\documentclass[11pt,a4paper]{article}
\usepackage{epsfig}
\usepackage[T1]{fontenc}    
\usepackage{graphics}
\usepackage{graphicx}
\usepackage{pstricks,pst-coil,pst-fill,pst-plot}
\usepackage[fleqn]{amsmath}    
\usepackage{amssymb}    
\usepackage{amsthm}    
\usepackage{amsfonts}   
\usepackage{verbatim}   
\usepackage{mathrsfs}   
\usepackage{dsfont}
\usepackage{euscript}
\usepackage{yfonts}
\usepackage{enumerate}     
\usepackage{txfonts}
\usepackage{marvosym}
\usepackage{stmaryrd}
\usepackage{vmargin}        
\usepackage{wasysym}		

\setmarginsrb{1.8cm}{2cm}{1.8cm}{2cm}{1cm}{1cm}{1cm}{1.6cm}
 \makeatletter
 \@addtoreset{equation}{section}
 \makeatother

\providecommand{\bysame}{\leavevmode\hbox to3em{\hrulefill}\thinspace}
\providecommand{\MR}{\relax\ifhmode\unskip\space\fi MR }

\providecommand{\href}[2]{#2}

\let\tend=\rightarrow

\long\def\symbolfootnote[#1]#2{\begingroup%
\def\thefootnote{\fnsymbol{footnote}}\footnote[#1]{#2}\endgroup}

\newtheorem{theorem}{Theorem}[section]
\newtheorem{prop}[theorem]{Proposition}
\newtheorem*{theorem*}{Theorem}

\newtheorem{defin}[theorem]{Definition}

\newtheorem{conj}[theorem]{Conjecture}

\newtheorem{lemme}[theorem]{Lemma}

\newtheorem*{axiomsFFFirst}{Bootstrap Axioms I-IV}
\newtheorem*{axiomsFFSecond}{Bootstrap Axiom V}

\def\Proof{\medskip\noindent {\it Proof --- \ }}

\def\qed{\hfill\rule{2mm}{2mm}}

\newcommand\beq{\begin{equation}}
\newcommand\enq{\end{equation}}
\newcommand\bem{\begin{multline}}
\newcommand\enm{\end{multline}}

\def\beqa{\begin{eqnarray}}
\def\eeqa{\end{eqnarray}}
\def\ba{\begin{array}}
\def\ea{\end{array}}
\def\det{\operatorname{det}}

\newcommand{\f}[2]{{\ensuremath{%
    \mathchoice%
    {\dfrac{#1}{#2}}
    {\dfrac{#1}{#2}}
    {\frac{#1}{#2}}
    {\frac{#1}{#2}}
}}}
\newcommand{\tf}[2]{\ensuremath{#1/#2}}

\def\a{\alpha}

\def\be{\beta}
\def\ga{\gamma}
\def\Ga{\Gamma}

\def\de{\delta}

\def\De{\Delta}
\def\eps{\epsilon}
\def\veps{\varepsilon}

\def\La{\Lambda}

\def\sg{\sigma}
\def\vsg{\varsigma}

\def\ups{\upsilon}
\def\th{\theta}

\def\vth{\vartheta}

\def\om{\omega}

\newcommand{\mc}[1]{\ensuremath{\mathcal{#1}}}
\newcommand{\mf}[1]{\ensuremath{\mathfrak{#1}}}
\newcommand{\msc}[1]{\ensuremath{\mathscr{#1}}}

\newcommand{\bs}[1]{\ensuremath{\boldsymbol{#1}}}

\DeclareFontFamily{OT1}{pzc}{}
\DeclareFontShape{OT1}{pzc}{m}{it}{<-> s * [1.10] pzcmi7t}{}
\DeclareMathAlphabet{\mathpzc}{OT1}{pzc}{m}{it}

\def \i{ \mathrm i}

\newcommand{\ov}[1]{\ensuremath{\overline{#1}}}
\newcommand{\wt}[1]{\ensuremath{\widetilde{#1}}}
\newcommand{\wh}[1]{\ensuremath{\widehat{#1}}}

\newcommand{\Int}[2]{\ensuremath{\int\limits_{#1}^{#2}}}
\newcommand{\Oint}[2]{\ensuremath{\oint\limits_{#1}^{#2}}}

\newcommand{\sul}[2]{\ensuremath{\sum\limits_{#1}^{#2}}}
\newcommand{\pl}[2]{\ensuremath{\prod\limits_{#1}^{#2}}}

\newcommand{\R}{\ensuremath{\mathbb{R}}}
\newcommand{\Cx}{\ensuremath{\mathbb{C}}}

\newcommand{\Dp}[1]{\ensuremath{\partial_{#1}}}

\newcommand{\limit}[2]{\ensuremath{\underset{#1 \tend #2}{\longrightarrow} }}

\newcommand{\ex}[1]{\ensuremath{\e{e}^{#1}}}

\newcommand{\op}[1]{ \boldsymbol{ \texttt{#1} } }

\newcommand{\norm}[1]{\ensuremath{  || #1 || }}

\newcommand{\dd}{\mathrm{d}}
\newcommand{\e}[1]{\ensuremath{\mathrm{#1}}}

\newcommand{\intff}[2]{\ensuremath{ [  #1 \,; #2 ] }}

\newcommand{\intoo}[2]{\ensuremath{ ]  #1 \,; #2 [ }}

\newcommand{\intn}[2]{\ensuremath{[\![ \, #1 \,;\, #2 \,]\!]}}

\begin{document}

\begin{center}
\begin{LARGE}
{\bf On multipoint correlation functions in the Sinh-Gordon 1+1 dimensional quantum field theory}
\end{LARGE}

\vspace{1cm}

\vspace{4mm}
{\large Karol K. Kozlowski \footnote{e-mail: karol.kozlowski@ens-lyon.fr}}%
\\[1ex]
Laboratoire de Physique (UMR CNRS 5672), ENS de Lyon, 46 Allée d'Italie, F-69364 LYON Cedex 07, France \\[2.5ex]

{\large Alex Simon \footnote{e-mail:alex.simon@ens-lyon.fr}}%
\\[1ex]
Laboratoire de Physique (UMR CNRS 5672), ENS de Lyon, 46 Allée d'Italie, F-69364 LYON Cedex 07, France \\[2.5ex]

\par 

\end{center}

\vspace{40pt}

\centerline{\bf Abstract} \vspace{1cm}
\parbox{12cm}{\small}

This work provides a closed, explicit and rigorous expression for the appropriately truncated $k$-point function of the integrable 1+1 dimensional
Sinh-Gordon quantum field theory. The results are obtained within the bootstrap program setting.

\vspace{20pt}


\vspace{40pt}

\tableofcontents

\section{Introduction}

 The boostrap program approach provides one with a path, based on the $\op{S}$-matrix formalism, for constructing
integrable quantum field theories in 1+1 dimensions, in an explicit and closed way. These correspond to quantisations of
classical integrable field theories in 1+1 dimensions, the most prominent examples being the Sinh and Sine-Gordon models.
The approach takes its root in the seminal work of Gryanik and Vergeles \cite{GryanikVergelesSMatrixAndOtherStuffForSinhGordon}
who proposed the exact $\op{S}$ matrix of the Sinh-Gordon 1+1 quantum field theory. Those ideas were
later extended and developed by Zamolodchikov \cite{ZalmolodchikovSMatrixSolitonAntiSolitonSineGordon},
Karowski and Thun \cite{KarowskiThunCompleteSMatrixThirring} what led to the full $\op{S}$-matrix for the
Sine-Gordon model. Later on, the reasoning were adapted so as to produce $\op{S}$-matrices for numerous other models
\cite{ArinshteinFateyevZamolodchikovSMatrixTodaChain,ZalZalBrosFactorizedSMatricesIn(1+1)QFT}.
This progress paved the way for the \textit{explicit} construction of the quantum
field theories for which $\op{S}$ is supposed to describe the scattering.
Traditionally, a quantum field theory is constructed by means of the path integral formalism -be it formal
as it appears in the physics literature or constructive as developed by mathematicians- in Euclidian space-type.
Then, the theory on the Minkowski space-time is reconstructed by means of the Osterwalder-Schrader theorem.
All-in-all, one may think of these constructions as a map allowing one to associate with a given classical Lagrangian
a collection of distributions on functions on Cartesian products of the space-time:
\beq
\mc{W}_{\a_1,\dots, \a_k}\big[G]   \; =  \hspace{-2mm} \Int{ \big( \R^{1,1} \big)^k }{} \pl{a=1}{k} \dd \bs{x}_a
\msc{W}_{\a_1,\dots, \a_k}\big(\bs{x}_1,\dots, \bs{x}_k \big) G\big(\bs{x}_1,\dots, \bs{x}_k\big) \;,
\qquad G \in \mc{S}\Big( \big( \R^{1,1} \big)^k \Big) \;,
\enq
in which $\msc{W}_{\a_1,\dots, \a_k}\big(\bs{x}_1,\dots, \bs{x}_k \big)$ are to be understood as generalised functions and $\mc{S}(X)$
refers to Schwarz functions on $X$. $\a_1,\dots, \a_k$ are indices labelling the possible operator content of the theory.
These generalised functions satisfy the Wightman axioms, see \cite{StreaterWightmanPCTStatisticsAllThat},
what ensures the existence of a Hilbert space $\mf{h}$, a vacuum vector $\bs{f}_{\e{vac}}$ and operator valued distributions $\bs{\Psi}_{\a_s}$, labelled by the indices $\a_s$,
called quantum fields. These are such that if $G(\bs{x}_1,\dots, \bs{x}_k)\, = \, \prod_{a=1}^{k}g_a(\bs{x}_a)$,  with $g$ in the Schwartz class, it holds
\beq
\big( \bs{f}_{\e{vac}}, \bs{\Psi}_{\a_1}[g_1]\cdots \bs{\Psi}_{\a_k}[g_k] \bs{f}_{\e{vac}} \big) \; = \;
\mc{W}_{\a_1,\dots, \a_k}\big[G]   \, .
\enq
This procedure fully constructs the quantum field theory. In fact, these are not the \textit{per se} quantum fields
which are of main interest to the physics at the root of a given quantum field theory, but rather the generalised functions
$\msc{W}_{\a_1,\dots, \a_k}\big(\bs{x}_1,\dots, \bs{x}_k \big)$ which are called correlation functions.
Indeed, for vast domains of the space-time $\big( \R^{1,1} \big)^k$, these are expected to
be \textit{bona fide} functions, whose Fourier transforms correspond to experimentally measurable quantities.

It seems fundamental to stress that any formalism, alternative to the path integral based approach, allowing one to construct the
correlation functions of quantum fields is fully acceptable. One of these roads is provided by the bootstrap program \cite{SmirnovFormFactors}
for integrable quantum field theories in $1+1$ dimensions.
One may summarise this program as follows. It takes as an input a Hilbert space $\mf{h}_{\e{in}}$,
typically a Fock space built over certain $L^2$-spaces,
and a postulated expression for the $\op{S}$ matrix which satisfies certain properties: unitarity,
crossing symmetry and the Yang-Baxter equation. These data are used to construct a vector-valued Riemann--Hilbert problem
on the space of meromorphic, vector-valued, functions in $0, 1, 2, \dots$ variables. Its solutions
provide one with integral kernels of operators -the quantum fields of the theory one aims to construct-.
From there, one may then deduce the correlation functions by computing the averages of operator products.
Expressions for two-point functions, \textit{viz}. for  $\msc{W}_{\a_1, \a_2}\big(\bs{x}_1, \bs{x}_2 \big)$, were obtained
in Euclidian regularisations starting from the early days of the theory and this for several integrable quantum field theories.
They were given as series over $N$ whose summand is given by a multiple-integral whose dimension grows linearly in $N$.
The convergence issue for such representation of two point functions was only solved recently and for the simplest instance of an integrable quantum field theory:
the Sinh-Gordon model in \cite{KozConvergenceFFSeriesSinhGordon2ptFcts}. The Euclidian regularisation corresponds, after going to the Minkowski space-time
to the space-like separation regime $(\bs{x}_1-\bs{x}_2)^2<0$. The representations  for
time-like separated points  $(\bs{x}_1-\bs{x}_2)^2>0$ are given by slightly different series  \cite{KozICMReviewConvergenceFFSeries} and this convergence is still, \textit{per se}, open.

However, the determination
of general $k$-point functions with $k\geq 3$ is basically absent from the literature in that only some partial results for $3$ and
$4$ point functions were obtained. Four point functions were considered for the first time in \cite{BalogNiedermaierNiedermayerPatrascioiuSeilerWeiszOnLown1stTermSeriesFourPtFct}
for the $\op{O}(n)$ model with $n=1,2,3$ while the work \cite{CaselleDelfinoGrinzaJahnMagnoli3PtFctsLowTermsO3PottsModel} investigated $3$ point functions in the $\op{O}(3)$ Potts model.
Finally, the work \cite{BabujianKarowskiTsvelikSomeMultiPtFcts} studied certain  3 and 4 point functions in the $\mathbb{Z}_2$-Ising, $\mathbb{Z}_3$-Potts and Sinh-Gordon models.
In a sense, those papers only provided a very partial answer for the $3$ and $4$ point functions. Indeed, they all addressed only
the calculation of the first or first and second terms that build up  the infinite series of multiple integrals
supposed to represent the correlation functions. Moreover, while explicit for the quantities considered, the obtained answers did not
unravel any specific structure.

The aim of the present work is to fill this gap by presenting closed and structured representations for the $\bs{r}\in \mathbb{N}^{k-1}$-truncated $k$-point functions
in the Sinh-Gordon model. In particular, we overcome all the technical combinatorial handlings involving distributions which are necessary so as to
obtain the final result in a closed, neat form. Moreover, the type of obtained answer suggest that very similar in spirit representations can be obtained for other, more involved, quantum
integrable field theories such as, for instance, the sine-Gordon model. Here, we focus on $\bs{r}\in \mathbb{N}^{k-1}$-truncated $k$-point functions
since these can be constructed on fully rigorous grounds from the axiomatics of the theory. The \textit{per se} $k$-point functions
are then obtained from our answer by summing up over  $\bs{r}\in \mathbb{N}^{k-1}$. However, since the question of
convergence of the associated series is open, we only state that final result in the form of a conjecture.

The paper is organised as follows. Section \ref{Section operator content} contains an overall recall of the bootstrap program approach to the Sinh-Gordon model.
Subsection \ref{SousSoussectionOperateursDeBase} discusses how the quantum fields are realised within the bootstrap program.
The Riemann--Hilbert problem characterising the main building blocks for the quantum fields, the form factors, is described in Subsection \ref{SubSection Bootstrap Program Zero part sector}.
The construction of the integral kernels which characterise the general action of the quantum fields are described in Subsection \ref{SubSection Bootstrap Program multi part sector}.
Section \ref{Section Towards multi pts fcts} develops the first steps of the computation of the multi-point functions.
Subsection \ref{Subsection Premilinary expression} establishes the expression one starts the calculation of the correlators with.
Subsection \ref{Subsection Auxiliary representation} proves a combinatorial representation for the multi-point densities in momentum space.
Then Subsection \ref{Subsection Smeared integral representation} establishes an intermediate representation for the $\bs{r}$-truncated $k$-point functions
in terms of multi-dimensional boundary values. Section \ref{Section Per se Correlation functions}
culminates the paper as it provides the constructions of various closed representations for the $\bs{r}$-truncated $k$-point functions. To achieved that,
Subsection \ref{SubSection Auxiliary Bounds} establishes a certain amount of preliminary estimates,
while Subsection \ref{Subsection Closed reps for r truncated pt fcts} gathers the mentioned results.
The conclusion is followed by several appendices.

Appendix \ref{Appendix Section Auxiliary Results} establishes an auxiliary combinatorial representation for
a generalisation of a Cauchy determinant in Subappendix \ref{Appendix Subsection auxiliary identities} and recalls some of the properties of the special
functions used in the bootstrap approach in Subappendix \ref{Appendix Fonctions Speciales}.
Appendix \ref{Appendix Section Master representation multi pt fcts} focuses on establishing a master combinatorial representation
in Subappendix \ref{Appendix SubSection Master representation}
which allows one to derive from it a large number of equivalent but functionally different representations for the  $\bs{r}$-truncated $k$-point functions.
Two representations are derived from it, one in Subappendix \ref{SousSection Preuve alternative representation totalement -}
and another one in Subappendix \ref{Appendix SubSection Intermediate Decomposition Multi pt fcts}.

\section{The operator content \& the Bootstrap program for the Sinh-Gordon model}
 \label{Section operator content}

The Sinh-Gordon 1+1 dimensional quantum field theory is realised on the Fock Hilbert space \cite{GryanikVergelesSMatrixAndOtherStuffForSinhGordon}
\beq
\mf{h}_{\e{ShG}} \, = \, \bigoplus\limits_{n=0}^{+\infty} L^2(\R^n_{>}) \qquad \e{with} \qquad \R^n_{>} \; = \; \Big\{ \bs{\be}_{n}=(\be_1,\dots, \be_n) \in \R^n \; : \; \be_1>\dots>\be_n  \Big\} \;.
\label{definition d eh in}
\enq
Vectors $\bs{f}\in \mf{h}_{\e{in}}$ will be written as $\bs{f}=(f^{(0)},\dots, f^{(n)},\dots)$ in which the superscript $(n)$ refers to the $L^2(\R^n_{>})$  subspace of the Fock space
to which the component belongs to. 
The component $f^{(n)}\in L^2( \R^n_{>} )$ has the physical heuristic interpretation of an incoming $n$-particle wave-packet density in rapidity space. More precisely, on physical grounds, one interprets elements of the
Hilbert space $\mf{h}_{\e{ShG}}$ as parameterised by $n$-particles states, $n\in \mathbb{N}$,  arriving, in the remote past, with well-ordered rapidities $\be_1>\dots>\be_n$
prior to any scattering which would be enforced by the interacting nature of the model.

For the 1+1 dimension quantum Sinh-Gordon model, the $\op{S}$-matrix proposed in \cite{GryanikVergelesSMatrixAndOtherStuffForSinhGordon} is purely diagonal and
thus fully described by one scalar function of the relative "in" rapidities of the two particles:
\beq
\op{S}(\beta)\, = \, \f{ \tanh\big[ \tfrac{1}{2}\beta - \i \pi  \mf{b}   \big]  }{ \tanh\big[ \tfrac{1}{2}\beta + \i \pi  \mf{b}   \big]   }
      \, = \, \f{ \sinh(\be) - \i \sin[ 2 \pi \mf{b}]   }{ \sinh(\be) +  \i \sin[ 2 \pi \mf{b}]  }
\qquad \e{with} \qquad  \mf{b}\, = \,   \f{1}{2} \f{ g^2  }{ 8\pi + g^{2}  } \, \in \intff{0}{\tfrac{1}{2}}\;.
\label{definition matrice S}
\enq
This $\op{S}$-matrix satisfies  unitarity $\op{S}(\be)\op{S}(-\be)=1$ and crossing  $\op{S}(\be)=\op{S}(\i \pi-\be)$ symmetries.
These are fundamental symmetry features of any $\op{S}$-matrix -describing an integrable or non-integrable theory- and,
properly generalised to the case of genuinely matrix valued $\op{S}$-matrix, arise in many other integrable quantum field theories.
In fact, the above $\op{S}$ matrix corresponds to the most elementary, scalar, solution to these equations.
The $\op{S}$-matrix  has no poles in the physical strip $0<\Im(\be)<\pi$, what is interpreted as an absence of bound states in the theory.

 Within the physical picture, throughout the flow of time, the "in" particles approach each other, interact, scatter and finally travel again as asymptotically free outgoing, \textit{viz}. "out", 
particles.
Within such a scheme, an "out" $n$-particle state is then paramaterised by $n$ well-ordered rapidities  $\be_1<\dots < \be_n$ and  can be seen as 
a component of a vector belonging to the Hilbert space 
\beq
 \mf{h}_{\e{out}} \, = \, \bigoplus\limits_{n=0}^{+\infty} L^2(\R^n_{<}) \qquad \e{with} \qquad \R^n_{<} \; = \; \Big\{ \bs{\be}_{n}=(\be_1,\dots, \be_n) \in \R^n \; : \; \be_1<\dots<\be_n  \Big\} \;.  
\enq
The $\op{S}$-matrix will allow one to express the "out" state $\bs{g}=(g^{(0)},\dots, g^{(n)},\dots)$ which results from the scattering of an "in" state 
$\bs{f}=(f^{(0)},\dots, f^{(n)},\dots)$ as 
\beq
g^{(n)}(\be_1,\dots,\be_n) \, = \,   \pl{a<b}{n} S(\be_{ab}) \cdot  f^{(n)}(\be_n,\dots,\be_1)   \qquad \e{with} \qquad \be_{ab} \, = \, \be_a-\be_b\;. 
\label{definition vecteur out plus contraction deux vars}
\enq
Note that in this integrable setting, there is \textit{no} particle production and that the scattering is a concatenation of two-body processes. 
See \textit{e.g.} \cite{ArinshteinFateyevZamolodchikovSMatrixTodaChain,ZalZalBrosFactorizedSMatricesIn(1+1)QFT} for examples of $\op{S}$-matrices 
related to other integrable quantum field theories.

\subsection{The basic operators}
\label{SousSoussectionOperateursDeBase}

In order to realise a quantum field theory of interest -the Sinh-Gordon one in this case of interest- on $\mf{h}_{\e{ShG}}$, one should construct the set of physically
pertinent operator valued distributions on $\mf{h}_{\e{ShG}}$ called quantum fields. The expectation values in the vacuum vector
\beq
\bs{f}_{\e{vac}} \, = \, \big(1,0,\dots, \big)
\label{definition vecteur f vacuum}
\enq
of their properly regularised products give rise to all physically measurable quantities -called correlation functions- which should be encapsulated by the given quantum field theory.

The quantum fields of the model should comply with various symmetries that one wants to impose on the quantum field theory, such as invariance under Lorentz boosts of the space-time 
coordinates or translational invariance. Thus, one endows $\mc{L}(\mf{h}_{\e{in}})$ with a unitary operator $\op{U}_{\op{T}_{\bs{y}}}$ -the translation operator by $\bs{y}=(y_0, y_1)$-
which acts  diagonally on $\mf{h}_{\e{ShG}}$, \textit{c.f.} \eqref{definition d eh in}:
\beq
\op{U}_{\op{T}_{\bs{y}}} \cdot \bs{f} \; = \; \Big( \op{U}_{\op{T}_{\bs{y}}}^{(0)}\cdot f^{(0)},\dots, \op{U}_{\op{T}_{\bs{y}}}^{(n)}\cdot f^{(n)},\dots \Big)
\quad \e{with} \quad \bs{f}=(f^{(0)},\dots, f^{(n)},\dots)
\enq
and where given $\bs{\be}_n \, = \, (\be_1,\dots, \be_n)$,
\beq
\op{U}_{\op{T}_{\bs{y}}}^{(n)} \cdot f^{(n)}(\bs{\be}_{n}) \; = \; \exp\bigg\{\i \, \ov{\bs{p}}( \bs{\be}_n )  \cdot \bs{y} \bigg\}  f^{(n)}(\bs{\be}_{n}) \;.
\label{ecriture action operateur de translation}
\enq
Above $\cdot$ stands for the Minkowski form with signature $(1,-1)$, \textit{viz}. $\bs{x}\cdot \bs{y}=x_0 y_0 \, - \, x_1y_1$, while
\beq
\ov{\bs{p}}( \bs{\be}_n )  \; = \; \sul{a=1}{n}\bs{p}(\be_a) \qquad \e{with} \qquad  \bs{p}(\be) \, = \,  \big( m\cosh(\be), m \sinh(\be) \big)
\label{definition de bar vect p}
\enq
being the relativistic $2$-momentum of a particle of mass $m$ having rapidity $\be$.

One imposes that a local quantum fields $ \op{O}(\bs{x})$ behave under the adjoint action of $\op{U}_{\op{T}_{\bs{y}}}$ as :
\beq
\op{U}_{\op{T}_{\bs{y}}} \cdot \op{O}(\bs{x}) \cdot \op{U}_{\op{T}_{\bs{y}}}^{-1} \, = \,  \op{O}(\bs{x}+\bs{y}) \;.
\label{ecriture action adjointe operateur de translation}
\enq
Recall that $ \op{O} $ is, \textit{per se}, an operator on $\mf{h}_{\e{ShG}}$ valued distribution on Schwartz functions $\msc{S}(\R^{1,1})$.
The symbol $ \op{O}(\bs{x}) $ in the above expression and those that will follow, is to be understood as its generalised, operator valued, symbol.
In other words, these expressions are to be understood in the distributional sense.

Next, one introduces the unitary boost operator $\op{U}_{\La_{\th}}$ which acts by overall translations on  $\mf{h}_{\e{ShG}}$:
\beq
\op{U}_{\La_{\th}} \cdot \bs{f} \; = \; \Big( \op{U}_{\La_{\th}}^{(0)}\cdot f^{(0)},\dots, \op{U}_{\La_{\th}}^{(n)}\cdot f^{(n)},\dots \Big)
\quad \e{with} \quad \bs{f}=(f^{(0)},\dots, f^{(n)},\dots)
\enq
and where 
\beq
\op{U}_{\La_{\th}}^{(n)} \cdot f^{(n)}(\bs{\be}_{n}) \; = \;    f^{(n)}(\bs{\be}_{n}+\th \, \ov{\bs{e}}_n) \qquad \e{with} \qquad \ov{\bs{e}}_n=(1,\dots, 1) \in \R^n\;.
\label{definition boost et vecteur uniforme}
\enq
One imposes that the quantum fields of the theory transform under under the adjoint action of $\op{U}_{\La_{\th}}$  as
\beq
\op{U}_{\La_{\th}} \cdot \op{O}(\bs{x}) \cdot \op{U}_{\La_{\th}}^{-1} \, = \,  \ex{ \th \op{s}_{ \op{O} } }\op{O}\big( \La_{\th} \cdot \bs{x} \big)  \qquad \e{with} \qquad
\La_{\th} \; = \; \left( \ba{cc}  \cosh(\th) & -\sinh(\th) \\ -\sinh(\th) & \cosh(\th) \ea \right) \;,
\enq
where $ \op{s}_{ \op{O} }$ is the spin of the quantum field $\op{O}(\bs{x})$.

 \subsection{The bootstrap program for the zero particle sector}
 \label{SubSection Bootstrap Program Zero part sector}
 
Taken the $L^2$-structure of the Fock Hilbert space $\mf{h}_{\e{ShG}}$, one may represent an operator $\op{O}(\bs{x})$
on $\mf{h}_{\e{ShG}}$ labelled by the space-time coordinate $\bs{x}$ as an integral operator
acting on the $L^2$-components of the Fock space
\beq
\op{O}(\bs{x}) \cdot \bs{f} \; = \; \Big( \, \op{O}^{(0)}(\bs{x})  \cdot \bs{f} , \cdots ,  \op{O}^{(n)}(\bs{x})\cdot \bs{f} , \cdots \Big) 
\enq
with $ \op{O}^{(n)}(\bs{x}) \, : \,  \mf{h}_{\e{ShG}} \rightarrow L^2(\R^{n}_{>})$. Later on, we will discuss more precisely the structure
of the operators $ \op{O}^{(n)}(\bs{x}) $ that one needs to impose so as to end up with a consistent quantum field theory. However, first, we focus our attention on the 
$0^{\e{th}}$ space operators whose action may be represented, whenever it makes sense, as 
\beq
\op{O}^{(0)}(\bs{x})\cdot \bs{f} \; = \; \sul{m \geq 0}{} \; \Int{ \be_1>\dots > \be_m}{} \hspace{-4mm}  \f{ \dd^m \be }{ (2\pi)^m }  \;
\mc{M}_{0;m}^{(\op{O})}(\bs{\be}_m)   \ex{- \i \ov{\bs{p}}( \bs{\be}_m)\cdot \bs{x} }
f^{(m)}\big( \bs{\be}_m \big)   \;. 
\label{ecriture chp quantique comme op integral secteur 0}
\enq
The oscillatory $\bs{x}$-dependence is a simple consequence of imposing the translation relation \eqref{ecriture action adjointe operateur de translation} along with the explicit form of the action of the translation operator \eqref{ecriture action operateur de translation}-\eqref{definition de bar vect p}.

In order for $\op{O}^{(0)}(\bs{x})$ to comply with the scattering data encoded by $\op{S}$, one needs to impose a certain amount of constraints on the integral kernels $\mc{M}_{0;m}^{(\op{O})}(\bs{\be}_m)$. 
First of all, general principles of quantum field theory lead to impose that, in order for these to give rise to kernels of quantum fields, $\mc{M}_{0;m}^{(\op{O})}(\bs{\be}_m)$
has to correspond to 
a  $+$ boundary value
\beq
\label{form factor BV}
\mc{F}_{m;+}^{(\op{O})}(\bs{\be}_m) \, = \, \lim_{ \substack{ \veps_1>\dots > \veps_m \\ \veps_a \tend 0^+}  }\mc{F}_{m}^{(\op{O})}(\bs{\be}_m+\i \bs{\veps}_m) \; , \quad
 \e{for} \quad \bs{\be}_m\in \R^m_{>}
\enq
of a meromorphic function $\mc{F}_{m}^{(\op{O})}(\bs{\be}_m)$
of the variables $\be_1, \dots, \be_m$ taken singly and belonging to the strip
\beq
\msc{S} \, = \,   \Big\{ z \in \Cx \; : \;  0 < \Im (z) < 2 \pi \Big\} .
\label{definition rapidity strip}
\enq
Traditionally, in the physics literature,  the functions $\mc{F}_{m}^{(\op{O})}(\bs{\be}_m)$
are called form factors. 
In the following, it will appear useful to consider partial boundary values  \textit{e.g.} given $\bs{\alpha}_n \in \R^n$ and $\bs{\be}_m \in \R^m$
\beq
\mc{F}_{n+m;-,+}^{(\op{O})}(\bs{\alpha}_n+\i\pi \ov{\bs{e}}_n ,\bs{\be}_m) \, = \, \lim_{ \substack{ \veps_1>\dots > \veps_m \\ \veps_a \tend 0^+}   }\lim_{ \substack{ \eta_1<\dots < \eta_m \\ \eta_a \tend 0^+}}\mc{F}_{n+m}^{(\op{O})}(\bs{\alpha}_n+\i\pi \ov{\bs{e}}_n -\i\bs{\eta}_n,\bs{\be}_m+\i \bs{\veps}_m) \; , \quad
\e{with} \quad \ov{\bs{e}}_n = (1,\dots,1)
\enq
will stand for a mixed boundary value, - for the first set of $n$ variables and + for the second set of $m$ variables. Similarly, given $\bs{\be}_m \in \R^m$ and $\bs{\alpha}_n \in \msc{S}^n$ in generic position, 
\beq
\mc{F}_{n+m;\emptyset,+}^{(\op{O})}(\bs{\alpha}_n,\bs{\be}_m) \, = \, \lim_{ \substack{ \veps_1>\dots > \veps_m \\ \veps_a \tend 0^+}   }\mc{F}_{n+m}^{(\op{O})}(\bs{\alpha}_n,\bs{\be}_m+\i \bs{\veps}_m) \; , 
\enq
as well as evident generalisations or variants thereof. \\
Further, one imposes a set of equations on the $\mc{F}_{m}^{(\op{O})}$s. These constitute the so-called form factor bootstrap program. On mathematical grounds, one should understand the form factor bootstrap program
as a set of \textit{axioms} that one imposes on the integral kernels of the model's quantum fields given the starting data $\big( \mf{h}_{\e{ShG}}, \op{S} \big)$. Upon solving them, one has to check
\textit{a posteriori} that their solutions do provide one, through \eqref{ecriture chp quantique comme op integral secteur 0} and \eqref{ecriture chp quantique comme op integral secteur general}, 
with a collection of operators satisfying all of the requirements of the theory discussed earlier on. In practical terms, this means checking that the model's correlation functions -which can be computed once that the
operators are constructed- satisfy the Wightman axioms.

The bootstrap program axioms take the form of a Riemann-Hilbert problem for a collection of functions in 
many variables, each of which lives on the strip $\msc{S}$, \textit{c.f.} \eqref{definition rapidity strip}.
In the case of the Sinh-Gordon model, since there are no bound states, these take the below form.

\begin{axiomsFFFirst}

 Find functions $\mc{F}_{n}^{(\op{O})}$,  $n \in \mathbb{N}$, such that, for each $k \in \intn{1}{n}$ and  $\be_a \in \msc{S}$, $a \in \intn{1}{n}\setminus \{k \}$ all being fixed,
the maps $\mf{f}_k^{(\op{O})}: \be_k \mapsto \mc{F}_{n}^{(\op{O})}(\bs{\be}_n)$ are
\begin{itemize}
 
 \item meromorphic on $\msc{S}$;
 
 \item admit $+$, resp. $-$, boundary values $\mf{f}_{k;+}^{(\op{O})}$ on $\R$, resp. $\mf{f}_{k;-}^{(\op{O})}$ on  $\R+2\i\pi$;
 
 \item are bounded at infinity by $ C\cdot \cosh\big( \mf{w}_{\op{O}} \Re(\be_k) \big)$ for some $n$ and $k$ independent real number $\mf{w}_{\op{O}} \in \R$.
 
\end{itemize}
The $\mc{F}_{n}^{(\op{O})}$ satisfy the multi-variable system of Riemann-Hilbert problems:
\begin{itemize}
\item[{ \bf I) } ]  $\mc{F}_{n}^{(\op{O})}(\bs{\be}_n) \; = \; \op{S}(\be_{a \, a+1}) \cdot \mc{F}_{n}^{(\op{O})}(\bs{\be}_n^{(a+1a)})$  where
$$
\bs{\be}_n^{(a+1a)} \; = \; \big(\be_1,\dots,\be_{a-1},\be_{a+1},\be_{a}, \be_{a+2},\dots, \be_n \big) \; ,
$$
while $\be_{ab}$ is as in \eqref{definition vecteur out plus contraction deux vars}.
\item[ { \bf II) }  ]  For $\be_1 \in \R$, and given generic $\bs{\be}_n^{\prime}=(\be_2,\dots, \be_n) \in \msc{S}^{n-1}$ and $\bs{e}_1=(1,0,\dots,0) \in \R^{n}$,
$$ \mc{F}_{n}^{(\op{O})}(\bs{\be}_n + 2\i\pi \bs{e}_1) \; = \; \ex{2\i\pi \om_{\op{O}} } \,  \mc{F}_{n}^{(\op{O})}(\bs{\be}_n^{\prime},\be_1)
 \, = \, \ex{2\i\pi \om_{\op{O}} } \, \pl{a=2}{n} \op{S}(\be_{a1})  \cdot \mc{F}_{n}^{(\op{O})}( \be_1,\bs{\be}_n^{\prime})\, , $$
for some $\om_{O} \in \R$.

\item[ { \bf III) }  ] The only poles of  $\mc{F}_{n}^{(\op{O})}$ are simple,  located at
$\i\pi$ shifted rapidities and
\beq
-\i \e{Res}\Big(\mc{F}_{n+2}^{(\op{O})}(\alpha+\i\pi, \be, \bs{\be}_n) \cdot \dd \alpha \, , \, \alpha=\be  \Big) \; = \; \Big\{ 1\, - \, \ex{2\i\pi \om_{\op{O}} }  \pl{a=1}{n}  \op{S}(\be-\be_{a}) \Big\} \cdot
\mc{F}_{n}^{(\op{O})}( \bs{\be}_n)  \; ,    
\nonumber
\enq
with $\bs{\be}_n\in \msc{S}^n$  such that  $\be_a \notin \be+\i\pi\mathbb{Z}$ for any $a \in \intn{1}{n}$
\item[ { \bf IV) }  ] $ \mc{F}_{n}^{(\op{O})}( \bs{\be}_n + \theta\,  \ov{\bs{e}}_n )  \; = \; \ex{\theta \op{s}_{\op{O}} } \cdot \mc{F}_{n}^{(\op{O})}(\bs{\be}_n)$ for some number $\op{s}_{\op{O}}$ and
$\ov{\bs{e}}_n$  as in \eqref{definition boost et vecteur uniforme}.
\end{itemize}

\end{axiomsFFFirst}
 
\noindent The form factor axioms involve three auxiliary, operator dependent, parameters:
\begin{itemize}
\item[$\a$)]  the phase $\ex{2\i\pi \om_{\op{O}} } $ called mutual locality index;
\item[$\be$)] the quantity $\op{s}_{\op{O}}$ called the spin;
\item[$\ga$)] the real number  $\mf{w}_{\op{O}}$ called operator's growth index.
\end{itemize}
 These  are intrinsic properties of a given operator. $\op{s}_{\op{O}}$  describes how the operator is modified through
a Lorentz boost, \textit{i.e.} change of Galilean reference frame, while $\om_{\op{O}}$ characterises how is the operator "local" in respect to the hypothetically
existing free asymptotic fields of the theory,  $\mf{w}_{\op{O}}$ captures the type of ultraviolet short-distance behaviour induced by the operator in a correlation.

We would also like to point out that the reduction occurring at the residues of the poles corresponding to $\be_{ab}=\i \pi$ can be readily inferred from axioms ${\bf I) }$ and ${\bf III) }$.

It seems pertinent to comment on the origin of the axioms. The first one translates how the scattering properties of the model manifest themselves at the level of the operator's kernel.
The second and third axioms may be interpreted heuristically as a consequence of the LSZ reduction \cite{LehmannSymanzikZimmermanLSZReductionFormulaOriginalPaper}, and locality of the operator,
see \textit{e.g.} \cite{BabujianFringKarowskiZapletalExactFFSineGordonBootsstrapI,SmirnovFormFactors} for heuristics on that matter. Finally, 
the last axiom is a manifestation of the Lorentz invariance of the theory. We point out that
 for more complex models, one would also need to add an additional axiom which would encapsulate the way how the presence of  bound states
in the model determines the \textit{locii} and residues of the additional poles in the form factors, \textit{c.f.} \cite{SmirnovFormFactors}.

The bootstrap axioms ${\bf I)}-{\bf IV)}$ can be reduced to solving  a simpler set of equations  through the $\mc{K}$-transform method
which was introduced in \cite{BabujianKarowskiBreatherFFSineGordon}, an analogous expression was also proposed in \cite{BrazhnikovLukyanovFreeFieldRepMassiveFFIntegrable}
by applying the angular quantisation method, first introduced in \cite{LukyanovFirstIntroFreeField}. The construction has several ingredients,
the first of which corresponds to the pole-free two-particle form factor $\op{F}$ which is a meromorphic function on $\Cx$:
\beq
\op{F}(\be) \, = \, \f{1}{  \Ga\Big( 1+\mf{z}, -\mf{z} \Big)}   G\left( \ba{cccc} 1-\mf{b} - \mf{z} \, ,  &   2 - \mf{b} + \mf{z} \,,  &   1- \hat{\mf{b}} - \mf{z}   \, ,  &   2 - \hat{\mf{b}}  + \mf{z}    \\
  \mf{b} - \mf{z}   \, ,  &   1+  \mf{b} + \mf{z}   \, ,  &   \hat{\mf{b}} - \mf{z}   \, ,  &   1+  \hat{\mf{b}} + \mf{z}    \ea \right)  \quad \e{with} \quad  \mf{z} \, = \, \f{\i \be }{2 \pi } \;.
\label{expression twpo body scattering via Barnes}
\enq
Above, $\Ga$ is the Gamma function, $G$ is the Barnes function, \textit{c.f.} Appendix \ref{Appendix Fonctions Speciales}, and we have adopted the product conventions
\beq
  \Ga\left(\ba{c} a_1,\dots, a_n \\ b_1,\dots, b_{\ell} \ea \right)  \, = \, \f{ \pl{k=1}{n} \Ga(a_k)  }{  \pl{k=1}{\ell} \Ga(b_k) } \qquad  \e{and}  \qquad
  G\left(\ba{c} a_1,\dots, a_n \\ b_1,\dots, b_{\ell} \ea \right)  \, = \, \f{ \pl{k=1}{n} G(a_k)  }{  \pl{k=1}{\ell} G(b_k) } \;.
\label{ecriture convention produit ratios fcts Gamma}
\enq
The $\mc{K}$-transform of  a function  $p_n$ on $\Cx^n\times \{0,1\}^n$ depending on $n$ complex variables $\bs{\be}_n \in \Cx^n$ and $n$ discrete variables $\bs{\ell}_n \in \{0,1\}^n$ is defined as
\beq
  \mc{K}_{n}\big[ p_n  \big]\big( \bs{\be}_{n} \big) \, =  \hspace{-2mm} \sul{ \bs{\ell}_n \in \{0,1\}^n }{} (-1)^{\ov{\bs{\ell}}_n}
\pl{k<s}{n} \bigg\{ 1 \, - \, \i \f{ \ell_{ks} \cdot \sin[2\pi \mf{b} ] }{ \sinh(\be_{ks})  }  \bigg\} \cdot p_n\big(\bs{\be}_n\mid \bs{\ell}_n\big)  \;,
\label{definition K transformee fct p}
\enq
in which $\ov{\bs{\ell}}_n \, = \, \sul{a=1}{n} \ell_k$.

We now state, without proof, one of the results obtained in \cite{BabujianKarowskiBreatherFFSineGordon} which provides an explicit representation for
$\mc{F}_n^{(\op{O})}$ in terms of the $\mc{K}$-transform of a function $p_{n}^{(\op{O})}$ satisfying a structurally simpler set of equations that
those provided by the axioms ${\bf I)}-{\bf IV)}$.

\begin{prop} \cite{BabujianKarowskiBreatherFFSineGordon}
\label{Proposition representation FF general en terme K transformee}

\noindent Let $\bs{\ell}_n\in \{0,1\}^n$ and $p_{n}^{(\op{O})}\big(\bs{\be}_n\mid \bs{\ell}_n\big)$ be a solution to the below constraints
\begin{itemize}

 \item[a)]  $\bs{\be}_n\mapsto p_{n}^{(\op{O})}\big(\bs{\be}_n\mid \bs{\ell}_n\big)$ is a collection of $2\i\pi$ periodic holomorphic functions on $\Cx$ that are symmetric in the two sets of variables jointly,  \textit{viz}.
 for any $\sg \in \mf{S}_n$ it holds $ p_{n}^{(\op{O})}\big(\bs{\be}_n^{\sg}\mid \bs{\ell}_n^{\sg} \big)=  p_{n}^{(\op{O})}\big(\bs{\be}_n\mid \bs{\ell}_n\big)$
with $\bs{\be}_n^{\sg}=\big( \be_{\sg(1)}, \dots, \be_{\sg(n)} \big)$;

  \item[b)] $p_{n}^{(\op{O})}\big(\be_2+\i\pi, \bs{\be}_n^{\prime}\mid \bs{\ell}_n\big)\, = \,g(\ell_1,\ell_2) p_{n-2}^{(\op{O})}\big(\bs{\be}_n^{\prime\prime}\mid \bs{\ell}_n^{\prime\prime}\big)
  \, + \, h(\ell_1,\ell_2\mid \bs{\be}_n^{\prime})$
where $\bs{\be}_n^{\prime}$ is as given in axiom {\bf II)}, $\bs{\be}_n^{\prime\prime} = (\be_3,\dots,\be_n)$, the function $h$ does not depend on the remaining set of variables $\bs{\ell}_n^{\prime\prime} = (\ell_3,\dots,\ell_n) $ and
\beq
g(0,1)\, = \, g(1,0) \, = \, \f{ -1 }{ \sin (2\pi \mf{b} ) \,  \op{F}(\i\pi)  } \, ;
\enq

 \item[c)]  $ p_{n}^{(\op{O})}\big( \bs{\be}_n + \theta \, \ov{\bs{e}}_n \mid \bs{\ell}_n \big)  \; = \; \ex{\theta \op{s}_{\op{O}} } \cdot p_{n}^{(\op{O})}\big(\bs{\be}_n\mid \bs{\ell}_n\big)$;

 \item[d)]  $ \big| p_{n}^{(\op{O})}\big(\bs{\be}_n\mid \bs{\ell}_n\big)  \big| \, \leq \, C \cdot  {\displaystyle \pl{a=1}{n} } \big| \cosh\big[ \Re(\be_a) \big]\big|^{ \mf{w}_{\op{O}} }$.

\end{itemize}
 Then, the sequence of meromorphic functions defined by means of the $\mc{K}$-transform \eqref{definition K transformee fct p}
\beq
\mc{F}_n^{(\op{O})}\big( \bs{\be}_n \big) \, = \, \pl{a<b}{n} \op{F}\big( \be_{ab} \big)  \cdot \mc{K}_{n}\big[ p_n^{(\op{O})} \big]\big( \bs{\be}_{n} \big)
\label{solution eqns bootstrap via K transformee}
\enq
solves the bootstrap axioms ${\bf I)} -{\bf IV)}$.

\end{prop}

To the best of our knowledge, it has not been established yet that every solution of the bootstrap axioms is given by \eqref{solution eqns bootstrap via K transformee}
for some solution $p_{n}^{(\op{O})}\big(\bs{\be}_n\mid \bs{\ell}_n\big)$ to $a)-d)$ above. This seems however a reasonable conjecture, and,
in the following, we shall only focus on this kind of solutions, \textit{viz}. form factors given by \eqref{solution eqns bootstrap via K transformee}
with $p_{n}^{(\op{O})}$ solving $a)-d)$. We refer to \cite{BabujianKarowskiBreatherFFSineGordon} for several examples of solutions $p_n^{(\op{O})}$.

\subsection{The bootstrap program for the multi-particle sector}
\label{SubSection Bootstrap Program multi part sector}

It is convenient to represent the action of the operators $\op{O}^{(n)}(\bs{x})$ in the form 
\beq
\Big(\op{O}^{(n)}(\bs{x})\cdot \bs{f} \Big)(\bs{\ga}_n) \; = \; \sul{m \geq 0}{} \;
  \op{M}_{\op{O}}^{(m)} \big[f^{(m)}\big]\big( \bs{x} \, \mid     \bs{\ga}_n \big) \;.
\label{ecriture chp quantique comme op integral secteur general}
\enq
There $\op{M}_{\op{O}}^{(m)}\big( \bs{x} \, \mid     \bs{\ga}_n \big)$ are distribution and linear form valued functions which act on appropriate spaces of sufficiently regular functions in $m$ variables. 
The regularity assumptions will clear out later on, once that we provide the explicit expressions \eqref{ecriture forme explicite de la distribution} for these distributions. In fact, it is convenient, in order to avoid heavy notations, to represent their action as 
generalised integral operators
\beq
  \op{M}_{\op{O}}^{(m)}\big[f^{(m)}\big] \big( \bs{x} \, \mid     \bs{\ga}_n \big)  \; = \hspace{-3mm} \Int{ \be_1>\dots > \be_m}{} \hspace{-4mm}  \f{ \dd^m \be  }{ (2\pi)^m } \;
  \mc{M}^{(\op{O})}_{n;m}\big(     \bs{\ga}_{n}  ; \bs{\be}_{m} \big) \cdot  \exp\Big\{ \i \, \big[ \ov{\bs{p}}( \bs{\ga}_n) - \ov{\bs{p}}( \bs{\be}_m) \big] \cdot \bs{x} \Big\} \cdot
f^{(m)}\big( \bs{\be}_m \big) \;, 
\label{definition action operatur dans secteur multipoints}
\enq
in which one understands the kernels $\mc{M}^{(\op{O})}_{n;m}\big( \bs{\ga}_{n}  ; \bs{\be}_{m} \big) $ as generalised functions.  
 
  The last axiom of the bootstrap program  provides one  with a way to compute these kernels. Heuristically, it can be seen as a consequence of the LSZ reduction \cite{LehmannSymanzikZimmermanLSZReductionFormulaOriginalPaper}. 
\begin{axiomsFFSecond}
\begin{itemize} 
 \item[ {\bf V)} ] For $\bs{\alpha}_n \in \R^{n}$ and $\bs{\be}_{m} \in \R^m$, one has
$$  \mc{M}^{(\op{O})}_{n;m}\big(  \bs{\alpha}_n ; \bs{\be}_{m} \big) \; = \;  \ex{- 2\i\pi \om_{\op{O}}} \cdot \bigg\{ \mc{M}^{(\op{O})}_{n-1;m+1}\big(  \bs{\alpha}_n^{\prime} ; (\alpha_1+\i \pi, \bs{\be}_m) \big)
\; + \; 2\pi \sul{a=1}{m}  \de_{\alpha_1;\be_a} \pl{k=1}{a-1} \op{S}(\be_{ka}) \cdot
\mc{M}^{(\op{O})}_{n-1;m-1}\big(  \bs{\alpha}_n^{\prime};  \wh{\bs{\be}}_m^{\, (a)}  \big)  \bigg\} \; .  $$
\end{itemize}
In this formula,  the evaluation at $\alpha_1+\i \pi$ is to be understood in the sense of the $-$ boundary value on $\R+\i\pi$ of the meromorphic continuation in the first secondary variable
from $\R$ to the strip $0<\Im(z)<\pi$ of the generalised function
$ \a \mapsto \mc{M}^{(\op{O})}_{n-1;m+1}\big(  \bs{\alpha}_n^{\prime} ; (\alpha, \bs{\be}_m) \big)   $.

The induction is complemented with the initialisation condition
\beq
\mc{M}^{(\op{O})}_{0;m}\big( \emptyset ; \bs{\be}_{m} \big)\; = \;   \mc{F}_{m;+}^{(\op{O})}(\bs{\be}_m) \quad when \quad  \bs{\be}_m \in \R^{m}_{>} \;.
\label{condition initialisation pour MO avec zero vecteur alpha}
\enq
\end{axiomsFFSecond}

In the above expression, we remind that $\bs{\alpha}_n^{\prime}=(\a_2,\dots,\a_n)$ while $ \wh{\bs{\be}}_m^{\, (a)} $ means that the variable $\be_a$ should be omitted in the vector, \textit{viz}.
\beq
\wh{\bs{\be}}_m^{\, (a)} =  \big( \be_1, \dots, \be_{a-1}, \be_{a+1} , \dots,  \be_m\big) \in \R^m \;.
\enq
Finally,  $\de_{x;y}$ refers to the Dirac mass distribution centred at $x$ and acting on functions of $y$.

It will appear convenient for later purposes to introduce a specific terminology for the two kinds of contributions that arise in the induction
provided by axiom ${ \bf V) } $. The first contribution will be called shifted concatenation and the second one, involving $\de_{\a;\be_a}$, will be called reduction.

\subsubsection{The direct representation for multi-particle matrix elements}

 With all these data
at hand, one may provide a fully explicit expression for $\mc{M}^{(\op{O})}_{n;m}\big(  \bs{\alpha}_n ; \bs{\be}_{m} \big)$ solely in terms of a linear combination of
form factors $\mc{F}_{q}^{(\op{O})}\big(\bs{\ga}_q\big)$, where $0\leq q \leq n+m$  and $\bs{\ga}_q$ is a vector whose entries are given by a subset of coordinates
of $\bs{\a}_n + \i \pi \ov{\bs{e}}_n$, with $\ov{\bs{e}}_n$ as introduced in \eqref{definition boost et vecteur uniforme}, and a subset of the coordinates of $\bs{\be}_m$.

\begin{prop}
\label{Proposition noyau integral generalise solution recursion}

The recursion associated with axiom ${\bf V)}$ may be solved in the below closed form:
\begin{multline}
\mc{M}^{(\op{O})}_{n;m}\big(  \bs{\alpha}_n ; \bs{\be}_{m} \big) \; = \; \sul{p=0}{ \e{min}(n,m) } \sul{  \substack{ k_1<\dots < k_p \\ 1 \leq k_a \leq n}  }{} 
\sul{  \substack{ i_1\not=\dots\not= i_p \\ 1 \leq i_a \leq m}  }{}  
\pl{a=1}{p}\Big\{2\pi \de_{\alpha_{k_a} ; \be_{i_a}}  \Big\}  \op{S}\big(\overleftarrow{\bs{\alpha}}_{n}  \mid \overleftarrow{\bs{\alpha}}_{n}^{(1)} \big)  \cdot \ex{ - 2\i\pi n \om_{\op{O}} } \\
\times \op{S}\big(\bs{\be}_{m}^{(1)}  \mid \bs{\be}_{m} \big)    \cdot \mc{F}_{n+m-2p;-,+ }^{(\op{O})}\big(  \overleftarrow{\bs{\alpha}}_{n}^{(2)} +\i\pi \ov{\bs{e}}_{n-p}, \bs{\be}_m^{(2)} \big)  \;.
\label{ecriture forme explicite de la distribution}
\end{multline}
There, $ \mc{F}_{n+m-2p;-,+ }^{(\op{O})}$ stands for the $-$ boundary value in respect to the first $(n-p)$ variables and $+$ boundary value in respect to the last $(m-p)$
variables. Also, the above formula builds on the shorthand notations 
\beq
 \left\{ \ba{c}   \bs{\alpha}_{n}^{(1)}=(\alpha_{k_1},\dots, \alpha_{k_p}) \vspace{2mm} \\  
 \bs{\be}_{m}^{(1)}=(\be_{i_1},\dots, \be_{i_p}) \ea \right.  \quad and \quad  
 \left\{ \ba{cc}  \bs{\alpha}_{n}^{(2)}=(\alpha_{\ell_1},\dots, \alpha_{\ell_{n-p}})  \vspace{2mm}  \\ 
\bs{\be}_{m}^{(2)}=(\be_{j_1},\dots, \be_{j_{m-p}}) \ea \right. 
\enq
while the new sets of indices are defined as 
\beq
\left\{ \ba{ccc} \{\ell_1,\dots, \ell_{n-p} \} \, = \, \intn{1}{n} \setminus \{ k_a\}_1^p \;\qquad  & with\qquad  & \ell_1<\dots < \ell_{n-p}   \vspace{2mm}\\ 
\{j_1,\dots, j_{m-p} \} \, = \, \intn{1}{m} \setminus \{ i_a\}_1^p \;\qquad & with \qquad & j_1<\dots < j_{m-p}  \ea \right. \;. 
\enq
Moreover, we have introduced 
\beq
 \op{S}\big(\overleftarrow{\bs{\alpha}}_{n}  \mid \overleftarrow{\bs{\alpha}}_{n}^{(1)} \big)  =    \pl{a=1}{p} \pl{ \substack{  b =1 \\  k_a> \ell_b}  }{ n-p } \op{S}\big( \alpha_{k_a} - \alpha_{\ell_b} \big)
\; , \quad 
\op{S}\big(\bs{\be}_{m}^{(1)}  \mid \bs{\be}_{m} \big)    =   \pl{a=1}{p}\pl{ \substack{ b =1 \\ b< i_a}  }{ m }  S(\be_b-\be_{i_a}) \cdot \pl{ \substack{ a>b \\  i_a>i_b} }{} S\big( \be_{i_a} - \be_{i_b} \big) \;.
\nonumber
\enq
Finally, we agree upon  $\overleftarrow{\bs{\ga}}_N = (\ga_N,\dots, \ga_1)$ for any $\bs{\ga}_N=(\ga_1,\dots, \ga_{N})$.

\end{prop}

\Proof

The expression follows from a direct inductive repetition of the recursive construction through Axiom {\bf V)}. \qed

\vspace{2mm}

Here, we omit the details of the proof since, later on, we shall present a more effective one that is furthermore easily extendable to the more involved cases of interest to the analysis.
We however need the explicit expression \eqref{ecriture forme explicite de la distribution} provided by Proposition \ref{Proposition noyau integral generalise solution recursion} so as to ensure that the solution
$\mc{M}^{(\op{O})}_{n;m}$  to the recursion given in Axiom {\bf V)} is a well-defined generalised function
defined through multiplications of distributions having disjoint supports. Notice that, because of the explicit expressions \eqref{solution eqns bootstrap via K transformee} for the form factors
which show a  dependence on the difference of the rapidities only and taken the order of the limits \eqref{form factor BV}, one can get rid of the $(m-p)$ + boundary values
altogether in \eqref{ecriture forme explicite de la distribution}. This means that
\beq
\mc{F}_{n+m-2p;-,+ }^{(\op{O})}\big(  \overleftarrow{\bs{\alpha}}_{n}^{(2)} +\i\pi \ov{\bs{e}}_{n-p}, \bs{\be}_m^{(2)} \big) \, = \,
\mc{F}_{n+m-2p;-, \emptyset }^{(\op{O})}\big(  \overleftarrow{\bs{\alpha}}_{n}^{(2)} +\i\pi \ov{\bs{e}}_{n-p}, \bs{\be}_m^{(2)} \big)
\enq
where the $\emptyset$ means that the form factor is evaluated on the real axis for the $\be_a$s variables.\\
The very form of the inductive construction of $\mc{M}^{(\op{O})}_{n;m}$ allows one to establish the behaviour under contiguous
permutations of coordinates which will play a central role later on.

\begin{prop}

For $\bs{\alpha}_n \in \R^n$ and $\bs{\be}_{m}\in \{ \R \cup (\R+\i\pi) \}^m$ generic,   $\mc{M}^{(\op{O})}_{n;m}$ enjoys the below exchange of contiguous coordinates property
\beq
\mc{M}^{(\op{O})}_{n;m}\big(  \bs{\alpha}_n ; \bs{\be}_{m} \big) \; = \;  \op{S}\big( \be_{a a+1} \big) \cdot \mc{M}^{(\op{O})}_{n;m}\big(  \bs{\alpha}_n ; \bs{\be}_{m}^{(a+1 a)} \big)
\quad and \quad
\mc{M}^{(\op{O})}_{n;m}\big(  \bs{\alpha}_n; \bs{\be}_{m} \big) \; = \;  \op{S}\big( \a_{a+1  a} \big) \cdot \mc{M}^{(\op{O})}_{n;m}\big(  \bs{\alpha}_n^{(a+1 a)}  ; \bs{\be}_{m} \big) \;,
\label{ecriture equation echange local pour M n m}
\enq
with $\bs{\be}_{m}^{(p+1p)}$ as introduced in axiom {\bf I)}.

\end{prop}

\Proof
We first prove the ${\bs\be}_{m}$ exchange property by induction over the dimensionality $n$ of $\bs{\a}_n$.

When $n=0$, since $\mc{M}^{(\op{O})}_{n;m}\big(  \bs{\alpha}_n; \bs{\be}_{m} \big) \; = \; \mc{F}_{m;+}^{(\op{O})}(\bs{\be}_m)$
by the initialisation property, there is only the exchange property of the $\bs{\be}_m$ variables to establish and the latter follows
from form factors axiom {\bf I}.

Now assume that the $\bs{\be}_{m}$ exchange property holds up to some dimensionality $n-1 \geq 0$ of the first set of variables. Looking
at the reduction equation provided in axiom {\bf V)},
one observes that the shifted concatenation contribution has already the sought transformation law:
\beq
\mc{M}^{(\op{O})}_{n-1;m+1}\big(\bs{\alpha}^{\prime}_{n} ; (\alpha_1 + \i\pi,~ \bs{\be}_{m})\big) \, = \,
\op{S}(\be_{pp+1}) \cdot \mc{M}^{(\op{O})}_{n-1;m+1}\big(\bs{\alpha}^{\prime}_{n} ; (\alpha_1 + \i\pi,  \bs{\be}_{m}^{(p+1p)})\big) \; , \quad p \in \intn{1}{m-1} \;.
\enq

It remains to establish the same property for the reduction contributions of axiom {\bf V)}. Let us fix $p \in \intn{1}{m}$ and split the latter as
\beq
\sul{a=1}{m}  \de_{\alpha_1;\be_a} \pl{k=1}{a-1} \op{S}(\be_{ka}) \cdot \mc{M}^{(\op{O})}_{n-1;m-1}\big(\bs{\alpha}^{\prime}_{n} ;   \wh{\bs{\be}}_{m}^{\,(a)}\big) \, = \, \mc{S}_1 \,  +\, \mc{S}_2 \;.
\label{ecriture contribution somme axiome V}
\enq
The first term, $\mc{S}_1$, avoids the contribution of $a=p, p+1$:
\beq
\mc{S}_1 \, = \, \sul{  \substack{a=1  \\ a \not=p, p+1} }{m}  \de_{\alpha_1;\be_a} \pl{k=1}{a-1} \op{S}(\be_{ka}) \cdot \mc{M}^{(\op{O})}_{n-1;m-1}\big(\bs{\alpha}^{\prime}_{n} ;   \wh{\bs{\be}}_{m}^{\, (a)}\big) \;.
\enq
As such, it directly enjoys the exchange property between $\be_p$ and $\be_{p+1}$ since its expression only involves variables which are swapped inside the
generalised function, what allows one to apply the induction hypothesis at $n-1$.
The second contribution in \eqref{ecriture contribution somme axiome V} only involves the $a=p$ and $a=p+1$ terms:
\beq
\mc{S}_2 \, = \, \de_{\alpha_1 ; \be_p} \prod_{k=1}^{p-1} \op{S}(\be_k - \alpha_1) \cdot \mc{M}^{(\op{O})}_{n-1;m-1}(\bs{\alpha}^{\prime}_n ; \widehat{\bs{\be}}^{\, (p)}_m)
+ \de_{\alpha_1 ; \be_{p+1}} \prod_{k=1}^{p} \op{S}(\be_k - \alpha_1) \cdot \mc{M}^{(\op{O})}_{n-1;m-1}(\bs{\alpha}^{\prime}_n ; \widehat{\bs{\be}}^{\,(p+1)}_m).
\enq
By using the property of Dirac masses, it can be rewritten and factorized:
\beq
\op{S}(\be_p-\be_{p+1}) \cdot \bigg\{ \de_{\alpha_1 ; \be_p}   \prod_{\substack{k=1 \\ k\neq p}}^{p+1} \op{S}(\be_k - \alpha_1) \cdot
\mc{M}^{(\op{O})}_{n-1;m-1}\big(  \bs{\alpha}_n^{\prime};  \wh{\bs{\be}}_m^{\, (p)}  \big) \; + \; \de_{\alpha_1 ; \be_{p+1}}
\prod_{k=1}^{p-1} \op{S}(\be_k - \alpha_1) \cdot \mc{M}^{(\op{O})}_{n-1;m-1}\big(  \bs{\alpha}_n^{\prime};  \wh{\bs{\be}}_m^{\, (p+1)}  \big)              \bigg\} \;.
\enq
The two terms in the brackets correspond to the contributions that one would get by writing down the inductive equation of axiom {\bf V)}
for the vector $\bs{\be}_{m}^{(p+1 p)}$. All-in-all, this entails the claim.

It remains to establish the exchange property for the $\bs{\a}_n$ variables which we will once again prove by induction on $n$. Since there is nothing to prove, the latter is obviously true for $n=0,1$.
Assume that the property holds up to some $n-1$.

First, we focus on the exchange of $\a_{p}$ and $\a_{p+1}$, with $2 \leq p \leq n-1$.
Upon applying the inductive reduction of axiom {\bf V)}, the property follows from the induction hypothesis.

Dealing with the exchange of $\a_{1}$ and $\a_{2}$, \textit{viz}. $p=1$, demands more care, in particular apply the axiom {\bf V)} reduction twice.
Upon observing that the reduction only applies to the setting where $\bs{\a}_n\in \R^n$, one has that the Dirac mass at $\a_2=\a_1+\i\pi$
has zero net contribution. Then recasting the contribution of exchanged $\a$'s by using the crossing symmetry
$ \op{S}(\alpha_1+\i\pi-\alpha_2)=\op{S}(\a_{2}-\a_{1})$, one obtains
\begin{align}
	\mc{M}^{(\op{O})}_{n;m}(\bs{\alpha}_n ; \bs{\be}_m) \, & = \, \mathrm{e}^{-4\pi\i\omega_0} \cdot \bigg\{ \mc{M}^{(\op{O})}_{n-2;m+2}(\bs{\alpha}_n^{\prime\prime} ; \alpha_2+\i\pi, \alpha_1+\i\pi, \bs{\be}_m)
	+ 2\pi\de_{\alpha_2;\alpha_1+\i\pi} \cdot \mc{M}^{(\op{O})}_{n-2;m}(\bs{\alpha}_n^{\prime\prime} ; \bs{\be}_m) \notag \\
	& + 2\pi \sum_{a=1}^m \de_{\alpha_2;\be_a}  \cdot\op{S}(\a_{2}-\a_{1}) \prod_{k=1}^{a-1} \op{S}(\be_k - \alpha_2)  \cdot \mc{M}^{(\op{O})}_{n-2;m}(\bs{\alpha}^{\prime\prime}_n ; \alpha_1+\i\pi,\widehat{\bs{\be}}^{\,(a)}_m) \notag \\
	& + 2\pi \sum_{a=1}^m \de_{\alpha_1;\be_a} \prod_{k=1}^{a-1} \op{S}(\be_k - \alpha_1)  \cdot \mc{M}^{(\op{O})}_{n-2;m}(\bs{\alpha}^{\prime\prime}_n ; \alpha_2+\i\pi, \widehat{\bs{\be}}^{\,(a)}_m) \notag \\
	& + 4\pi^2\sum_{a=1}^m\sum_{b=1}^{a-1} \de_{\alpha_1;\be_a}\de_{\alpha_2;\be_b}\prod_{k=1}^{a-1}\op{S}(\be_k-\alpha_1)\prod_{j=1}^{b-1} \op{S}(\be_j - \alpha_2)
	\cdot 	\mc{M}^{(\op{O})}_{n-2;m-2}(\bs{\alpha}_n^{\prime\prime} ; \widehat{\bs{\be}}_m^{\,(b,a)})\bigg\} \notag\\
	& + 4\pi^2\sum_{a=1}^m\sum_{b=a+1}^m \de_{\alpha_1;\be_a}\de_{\alpha_2;\be_b}\prod_{k=1}^{a-1}\op{S}(\be_k-\alpha_1) \prod_{ \substack{j=1 \\  j\neq a} }^{b-1} \op{S}(\be_j - \alpha_2)
	\cdot 	\mc{M}^{(\op{O})}_{n-2;m-2}(\bs{\alpha}_n^{\prime\prime} ; \widehat{\bs{\be}}_m^{\,(a,b)})\bigg\} \;.
	\label{double reduction}
\end{align}
Above, we agree that $\bs{\a}_n^{\prime\prime}=(\a_3,\dots, \a_n)$ and have split the double Dirac mass contributions in two.
The exchange property of the second set of variables proven previously applied to the first two components of $(\alpha_2+\i\pi, \alpha_1+\i\pi, \bs{\be}_m)$ in the first term lead to
\begin{align}
\mc{M}^{(\op{O})}_{n;m}(\bs{\alpha}_n ; \bs{\be}_m) \,& = \,
\op{S}(\alpha_2-\alpha_1) \cdot \mathrm{e}^{-4\pi\i\omega_0} \cdot \bigg\{ \mc{M}^{(\op{O})}_{n-2;m+2}(\bs{\alpha}_n^{\prime\prime} ; \alpha_1+\i\pi, \alpha_2+\i\pi, \bs{\be}_m) \notag \\
& + 2\pi \sum_{a=1}^m \de_{\alpha_2;\be_a} \prod_{k=1}^{a-1} \op{S}(\be_k - \alpha_2)  \cdot \mc{M}^{(\op{O})}_{n-2;m}(\bs{\alpha}^{\prime\prime}_n ; \alpha_1+\i\pi,\widehat{\bs{\be}}^{\,(a)}_m) \notag \\
& + 2\pi \op{S}(\alpha_1 - \alpha_2) \sum_{a=1}^m \de_{\alpha_1;\be_a}  \prod_{k=1}^{a-1} \op{S}(\be_k - \alpha_1)  \cdot \mc{M}^{(\op{O})}_{n-2;m}(\bs{\alpha}^{\prime\prime}_n ; \alpha_2+\i\pi, \widehat{\bs{\be}}^{\,(a)}_m) \notag \\
& + 4\pi^2 \op{S}(\alpha_1 - \alpha_2)\sum_{a=1}^m\sum_{b=1}^{a-1} \de_{\alpha_1;\be_a}\de_{\alpha_2;\be_b}\prod_{k=1}^{a-1}\op{S}(\be_k-\alpha_1)\prod_{j=1}^{b-1} \op{S}(\be_j - \alpha_2)  \cdot 	\mc{M}^{(\op{O})}_{n-2;m-2}
(\bs{\alpha}_n^{\prime\prime} ; \widehat{\bs{\be}}_m^{\,(b,a)})\bigg\} \notag \\
& + 4\pi^2 \op{S}(\alpha_1 - \alpha_2)\sum_{a=1}^m\sum_{b=a+1}^m \de_{\alpha_1;\be_a}\de_{\alpha_2;\be_b}\prod_{k=1}^{a-1}\op{S}(\be_k-\alpha_1)
\prod_{ \substack{ j=1 \\ j\neq a} }^{b-1} \op{S}(\be_j - \alpha_2)  \cdot 	\mc{M}^{(\op{O})}_{n-2;m-2}
(\bs{\alpha}_n^{\prime\prime} ; \widehat{\bs{\be}}_m^{\,(a,b)})\bigg\}.
\end{align}
One can use the product of Dirac masses in the two last terms to get:
\begin{align}
\mc{M}^{(\op{O})}_{n;m}(\bs{\alpha}_n ; \bs{\be}_m) & = \op{S}(\alpha_2-\alpha_1) \cdot \mathrm{e}^{-4\pi\i\omega_0} \cdot \bigg\{ \mc{M}^{(\op{O})}_{n-2;m+2}(\bs{\alpha}_n^{\prime\prime} ; \alpha_1+\i\pi, \alpha_2+\i\pi, \bs{\be}_m) \notag \\
& + 2\pi \sum_{a=1}^m \de_{\alpha_2;\be_a} \prod_{k=1}^{a-1} \op{S}(\be_k - \alpha_2) \cdot  \mc{M}^{(\op{O})}_{n-2;m}(\bs{\alpha}^{\prime\prime}_n ; \alpha_1+\i\pi,\widehat{\bs{\be}}^{\,(a)}_m) \notag \\
& + 2\pi \sum_{a=1}^m \de_{\alpha_1;\be_a} \cdot \op{S}(\alpha_1-\alpha_2)\prod_{k=1}^{a-1} \op{S}(\be_k - \alpha_1)  \cdot \mc{M}^{(\op{O})}_{n-2;m}(\bs{\alpha}^{\prime\prime}_n ; \alpha_2+\i\pi, \widehat{\bs{\be}}^{\,(a)}_m) \notag \\
& + 4\pi^2\sum_{a=1}^m\sum_{b=1}^{a-1} \de_{\alpha_1;\be_a}\de_{\alpha_2;\be_b}\prod_{k\neq b}^{a-1}\op{S}(\be_k-\alpha_1)\prod_{j=1}^{b-1} \op{S}(\be_j - \alpha_2) \cdot
                                        \mc{M}^{(\op{O})}_{n-2;m-2}(\bs{\alpha}_n^{\prime\prime} ; \widehat{\bs{\be}}_m^{\,(b,a)}) \notag \\
& + 4\pi^2\sum_{a=1}^m\sum_{b=a+1}^m \de_{\alpha_1;\be_a}\de_{\alpha_2;\be_b}\prod_{k=1}^{a-1}\op{S}(\be_k-\alpha_1)\prod_{j=1}^{b-1} \op{S}(\be_j - \alpha_2) \cdot
\mc{M}^{(\op{O})}_{n-2;m-2}(\bs{\alpha}_n^{\prime\prime} ; \widehat{\bs{\be}}_m^{\,(a,b)})\bigg\}.
	\label{switch double reduction}
\end{align}
At this stage, it remains to observe that exchanging the $\a_1$ with $\a_2$ and the indices $a,b$ in the double sums
in \eqref{double reduction} exactly yields \eqref{switch double reduction}, up to the overall factor $\op{S}(\a_2-\a_1)$. \qed

\vspace{4mm}

In the remainder of this section, we shall prove several equivalent representations for $\mc{M}^{(\op{O})}_{n;m}\big(  \bs{\alpha}_n ; \bs{\be}_{m} \big)$. For that purpose, we need to
introduce a few notations. Given vectors
\beq
\bs{\a}_n \, = \, \big( \a_1,\dots,\a_n \big)  \qquad  \e{and} \qquad  \bs{\be}_m \, = \, \big( \be_1,\dots, \be_m \big)  \; ,
\enq
one introduces two sets built out of their coordinates
\beq
A \; = \; \big\{ \a_a \big\}_{a=1}^{n} \qquad \e{and} \qquad B \; = \; \big\{ \be_a \big\}_{a=1}^{m} \;.
\enq
At this stage one may consider an arbitrary partition
\beq
A\, = \, A_1 \cup A_2 \qquad  \e{where}  \qquad  A_1 \, = \, \{ \a_{k_a} \}_{a=1}^{p} \quad  \e{and} \quad
 A_2 \, = \, \{ \a_{\ell_a} \}_{a=1}^{n-p} \;.
\enq
In such a writing, one assumes that the indices $k_a$ and $\ell_b$ are taken in the strictly increasing order
\beq
1 \leq \ell_1<\dots < \ell_{n-p} \leq n\; ,\qquad   1 \leq k_1<\dots < k_{p} \leq n
\enq
and are such that $\{\ell_1,\dots, \ell_{n-p} \} \, = \, \intn{1}{n} \setminus \{ k_a\}_1^p$.

One then associates with such partitions the vectors
\beq
\bs{A}_1 \; = \; \big(\a_{k_1},\dots, \a_{k_p}  \big)  \qquad \e{and} \qquad
\overleftarrow{\bs{A}_1} \; = \; \big(\a_{k_p},\dots, \a_{k_1}  \big)  \;,
\enq
and likewise for $\bs{A}_2$ and $\overleftarrow{\bs{A}_2}$.
It will also be useful to consider non-ordered partitions: $B=B_{1} \underset{ 1}{ \cup } B_2 $. The notation means that the elements of $B$
are split into two sets
\beq
B_1 \, = \, \{ \, \be_{i_a} \}_{a=1}^{p} \quad  \e{and} \quad
B_2 \, = \, \{ \, \be_{j_a} \}_{a=1}^{n-p}
\enq
in which
\beq
1\leq i_1\not= \cdots \not= i_p \leq n \qquad \e{and} \qquad   1 \leq j_1<\dots < j_{n-p} \leq n
\enq
are such that $\{j_1,\dots, j_{n-p} \} \, = \, \intn{1}{n} \setminus \{ i_a\}_1^p$. Hence, in such a writing, one considers partitions
with the additional data relative to a permutation of the strict order usually taken for the labelling of the elements of the first partition.
We stress that in the writing, $B_{1} \underset{ 1}{ \cup } B_2$, the subscript $1$
indicates the sets whose elements are labelled with unordered, \textit{i.e.} possible permuted, indices. Then, one associates the vectors
\beq
\left\{ \ba{ccc}  \bs{B}_1 & = & \big(\be_{i_1},\dots, \be_{i_p}  \big)  \vspace{2mm}  \\
\overleftarrow{ \bs{B}}_1 & = & \big(\be_{i_p},\dots, \be_{i_1}  \big)   \ea \right.
\qquad \e{and} \qquad
\left\{ \ba{ccc}  \bs{B}_2 & = & \big(\be_{j_1},\dots, \be_{j_{n-p}}  \big)   \vspace{2mm} \\
\overleftarrow{ \bs{B} }_2  & = & \big(\be_{j_{n-p} },\dots, \be_{j_1}  \big)   \ea \right.     \;.
\enq

Finally, given $A_1 \, = \, \{ \a_{k_a} \}_{a=1}^{p}$ and $A_2 \, = \, \{ \a_{r_a} \}_{a=1}^{q}$ with $r_a \not= k_b$ $\forall a\not=b$,
one may write
\beq
A_1\cup A_2 \; = \; \{ \a_{s_a} \}_{a=1}^{p+q} \qquad \e{with} \qquad  s_1 < \dots < s_{p+q}\; .
\enq
In the following, we denote by $\bs{A}_1\cup  \bs{A}_2$
the vector obtained by concatenating the coordinates of $\bs{A}_1$ and $\bs{A}_2$, while
$\overrightarrow{\bs{A}_1\cup \bs{A}_2}$ the vector built out of the set $A_1\cup A_2$, namely
\beq
\bs{A}_1\cup  \bs{A}_2 \; = \; \big( \a_{k_1},\dots, \a_{k_p}  , \, \a_{r_1},\dots, \a_{r_q}  \big)
\qquad \e{and} \qquad
 \overrightarrow{ \bs{A}_1\cup \bs{A}_2 } \; = \; \big(  \a_{s_1},\dots, \a_{s_{p+q}}  \big)  \;.
\enq
Analogously, 
\beq
\overleftarrow{\bs{A}_1}\cup  \overleftarrow{\bs{A}_2 }\; = \;
\big( \a_{k_p},\dots, \a_{k_1}  , \, \a_{r_q},\dots, \a_{r_1}  \big) \qquad \e{and} \qquad \overleftarrow{ \bs{A}_1\cup \bs{A}_2 } \; = \; \big(  \a_{s_{p+q}},\dots, \a_{s_{1}}  \big)  \;.
\enq

By applying the local exchange relations \eqref{ecriture equation echange local pour M n m}, given partitions $A \, =\, A_1 \cup A_2$ and $B=B_1\underset{1}{\cup} B_2$, one then defines the
product of $S$ matrices subordinate to the permutation of coordinates
\beq
\bs{B} \hookrightarrow \bs{B}_1\cup  \bs{B}_2 \;, \quad \e{resp}. \qquad
\overleftarrow{\bs{B}} \hookrightarrow \overleftarrow{\bs{B}}_1\cup  \overleftarrow{\bs{B}}_2
\enq
as
\beqa
\mc{M}^{(\op{O})}_{n;m}\Big( \,  \bs{A} ; \bs{B} \Big) & = & \op{S}\Big( \, \bs{B} \mid \bs{B}_1\cup  \bs{B}_2 \Big) \cdot
          \mc{M}^{(\op{O})}_{n;m}\Big( \,  \bs{A} ;\bs{B}_1\cup  \bs{B}_2 \Big) \label{ecriture permutation 1 dans M}\\
\mc{M}^{(\op{O})}_{n;m}\Big( \,  \bs{A} ;\overleftarrow{ \bs{B} } \Big) & = & \op{S}\Big( \, \overleftarrow{ \bs{B} }  \mid \overleftarrow{ \bs{B}_1}\cup  \overleftarrow{ \bs{B}_2} \Big) \cdot
          \mc{M}^{(\op{O})}_{n;m}\Big( \,  \bs{A} ;\overleftarrow{ \bs{B}_1 }\cup  \overleftarrow{ \bs{B}_2 } \Big)
\label{ecriture permutation 2 dans M}
\eeqa
and likewise for more complex permutations. Then, it follows from the local exchange relations \eqref{ecriture equation echange local pour M n m}
for the $\bs{\a}_n$ coordinates and the unitarity of the S-matrix (\ref{definition matrice S}) that
\beqa
\mc{M}^{(\op{O})}_{n;m}\Big( \,  \bs{A} ;  \bs{B} \Big) & = &\op{S}\Big( \,\bs{A}_1\cup  \bs{A}_2  \mid  \bs{A} \Big) \cdot
          \mc{M}^{(\op{O})}_{n;m}\Big( \, \bs{A}_1 \cup  \bs{A}_2 ;   \bs{B} \Big) \label{ecriture permutation 3 dans M}\\
\mc{M}^{(\op{O})}_{n;m}\Big( \, \overleftarrow{ \bs{A} } ; \bs{B} \Big) & = & \op{S}\Big( \, \overleftarrow{ \bs{A}_1}\cup  \overleftarrow{ \bs{A}_2} \mid  \overleftarrow{ \bs{A} }  \,  \Big) \cdot
          \mc{M}^{(\op{O})}_{n;m}\Big( \, \overleftarrow{ \bs{A}_1 }\cup  \overleftarrow{ \bs{A}_2 } ;  \bs{B} \Big) \;.
\label{ecriture permutation 4 dans M}
\eeqa
Obviously, these definitions generalise straightforwardly to more complex permutations.

We close this preliminary discussion by observing that a direct calculation shows that
\beq
\op{S}\Big( \,\bs{A}_1\cup  \bs{A}_2  \mid  \bs{A} \Big) \; = \; \op{S}\Big( \, \overleftarrow{\bs{A}}  \mid \overleftarrow{\bs{A}}_2\cup  \overleftarrow{ \bs{A}}_1 \Big) \;.
\label{ecriture identite inversion coordonnee dans produit S permutationnel}
\enq

In order to lighten the notations, from now on, we shall drop the number of variables index from the form factors; the latter may always be inferred from the dimensionality of the vector
appearing in the argument, \textit{i.e.}
\beq
\mc{F}^{(\op{O})}(\bs{A}) \, \equiv \,  \mc{F}_n^{(\op{O})}(\bs{A}) \quad \e{for} \quad \bs{A} \in \Cx^n \;.
\enq
The dimensionality of eventual overall variable shifts will also always be undercurrent by the dimensionality of the vectors, \textit{e.g.}
\beq
\mc{F}^{(\op{O})}(\bs{A},\bs{B}+\a \bs{e}) \, \equiv \,  \mc{F}^{(\op{O})}_{n+m}(\bs{A},\bs{B}+\a \bs{e}_m) \quad \e{if} \quad \bs{A}\in \Cx^n \; \e{and} \; \bs{B}\in \Cx^m \;.
\enq

We now re-express \eqref{ecriture forme explicite de la distribution} with the help of these new notations and present a very simple proof of the decomposition
formula for $\mc{M}^{(\op{O})}_{n;m}\big(  \bs{\alpha}_n ; \bs{\be}_{m} \big)$.

\begin{lemme}
\label{Lemma rep direct pour noyau integral generalise secteur n m particules}

Let $\bs{\alpha}_n \in \R^{n}$ and $\bs{\be}_{m} \in \R^m$ and $A \, = \, \{\a_a\}_1^n$, $B\, = \, \{ \be_a\}_1^m$. Axiom ${\bf V)}$ implies that
\bem
\mc{M}^{(\op{O})}_{n;m}\big(  \bs{\alpha}_n ; \bs{\be}_{m} \big) \; = \;  \ex{-2\i\pi \om_{\op{O}} |A| }  \hspace{-2mm}  \sul{ A= A_1 \cup A_2  }{}  \sul{ B = B_1 \underset{1}{\cup} B_2  }{}
 \hspace{-2mm}  \De\big( \bs{A}_1 \mid \bs{B}_1  \big) \cdot \op{S}\Big( \, \overleftarrow{ \bs{A} }  \mid \overleftarrow{ \bs{A}_2 }\cup  \overleftarrow{ \bs{A}_1 } \Big)
\cdot \op{S}\big(  \bs{B} \mid \bs{B}_1\cup  \bs{B}_2 \big) \\
\times \mc{F}_{ - , 0}^{(\op{O})} \Big( \overleftarrow{ \bs{A}_2 } + \i \pi \ov{\bs{e}}, \bs{B}_2  \Big) \;.
\label{ecriture representation combinatoire pour MO via reduction de type 1}
\end{multline}
There $-,0$ means that the first set of variables should be understood in the sense of distributional $-$ boundary values while there is no per se prescription for the second set which simply takes real values:
\beq
 \mc{F}_{ - , 0}^{(\op{O})} \Big( \overleftarrow{ \bs{A} } + \i \pi \ov{\bs{e}},  \bs{B}  \Big) \, \equiv \,
 \lim_{ \substack{ \veps_1<\dots < \veps_n \\ \veps_a \tend 0^+}  }\mc{F}^{(\op{O})}(\bs{A}  + \i \pi \ov{\bs{e}}_n  - \i \bs{\veps}_n, \bs{B})
\quad for \quad \bs{A} \in \Cx^n \; and \; \bs{B}\in \Cx^m \;.
\enq

The summation runs through all ordered partitions $A_1 \cup A_2$ of $A$ and all partitions $B_1 \underset{1}{\cup} B_2$  of $B$ such that elements of $B_1$ appear in any order while those of
$B_2$ are ordered. The choices of partitions are constrained  to the condition
\beq
|A_1| \; = \; |B_1| \;.
\enq
Finally, we have set
\beq
\De\big( \bs{A}_1 \mid \bs{B}_1  \big) \; = \; \pl{a=1}{|A_1|} \Big\{ 2\pi \de_{ \a_{k_a},\,  \be_{i_a}} \Big\} \qquad where \qquad
\left\{\ba{ccc}  A_1 & =& \big\{ \a_{k_a} \big\}_{a=1}^{|A_1|}  \vspace{2mm}  \\
        B_1 &  = & \big\{ \be_{i_a} \big\}_{a=1}^{|B_1|}   \ea \right.  \; .
\label{definition facteur masse dirac globale}
\enq

\end{lemme}

\Proof

The proof is based on the completely direct/indirect action method, see \cite{SlavnovBookCorrelationFunctions}. It is clear that the overall form
$\mc{M}^{(\op{O})}_{n;m}\big(  \bs{\alpha}_n ; \bs{\be}_{m} \big)$ issuing from the reduction provided by axiom ${\bf V)}$
takes the general form
\beq
\mc{M}^{(\op{O})}_{n;m}\big(  \bs{\alpha}_n ; \bs{\be}_{m} \big) \; =   \hspace{-2mm}  \sul{ A= A_1 \cup A_2  }{}  \sul{ B = B_1 \underset{1}{\cup} B_2  }{}
 \hspace{-2mm}  \De\big( \bs{A}_1 \mid \bs{B}_1  \big) \cdot \op{C}\big( \bs{A}_1    ,    \bs{A}_2;
\bs{B}_1   ,     \bs{B}_2 \big)
\cdot \mc{F}_{ - , 0}^{(\op{O})} \Big( \overleftarrow{ \bs{A}_2} + \i \pi \ov{\bs{e}},  \bs{B}_2 \Big) \;,
\enq
for some unknown coefficient $\op{C}\big( \bs{A}_1    ,    \bs{A}_2; \bs{B}_1   ,     \bs{B}_2 \big)$.
Hence, first, consider the permutation
\beq
\bs{A}  \hookrightarrow \bs{A}_1\cup  \bs{A}_2 \qquad \e{and} \qquad \bs{B}  \hookrightarrow \bs{B}_1\cup  \bs{B}_2
\enq
which, upon invoking \eqref{ecriture identite inversion coordonnee dans produit S permutationnel}, leads to
\beq
\mc{M}^{(\op{O})}_{n;m}\big(   \bs{A} ;\bs{B} \big) \; = \; \op{S}\Big( \, \overleftarrow{ \bs{A} }  \mid \overleftarrow{ \bs{A}_2}\cup  \overleftarrow{ \bs{A}_1 } \Big)
\cdot \op{S}\big( \bs{B} \mid \bs{B}_1\cup  \bs{B}_2 \big) \cdot
          \mc{M}^{(\op{O})}_{n;m}\big(  \bs{A}_1\cup  \bs{A}_2 ;  \bs{B}_1\cup  \bs{B}_2   \big)
\enq
We now reduce variables in $ \mc{M}^{(\op{O})}_{n;m}\big(  \bs{A}_1\cup  \bs{A}_2 ;  \bs{B}_1\cup  \bs{B}_2   \big)$
by applying axiom $\bf{V)}$. Starting from this new ordering, the only way to produce the distribution  $\De\big( \bs{A}_1 \mid \bs{B}_1  \big)$ is to
reduce, by means of the reduction contributions present in axiom $\bf{V)}$, the first component of  $\bs{A}_1$ with the first component of $\bs{B}_1$ and so on,
until all vectors are reduced. This generates \textit{no} additional $\op{S}$-matrix products and only produces the additional factor $\ex{-2\i\pi \om_{\op{O}}|A_1| }$, \textit{viz}.
\beq
\mc{M}^{(\op{O})}_{n;m}\big(   \bs{A}_1\cup  \bs{A}_2 ;  \bs{B}_1\cup  \bs{B}_2   \big) \hookrightarrow
\ex{-2\i\pi \om_{\op{O}}|A_1| } \cdot \De\big( \bs{A}_1 \mid \bs{B}_1  \big) \cdot \mc{M}^{(\op{O})}_{|A_2|;|B_2|}\big(    \bs{A}_2 ;   \bs{B}_2   \big) \;.
\enq
After that first reduction, one "moves" the coordinates of $\bs{A}_2$ in the "$B$"-coordinate type slot, by means of the shifted
concatenation contributions in axiom $\bf{V)}$. These  produce $-$ boundary value prescriptions for the concatenated coordinates.
Also, the operation  induces a flip in the vector's orientation  $\bs{A}_2 \hookrightarrow  \overleftarrow{ \bs{A}_2 }$ and a prefactor $\ex{-2\i\pi \om_{\op{O}}|A_2| }$, \textit{viz}.
\beq
\mc{M}^{(\op{O})}_{|A_2|;|B_2|}\Big( \,  \bs{A}_2 ;   \bs{B}_2   \Big) \hookrightarrow
\ex{-2\i\pi \om_{\op{O}}|A_2| } \cdot  \mc{M}_{0 ; |A_2| + |B_2|; - , 0}^{(\op{O})} \Big( \emptyset; \big(\,  \overleftarrow{\bs{A}_2} + \i \pi \ov{\bs{e}}_{|A_2|},   \bs{B}_2 \big)\Big) \;.
\enq
This is the only possible reduction that does not give rise to more Dirac masses. This last quantity may be evaluated by using the initialisation condition \eqref{condition initialisation pour MO avec zero vecteur alpha}. Thus, all-in-all, the reasoning allows one to
identify
\beq
 \op{C}\big(   \bs{A}_1  , \,   \bs{A}_2 ; \bs{B}_1 \, , \,  \bs{B}_2 \big)  \; = \;
 \op{S}\Big( \, \overleftarrow{\bs{A} }  \mid \overleftarrow{ \bs{A}_2 }\cup  \overleftarrow{ \bs{A}_1} \Big)
\cdot \op{S}\big( \bs{B} \mid \bs{B}_1\cup  \bs{B}_2 \big) \;.
\enq
This entails the claim. \qed

\subsubsection{Dual representation and its equivalence with the direct one}

It so happens that the form factor axioms allow one to produce quite a deal of equivalent although structurally different combinatorial representations for the generalised integral
kernels  $\mc{M}^{(\op{O})}_{n;m}\big(  \bs{\alpha}_n ; \bs{\be}_{m} \big)$. Their existence plays a crucial role in providing well-defined
expressions for the multi-point correlation functions of local operators located at generic space-time positions.

For now, we shall introduce a recursion equation that is, in some sense,  dual to axiom ${\bf V)}$.
This recursion will hold for an auxiliary quantity $\wt{\mc{M}}^{(\op{O})}_{n;m}\big(  \bs{\alpha}_n ; \bs{\be}_{m} \big)$.
It reads
\bem
\wt{\mc{M}}^{(\op{O})}_{n;m}\big(  \bs{\alpha}_n ; \bs{\be}_{m} \big) \; = \;   \pl{k=2}{n}\op{S}(\a_{k1})\cdot \wt{\mc{M}}^{(\op{O})}_{n-1;m+1}\big(  \bs{\alpha}_n^{\prime} ; (\bs{\be}_m,\alpha_1-\i \pi ) \big) \\
\; + \; 2\pi \sul{a=1}{m}  \de_{\alpha_1;\be_a}  \pl{k=2}{n}\op{S}(\a_{k1})\cdot  \pl{k=a+1}{m} \op{S}(\be_{ak}) \cdot
\wt{\mc{M}}^{(\op{O})}_{n-1;m-1}\big(  \bs{\alpha}_n^{\prime};  \wh{\bs{\be}}_m^{\,(a)}  \big)   \; .
\label{ecriture relation reduction forme alternative}
\end{multline}
The induction holds for $\bs{\alpha}_n \in \R^{n}$, $\bs{\be}_{m} \in \R^m$ and the evaluation at $\alpha_1-\i \pi$ is to be understood in the sense of the $+$ boundary value on
$\R-\i\pi$ of the analytic continuation  from $\R$ to the strip $-\pi <\Im(z)<0$ of the generalised function
$ \a \mapsto \wt{\mc{M}}^{(\op{O})}_{n-1;m+1}\big(  \bs{\alpha}_n^{\prime} ; (\bs{\be}_m, \alpha) \big)   $.
The dual induction is complemented with the initialisation condition
\beq
 \wt{\mc{M}}^{(\op{O})}_{0;m}\big( \emptyset ; \bs{\be}_{m} \big)\; = \;   \mc{F}_{m;+}^{(\op{O})}(\bs{\be}_m) \quad \e{when} \quad  \bs{\be}_m \in \R^{m}_{>} \;.
\enq
On that basis, one may show, exactly as for  $\mc{M}^{(\op{O})}_{n;m}\big(  \bs{\alpha}_n ; \bs{\be}_{m} \big)$, that  $\wt{\mc{M}}^{(\op{O})}_{n;m}\big(  \bs{\alpha}_n ; \bs{\be}_{m} \big)$
enjoys the symmetry properties
\beq
\wt{\mc{M}}^{(\op{O})}_{n;m}\big(  \bs{\alpha}_n ; \bs{\be}_{m} \big) \; = \;  \op{S}\big( \be_{a a+1} \big) \cdot \wt{\mc{M}}^{(\op{O})}_{n;m}\big(  \bs{\alpha}_n ; \bs{\be}_{m}^{(a+1 a)} \big)
\quad \e{and} \quad
\wt{\mc{M}}^{(\op{O})}_{n;m}\big(  \bs{\alpha}_n ; \bs{\be}_{m} \big) \; = \;  \op{S}\big( \a_{a+1  a} \big) \cdot \wt{\mc{M}}^{(\op{O})}_{n;m}\big(  \bs{\alpha}_n^{(a+1 a)} ; \bs{\be}_{m} \big) \;.
\label{ecriture equation echange local pour tilde M n m}
\enq
This entails that $\wt{\mc{M}}^{(\op{O})}_{n;m}$ satisfies the very same equations as \eqref{ecriture permutation 1 dans M}-\eqref{ecriture permutation 4 dans M}
under general permutations.

These imply an equivalent form of the reduction equation \eqref{ecriture relation reduction forme alternative}:
\beq
\wt{\mc{M}}^{(\op{O})}_{n;m}\big(  \overleftarrow{\bs{\alpha}}_n; \bs{\be}_{m} \big) \; = \;    \wt{\mc{M}}^{(\op{O})}_{n-1;m+1}\big(  \overleftarrow{\bs{\alpha}}_n^{\, \prime} ; (\bs{\be}_m,\alpha_1-\i \pi ) \big) \\
\; + \; 2\pi \sul{a=1}{m}  \de_{\alpha_1;\be_a}  \pl{k=1+a}{m} \op{S}(\be_{ak}) \cdot
\wt{\mc{M}}^{(\op{O})}_{n-1;m-1}\big( \overleftarrow{\bs{\alpha}}_n^{\, \prime} ;  \wh{\bs{\be}}_m^{(a)}  \big)   \; .
\label{ecriture formule recursive de type 2 equivalente a l originale}
\enq

\begin{lemme}
For $\bs{\alpha}_n \in \R^{n}$ and $\bs{\be}_{m} \in \R^m$, it holds
\bem
\wt{\mc{M}}^{(\op{O})}_{n;m}\big(  \bs{\alpha}_n ; \bs{\be}_{m} \big) \; =   \hspace{-2mm}  \sul{ A= A_1 \cup A_2  }{}  \sul{ B = B_1 \underset{1}{\cup} B_2  }{}
 \hspace{-2mm}  \De\big( \bs{A}_1 \mid \bs{B}_1  \big) \cdot \op{S}\Big( \, \overleftarrow{ \bs{A} }  \mid \bs{A}_1\cup  \bs{A}_2 \Big)
\cdot \op{S}\Big( \, \bs{B} \mid \bs{B}_2\cup  \overleftarrow{ \bs{B}_1} \Big) \\
\times \mc{F}_{ 0, +}^{(\op{O})} \Big(   \bs{B}_2, \bs{A}_2 - \i \pi \ov{\bs{e}}  \Big) \;.
\label{ecriture representation combinatoire pour MO via reduction de type 2}
\end{multline}
There $0,+$ means that the first set of variables  is simply taking real values while the second set should be understood in the sense of $+$ boundary values.
The summation runs through all ordered partitions $A_1 \cup A_2$ of $A$ and all partitions $B_1 \cup B_2$  of $B$ such that elements of $B_1$ appear in any order while those of
$B_2$ are ordered. The choices of partitions are constrained  to the condition
\beq
|A_1| \; = \; |B_1| \;.
\enq
Finally, $\De\big( * \mid  * \big)$ is as defined in \eqref{definition facteur masse dirac globale}.

\end{lemme}

\Proof

The proof's strategy is similar to Lemma \ref{Lemma rep direct pour noyau integral generalise secteur n m particules}. First of all, given partitions $ A= A_1 \cup A_2$ and $B = B_1 \underset{1}{\cup} B_2$, one has that
\begin{align}
\wt{\mc{M}}^{(\op{O})}_{n;m}\big(   \bs{A} ;\bs{B} \big) \; & = \; \op{S}\Big( \, \bs{B} \mid \bs{B}_2\cup  \overleftarrow{\bs{B}_1} \Big) \cdot
\op{S}\Big( \, \overleftarrow{\bs{A}_2}\cup  \overleftarrow{\bs{A}_1} \mid \bs{A}   \Big) \cdot
          \wt{\mc{M}}^{(\op{O})}_{n;m}\Big( \, \overleftarrow{\bs{A}_2}\cup   \overleftarrow{\bs{A}_1} ;  \bs{B}_2\cup  \overleftarrow{\bs{B}_1}   \Big) \notag \\
& = \;\op{S}\Big( \, \overleftarrow{\bs{A}}  \mid \bs{A}_1\cup  \bs{A}_2 \Big) \cdot
 \op{S}\Big( \, \bs{B} \mid \bs{B}_2\cup  \overleftarrow{\bs{B}_1} \Big) \cdot
          \wt{\mc{M}}^{(\op{O})}_{n;m}\Big( \,  \overleftarrow{\bs{A}_2}\cup   \overleftarrow{\bs{A}_1} ;  \bs{B}_2\cup  \overleftarrow{\bs{B}_1}   \Big) \;,
\end{align}
where we have used (\ref{ecriture identite inversion coordonnee dans produit S permutationnel}) in the last line. After such a permutation of the variables, one reduces the last coordinate of $\overleftarrow{\bs{A}_1}$ with the last one of $\overleftarrow{\bs{B}_1}$
by means of the reduction contribution in  \eqref{ecriture formule recursive de type 2 equivalente a l originale}, and so-on until all vectors of type $1$ are reduced.
Further, one "moves", to the right of $ \bs{B}_2 $, one-by-one, the components of  $\overleftarrow{\bs{A}_2}$ by means of the shifted concatenation contribution in  \eqref{ecriture relation reduction forme alternative}.
This yields
\beq
 \wt{\mc{M}}^{\,(\op{O})}_{n;m}\Big( \,  \overleftarrow{\bs{A}_2}\cup   \overleftarrow{\bs{A}_1} ;  \bs{B}_2\cup  \overleftarrow{\bs{B}_1}   \Big)
\quad \hookrightarrow  \quad \De\big(   \bs{A}_1  \mid  \bs{B}_1   \big) \cdot
\mc{F}^{(\op{O})}_{ 0, + }\Big( \,   \bs{B}_2,  \bs{A}_2  - \i \pi  \ov{\bs{e}}   \Big) \;.
\enq
This entails the claim. \qed

\vspace{2mm}

We shall now establish that the expansions \eqref{ecriture representation combinatoire pour MO via reduction de type 1} and \eqref{ecriture representation combinatoire pour MO via reduction de type 2}
are, in fact, equivalent, \textit{viz}. that
\beq
\mc{M}^{(\op{O})}_{n;m}\big(  \bs{\alpha}_n; \bs{\be}_{m} \big) \; = \;\wt{\mc{M}}^{(\op{O})}_{n;m}\big(  \bs{\alpha}_n; \bs{\be}_{m} \big)
\enq

For that purpose, we need the below auxiliary lemma. An analogous result, in the case of the Massive Thirring model, has been established in \cite{KirillovSmirnovUseOfBootstrapAxiomsForQIFTToGetMassiveThirringFF}
by means of more combinatorial handlings.
\begin{lemme}
\label{Lemme relation entre valeurs plus et moins des FF}
Given two sets of real valued variables $A=\{\a_a\}_{a=1}^{n}$ and $B=\{\be_a\}_{a=1}^{m}$, one has the equality
\bem
\mc{F}_{ +, 0}^{(\op{O})}\Big( \, \overleftarrow{\bs{A}}  + \i \pi  \ov{\bs{e}},   \bs{B}    \Big) \; = \; \sul{ A =   \cup\, _{a=1}^{3}  A_a}{}   \sul{   B =   \underset{23}{\cup} \, _{a=1}^{3} B_a  }{}
\op{S}\Big( \, \overleftarrow{\bs{A}}  \mid \bs{A}_3\cup  \overleftarrow{\bs{A}_1} \cup  \overleftarrow{\bs{A}_2} \Big)
\cdot \op{S}\Big( \, \bs{B} \mid \bs{B}_2\cup  \bs{B}_1 \cup  \overleftarrow{ \bs{B}_3} \Big)  \cdot  \Big( - \ex{2\i\pi \om_{\op{O}}} \Big)^{|A_3|} \\
\times  \De\big(   \bs{A}_2 \cup  \bs{A}_3 \mid  \bs{B}_2 \cup  \bs{B}_3  \big) \cdot
\mc{F}^{(\op{O})}_{  -, 0 }\Big( \, \overleftarrow{\bs{A}_1}  + \i \pi  \ov{\bs{e}} ,   \bs{B}_1    \Big)
\end{multline}
The summation is subject to the constraint
\beq
|A_2| \, = \, |B_2| \qquad and \qquad  |A_3| \, = \, |B_3|
\enq
and $\De\big( * \mid *  \big)$ is as defined in \eqref{definition facteur masse dirac globale}.
\end{lemme}

\Proof

To start with, one observes that owing to
\beq
\mc{F}^{(\op{O})}\big( \a+\i\pi , \be , \bs{\be}_n \big)  \; = \; \pl{k=1}{a-1} \op{S}\big( \be - \be_k \big) \cdot
\mc{F}^{(\op{O})}\big( \a+\i\pi , \bs{\be}_{a-1} , \be , \wh{\bs{\be}}_{n;a-1} \big)  \quad \e{with} \quad
 \wh{\bs{\be}}_{n;a} \; = \; \big( \be_{a+1}, \dots, \be_{n} \big)  \;,
\enq
the pole axiom $\bs{III)}$ may be recast as
\beq
-\i \e{Res}\Big(\mc{F}^{(\op{O})}\big( \a+\i\pi , \bs{\be}_{a-1} , \be , \wh{\bs{\be}}_{n;a-1} \big) \cdot \dd \alpha \, , \, \alpha=\be  \Big) \; = \;
\bigg\{  \pl{k=1}{a-1} \op{S}\big( \be_k - \be \big) \, - \, \ex{2\i\pi \om_{\op{O}} }  \pl{k=a}{n}  \op{S}(\be-\be_{k}) \bigg\} \cdot
\mc{F}^{(\op{O})}( \, \bs{\be}_n)   \;.
\enq
Therefore, taken  $\bs{\be}_n$ having pairwise distinct coordinates, one has:
\beq
-\i \e{Res}\Big(\mc{F}^{(\op{O})}\big( \a+\i\pi , \bs{\be}_{n}   \big) \cdot \dd \alpha \, , \, \alpha=\be_a  \Big) \; = \;
\bigg\{  \pl{k=1}{a-1} \op{S}\big( \be_{ka} \big) \, - \, \ex{2\i\pi \om_{\op{O}} }  \pl{k=1+a}{n}  \op{S}(\be_{ak}) \bigg\} \cdot
\mc{F}^{(\op{O})}\big( \,   \wh{\bs{\be}}_n^{\, (a)} \big)   \;.
\enq
From the above, since all poles are simple for pairwise different $\be_a$s, one infers the jump conditions
\bem
 \mc{F}_{+,\emptyset}^{(\op{O})}\big( \a + \i\pi , \bs{\be}_{n}   \big)  \; = \;  \mc{F}_{-,\emptyset}^{(\op{O})}\big( \a + \i\pi , \bs{\be}_{n}   \big)
\; + \; \sul{a=1}{n} 2\pi \de_{\a;\be_a} \cdot   \pl{k=1}{a-1} \op{S}\big( \be_{ka} \big) \cdot \mc{F}^{(\op{O})}\big( \,   \wh{\bs{\be}}_n^{\, (a)} \big) \\
\, - \, \ex{2\i\pi \om_{\op{O}} }   \sul{a=1}{n} 2\pi \de_{\a;\be_a} \cdot  \pl{k=1+a}{n}  \op{S}(\be_{ak})   \cdot \mc{F}^{(\op{O})}\big( \,   \wh{\bs{\be}}_n^{\, (a)} \big)  \;.
\label{ecriture equation saut avec reduction gauche}
\end{multline}
Further, upon permuting some of the variables, one infers that
\bem
 \mc{F}_{\emptyset,+,0}^{(\op{O})}\big( \bs{\ga}_m + \i\pi \ov{\bs{e}}_m, \a + \i\pi, \bs{\be}_{n}   \big)  \; = \;  \mc{F}_{\emptyset,-,0}^{(\op{O})}\big( \bs{\ga}_m + \i\pi \ov{\bs{e}}_m, \a + \i\pi, \bs{\be}_{n}   \big) \\
\; + \; \sul{a=1}{n} 2\pi \de_{\a;\be_a} \cdot \pl{k=1}{a-1} \op{S}\big( \be_{ka} \big) \cdot \mc{F}^{(\op{O})}_{\emptyset,0}\big(    \bs{\ga}_m + \i\pi \ov{\bs{e}}_m,  \wh{\bs{\be}}_n^{\, (a)} \big) \\
\, - \, \ex{2\i\pi \om_{\op{O}} } \sul{a=1}{n} 2\pi \de_{\a;\be_a} \cdot \pl{k=1}{m}\op{S}(\ga_k-\a) \pl{k=1+a}{n}  \op{S}(\be_{ak})  \cdot
\mc{F}^{(\op{O})}_{\emptyset,0}\big(  \bs{\ga}_m + \i\pi \ov{\bs{e}}_m,  \wh{\bs{\be}}_n^{\, (a)} \big)  \;.
\label{ecriture equation saut avec reduction droite}
\end{multline}
Above, the boundary values pertain to the $\a\in \R$ variable, while $\bs{\be}_n\in \R^n$ and $\bs{\ga}_m \in \Cx^m$ are such that $\Im(\bs{\ga}_a)\not=0$ for any $a\in \intn{1}{m}$
as well as $|\Im(\bs{\ga}_a)| \ll 1$. Note that at the level of \eqref{ecriture equation saut avec reduction droite}, one may take the $+$ or $-$ boundary values for the $\ga_a$s,
possibly varying the sign of the boundary value with $a$, to get the relation valid for $\bs{\ga}_m \in \R^m$ in the distributional sense. This yields for $\bs{\vsg}_m\in \{\pm\}^m$
\bem
 \mc{F}_{ \bs{\vsg}_{m}, +, 0}^{(\op{O})}\big( \bs{\ga}_m + \i\pi \ov{\bs{e}}_m, \a + \i\pi, \bs{\be}_{n}   \big)
                                              \; = \;  \mc{F}_{ \bs{\vsg}_{m}, -, 0 }^{(\op{O})}\big( \bs{\ga}_m + \i\pi \ov{\bs{e}}_m, \a + \i\pi, \bs{\be}_{n}   \big) \\
\; + \; \sul{a=1}{n} 2\pi \de_{\a;\be_a} \cdot \pl{k=1}{a-1} \op{S}\big( \be_{ka} \big) \cdot \mc{F}^{(\op{O})}_{ \bs{\vsg}_{m},   0 }\big(    \bs{\ga}_m + \i\pi \ov{\bs{e}}_m,  \wh{\bs{\be}}_n^{\, (a)} \big) \\
\, - \, \ex{2\i\pi \om_{\op{O}} } \sul{a=1}{n} 2\pi \de_{\a;\be_a} \cdot \pl{k=1}{m}\op{S}(\ga_k-\a) \pl{k=1+a}{n}  \op{S}(\be_{ak})  \cdot
\mc{F}^{(\op{O})}_{ \bs{\vsg}_{m},   0 }\big(  \bs{\ga}_m + \i\pi \ov{\bs{e}}_m,  \wh{\bs{\be}}_n^{\, (a)} \big)  \;.
\label{ecriture equation saut general bv}
\end{multline}
Now introduce a new set functions $\widehat{\mc{F}}^{(\op{O})}$ satisfying
$\widehat{\mc{F}}^{(\op{O})}\big(\bs{B}\big) \, = \, \op{S}\big( \bs{B} \mid \bs{B}_1\cup \bs{B}_2\big) \widehat{\mc{F}}^{(\op{O})}\big( \bs{B}_1\cup \bs{B}_2 \big)$
as well as equations \eqref{ecriture equation saut avec reduction gauche} and \eqref{ecriture equation saut general bv} in which one implements the substitution
\beq
	\om_{\op{O}} \rightarrow \psi(\alpha)
\enq
where $\psi$ is some function of the variable $\alpha$. In other terms, we have the properties:
\bem
\widehat{\mc{F}}_{+,0}^{(\op{O})}\big( \a + \i\pi , \bs{\be}_{n}   \big)  \; = \;  \widehat{\mc{F}}_{-,0}^{(\op{O})}\big( \a + \i\pi , \bs{\be}_{n}   \big)
\; + \; \sul{a=1}{n} 2\pi \de_{\a;\be_a} \cdot   \pl{k=1}{a-1} \op{S}\big( \be_{ka} \big) \cdot \widehat{\mc{F}}^{(\op{O})}_0\big( \,   \wh{\bs{\be}}_n^{\, (a)} \big) \\
\, - \, \ex{2\i\pi \psi(\alpha) }   \sul{a=1}{n} 2\pi \de_{\a;\be_a} \cdot  \pl{k=1+a}{n}  \op{S}(\be_{ak})   \cdot \widehat{\mc{F}}^{(\op{O})}_0\big( \,   \wh{\bs{\be}}_n^{\, (a)} \big)
\label{ecriture equation saut avec reduction gauche mod}
\end{multline}
and 
\bem
\widehat{\mc{F}}_{ \bs{\vsg}_{m}, +, 0}^{(\op{O})}\big( \bs{\ga}_m + \i\pi \ov{\bs{e}}_m, \a + \i\pi, \bs{\be}_{n}   \big)
\; = \;  \widehat{\mc{F}}_{ \bs{\vsg}_{m}, -, 0 }^{(\op{O})}\big( \bs{\ga}_m + \i\pi \ov{\bs{e}}_m, \a + \i\pi, \bs{\be}_{n}   \big) \\
\; + \; \sul{a=1}{n} 2\pi \de_{\a;\be_a} \cdot \pl{k=1}{a-1} \op{S}\big( \be_{ka} \big) \cdot \widehat{\mc{F}}^{(\op{O})}_{ \bs{\vsg}_{m},   0 }\big(    \bs{\ga}_m + \i\pi \ov{\bs{e}}_m,  \wh{\bs{\be}}_n^{\, (a)} \big) \\
\, - \, \ex{2\i\pi\psi(\alpha) } \sul{a=1}{n} 2\pi \de_{\a;\be_a} \cdot \pl{k=1}{m}\op{S}(\ga_k-\a) \pl{k=1+a}{n}  \op{S}(\be_{ak})  \cdot
\widehat{\mc{F}}^{(\op{O})}_{ \bs{\vsg}_{m},   0 }\big(  \bs{\ga}_m + \i\pi \ov{\bs{e}}_m,  \wh{\bs{\be}}_n^{\, (a)} \big)  \;.
\label{ecriture equation saut general bv mod}
\end{multline}

At this stage we implement the permutation of coordinates
\bem
\widehat{\mc{F}}_{ +, 0}^{(\op{O})}\Big( \, \overleftarrow{\bs{A}}  + \i \pi  \ov{\bs{e}} ,   \bs{B}    \Big) \; = \;
\op{S}\Big( \, \overleftarrow{\bs{A}}  \mid \bs{A}_3\cup  \overleftarrow{\bs{A}_1} \cup  \overleftarrow{\bs{A}_2} \Big)
\cdot \op{S}\Big( \, \bs{B} \mid \bs{B}_2\cup  \bs{B}_1 \cup  \overleftarrow{ \bs{B}_3} \Big) \\
\times \widehat{\mc{F}}_{ +, 0}^{(\op{O})}\Big( \, \bs{A}_3\cup  \overleftarrow{\bs{A}_1} \cup  \overleftarrow{\bs{A}_2}  + \i \pi  \ov{\bs{e}} ,
        \bs{B}_2\cup  \bs{B}_1 \cup  \overleftarrow{ \bs{B}_3}   \Big) \;.
\end{multline}
Then,
\begin{itemize}

 \item we reduce $\bs{A}_3$ starting from the first entry and so on, with the last entry, and so on, of $\overleftarrow{ \bs{B}_3}$
by using the second kind of reduction contributions present in \eqref{ecriture equation saut avec reduction gauche mod},

 \item we reduce $\overleftarrow{\bs{A}_2}$ starting from the last entry and so on, with the first entry, and so on, of $\bs{B}_2$
by using the first kind of reduction contributions present in \eqref{ecriture equation saut general bv mod},

\item we trade the $+$ boundary value for $\overleftarrow{\bs{A}_1} + \i \pi  \ov{\bs{e}}_{|A_1|}$ for the $-$ boundary value by means of the first contribution in \eqref{ecriture equation saut avec reduction gauche mod}.

\end{itemize}
All of this leads to
\bem
\widehat{\mc{F}}_{ +, 0}^{(\op{O})}\Big( \, \bs{A}_3\cup  \overleftarrow{\bs{A}_1} \cup  \overleftarrow{\bs{A}_2}  + \i \pi  \ov{\bs{e}} ,
        \bs{B}_2\cup  \bs{B}_1 \cup  \overleftarrow{ \bs{B}_3}   \Big)  \\
\quad \hookrightarrow \quad
\widehat{\mc{F}}_{ -, 0}^{(\op{O})}\Big(  \,  \overleftarrow{\bs{A}_1}    + \i \pi  \ov{\bs{e}}, \bs{B}_1   \Big) \cdot
\pl{\alpha \in \bs{A}_3}{}\Big( - \ex{2\i\pi \psi(\alpha) } \Big)  \De\big(   \bs{A}_2 \cup  \bs{A}_3 \mid  \bs{B}_2 \cup  \bs{B}_3  \big) \;.
\end{multline}
Thus, under the constraints $|A_2| \, = \, |B_2|$ and $|A_3| \, = \, |B_3|$,
\bem
\mc{F}_{ +, 0}^{(\op{O})}\Big( \, \overleftarrow{\bs{A}}  + \i \pi  \ov{\bs{e}},   \bs{B}    \Big) \; = \; \sul{ A =   \cup\, _{a=1}^{3}  A_a}{}   \sul{   B =   \underset{23}{\cup} \, _{a=1}^{3} B_a  }{}
\op{S}\Big( \, \overleftarrow{\bs{A}}  \mid \bs{A}_3\cup  \overleftarrow{\bs{A}_1} \cup  \overleftarrow{\bs{A}_2} \Big)
\cdot \op{S}\Big( \, \bs{B} \mid \bs{B}_2\cup  \bs{B}_1 \cup  \overleftarrow{ \bs{B}_3} \Big)  \cdot \pl{\a \in A_3}{} \Big\{ - \ex{2\i\pi \psi(\a) } \Big\} \\
\times  \De\big(   \bs{A}_2 \cup  \bs{A}_3 \mid  \bs{B}_2 \cup  \bs{B}_3  \big) \cdot
\mc{F}^{(\op{O})}_{  -, 0 }\Big( \, \overleftarrow{\bs{A}_1}  + \i \pi  \ov{\bs{e}} ,   \bs{B}_1    \Big) \;.
\end{multline}
The above then entails the claim upon specialising to the case of a constant function $\psi = \om_{\op{O}}$. \qed

\begin{prop}

The expansions \eqref{ecriture representation combinatoire pour MO via reduction de type 1} and \eqref{ecriture representation combinatoire pour MO via reduction de type 2}
are compatible with each other, \textit{viz}. $\mc{M}^{(\op{O})}_{n;m} \; = \;\wt{\mc{M}}^{(\op{O})}_{n;m}$.

\end{prop}

\Proof

The strategy of the proof consists in starting from the combinatorial sum given in \eqref{ecriture representation combinatoire pour MO via reduction de type 2} and then
using Lemma \ref{Lemme relation entre valeurs plus et moins des FF} so as to trade the $+$ boundary value appearing there for a $-$ boundary value and then resum the resulting expression into
\eqref{ecriture representation combinatoire pour MO via reduction de type 1}. By virtue of axiom $\bf{II)}$ and the permutation properties of the form factors stated in axiom $\bf{I)}$,
one may recast \eqref{ecriture representation combinatoire pour MO via reduction de type 2} in the form
\bem
\wt{\mc{M}}^{\,(\op{O})}_{n;m}\big(  \bs{\alpha}_n ; \bs{\be}_{m} \big) \; =      \hspace{-2mm}  \sul{ A= A_1 \cup A_2  }{}  \sul{ B = B_1 \underset{1}{\cup} B_2  }{}
 \hspace{-2mm}  \De\big( \bs{A}_1 \mid \bs{B}_1  \big) \cdot \op{S}\Big( \, \overleftarrow{\bs{A}}  \mid \bs{A}_1\cup   \overleftarrow{\bs{A}_2}  \Big)
\cdot \op{S}\Big( \, \bs{B} \mid \bs{B}_2\cup  \overleftarrow{\bs{B}_1} \Big) \\
\times \ex{-2\i\pi \om_{\op{O}} |A_2| } \cdot  \mc{F}_{ +, 0}^{(\op{O})} \Big(   \overleftarrow{\bs{A}_2} + \i \pi \ov{\bs{e}},  \bs{B}_2  \Big) \;.
\end{multline}
Above, we made use of
\beq
 \op{S}\Big( \, \overleftarrow{\bs{A}}  \mid \bs{A}_1\cup  \bs{A}_2 \Big) \, \op{S}\Big(\, \bs{A}_2 \mid \overleftarrow{\bs{A}_2}  \Big)
          \, = \, \op{S}\Big( \, \overleftarrow{\bs{A}}  \mid \bs{A}_1\cup   \overleftarrow{\bs{A}_2}  \Big) \;.
\enq
Thus, inserting the expansion obtained in Lemma \ref{Lemme relation entre valeurs plus et moins des FF}, one gets
\bem
\wt{\mc{M}}^{\,(\op{O})}_{n;m}\big(  \bs{\alpha}_n ; \bs{\be}_{m} \big) \; =      \hspace{-2mm}  \sul{ A=   \cup_{a=1}^{4} A_a  }{}  \sul{ B = \underset{134}{\cup} \, _{a=1}^{4} B_a  }{}
 \hspace{-2mm}  \De\big( \bs{A}_1 \cup \bs{A}_3 \cup \bs{A}_4 \mid \bs{B}_1 \cup \bs{B}_3 \cup \bs{B}_4  \big)
  \cdot \Big( - \ex{2\i\pi \om_{\op{O}}} \Big)^{|A_4|}  \, \ex{-2\i\pi \om_{\op{O}} |A_2\cup A_3 \cup A_4| }  \\
\times \op{S}\Big( \, \overleftarrow{\bs{A}}  \mid \bs{A}_1\cup   \overleftarrow{ \bs{A}_2\cup \bs{A}_3\cup \bs{A}_4}  \Big)
\cdot \op{S}\Big( \, \bs{B} \mid \overrightarrow{\bs{B}_2 \cup \bs{B}_3\cup \bs{B}_4 }\cup  \overleftarrow{\bs{B}_1} \Big)
\cdot \op{S}\Big( \,     \overleftarrow{ \bs{A}_2 \cup \bs{A}_3 \cup \bs{A}_4 } \mid  \bs{A}_4 \cup \overleftarrow{\bs{A}_2 } \cup \overleftarrow{ \bs{A}_3 } \Big)  \\
\times \op{S}\Big( \,  \overrightarrow{ \bs{B}_2 \cup \bs{B}_3 \cup \bs{B}_4 }\mid  \bs{B}_3 \cup \bs{B}_2 \cup \overleftarrow{ \bs{B}_4 }  \Big)
\cdot
\mc{F}_{ -, 0}^{(\op{O})} \Big(   \overleftarrow{\bs{A}_2} + \i \pi \ov{\bs{e}},  \bs{B}_2  \Big) \;.
\end{multline}
Above, the summation over partitions is constrained to ensembles satisfying
\beq
|A_1|=|B_1|\, , \quad |A_3|=|B_3|\, , \quad  |A_4|=|B_4| \; .
\enq
Upon reducing the products of $\op{S}$-matrices, one ends up with the sum
\bem
\wt{\mc{M}}^{\,(\op{O})}_{n;m}\big(  \bs{\alpha}_n ; \bs{\be}_{m} \big) \; =      \hspace{-2mm}  \sul{ A=   \cup_{a=1}^{4} A_a  }{}  \sul{ B = \underset{134}{\cup} \, _{a=1}^{4} B_a  }{}
 \hspace{-2mm}  \De\big( \bs{A}_1 \cup \bs{A}_3 \cup \bs{A}_4 \mid \bs{B}_1 \cup \bs{B}_3 \cup \bs{B}_4  \big)
  \cdot \big( - 1 \big)^{|A_4|} \,  \ex{-2\i\pi \om_{\op{O}} |A_2\cup A_3 | }  \\
\times \op{S}\Big( \, \overleftarrow{\bs{A}}  \mid \bs{A}_1\cup    \bs{A}_4 \cup \overleftarrow{ \bs{A}_2 } \cup \overleftarrow{ \bs{A}_3 }  \Big)
\cdot \op{S}\Big( \, \bs{B} \mid \bs{B}_3 \cup \bs{B}_2 \cup \overleftarrow{ \bs{B}_4 }  \cup  \overleftarrow{\bs{B}_1} \Big)
\cdot
\mc{F}_{-, 0}^{(\op{O})} \Big(   \overleftarrow{\bs{A}_2} + \i \pi \ov{\bs{e}},  \bs{B}_2  \Big) \;.
\label{ecriture formule intermediare pour tilde MO apres reduction droite}
\end{multline}
At this stage one observes the identity
\bem
 \op{S}\Big( \, \overleftarrow{ \bs{B}_1 \cup \bs{B}_4 }   \mid   \overleftarrow{ \bs{B}_4 }  \cup  \overleftarrow{\bs{B}_1} \Big) \cdot
 \De\big( \bs{A}_1 \cup  \bs{A}_4 \mid \bs{B}_1   \cup \bs{B}_4  \big) \cdot
 \op{S}\Big( \, \overrightarrow{ \bs{A}_1 \cup \bs{A}_4 }   \mid   \bs{A}_1  \cup  \bs{A}_4 \Big)   \\
\; = \;  \De\big( \bs{A}_1 \cup  \bs{A}_4 \mid \bs{B}_1   \cup \bs{B}_4  \big) \cdot
\op{S}\Big( \, \bs{A}_1   \cup  \bs{A}_4    \mid  \overrightarrow{ \bs{A}_1 \cup \bs{A}_4 } \Big) \cdot
 \op{S}\Big( \,   \overrightarrow{ \bs{A}_1 \cup \bs{A}_4} \mid   \bs{A}_1   \cup  \bs{A}_4 \Big) \; = \;
 \De\big( \bs{A}_1 \cup  \bs{A}_4 \mid \bs{B}_1   \cup \bs{B}_4  \big) \;,
\nonumber
\end{multline}
where we made use of \eqref{ecriture identite inversion coordonnee dans produit S permutationnel}.
The latter allows one to replace in \eqref{ecriture formule intermediare pour tilde MO apres reduction droite} some of the concatenated vectors by the fully ordered ones:
\beq
 \overleftarrow{ \bs{B}_4 }  \cup  \overleftarrow{\bs{B}_1}  \quad \hookrightarrow \quad  \overleftarrow{ \bs{B}_1 \cup \bs{B}_4 } \qquad \e{and} \qquad
 \bs{A}_1   \cup  \bs{A}_4  \quad \hookrightarrow \quad  \overrightarrow{ \bs{A}_1 \cup \bs{A}_4 }  \;,
\enq
leading to
\bem
\wt{\mc{M}}^{\,(\op{O})}_{n;m}\big(  \bs{\alpha}_n ; \bs{\be}_{m} \big) \; =      \hspace{-2mm}  \sul{ A=   \cup_{a=1}^{4} A_a  }{}  \sul{ B = \underset{134}{\cup} \, _{a=1}^{4} B_a  }{}
 \hspace{-2mm}  \De\big( \bs{A}_1 \cup \bs{A}_3 \cup \bs{A}_4 \mid \bs{B}_1 \cup \bs{B}_3 \cup \bs{B}_4  \big)
  \cdot \big( - 1 \big)^{|A_4|} \,  \ex{-2\i\pi \om_{\op{O}} |A_2\cup A_3 | }  \\
\times \op{S}\Big( \, \overleftarrow{\bs{A}}  \mid \overrightarrow{ \bs{A}_1 \cup \bs{A}_4 }  \cup \overleftarrow{ \bs{A}_2 } \cup \overleftarrow{ \bs{A}_3 }  \Big)
\cdot \op{S}\Big( \, \bs{B} \mid \bs{B}_3 \cup \bs{B}_2 \cup \overleftarrow{ \bs{B}_1 \cup \bs{B}_4}  \Big)
\cdot
\mc{F}_{ -, 0}^{(\op{O})} \Big(   \overleftarrow{\bs{A}_2} + \i \pi \ov{\bs{e}},  \bs{B}_2  \Big) \;.
\end{multline}
This then allows one, taken that the partitions over the $B_a$s, $a=1,3,4$ are not ordered, to change the variables in the partitioning to
\beq
A_1^{\prime}\, = \, A_1\cup A_4\;, \quad  A_2^{\prime}\, = \, A_2 \;,\quad  A_3^{\prime}\, = \, A_3
\quad \e{and} \quad
B_1^{\prime}\, = \, B_1\cup B_4\;, \quad  B_2^{\prime}\, = \, B_2 \;,\quad  B_3^{\prime}\, = \, B_3 \;,
\enq
what eventually leads to
\bem
\wt{\mc{M}}^{\,(\op{O})}_{n;m}\big(  \bs{\alpha}_n ; \bs{\be}_{m} \big) \; =      \hspace{-2mm}  \sul{ A=   \cup_{a=1}^{3} A_a  }{}  \sul{ B = \underset{13}{\cup} \, _{a=1}^{3} B_a  }{}
 \hspace{-2mm}  \De\big( \bs{A}_1 \cup \bs{A}_3   \mid \bs{B}_1 \cup \bs{B}_3   \big)
 \cdot \ex{-2\i\pi \om_{\op{O}} |A_2\cup A_3 | } \cdot \op{S}\Big( \, \overleftarrow{\bs{A}}  \mid \bs{A}_1  \cup \overleftarrow{ \bs{A}_2 } \cup \overleftarrow{ \bs{A}_3 }  \Big)  \\
\times
 \op{S}\Big( \, \bs{B} \mid \bs{B}_3 \cup \bs{B}_2 \cup \overleftarrow{\bs{B}_1}  \Big)
\cdot
\mc{F}_{-, 0}^{(\op{O})} \Big(   \overleftarrow{\bs{A}_2} + \i \pi \ov{\bs{e}},  \bs{B}_2  \Big)
\cdot \sul{ A^{\prime} \subset A_1 }{}    \big( - 1 \big)^{|A^{\prime}|}   \;.
\end{multline}
Then, ones observes that
\beq
 \sul{ A^{\prime} \subset A_1 }{}    \big( - 1 \big)^{|A^{\prime}|} \; = \; \left\{ \ba{ccc} 1 & \e{if} & A_1=\emptyset  \\
                                                                                              0 & \e{if} & A_1 \not=\emptyset \ea \right. \;.
\enq
This reduces the combinatorial expression to
\bem
\wt{\mc{M}}^{\,(\op{O})}_{n;m}\big(  \bs{\alpha}_n ; \bs{\be}_{m} \big) \; =     \ex{-2\i\pi \om_{\op{O}} |A | }  \hspace{-2mm}  \sul{ A=   A_2\cup A_3  }{}  \sul{ B = B_2 \underset{3}{\cup} B_3  }{}
 \hspace{-2mm}  \De\big(  \bs{A}_3   \mid  \bs{B}_3   \big)
 \cdot \op{S}\Big( \, \overleftarrow{\bs{A}}  \mid   \overleftarrow{ \bs{A}_2 } \cup \overleftarrow{ \bs{A}_3 }  \Big)  \\
\times
 \op{S}\big(  \bs{B} \mid \bs{B}_3 \cup \bs{B}_2    \big)
\cdot
\mc{F}_{ -, 0}^{(\op{O})} \Big(   \overleftarrow{\bs{A}_2} + \i \pi \ov{\bs{e}},  \bs{B}_2  \Big) \;.
\end{multline}
Upon setting $A_3\hookrightarrow A_1$ and $B_3\hookrightarrow B_1$, one exactly recovers \eqref{ecriture representation combinatoire pour MO via reduction de type 1}. \qed

To close this section, we provide a formula that represents $\mc{M}^{\,(\op{O})}_{n;m}\big(  \bs{\alpha}_n ; \bs{\be}_{m} \big)$ as a mixture of $+$ and $-$
boundary values of linear combinations of form factors.

\begin{lemme}
\label{Lemme ecriture generale densite FF}

 Given $\bs{\alpha}_n \in \R^{n}$ and $\bs{\be}_{m} \in \R^m$, let $A=\{\a_{a}\}_1^{n}$ and consider $A=A_1\cup A_2$ some partition of $A$.
 Further, let $B=\{\be_a\}_{1}^{m}$. Then, one has the expansion
\bem
\mc{M}^{(\op{O})}_{n;m}\big(  \bs{\alpha}_n ; \bs{\be}_{m} \big) \; =    \ex{ - 2 \i\pi \om_{\op{O}} |A_1| } \hspace{-2mm}  \sul{   A_1 = C_1 \cup C_2  }{}
\sul{   A_2 = D_1 \cup D_2  }{} \sul{ B =  \underset{13}{\cup}\, _{a=1}^{3} B_a  }{}
 \hspace{-2mm}  \De\big( \bs{C}_1\cup \bs{D}_1 \mid \bs{B}_1 \cup \bs{B}_3 \big)
\cdot \op{S}\Big( \, \overleftarrow{\bs{A}}  \mid \overleftarrow{\bs{A}_2}\cup  \overleftarrow{\bs{A}_1} \Big)  \\
 \times \op{S}\Big( \, \overleftarrow{\bs{A}_1}  \mid \overleftarrow{ \bs{C}_2}\cup  \overleftarrow{ \bs{C}_1} \Big)
\cdot \op{S}\Big( \, \overleftarrow{\bs{A}_2}  \mid \overleftarrow{ \bs{D}_1 }\cup  \overleftarrow{ \bs{D}_2 } \Big)
\cdot \op{S}\big(   \bs{B} \mid \bs{B}_1 \cup \bs{B}_2\cup  \bs{B}_3 \big) \\
\times  \mc{F}_{ -, 0, +}^{(\op{O})} \Big( \,  \overleftarrow{ \bs{C}_2} + \i \pi \ov{\bs{e}} ,  \bs{B}_2, \overleftarrow{ \bs{D}_2} - \i \pi \ov{\bs{e}}  \Big) \;.
\label{ecriture representation combinatoire pour MO via reduction mixte entre 1 et 2}
\end{multline}
The $-,0,+$ subscript means that the first set of variables is to be understood as a $-$ boundary value, the central set of variables simply takes real values
while the third set of variables ought to be taken in the sense of a $+$ boundary value.
The summation runs through all ordered partitions $C_1 \cup C_2$ of $A_1$, $D_1 \cup D_2$ of $A_2$, and all partitions $B_1 \cup B_2 \cup B_3$  of $B$ such that elements of $B_1$ and $B_3$ appear in any order while those of
$B_2$ are ordered. The summation over the various partitions is constrained  by the condition
\beq
|C_1| \; = \; |B_1| \qquad and \qquad  |D_1| \; = \; |B_3| \;.
\enq
Finally, $\De\big( * \mid *  \big)$ is as defined in \eqref{definition facteur masse dirac globale}.

\end{lemme}

\Proof

Focusing on a given partition $A_1 = C_1 \cup C_2$, $A_2 = D_1 \cup D_2$ and $B = B_1 \cup B_2 \cup B_3$ as given in the summation in \eqref{ecriture representation combinatoire pour MO via reduction mixte entre 1 et 2},
one has that
\bem
\mc{M}^{(\op{O})}_{n;m}\Big(\,   \bs{A} ; \bs{B} \Big) \; = \; \op{S}\Big( \, \overleftarrow{\bs{A}}  \mid \overleftarrow{\bs{A}_2}\cup  \overleftarrow{\bs{A}_1} \Big)
\cdot \op{S}\Big( \, \overleftarrow{\bs{A}_1}  \mid \overleftarrow{ \bs{C}_2 }\cup  \overleftarrow{ \bs{C}_1 } \Big)
\cdot \op{S}\Big( \, \overleftarrow{\bs{A}_2}  \mid \overleftarrow{ \bs{D}_1 }\cup  \overleftarrow{ \bs{D}_2 } \Big)
\cdot \op{S}\big(   \bs{B} \mid \bs{B}_1 \cup \bs{B}_2\cup  \bs{B}_3 \big)  \\
\mc{M}^{(\op{O})}_{n;m}\big(     \bs{C}_1 \cup  \bs{C}_2  \cup \bs{D}_2 \cup  \bs{D}_1   ; \bs{B}_1 \cup \bs{B}_2\cup  \bs{B}_3  \big) \; .
\end{multline}
At this stage, one reduces the first variable in $\bs{C}_1$ with the first one occurring in $\bs{B}_1$ by means of the reduction contributions occurring in the recursive formula
provided by axiom $\bf{V)}$, and so on until having all of $\bs{C}_1$ and $\bs{B}_1$ disappears. Next, one reduces the last variable in $\bs{D}_1$ with the last on occurring in
$\bs{B}_3$ by means of the reduction contributions occurring in the recursive formula \eqref{ecriture formule recursive de type 2 equivalente a l originale}, and so on until   all of
$\bs{D}_1$ and $\bs{B}_3$ disappears. Then, one moves, starting from the first entry,  $\bs{C}_2$ into the "$B$" variables by means of the shifted concatenation
occurring in axiom $\bf{V)}$. Finally, one moves, starting from the last entry,  $\bs{D}_2$ into the "$B$" variables by means of the shifted concatenation
occurring in  \eqref{ecriture formule recursive de type 2 equivalente a l originale}. This yields the reduction
\bem
\mc{M}^{(\op{O})}_{n;m}\big(    \bs{C}_1 \cup  \bs{C}_2  \cup \bs{D}_2 \cup  \bs{D}_1     ; \bs{B}_1 \cup \bs{B}_2\cup  \bs{B}_3  \big)
 \\
 \qquad \hookrightarrow \qquad  \ex{ - 2 \i\pi \om_{\op{O}} |A_1| }  \cdot \De\big( \bs{C}_1\cup \bs{D}_1 \mid \bs{B}_1 \cup \bs{B}_3 \big)
\cdot \mc{F}_{-, 0, +}^{(\op{O})} \Big(  \, \overleftarrow{ \bs{C}_2 } + \i \pi \ov{\bs{e}} ,  \bs{B}_2, \overleftarrow{ \bs{D}_2 } - \i \pi \ov{\bs{e}}  \Big) \;.
\end{multline}
This entails the claim. \qed

We now establish explicitly the equivalence of \eqref{ecriture representation combinatoire pour MO via reduction mixte entre 1 et 2} with
\eqref{ecriture representation combinatoire pour MO via reduction de type 1}. First, we start with an auxiliary lemma

\begin{lemme}

It holds
\bem
 \mc{F}_{ +, -, 0}^{(\op{O})} \Big( \,  \overleftarrow{ \bs{D} } + \i \pi \ov{\bs{e}} ,  \overleftarrow{  \bs{C} } + \i \pi \ov{\bs{e}} ,  \bs{B} \Big)
 \; = \hspace{-3mm}
\sul{ D =   \cup\, _{a=1}^{3}  D_a}{}   \sul{   B =   \underset{23}{\cup} \, _{a=1}^{3} B_a  }{}  \hspace{-3mm}
\op{S}\Big( \, \overleftarrow{ \bs{D} } \cup   \overleftarrow{ \bs{C} }  \mid \bs{D}_3\cup  \overleftarrow{ \bs{D}_1 } \cup  \overleftarrow{ \bs{C} } \cup  \overleftarrow{\bs{D}_2} \Big)  \\
\times \op{S}\big(   \bs{B} \mid \bs{B}_2\cup  \bs{B}_1 \cup  \overleftarrow{\bs{B}_3} \big)
\cdot \Big( - \ex{2\i\pi \om_{\op{O}}} \Big)^{|D_3|}  \cdot \De\big(    \bs{D}_2 \cup  \bs{D}_3 \mid  \bs{B}_2 \cup  \bs{B}_3  \big)
\cdot \mc{F}^{(\op{O})}_{  -, -, 0 }\Big( \, \overleftarrow{ \bs{D}_1 }  + \i \pi  \ov{\bs{e}}   ,  \overleftarrow{ \bs{C} } + \i \pi \ov{\bs{e}} ,  \bs{B}_1    \Big) \;.
\end{multline}

\end{lemme}

\Proof

First, observe that given a partition $D=D_1\cup D_2 \cup D_3$ and $B = B_1\cup B_2 \cup B_3$, one may reorder the entries of $\mc{F}^{(\op{O})}$
in the form
\bem
 \mc{F}_{+, -, 0}^{(\op{O})} \Big( \,  \overleftarrow{ \bs{D} } + \i \pi \ov{\bs{e}} ,  \overleftarrow{ \bs{C} } + \i \pi \ov{\bs{e}} ,  \bs{B} \Big) \; =  \;
\op{S}\big(  \bs{B} \mid \bs{B}_2\cup  \bs{B}_1 \cup  \overleftarrow{\bs{B}_3} \big)
 \cdot \op{S}\Big( \, \overleftarrow{ \bs{D} } \cup   \overleftarrow{ \bs{C} }  \mid \bs{D}_3\cup  \overleftarrow{ \bs{D}_1 } \cup  \overleftarrow{ \bs{C} } \cup  \overleftarrow{ \bs{D}_2 } \Big)  \\
\times
\mc{F}^{(\op{O})}_{  +, -, +, 0 }\Big( \bs{D}_3  \cup \overleftarrow{ \bs{D}_1 }  + \i \pi  \ov{\bs{e}} ,
 \overleftarrow{ \bs{C} } + \i \pi \ov{\bs{e}} ,\overleftarrow{ \bs{D}_2}  + \i \pi  \ov{\bs{e}} , \bs{B}_2  \cup \bs{B}_1   \cup \overleftarrow{ \bs{B}_3}  \Big) \;.
\end{multline}
This then allows to
\begin{itemize}

 \item reduce $\bs{D}_3$, starting from the first entry and so on, with the last entry, and so on, of $\overleftarrow{ \bs{B}_3}$
by using the second reduction contributions present in \eqref{ecriture equation saut avec reduction gauche},

 \item reduce $\overleftarrow{ \bs{D}_2 }$, starting from the last entry and so on, with the first entry, and so on, of $\bs{B}_2$
by using the first reduction  contributions present in \eqref{ecriture equation saut general bv},

 \item change the boundary value in $\overleftarrow{ \bs{D}_1 } + \i \pi  \ov{\bs{e}}_{|D_1|} $ by taking into account the first term in \eqref{ecriture equation saut avec reduction gauche}.

\end{itemize}
These reductions generate a factor
\beq
 \Big( - \ex{2\i\pi \om_{\op{O}}} \Big)^{|D_3|} \cdot  \De\big(  \bs{D}_2  \cup  \bs{D}_3 \mid  \bs{B}_2 \cup  \bs{B}_3  \big)
\mc{F}^{(\op{O})}_{   -, -, 0 }\Big(  \overleftarrow{ \bs{D}_1 }  + \i \pi  \ov{\bs{e}} ,
 \overleftarrow{ \bs{C} } + \i \pi \ov{\bs{e}}  ,   \bs{B}_1    \Big)  \, ,
\enq
what entails the claim. \qed

\begin{prop}

The expansions \eqref{ecriture representation combinatoire pour MO via reduction mixte entre 1 et 2} and \eqref{ecriture representation combinatoire pour MO via reduction de type 1}
coincide.

\end{prop}

\Proof

Axiom $\bs{II)}$ allows one to transform
\beq
\mc{F}_{ -, 0, +}^{(\op{O})} \Big( \,  \overleftarrow{ \bs{C}_2 } + \i \pi \ov{\bs{e}} ,  \bs{B}_2, \overleftarrow{ \bs{D}_2} - \i \pi \ov{\bs{e}}  \Big)
\; = \; \mc{F}_{ +, -, 0}^{(\op{O})} \Big( \,  \overleftarrow{ \bs{D}_2 } + \i \pi \ov{\bs{e}} ,  \overleftarrow{ \bs{C}_2 } + \i \pi \ov{\bs{e}} ,  \bs{B}_2 \Big) \cdot \ex{-2\i\pi \om_{\op{O}}|D_2|}\;.
\enq
Hence, implementing the latter transform at the level of \eqref{ecriture representation combinatoire pour MO via reduction mixte entre 1 et 2} leads to
\bem
\mc{M}^{(\op{O})}_{n;m}\big(  \bs{\alpha}_n ; \bs{\be}_{m} \big) \; =    \hspace{-2mm}  \sul{   A_1 = C_1 \cup C_2  }{}
\sul{   A_2 =  \cup\, _{a=1}^{4} D_a  }{} \sul{ B =  \underset{1345}{\cup}\, _{a=1}^{5} B_a  }{}
 \hspace{-2mm}  \De\big(  \bs{C}_1  \cup \bs{D}_1 \mid \bs{B}_1 \cup \bs{B}_5 \big)
\cdot \op{S}\Big( \, \overleftarrow{\bs{A}}  \mid \overleftarrow{\bs{A}_2}\cup  \overleftarrow{\bs{A}_1} \Big)  \\
 \times \op{S}\Big( \, \overleftarrow{\bs{A}_1}  \mid \overleftarrow{ \bs{C}_2 }\cup  \overleftarrow{ \bs{C}_1 } \Big)
\cdot \op{S}\Big( \, \overleftarrow{\bs{A}_2}  \mid \overleftarrow{ \bs{D}_1 }\cup  \overleftarrow{ \bs{D}_2 \cup \bs{D}_3 \cup \bs{D}_4} \Big)
\cdot \op{S}\Big( \, \bs{B} \mid \bs{B}_1 \cup \overrightarrow{ \bs{B}_2 \cup \bs{B}_3 \cup  \bs{B}_4}\cup  \bs{B}_5 \Big) \\
\times \op{S}\Big( \,   \overrightarrow{ \bs{B}_2\cup \bs{B}_3 \cup \bs{B}_4 }  \mid  \bs{B}_3 \cup \bs{B}_2\cup  \overleftarrow{ \bs{B}_4 }  \Big)
\cdot \op{S}\Big( \,  \overleftarrow{ \bs{D}_2 \cup \bs{D}_3 \cup \bs{D}_4 }  \cup   \overleftarrow{ \bs{C}_2 }  \mid  \bs{D}_4\cup  \overleftarrow{ \bs{D}_2 } \cup  \overleftarrow{ \bs{C}_2 }  \cup
\overleftarrow{ \bs{D}_3 }  \Big) \\
\times  \ex{ - 2 \i\pi \om_{\op{O}} ( |A_1| + |D_2\cup D_3 \cup D_4|) } \cdot \Big( - \ex{ 2 \i\pi \om_{\op{O}} } \Big)^{|D_4|}
\cdot  \De\big( \bs{D}_3\cup \bs{D}_4 \mid \bs{B}_3 \cup \bs{B}_4 \big) \\
\times \mc{F}_{ -, -, 0}^{(\op{O})} \Big( \,  \overleftarrow{ \bs{D}_2 } + \i \pi \ov{\bs{e}} ,  \overleftarrow{ \bs{C}_2 } + \i \pi \ov{\bs{e}} ,  \bs{B}_2 \Big) \;.
\end{multline}
There, the summations run under the constraints $|C_1|=|B_1|$, $|D_1|=|B_5|$, $|D_3|=|B_3|$ and $|D_4|=|B_4|$. Simplifying the products over the various $\op{S}$-matrices,
one gets that
\bem
\mc{M}^{(\op{O})}_{n;m}\big(  \bs{\alpha}_n ; \bs{\be}_{m} \big) \; =     \hspace{-2mm}  \sul{   A_1 = C_1 \cup C_2  }{}
\sul{   A_2 =  \cup\, _{a=1}^{4} D_a  }{} \sul{ B =  \underset{1345}{\cup}\, _{a=1}^{5} B_a  }{}
 \hspace{-2mm}
\De\big( \bs{C}_1\cup \bs{D}_1 \cup \bs{D}_3\cup \bs{D}_4 \mid
\bs{B}_1 \cup \bs{B}_5 \cup \bs{B}_3 \cup \bs{B}_4 \big) \\
\times  \op{S}\Big( \, \overleftarrow{\bs{A}}  \mid \overleftarrow{\bs{A}_2}\cup  \overleftarrow{\bs{A}_1} \Big)  \cdot  \op{S}\Big( \, \overleftarrow{\bs{A}_1}
                          \mid \overleftarrow{ \bs{C}_2 }\cup  \overleftarrow{ \bs{C}_1 } \Big)
\cdot \op{S}\Big( \, \overleftarrow{\bs{A}_2}  \cup   \overleftarrow{ \bs{C}_2 }
                                                    \mid \overleftarrow{ \bs{D}_1 }\cup  \bs{D}_4\cup  \overleftarrow{\bs{D}_2}\cup  \overleftarrow{ \bs{C}_2}  \cup   \overleftarrow{ \bs{D}_3}  \Big) \\
\times \op{S}\Big( \, \bs{B} \mid \bs{B}_1 \cup  \bs{B}_3 \cup \bs{B}_2\cup  \overleftarrow{ \bs{B}_4 }  \cup   \bs{B}_5  \Big)
\cdot  \ex{ - 2 \i\pi \om_{\op{O}} ( |A_1| + |D_2\cup D_3| ) } \cdot \big( - 1 \big)^{|D_4|} \\
\times  \mc{F}_{-, -, 0}^{(\op{O})} \Big( \,  \overleftarrow{ \bs{D}_2 } + \i \pi \ov{\bs{e}} ,  \overleftarrow{\bs{C}_2} + \i \pi \ov{\bs{e}} ,  \bs{B}_2 \Big)  \;.
\end{multline}
One now observes the identity
\bem
\op{S}\Big(\,   \overleftarrow{ \bs{D}_1 }\cup  \bs{D}_4\mid  \overleftarrow{ \bs{D}_1 \cup \bs{D}_4}   \Big) \cdot
\De\big( \bs{D}_1 \cup \bs{D}_4 \mid \bs{B}_5  \cup \bs{B}_4 \big) \cdot
\op{S}\Big(\,   \overleftarrow{ \bs{B}_4}\cup  \bs{B}_5\mid  \overrightarrow{ \bs{B}_4 \cup \bs{B}_5}   \Big) \\  \; = \;
\op{S}\Big(\,  \overrightarrow{ \bs{B}_4 \cup \bs{B}_5} \mid \overleftarrow{ \bs{B}_4}\cup  \bs{B}_5  \Big) \cdot
\De\big( \bs{D}_1 \cup \bs{D}_4 \mid \bs{B}_5  \cup \bs{B}_4 \big) \cdot
\op{S}\Big( \,  \overleftarrow{ \bs{B}_4}\cup  \bs{B}_5\mid  \overrightarrow{ \bs{B}_4 \cup \bs{B}_5 }   \Big) \; = \;
\De\big( \bs{D}_1 \cup \bs{D}_4 \mid \bs{B}_5  \cup \bs{B}_4 \big) \;.
\end{multline}
The latter allows one to carry out the replacement $ \overleftarrow{ \bs{D}_1 }\cup  \bs{D}_4 \hookrightarrow  \overleftarrow{ \bs{D}_1 \cup \bs{D}_4}  $
and $\overleftarrow{ \bs{B}_4}\cup  \bs{B}_5\hookrightarrow  \overrightarrow{ \bs{B}_4 \cup \bs{B}_5}$ in the above formula. In its turn,
this allows one to implement a change of summation set $D^{\prime}_1\, = \, D_1\cup D_4$ and $B^{\prime}_4 \, = \, B_4\cup B_5$, what recast the expansion in the form
\bem
\mc{M}^{(\op{O})}_{n;m}\big(  \bs{\alpha}_n ; \bs{\be}_{m} \big) \; =     \hspace{-2mm}  \sul{   A_1 = C_1 \cup C_2  }{}
\sul{   A_2 =  \cup\, _{a=1}^{3} D_a  }{} \sul{ B =  \underset{134}{\cup}\, _{a=1}^{4} B_a  }{}
 \hspace{-2mm}
\De\big( \bs{C}_1\cup \bs{D}_1 \cup \bs{D}_3  \mid
\bs{B}_1   \cup \bs{B}_4 \cup \bs{B}_3 \big) \\
\times  \op{S}\Big( \, \overleftarrow{\bs{A}}  \mid \overleftarrow{\bs{A}_2}\cup  \overleftarrow{\bs{A}_1} \Big)
\cdot  \op{S}\Big( \, \overleftarrow{\bs{A}_1}  \mid \overleftarrow{ \bs{C}_2 }\cup  \overleftarrow{ \bs{C}_1} \Big)
\cdot \op{S}\Big( \, \overleftarrow{\bs{A}_2}  \cup   \overleftarrow{ \bs{C}_2 }  \mid \overleftarrow{ \bs{D}_1 } \cup  \overleftarrow{ \bs{D}_2 }\cup  \overleftarrow{ \bs{C}_2 }  \cup   \overleftarrow{ \bs{D}_3}  \Big) \\
\times \op{S}\big(   \bs{B} \mid \bs{B}_1 \cup  \bs{B}_3 \cup \bs{B}_2\cup  \bs{B}_4    \big)
\cdot  \ex{ - 2 \i\pi \om_{\op{O}} ( |C_1\cup C_2| + |D_2\cup D_3| ) }    \\
\times  \mc{F}_{ -, -, 0}^{(\op{O})} \Big( \,  \overleftarrow{ \bs{D}_2 } + \i \pi \ov{\bs{e}} ,  \overleftarrow{ \bs{C}_2 } + \i \pi \ov{\bs{e}} ,  \bs{B}_2 \Big)
 \cdot \sul{ D_4 \subset D_1  }{} \big( - 1 \big)^{|D_4|}  \;.
\end{multline}
Above, the summation is constrained to $|D_1|=|B_4|$, $|C_1|=|B_1|$ and $|D_3|=|B_3|$. The last sum vanishes unless $D_1=\emptyset$, what thus also imposes that $B_4=\emptyset$.
Thus,
\bem
\mc{M}^{(\op{O})}_{n;m}\big(  \bs{\alpha}_n ; \bs{\be}_{m} \big) \; =     \hspace{-2mm}  \sul{   A_1 = C_1 \cup C_2  }{}
\sul{   A_2 =  D_2\cup D_3  }{} \sul{ B =  \underset{13}{\cup}\, _{a=1}^{3} B_a  }{}
 \hspace{-2mm}
\De\big( \bs{C}_1\cup \bs{D}_3  \mid \bs{B}_1   \cup \bs{B}_3  \big) \\
\times  \op{S}\Big( \, \overleftarrow{\bs{A}}  \mid \overleftarrow{\bs{A}_2}\cup  \overleftarrow{\bs{A}_1} \Big)
\cdot  \op{S}\Big( \, \overleftarrow{\bs{A}_1}  \mid \overleftarrow{ \bs{C}_2}\cup  \overleftarrow{ \bs{C}_1} \Big)
\cdot \op{S}\Big( \, \overleftarrow{\bs{A}_2}  \cup   \overleftarrow{ \bs{C}_2}  \mid   \overleftarrow{ \bs{D}_2 }\cup  \overleftarrow{ \bs{C}_2 }  \cup   \overleftarrow{ \bs{D}_3 }  \Big)
\cdot \op{S}\big(   \bs{B} \mid \bs{B}_1 \cup  \bs{B}_3 \cup \bs{B}_2    \big) \\
\times  \ex{ - 2 \i\pi \om_{\op{O}} ( |C_1\cup C_2| + |D_2\cup D_3| ) }
\cdot  \mc{F}_{ -, -, 0}^{(\op{O})} \Big( \,  \overleftarrow{ \bs{D}_2 } + \i \pi \ov{\bs{e}} ,  \overleftarrow{ \bs{C}_2 } + \i \pi \ov{\bs{e}} ,  \bs{B}_2 \Big) \;.
\end{multline}
One may then combine the products of $\op{S}$ matrices:
\bem
\op{S}\Big( \, \overleftarrow{\bs{A}}  \mid \overleftarrow{\bs{A}_2}\cup  \overleftarrow{\bs{A}_1} \Big)
\cdot  \op{S}\Big( \, \overleftarrow{\bs{A}_1}  \mid \overleftarrow{ \bs{C}_2 }\cup  \overleftarrow{ \bs{C}_1 } \Big)
\cdot  \op{S}\Big( \, \overleftarrow{\bs{A}_2}  \cup   \overleftarrow{ \bs{C}_2 }  \mid   \overleftarrow{ \bs{D}_2 }\cup  \overleftarrow{ \bs{C}_2 }  \cup   \overleftarrow{ \bs{D}_3 }  \Big)\\
\; = \; \op{S}\Big( \, \overleftarrow{\bs{A}}  \mid \overleftarrow{\bs{A}_2}\cup  \overleftarrow{ \bs{C}_2 }\cup  \overleftarrow{ \bs{C}_1 } \Big)  \cdot
\op{S}\Big( \, \overleftarrow{\bs{A}_2}  \cup   \overleftarrow{ \bs{C}_2 }  \mid   \overleftarrow{ \bs{D}_2 }\cup  \overleftarrow{ \bs{C}_2 }  \cup   \overleftarrow{ \bs{D}_3 }  \Big)
\; = \; \op{S}\Big( \, \overleftarrow{\bs{A}}  \mid   \overleftarrow{ \bs{D}_2 }\cup  \overleftarrow{ \bs{C}_2 }  \cup   \overleftarrow{ \bs{D}_3 } \cup  \overleftarrow{ \bs{C}_1 } \Big) \;.
\end{multline}
Further, one has
\bem
 \op{S}\Big( \, \overleftarrow{\bs{A}}  \mid   \overleftarrow{ \bs{D}_2 }\cup  \overleftarrow{ \bs{C}_2 }  \cup   \overleftarrow{ \bs{D}_3 } \cup  \overleftarrow{ \bs{C}_1 } \Big)
\cdot \mc{F}_{ -, -, 0}^{(\op{O})} \Big( \,  \overleftarrow{ \bs{D}_2 } + \i \pi \ov{\bs{e}} ,  \overleftarrow{ \bs{C}_2 } + \i \pi \ov{\bs{e}} ,  \bs{B}_2 \Big) \\
\; = \; \op{S}\Big( \, \overleftarrow{\bs{A}}  \mid   \overleftarrow{ \bs{C}_2\cup \bs{D}_2 }  \cup   \overleftarrow{ \bs{D}_3 } \cup  \overleftarrow{ \bs{C}_1 } \Big)
\cdot \mc{F}_{ -, 0}^{(\op{O})} \Big( \,   \overleftarrow{ \bs{C}_2\cup \bs{D}_2 }  + \i \pi \ov{\bs{e}}  ,  \bs{B}_2 \Big) \;.
\end{multline}
Finally, one observes that
\bem
\op{S}\Big(\,   \overleftarrow{ \bs{D}_3 }\cup  \overleftarrow{ \bs{C}_1 }\mid  \overleftarrow{ \bs{D}_3 \cup \bs{C}_1 }   \Big) \cdot
\De\big( \bs{C}_1 \cup \bs{D}_3 \mid \bs{B}_1  \cup \bs{B}_3 \big) \cdot
\op{S}\Big(\,   \bs{B}_1\cup  \bs{B}_3\mid  \overrightarrow{ \bs{B}_1 \cup \bs{B}_3}   \Big) \\  \; = \;
\op{S}\Big(\,  \overrightarrow{ \bs{B}_1 \cup \bs{B}_3}   \mid  \bs{B}_1\cup  \bs{B}_3  \Big) \cdot
\De\big( \bs{C}_1 \cup \bs{D}_3 \mid \bs{B}_1  \cup \bs{B}_3 \big) \cdot
\op{S}\Big(\,   \bs{B}_1\cup  \bs{B}_3\mid  \overrightarrow{ \bs{B}_1 \cup \bs{B}_3}   \Big)  \; = \;
\De\big( \bs{C}_1 \cup \bs{D}_3 \mid \bs{B}_1  \cup \bs{B}_3 \big) \;.
\end{multline}
All of the above allows one to recast the sum as
\bem
\mc{M}^{(\op{O})}_{n;m}\big(  \bs{\alpha}_n ; \bs{\be}_{m} \big) \; =     \hspace{-2mm}  \sul{   A_1 = C_1 \cup C_2  }{}
\sul{   A_2 =  D_2\cup D_3  }{} \sul{ B =  \underset{13}{\cup}\, _{a=1}^{3} B_a  }{}
 \hspace{-2mm}
\De\big( \bs{C}_1\cup \bs{D}_3  \mid \bs{B}_1   \cup \bs{B}_3  \big) \, \ex{ - 2 \i\pi \om_{\op{O}}  |A|  } \\
\times\op{S}\Big( \, \overleftarrow{\bs{A}}  \mid   \overleftarrow{ \bs{C}_2\cup \bs{D}_2 }  \cup   \overleftarrow{ \bs{D}_3 \cup \bs{C}_1 }  \Big)
\cdot \op{S}\Big( \, \bs{B} \mid \overrightarrow{ \bs{B}_1 \cup \bs{B}_3 } \cup \bs{B}_2    \Big)
\cdot \mc{F}_{  -, 0}^{(\op{O})} \Big( \,   \overleftarrow{ \bs{C}_2\cup \bs{D}_2 }  + \i \pi \ov{\bs{e}}  ,  \bs{B}_2 \Big) \;.
\end{multline}
At this stage, one may set
\beq
\wt{A}_1 \, = \, D_3 \cup C_1 \;, \quad \wt{A}_2 \, = \, D_2 \cup C_2 \;, \quad
\wt{B}_1 \, = \, B_1 \cup B_3 \;, \quad \wt{B}_2 \, = \, B_2 \;.
\enq
The summation over partitions $A_1 = C_1 \cup C_2 $, $ A_2 =  D_2\cup D_3 $, $B =  \underset{13}{\cup}\, _{a=1}^{3} B_a$ with $A=A_1\cup A_2$ and $A_1, A_2$
fixed is then fully equivalent to summing up over partitions $A= \wt{A}_1\cup \wt{A}_2$ and $B= \wt{B}_1\cup \wt{B}_2$ with $\wt{B}_1$ unordered.
Since $B_1, B_3$ were unordered sets, one may replace as well
\beq
\De\Big( \bs{C}_1\cup \bs{D}_3  \mid \bs{B}_1   \cup \bs{B}_3  \Big)  \hookrightarrow
\De\Big( \bs{\wt{A}}_1  \mid  \bs{\wt{B}}_1  \Big) \;.
\enq
This recasts the sum as
\bem
\mc{M}^{(\op{O})}_{n;m}\big(  \bs{\alpha}_n ; \bs{\be}_{m} \big) \; =    \ex{ - 2 \i\pi \om_{\op{O}}  |A|   }  \hspace{-2mm}  \sul{   A = A_1 \cup A_2  }{}\sul{ B =  B_1 \underset{1}{\cup} B_2  }{}
 \hspace{-2mm}
\De\big( \bs{A}_1  \mid \bs{B}_1   \big) \cdot \op{S}\Big( \, \overleftarrow{\bs{A}}  \mid   \overleftarrow{\bs{A}_2}  \cup   \overleftarrow{\bs{A}_1}  \Big)
\cdot \op{S}\big(   \bs{B} \mid \bs{B}_1 \cup \bs{B}_2    \big) \\
\times \mc{F}_{  -, 0}^{(\op{O})} \Big( \,   \overleftarrow{\bs{A}_2}  + \i \pi \ov{\bs{e}} ,  \bs{B}_2 \Big) \;,
\end{multline}
and thus exactly reproduces \eqref{ecriture representation combinatoire pour MO via reduction de type 1}. \qed

\section{Toward truncated  multi-point correlation functions: the smeared integral representation}
\label{Section Towards multi pts fcts}

\subsection{A premilinary expression}
\label{Subsection Premilinary expression}

In this subsection, we shall obtain a first integral representation for a smeared $k$-point function. In the following, we shall always consider a smearing function $g$
belonging to the Schwarz class $\mc{S}(\R^{1,1})$
%
%
%
%
%
%
%
%
%
%
Further, here and in the following, we denote by $\op{O}[g]$ the result of smearing-out the operator $\op{O}$ versus a smooth compactly supported function $g$ on $\R^{1,1}$.
Following \eqref{definition action operatur dans secteur multipoints}, given a sufficiently regular function
\beq
\bs{f} \, = \, \big( f^{(0)}, \dots, f^{(n)}, \dots \big) \, \in \, \mf{h}_{ \e{ShG} } \, ,
\enq
one has $\op{O}[g] \, = \, \big( \op{O}^{(0)}[g]\cdot \bs{f}, \dots, \op{O}^{(n)}[g]\cdot \bs{f}, \dots \big)$, in which
\beq
 \Big( \op{O}^{(n)}[g]\cdot \bs{f} \Big)\big(  \bs{\a}_n \big)   \; = \; \sul{m \geq 0}{}  \op{M}_{\op{O}}^{(m)} \big[g , f^{(m)}\big]  \big(  \bs{\a}_n \big) \;.
\enq
There, defining $A = \{\a_a \}_{a=1}^{n}$ with $\bs{\a}_n$ generic, as follows from \eqref{ecriture representation combinatoire pour MO via reduction de type 1}
of Lemma \ref{Lemma rep direct pour noyau integral generalise secteur n m particules}, one has the representation
\bem
\op{M}_{\op{O}}^{(m)} \big[g , f^{(m)}\big]  \big(  \bs{\a}_n \big)   \; = \;  \ex{-2 n \i\pi \om_{\op{O}}    } \lim_{ \veps \tend 0^+} \sul{A = A_1 \cup A_2 }{}
 \Int{ \R_{>}^m }{}    \f{ \dd^m \be }{ (2\pi)^m }  \;\sul{ B = B_1 \underset{1}{\cup} B_2  }{}
 \De\big( \bs{A}_1 \mid \bs{B}_1  \big) \cdot \op{S}\Big( \, \overleftarrow{\bs{A}}  \mid \overleftarrow{\bs{A}_2}\cup  \overleftarrow{\bs{A}_1} \Big) \\
\times \op{S}\big(  \bs{B} \mid \bs{B}_1\cup  \bs{B}_2 \big) \cdot
\mc{F}^{(\op{O})}  \Big( \, \overleftarrow{\bs{A}_2} + \i \pi \ov{\bs{e}}_{\veps};  \bs{B}_2 \, \Big)
    \cdot \mc{R}[g]\big(    \bs{A} ,  \bs{B} \big)   \cdot  f^{(m)}\big(   \bs{B}   \big) \;,
\label{definition action operatur dans secteur multipoints}
\end{multline}
in which  $B \, = \, \big\{ \be_a \big\}_{a=1}^{m}$ and we have introduced
\beq
\mc{R}[g]\big(    \bs{A} ,  \bs{B} \big)   \; = \;  \Int{ \R^{1,1} }{} \hspace{-1mm} \dd^2 \bs{x} \;
\ex{ \i \,  [ \ov{\bs{p}}( \bs{A} ) - \ov{\bs{p}}( \bs{B})  ] \cdot \bs{x} } \, g(\bs{x}) \;,
\label{definition facteur regularisant espace reel}
\enq
where $ \ov{\bs{p}}( \bs{A} )$ is as given in \eqref{definition de bar vect p}. Furthermore, we agree upon
\beq
\ov{\bs{e}}_{\veps} \; = \; \Big( 1 \, - \, \f{\veps}{\pi} \Big) \cdot \big( 1,\dots, 1 \big) \: .
\label{definition ov bs eps avec indice regulateur}
\enq
Above, the $\veps \tend 0^+$ limit issues from the distributional $-$ boundary values regularisation of the poles at
\beq
(\bs{A}_2)_k=(\bs{B}_2)_{\ell} \, .
\enq
The techniques developed below ensure that for generic values of $\bs{\a}_n\in \R^n_{>}$, \eqref{definition action operatur dans secteur multipoints}
is well-defined as soon as $f^{(m)}$ is regular enough.

Thus, provided that the integrations and limits make sense -and this shall be established at a later stage-, given
$r_0, \dots, r_k \in \mathbb{N}$ and $g_1,\dots, g_k \in \mc{S}(\R^{1,1})$, one has the below representation for the concatenation of operator actions
\bem
\Big( \op{O}^{(r_0)}_1[g_1] \cdot  \op{O}^{(r_1)}_2[g_2] \cdots   \op{O}^{(r_{k-1})}_k[g_k]  \cdot \bs{f}_{r_k} \Big) \big( \bs{\a}^{(0)}_{r_0} \big) \; = \;
\lim_{\veps_1 \tend 0^+} \Int{  \R_{>}^{r_1}   }{} \f{ \dd^{ r_1 } \a^{(1)} }{ (2\pi)^{r_1} } \cdots \lim_{\veps_{k} \tend 0^+} \Int{  \R_{>}^{r_k}   }{} \f{ \dd^{ r_k } \a^{(k)}  }{ (2\pi)^{r_k} }
f^{(r_k)} \big( \bs{A}^{(k)} \big)\\
 \times  \mc{G}\Big( \big\{  \bs{A}^{(s)}   \big\}^{k}_{0} ; \bs{\veps}_k \Big)  \cdot  \mc{R}[ G_k]\Big(   \big\{  A^{(s)}     \big\}^{k}_{0} \Big)  \; .
\label{ecriture rep int fct k pts forme primordiale}
\end{multline}
There are several ingredients to the formula. First of all,
\beq
 \bs{f}_{r} \; = \; \Big( 0,\dots, 0, f^{(r)},0,\dots \Big)\qquad \e{with} \qquad f^{(r)} \in \mc{C}^{\infty}_{\e{c}}(\R^{r}) \;.
\label{definition vecteur fr dans secteur r particule pure}
\enq
Also,  we have introduced $ A^{(s)} = \{  \a^{(s)}_{a}  \}_{a=1}^{r_s}$ and have set
\beq
G_k\big(\bs{x}_1,\dots, \bs{x}_k \big) \, = \,  \pl{s=1}{k} g_s\big( \bs{x}_s \big)
\label{definition fonction Gk}
\enq
Further,
for any Schwarz function $G \in \mc{S}\Big( \big( \R^{1,1} \big)^{k}  \Big)$, we denote
\beq
 \mc{R}[ G]\Big(   \big\{  A^{(s)}     \big\}^{k}_{0} \Big)   \; = \; \Int{ \big( \R^{1,1} \big)^{k} }{} \pl{s=1}{k}\dd \bs{x}_s
\cdot  G\big(\bs{x}_1,\dots, \bs{x}_k \big) \cdot
 \ex{ \i \mc{P} \big( \{ A^{(s)} \}^{k}_{0} ; \{\bs{x}_a \}_{1}^{k} \big) } \; ,
\label{definition Fourier direct de G a k veriable Minkowski}
\enq
where
\beq
 \mc{P} \big( \{ A^{(s)}\}^{k}_{0} ; \{\bs{x}_a \}_{1}^{k} \big) \; = \; \ov{\bs{p}} \, \big(   \bs{A}^{(0)}   \big) \cdot \bs{x}_{1} \, + \,   \sul{p=1}{k-1} \ov{\bs{p}}\,\big(   \bs{A}^{(p)}  \big) \cdot \bs{x}_{p+1 p}
 \; - \; \ov{\bs{p}} \, \big(   \bs{A}^{(k)}   \big) \cdot \bs{x}_{k} \;.
\label{defintion impulsion globale pour produit elements matrice}
\enq
Finally, we agree upon

\beq
\mc{G}\Big( \big\{  \bs{A}^{(s)}   \big\}^{k}_{0} ; \bs{\veps}_k \Big) \; = \; \mc{M}^{(\op{O}_1)}_{r_0;r_1}\big(  \bs{\alpha}_{r_0}^{(0)} ; \bs{\alpha}_{r_1}^{(1)} \big)_{\veps_1}
\cdot \mc{M}^{(\op{O}_2)}_{r_1;r_2}\big(  \bs{\alpha}_{r_1}^{(1)} ; \bs{\alpha}_{r_2}^{(2)} \big)_{\veps_2}  \cdots
\mc{M}^{(\op{O}_k)}_{r_{k-1};r_k}\big(  \bs{\alpha}_{r_{k-1}}^{(k-1)} ; \bs{\alpha}_{r_k}^{(k)} \big)_{\veps_k} \;,
\label{definition G produit noyaux integraux operateurs}
\enq
in which the fundamental building block takes the form
\bem
\mc{M}^{(\op{O}_{\ell})}_{ |A^{(\ell-1)}| ; |A^{(\ell)}| }\big(  \bs{A}^{(\ell-1)}     ; \bs{A}^{(\ell)}  \big)_{\veps} \; = \;
\ex{-2\i\pi \om_{\op{O}_{\ell}}|A^{(\ell-1)}| }   \hspace{-5 mm}  \sul{ A^{(\ell-1)}  =  A_1^{(\ell-1)} \cup A_2^{(\ell-1)}  }{}
\sul{ A^{(\ell)} = B_1^{(\ell)} \underset{1}{\cup} B_2^{(\ell)}  }{}
 \hspace{-2mm}  \De\Big( \bs{A}_1^{(\ell-1)} \mid \bs{B}_1^{(\ell)}  \Big)  \\
\times \op{S}\Big( \, \overleftarrow{\bs{A}^{(\ell-1)}}  \mid \overleftarrow{\bs{A}_2^{(\ell-1)}}\cup  \overleftarrow{\bs{A}_1^{(\ell-1)}} \Big)
\cdot  \op{S}\big(  \bs{A}^{(\ell)} \mid \bs{B}_1^{(\ell)}\cup  \bs{B}_2^{(\ell)} \big) \cdot \mc{F}^{(\op{O}_{\ell})} \Big( \overleftarrow{\bs{A}_2^{(\ell-1)}} + \i \pi \ov{\bs{e}}_{\veps};  \bs{B}_2^{(\ell)} \Big) \;.
\label{ecriture representation combinatoire pour noyau integral}
\end{multline}

\subsection{An auxiliary representation}
\label{Subsection Auxiliary representation}

In this subsection, we shall focus on the product
\beq
 \mc{G}_{\e{tot}}[ G] \Big( \big\{  \bs{A}^{(s)}   \big\}^{k}_{0} ; \bs{\veps}_k \Big)\; = \;
 \mc{G}\Big( \big\{  \bs{A}^{(s)}   \big\}^{k}_{0} ; \bs{\veps}_k \Big) \,  \mc{R}[ G]\Big(   \big\{  A^{(s)} \big\}^{k}_{0} \Big)
\label{definition G tot de eps}
\enq
and provide a closed combinatorial expression for this product of generalised integral kernels in the case where $A^{(0)}=\emptyset$, \textit{viz}. $\bs{\a}_{r_0}^{(0)}$ is the empty vector,
which allows one to immediately compute the effect of $\De$-distributions present in the "raw" representation for $\mc{G}$ which can be obtained by
simply taking the products over the kernels given in \eqref{ecriture representation combinatoire pour noyau integral}.

Prior to stating the representation for \eqref{definition G tot de eps}, we introduce a convenient notation. Given an index ordered set $A=\{\a_{j_a} \}_{a=1}^{k}$
with $j_a$ being pairwise distinct, and $\sg \in \mf{S}_{k}$, the ordered set $A^{\sg}$ and the vector $\vec{\bs{A}}\,\!^{\sg}$ corresponds to
\beq
A^{\sg} \, = \, \{ \a_{ j_{\sg(a)} } \}_{a=1}^{k} \;,   \qquad   \vec{\bs{A}}\,\!^{\sg} \, = \, (\a_{j_{\sg(1)}},\dots, \a_{j_{\sg(k)} }  )  \;.
\enq

\begin{prop}
\label{Proposition rep combinatoire pour G tot de epsk}

Let $A^{(0)}=\emptyset$, then it holds
\bem
 \mc{G}_{\e{tot}}[ G] \Big( \big\{  \bs{A}^{(s)}   \big\}^{k}_{0} ; \bs{\veps}_k \Big) \; = \;\pl{s=1}{k} \ex{-2\i\pi \om_{\op{O}_s}|A^{(s-1)}| } \cdot   \pl{p=1}{k-1} \bigg\{   \sul{ \op{P}_p[  A^{(p)} ]  }{}
\pl{s=1}{p-1} \sul{ \sg_s^{(p-1)} \in \mf{S}_{  |  A^{(p-1)}_s |  } }{}   \bigg\} \sul{ \op{P}_k[  A^{(k)} ]  }{}
\msc{S}\Big( \big\{ B_{s}^{(k)}; \ga^{(ba)} \big\}  \Big) \\
\times \pl{p=2}{k-1} \pl{s=1}{p-1} \De\Big( \bs{A}_s^{(p-1)}  \mid \overrightarrow{ \big( \bs{A}^{(p)}_s  \cup \bs{\ga}^{(p+1s)} \big) }^{ \sg_s^{(p-1)} } \Big)
\cdot \pl{s=1}{k-1} \De\big( \bs{ A}_s^{(k-1)}  \mid \bs{B}_s^{(k)}  \big) \\
\times \pl{p=1}{k} \mc{F}^{(\op{O}_p)}\Big( \overleftarrow{ \bs{\ga}^{(pp-1)} } \cup \cdots \cup \overleftarrow{ \bs{\ga}^{(p1)} }  + \i \pi \ov{\bs{e}}_{\veps_{p}} ,
\bs{B}_p^{(k)}  \cup \bs{\ga}^{(kp)}  \cup \cdots \cup \bs{\ga}^{(p+1p)}   \Big)
\cdot  \msc{R}[ G]\Big(   \{ B_s^{(k)} \}; \{\ga^{(ba)} \}  \Big)   \;.
\label{ecriture formule combinatoire pour G tot totalement moins}
\end{multline}
Above,  the sums run through partitions  $\op{P}_p[  A^{(p)} ]$ of $A^{(p)}$ such that
\beq
 A^{(p)} \, = \, \bigcup\limits_{s=1}^{p}A_s^{(p)}  \, \bigcup\limits_{s=1}^{p}\ga^{(p+1s)} \quad p=1 , \dots, k-1 \quad and \quad
A^{(k)} \, = \, \bigcup\limits_{1, \dots,\,  k-1}^{k} \hspace{-2mm} B_{s}^{(k)} \;.
\label{ecriture sous partitions des Ap et Ak pour calcul multipoints}
\enq
Further, one sums over permutations $\sg_s^{(p-1)}$, $s=1,\dots, p-1$ and $p = 2, \dots, k-1$. The summations are constrained so that
\beq
 \big|  A^{(p-1)}_s \big| \; = \, \big| A^{(p)}_s  \cup \ga^{(p+1s)} \big| \qquad for \quad s=1,\dots, p-1\;,  \quad p=1,\dots, k-1
\enq
and
\beq
\big|  A^{(k-1)}_s \big| \; = \, \big| B^{(k)}_s  \big| \qquad s=1,\dots, k-1 \;.
\enq
Next,
\beq
 \msc{R}[ G]\Big(   \{ B_s^{(k)} \}; \{\ga^{(ba)} \}  \Big)   \; = \hspace{-2mm} \Int{ \big( \R^{1,1} \big)^{k} }{}  \hspace{-1mm} \pl{s=1}{k}\dd \bs{x}_s
\cdot  G\big(\bs{x}_1,\dots, \bs{x}_k \big) \cdot
 \ex{\i \msc{P}\big(  \{ B_s^{(k)} \}; \{\ga^{(ba)} \}; \{\bs{x}_s\} \big)  } \; ,
\enq
where we have used the shorthand notation
\beq
\msc{P}\big(  \{ B_s^{(k)} \}; \{\ga^{(ba)} \}; \{\bs{x}_s\} \big) \; = \; \sul{u>s}{k} \ov{\bs{p}}\, \big(   \bs{\ga}^{(us)}  \big)\cdot \bs{x}_{us}
\, - \, \sul{s=1}{k} \ov{\bs{p}}\, \big(   \bs{B}_s^{(k)}  \big)\cdot \bs{x}_{s} \;.
\label{definition curly P}
\enq
Finally, one has the product representation
\bem
\msc{S}\Big( \big\{ B_{s}^{(k)} ; \ga^{(ba)} \big\}  \Big) \; = \;
S\big( \bs{A}^{(k)}  \mid \bs{B}_1^{(k)}  \cup \cdots \cup \bs{B}_k^{(k)}  \big)
 \cdot \pl{p=2}{k-1}\pl{s=1}{p-1} \pl{v=s+1}{p} S\big( \bs{B}_v^{(k)} \cup  \bs{\ga}^{(p+1s)} \mid  \bs{\ga}^{(p+1s)}  \cup  \bs{B}_v^{(k)}  \big) \\
\times  \pl{  \substack{ v >p \\ p \geq 3 } }{ k }  \pl{u>s}{p-1}
S\big(  \bs{\ga}^{(vu)}  \cup  \bs{\ga}^{(ps)}   \mid   \bs{\ga}^{(ps)}   \cup \bs{\ga}^{(vu)}   \big)  \;.
\label{definition matrice S cas totalement -}
\end{multline}

\end{prop}

\Proof

We first focus on re-expressing the pure product of regularised kernels $ \mc{M}^{(\op{O}_{\ell})}_{\veps}$:
\beq
 \wt{\mc{G}}\Big( \big\{  \bs{A}^{(s)}   \big\}^{k}_{0} ; \bs{\veps}_k \Big)
 \; = \;  \pl{s=1}{k} \ex{2\i\pi \om_{\op{O}_s}|A^{(s-1)}| } \cdot  \mc{M}^{(\op{O}_1)}_{n_0;n_1}\big(  \bs{\alpha}_{n_0}^{(0)} ; \bs{\alpha}_{n_1}^{(1)} \big)_{\veps_1}
\cdot \mc{M}^{(\op{O}_2)}_{n_1;n_2}\big(  \bs{\alpha}_{n_1}^{(1)} ; \bs{\alpha}_{n_2}^{(2)} \big)_{\veps_2}  \cdots
\mc{M}^{(\op{O}_k)}_{n_{k-1};n_k}\big(  \bs{\alpha}_{n_{k-1}}^{(k-1)} ; \bs{\alpha}_{n_k}^{(k)} \big)_{\veps_k} \, ,
\enq
in a more convenient way.

Each of the building blocks may be represented through  \eqref{ecriture representation combinatoire pour MO via reduction de type 1} what yields to
\bem
 \wt{\mc{G}}\Big( \big\{  \bs{A}^{(s)}   \big\}^{k}_{0} ; \bs{\veps}_k \Big) \; = \; \pl{p=1}{k-1} \Bigg\{  \sul{ A^{(p)}= C_1^{(p)} \cup C_2^{(p)}  }{}  \sul{ A^{(p)} = D_1^{(p)} \underset{1}{\cup} D_2^{(p)}  }{}  \Bigg\}
 \sul{ A^{(k)} = D_1^{(k)} \underset{1}{\cup} D_2^{(k)}  }{}
 \pl{p=1}{k} \De\big( \bs{C}_1^{(p-1)}  \mid \bs{D}_1^{(p)}  \big) \\
\times \pl{p=1}{k-1}\op{S}\Big( \, \overleftarrow{ \bs{A}^{(p)} }  \mid \overleftarrow{ \bs{C}_2^{(p)} }\cup  \overleftarrow{ \bs{C}_1^{(p)} } \Big)
\cdot  \pl{p=1}{k}\op{S}\big(  \bs{A}^{(p)} \mid \bs{D}_1^{(p)} \cup  \bs{D}_2^{(p)} \big)
\cdot  \pl{p=1}{k}\mc{F}^{(\op{O}_p)} \Big( \overleftarrow{ \bs{C}_2^{(p-1)} } + \i \pi \ov{\bs{e}}_{\veps_{p}};  \bs{D}_2^{(p)}  \Big) \;.
\label{ecriture tilge G via somme sur les C et D partitions}
\end{multline}
The sub-partitions arising above are constrained so that
\beq
C_{1}^{(0)} \; = \; \emptyset  \qquad \e{and} \qquad |C_1^{(p-1)}| \, = \, |D_1^{(p)}| \;,  \qquad p=1,\dots, k \, .
\enq
We now define a sub-partitioning of the sets $A^{(p)}$ as in \eqref{ecriture sous partitions des Ap et Ak pour calcul multipoints}. Then, for each such a sub-partition, we define
\beq
C_1^{(p)} \; = \; \bigcup\limits_{s=1}^{p}A_s^{(p)} \; , \qquad  C_2^{(p)} \; = \; \bigcup\limits_{s=1}^{p} \ga^{(p+1s)}
\label{ecriture partition des Cp via les Asp et gamma p s}
\enq
for $p=1 , \dots, k-1$, as well as
\beq
D_1^{(p)} \; = \; \bigcup\limits_{s=1}^{p-1} \Big\{ A_s^{(p)} \cup \ga^{(p+1s)}   \Big\}^{ \sg_s^{(p-1)} }\qquad \e{and} \qquad D_2^{(p)} \; = \; A_p^{(p)} \cup \ga^{(p+1p)}
\label{ecriture partition des Dp via les Asp et gamma p s}
\enq
for $p=1 , \dots, k-1$. Finally, we set
\beq
D_1^{(k)} \; = \; \bigcup\limits_{1,\dots, k-1 }^{k-1} B_s^{(k)} \qquad \e{and} \qquad D_2^{(k)} \; = \; B_k^{(k)}    \;.
\label{ecriture D1 et D2 comme B partitions}
\enq

We now establish that that the original summation over the double partitioning of the sets $A^{(p)}$ into $C$ and $D$ type subsets
is equivalent to a summation over the partitions that we have just described. This is implemented by induction. The induction hypothesis at level $r\leq k-1$
is formulated as follows. For given choices of partitions
\beq
A^{(p)} \, =\, \left\{ \ba{ccc}   C_1^{(p)} \cup C_2^{(p)}  \vspace{2mm} \\
                        D_1^{(p)} \underset{1}{\cup} D_2^{(p)}   \ea \right. p=1,\dots, r
  \qquad \e{constrained} \, \e{as} \qquad |C_1^{(p-1)}| \, = \, |D_1^{(p)}| \;,  \qquad p=1,\dots, r \,
\enq
with $C_{1}^{(0)} \; = \; \emptyset$, there exists a unique choice of $2p$-fold partitions of $A^{(p)}$, $p=1 , \dots, r$,
\beq
 A^{(p)} \, = \, \bigcup\limits_{s=1}^{p}A_s^{(p)}  \, \bigcup\limits_{s=1}^{p}\ga^{(p+1s)} \;
\; \e{constrained}\; \e{as} \quad
\big|  A^{(p-1)}_s \big| \; = \, \big| A^{(p)}_s  \cup \ga^{(p+1s)} \big| \qquad \e{with} \quad s=1,\dots, p-1\;,
\enq
and of permutations $\sg_{s}^{(p-1)} \in \mf{S}_{ | A^{(p-1)}_s | }$, $s=1,\dots, p-1$ such that
\eqref{ecriture partition des Cp via les Asp et gamma p s}-\eqref{ecriture partition des Dp via les Asp et gamma p s} hold for $p=1,\dots,r$.

Indeed, for $r=1$, one sets $A_{1}^{(1)}\, = \, D_{2}^{(1)}\cap C_{1}^{(1)}$ and $\ga^{(21)}\, = \, D_{2}^{(1)}\cap C_{2}^{(1)}$
what defines the building blocks unambiguously and is consistent with $D_{1}^{(1)}=\emptyset$. At the initialisation step,
there are no permutations involved in the construction.

We now assume that the induction hypothesis holds up to some $r-1 \leq k-2$ and we are given the $D-C$ partitions up to subscript $p=r$.
Then, we set
\beq
A^{(r)}_{r} \, = \, D_{2}^{(r)}\cap C_{1}^{(r)} \quad \e{and} \quad  \ga^{(r+1r)} \, = \, D_{2}^{(r)}\cap C_{2}^{(r)} \;.
\enq
Since $C_{1}^{(r)}\cup C_{2}^{(r)}=A^{(r)}$, this implies that $D_{2}^{(r)}=A^{(r)}_{r} \cup \ga^{(r+1r)}$. The numbers
$|A_{s}^{(r-1)}|$, $s=1,\dots, r-1$ are given and sum up as
\beq
\sul{s=1}{r-1}|A_{s}^{(r-1)}| \, = \, |C_1^{(r-1)}| \, = \, |D_{1}^{(r)}| \;.
\enq
One then looks at the positions $p_1^{(s)}, \dots, p_{|A_{s}^{(r-1)}|}^{(s)}$ of appearance of the elements building up  the set $A_{s}^{(r-1)}$
in the full vector $\bs{C}^{(r-1)}_1$ and gathers the index-ordered elements building up $D_{1}^{(r)}$ with index labels  $p_1^{(s)}, \dots, p_{|A_{s}^{(r-1)}|}^{(s)}$
into the index-ordered sets $G_{s}^{(r)}$ of cardinality $|G_{s}^{(r)}|=|A_{s}^{(r-1)}|$, namely starting from
\beq
D_{1}^{(r)} \, = \, \big\{ d_{i_a} \big\}_{1}^{ | D_{1}^{(r)} | } \quad \e{one} \, \e{has} \quad
G_{s}^{(r)} \, = \, \Big\{ d_{i_{p_{a}^{(s)}}  } \Big\}_{a = 1 }^{    |A_{s}^{(r-1)}|   } \;.
\enq
This allows one to further define the \textit{per-se} sets
\beq
A_{s}^{(r)} \, = \, G_{s}^{(r)} \cap C_{1}^{(r)} \qquad \e{and} \qquad
\ga^{(r+1s)} \, = \, G_{s}^{(r)} \cap C_{2}^{(r)} \;.
\enq
By construction, the union set is given by  $A_{s}^{(r)}\cup \ga^{(r+1s)} = \big\{ d_{k_a} \big\}_{ a = 1 }^{ |G_s^{(r)}| }$ for some $k_1 < \cdots < k_{ |G_s^{(r)}| }$
such that $k_a \in \{ i_a \}_{1}^{ | D_{1}^{(r)} | } $.  The permutation $\sg_{s}^{(r-1)}$ is then uniquely defined as the one realising the correspondence of index-ordered ensembles
\beq
\big( A_{s}^{(r)}\cup \ga^{(r+1s)} \big)^{ \sg_{s}^{(r-1)}  } \,  =  \, G_s^{(r)} \;.
\enq
This completes the construction for $r$. Setting $r=k-1$ completes the first step of the construction.

The procedure for the construction of the $B$-partitions is quite similar. One first identifies the positions  $q_1^{(s)}, \dots, q_{|A_{s}^{(k-1)}|}^{(s)}$ of the
appearance if the elements building up the sets $A_{s}^{(k-1)}$ in the vector $\bs{C}_1^{(k-1)}$ and then gathers the index ordered elements building up $\bs{D}_1^{(k)}$
with label indices $q_1^{(s)}, \dots, q_{|A_{s}^{(k-1)}|}^{(s)}$ in the index ordered set $B_s^{(k)}$ of cardinality $|B_s^{(k)}| = |A_s^{(k-1)}|$, namely given
\beq
D_{1}^{(k)} \, = \, \big\{ \wt{d}_{i_a} \big\}_{1}^{ | D_{1}^{(k)} | } \quad \e{one} \, \e{has} \quad
B_{s}^{(k)} \, = \, \Big\{ \wt{d}_{i_{q_{a}^{(s)}}  } \Big\}_{a = 1 }^{    |A_{s}^{(k-1)}|   } \;.
\enq
This realises $B_s^{(k)}$, $s=1,\dots, k-1$ as an index ordered set $B_s^{(k)}=\big\{ \wt{d}_{j_{a;s}} \big\}_{a=1}^{ | A_{1}^{(k-1)} | } $ with $j_{a;s} = i_{ q_{a}^{(s)} }$.
Finally, one sets $B_{k}^{(k)}=D_2^{(k)}$.

It is clear that for any choice of partitions and permutations as given in the statement of the Proposition
one obtains through the identities \eqref{ecriture partition des Cp via les Asp et gamma p s}-\eqref{ecriture D1 et D2 comme B partitions}
the $C_u^{(p)}$ and $D_u^{(p)}$ partitions as appearing in the sums \eqref{ecriture tilge G via somme sur les C et D partitions}.
Moreover, it is clear from the discussion that the correspondence is bijective.
This completes the construction.

The previous construction ensures that the positions of the coordinates of the vector $\bs{A}^{(p-1)}_{s}$ inside the vector $\bs{C}^{(p-1)}_1$
are exactly located at the positions of the coordinates of the vector $\overrightarrow{ \big( \bs{A}_s^{(p)} \cup \bs{\ga}^{(p+1s)}   \big)}^{ \sg_s^{(p-1)} } $
inside of the vector $\bs{D}_1^{(p)}$, this for $p=2,\dots, k-1$. Similarly, the coordinates of the vector $\bs{A}^{(k-1)}_{s}$ inside the vector $\bs{C}^{(k-1)}_1$
are exactly located at the positions of the coordinates of the vector $\bs{B}_s^{(k)}$ inside of the vector $\bs{D}_1^{(k)}$. This ensures that it holds
\beq
 \De\big( \bs{C}_1^{(p-1)}  \mid \bs{D}_1^{(p)}  \big) \; = \;
 \pl{s=1}{p-1}  \De\Big( \bs{A}_s^{(p-1)}  \mid \overrightarrow{ \big( \bs{A}_s^{(p)} \cup \bs{\ga}^{(p+1s)}   \big)}^{ \sg_s^{(p-1)} }  \Big) \qquad \e{for} \qquad  p = 2, \dots, k-1
\label{ecriture contraintes Delta sur entrees As autres que k-1}
\enq
as well as
\beq
 \De\big( \bs{C}_1^{(k-1)}  \mid \bs{D}_1^{(k)}  \Big) \; = \;
 \pl{s=1}{k-1}  \De\big( \bs{A}_s^{(k-1)}  \mid \bs{B}_s^{(k)}   \big) \;.
\label{ecriture contraintes Delta sur entrees As de type k-1}
\enq

As a consequence, one ends up with the following formula
\beq
 \wt{\mc{G}}\Big( \big\{  \bs{A}^{(s)}   \big\}^{k}_{0} ; \bs{\veps}_k \Big)  \; = \; \pl{p=1}{k-1} \bigg\{   \sul{ \op{P}_p[  A^{(p)} ]  }{}
\pl{s=1}{p-1} \sul{ \sg_s^{(p-1)} \in \mf{S}_{  |  A^{(p-1)}_s |  } }{}   \bigg\} \sul{ \op{P}_k[  A^{(k)} ]  }{}
  \pl{p=1}{k} \De\big( \bs{C}_1^{(p-1)}  \mid \bs{D}_1^{(p)}  \big) \cdot \mc{W}\Big( \big\{ C_a^{(p)}, D_a^{(p)} \big\} \Big)
\label{ecriture représentation pour wt G apres changement parametrisation partition}
\enq
in which $C_a^{(p)}$s and $D_a^{(p)}$s are defined as above while
\beq
 \mc{W}\Big( \big\{ C_a^{(p)}, D_a^{(p)} \big\} \Big) \, = \,  \pl{p=1}{k-1}\op{S}\Big( \, \overleftarrow{ \bs{A}^{(p)} }  \mid \overleftarrow{ \bs{C}_2^{(p)} }\cup  \overleftarrow{ \bs{C}_1^{(p)} } \Big)
\cdot  \pl{p=2}{k}\op{S}\big( \bs{A}^{(p)} \mid \bs{D}_1^{(p)}\cup  \bs{D}_2^{(p)} \big)
\cdot  \pl{p=1}{k}\mc{F}^{(\op{O}_p)} \Big( \overleftarrow{ \bs{C}_2^{(p-1)} } + \i \pi \ov{\bs{e}}_{\veps_{p}};  \bs{D}_2^{(p)}  \Big) \;.
\label{definition sommant W}
\enq

The main advantage of partitioning the sets as above is that one may easily resolve
the constraints imposed by the $\De$ factors.
To start with, by \eqref{ecriture contraintes Delta sur entrees As de type k-1}, it holds that
\beq
 \bs{A}_s^{(k-1)}  \, = \,  \bs{B}_s^{(k)}  \qquad \e{with} \qquad s=1,\dots, k-1 \;.
\enq
Further, by \eqref{ecriture contraintes Delta sur entrees As de type k-1},
\beq
 \bs{A}_s^{(k-2)}  \; = \; \overrightarrow{ \big( \bs{A}_s^{(k-1)} \cup \bs{\ga}^{(ks)}   \big)}^{ \sg_s^{(k-2)} }  \qquad \e{for} \qquad s=1, \dots, k-2
\enq
so that one may substitute the expression for the entries of $ \bs{A}_s^{(k-1)} $ so as to get that
\beq
 \bs{A}_s^{(k-2)}  \; = \; \overrightarrow{ \big( \bs{B}_s^{(k)} \cup \bs{\ga}^{(ks)}   \big)}^{ \sg_s^{(k-2)} }  \qquad s=1, \dots, k-2
\enq
where one should understand the resulting vector as being obtained from a direct concatenation of the elements of the sets $B_s^{(k)}$ and $\ga^{(ks)}$
followed by a global permutation of the entries which reshuffles the order and finally, by producing the index ordered vector out of such set.

Similarly, the very same handlings yield that
\beq
D_1^{(k-1)} \; = \; \bigcup\limits_{s=1}^{k-2} \Big\{ B_s^{(k)} \cup \ga^{(ks)}   \Big\}^{ \sg_s^{(k-2)} }\;,
\enq
where one concatenates the sets $B_s^{(k)}$ and $\ga^{(ks)}$
and then applies the permutation $\sg_s^{(k-2)}$ so as to re-shuffle the elements and obtain the ordered set $D_1^{(k-1)}$.

We now assume that we have already established that, upon performing a translation in the various sums over permutations, the $\De$
constraint implies the relations for all $p \geq p_0$ and $s=1,\dots, p$
\beq
\bs{A}_s^{(p)}  \; = \; \overrightarrow{ \big( \bs{B}_s^{(k)} \cup \bs{\ga}^{(ks)} \cup \cdots \cup \bs{\ga}^{(p+2 s)}  \big)}^{ \sg_s^{(p)} }  \qquad \quad \e{and} \qquad \quad
D_1^{(p+1)} \; = \; \bigcup\limits_{s=1}^{p} \Big\{ B_s^{(k)} \cup \ga^{(p+2s)}   \Big\}^{ \sg_s^{(p)} }\;.
\label{ecriture vecteurs Aps et Dp+11 apres resolution contrainte}
\enq
There all index ordered sets are to be understood as a concatenation of the elementary sets which realise the union.
This being settled, one then gets that
\beq
\bs{A}_s^{(p_0-1)}  \; = \; \overrightarrow{ \Big(\big( \bs{B}_s^{(k)} \cup \bs{\ga}^{(ks)} \cup \cdots \cup \bs{\ga}^{(p_0+2 s)}  \big)^{ \sg_s^{(p_0)} } \cup \bs{\ga}^{(p_0+1 s)} \Big)}^{ \sg_s^{(p_0-1)} }
\enq
and
\beq
D_1^{(p_0)} \; = \; \bigcup\limits_{s=1}^{p_0-1} \Big\{ \big( B_s^{(k)} \cup \ga^{(ks)} \cup \cdots \cup \ga^{(p_0+2 s)}  \big)^{ \sg_s^{(p_0)} } \cup \ga^{(p_0+1 s)} \Big\}^{ \sg_s^{(p_0-1)} }  \;.
\enq
Then, one proceeds with the change of summation over permutations  $\wt{\sg}_s^{(p_0-1)}=  \sg_s^{(p_0-1)} \cdot \big( \sg_s^{(p_0)} \times \e{id} \big)$, what
establishes the induction hypothesis down to $p_0-1$.

Now observe that the substitution of the $\De$ constraints into the summand given in \eqref{definition sommant W} leads to
\bem
 \mc{W}\Big( \big\{ C_a^{(p)}, D_a^{(p)} \big\} \Big) \, = \,
 \op{S}\big( \bs{A}^{(k)} \mid \bs{D}_1^{(k)} \cup  \bs{D}_2^{(k)} \Big)
\cdot  \pl{p=2}{k-1}\op{S}\big( \bs{D}_1^{(p+1)} \cup  \bs{C}_2^{(p)}  \mid \bs{D}_1^{(p)}\cup  \bs{D}_2^{(p)} \Big) \\
\times \mc{F}_{ }^{(\op{O}_1)} \Big( \bs{D}_1^{(2)}\cup  \bs{C}_2^{(1)}  \Big) \cdot
\pl{p=2}{k}\mc{F}^{(\op{O}_p)} \Big( \overleftarrow{ \bs{C}_2^{(p-1)} } + \i \pi \ov{\bs{e}}_{\veps_{p}};  \bs{D}_2^{(p)}  \Big) \;.
\end{multline}
The above expression is readily seen to be invariant under any permutation of coordinates of any of the vectors
\beq
\bs{C}_2^{(p)} \;, \quad p=1,\dots k-1 \;, \quad    \bs{D}_2^{(p)}  \;, \quad p=2,\dots k \; \quad \e{or} \quad
  \bs{D}_1^{(p)}  \;, \quad p=2,\dots k \; .
\enq
This thus means that one may directly substitute above, using this symmetry
\beq
\bs{C}_2^{(p)} \, \hookrightarrow \, \bs{\ga}^{(p+1 1)}  \cup  \cdots \cup \bs{\ga}^{(p+1 p)}  \;, \quad \e{for} \quad p=1,\dots k-1
\enq
as well as
\beq
\bs{D}_1^{(k)}  \; \hookrightarrow \;  \bs{B}_1^{(k)} \cup \cdots \cup \bs{B}_{k-1}^{(k)}  \qquad \e{and} \qquad
\bs{D}_2^{(k)}  \; = \;  \bs{B}_k^{(k)} \;.
\enq
Further, one also may substitute
\beq
\bs{D}_2^{(p)} \, \hookrightarrow \, \bs{B}_p^{(k)} \cup  \bs{\ga}^{(k p)}  \cup  \cdots \cup \bs{\ga}^{( p+1 p)}  \;, \quad \e{for} \quad p=2,\dots k-1
\enq
as well as, for $p=2,\dots k-1$,
\beq
\bs{D}_1^{(p)} \, \hookrightarrow \, \bs{B}_1^{(k)} \cup  \bs{\ga}^{(k 1)}  \cup  \cdots \cup \bs{\ga}^{( p+1 1)}
\cup  \bs{B}_2^{(k)} \cup  \bs{\ga}^{(k 2)}  \cup  \cdots \cup \bs{\ga}^{( p+1 2)}   \cup \cdots \cup
\bs{B}_{p-1}^{(k)} \cup  \bs{\ga}^{(k p-1)}  \cup  \cdots \cup \bs{\ga}^{( p+1  p-1)}  \;.
\enq
This leads to the substitution relative to the various building blocks of $\mc{W}$
\beq
\op{S}\big( \bs{A}^{(k)} \mid \bs{D}_1^{(k)}\cup  \bs{D}_2^{(k)} \Big) \; \hookrightarrow \;
S\big( \bs{A}^{(k)}  \mid \bs{B}_1^{(k)}  \cup \cdots \cup \bs{B}_k^{(k)}  \Big) \;,
\label{ecriture substitution pour S Ak slash D1k D2k}
\enq
just as
\bem
 \mc{F}_{ }^{(\op{O}_1)} \Big( \bs{D}_1^{(2)}\cup  \bs{C}_2^{(1)}  \Big) \cdot
\pl{p=2}{k}\mc{F}^{(\op{O}_p)} \Big( \overleftarrow{ \bs{C}_2^{(p-1)} } + \i \pi \ov{\bs{e}}_{\veps_{p}};  \bs{D}_2^{(p)}  \Big)  \\
\; \hookrightarrow \;  \pl{p=1}{k} \mc{F}^{(\op{O}_p)}\Big( \overleftarrow{ \bs{\ga}^{(pp-1)} } \cup \cdots \cup \overleftarrow{ \bs{\ga}^{(p1)} }  + \i \pi \ov{\bs{e}}_{\veps_{p}} ,
\bs{B}_p^{(k)}  \cup \bs{\ga}^{(kp)}  \cup \cdots \cup \bs{\ga}^{(p+1p)}   \Big) \;.
\end{multline}
Implementing effectively the substitution at the level of the $S$-matrix product demands some more investigations. First, one has
\bem
\op{S}\big( \bs{D}_1^{(p+1)} \cup  \bs{C}_2^{(p)}  \mid \bs{D}_1^{(p)} \cup  \bs{D}_2^{(p)} \big)  \; \hookrightarrow  \;
\op{S}\big( \bs{B}_1^{(k)} \cup  \bs{\ga}^{(k 1)}  \cup  \cdots \cup \bs{\ga}^{( p+2 1)}
\cup  \bs{B}_2^{(k)} \cup  \bs{\ga}^{(k 2)}  \cup  \cdots \cup \bs{\ga}^{( p+2 2)}   \cup \cdots  \\
\cup \bs{B}_{p}^{(k)} \cup  \bs{\ga}^{(k p)}  \cup  \cdots \cup \bs{\ga}^{( p+2  p)}   \cup
 \bs{\ga}^{(p+1 1)}  \cup  \cdots \cup \bs{\ga}^{(p+1 p)}   \mid
 \bs{B}_1^{(k)} \cup  \bs{\ga}^{(k 1)}  \cup  \cdots \cup \bs{\ga}^{( p+1 1)}  \\
\cup  \bs{B}_2^{(k)} \cup  \bs{\ga}^{(k 2)}  \cup  \cdots \cup \bs{\ga}^{( p+1 2)}   \cup \cdots
\cup \bs{B}_{p}^{(k)} \cup  \bs{\ga}^{(k p)}  \cup  \cdots \cup \bs{\ga}^{( p+1  p)}  \big) \;.
\label{ecriture substitution pour S D1p+1 C2p slash D1p D2p}
\end{multline}
One may then reduce the complicated permutation issued $S$-factor into elementary ones as follows. First one "moves"
$ \bs{\ga}^{(p+1 1)}  $ through the chain
\beq
\bs{B}_2^{(k)} \cup  \bs{\ga}^{(k 2)}  \cup  \cdots \cup \bs{\ga}^{( p+2 2)}   \cup \cdots
\cup \bs{B}_{p}^{(k)} \cup  \bs{\ga}^{(k p)}  \cup  \cdots \cup \bs{\ga}^{( p+2  p)}
\enq
appearing in the right argument, what owing to the identity valid for any $\bs{X}$
\beq
S\big( \bs{A} \cup \bs{B}  \cup \bs{C} \cup \bs{D}  \mid \bs{X}  \big)  \; = \;
S\big(  \bs{B}  \cup \bs{C}   \mid   \bs{C}  \cup \bs{B}    \big) \cdot
S\big( \bs{A} \cup \bs{C}  \cup \bs{B} \cup \bs{D}  \mid \bs{X}  \big)
\enq
produces the factor
\beq
\pl{s=2}{p} S\big(  \bs{B}_{s}^{(k)}   \cup \bs{\ga}^{( p+1 1)}   \mid  \bs{\ga}^{( p+1 1)}  \cup  \bs{B}_{s}^{(k)}   \big)
\cdot \pl{s=p+2}{k} \pl{u=2}{p}
S\big(  \bs{\ga}^{( s u )}    \cup \bs{\ga}^{( p+1 1)}   \mid  \bs{\ga}^{( p+1 1)}  \cup   \bs{\ga}^{( s u )}   \big) \;.
\enq
More generally, for $s=2, \dots, p-1$, one permutes $\bs{\ga}^{(p+1 s)}  $ through the chain
\beq
\bs{B}_{s+1}^{(k)} \cup  \bs{\ga}^{(k s+1)}  \cup  \cdots \cup \bs{\ga}^{( p+1 s+1)}   \cup \cdots
\cup \bs{B}_{p}^{(k)} \cup  \bs{\ga}^{(k p)} \cup  \cdots \cup \bs{\ga}^{( p+2  p)}
\enq
what results in the factor
\beq
\pl{v=s+1}{p} S\big(  \bs{B}_{v}^{(k)}   \cup \bs{\ga}^{( p+1 s)}   \mid  \bs{\ga}^{( p+1 s)}  \cup  \bs{B}_{v}^{(k)}   \big)
\cdot \pl{v=p+2}{k} \pl{u=s+1}{p}
S\big(    \bs{\ga}^{( v u )}   \cup \bs{\ga}^{( p+1s)}   \mid  \bs{\ga}^{( p+1 s)}  \cup   \bs{\ga}^{( v u )}   \big) \;.
\enq
As a consequence, one gets that
\bem
\op{S}\big( \bs{D}_1^{(p+1)} \cup  \bs{C}_2^{(p)}  \mid \bs{D}_1^{(p)} \cup  \bs{D}_2^{(p)} \big)  \; \hookrightarrow  \;
\pl{s=1}{p-1} \pl{v=s+1}{p} S\big(  \bs{B}_{v}^{(k)}   \cup \bs{\ga}^{( p+1 s)}   \mid  \bs{\ga}^{( p+1 s)}  \cup  \bs{B}_{v}^{(k)}   \big) \\
\times \pl{s=1}{p-1} \pl{v=p+2}{k} \pl{u=s+1}{p}
S\big(   \bs{\ga}^{( v u )}   \cup \bs{\ga}^{( p+1s)}   \mid  \bs{\ga}^{( p+1 s)}  \cup   \bs{\ga}^{( v u )}   \big) \;.
\end{multline}

Thus, all-in-all, one has the substitution
\beq
 \mc{W}\Big( \big\{ C_a^{(p)}, D_a^{(p)} \big\} \Big) \, \hookrightarrow \,
 \msc{S}\Big( \big\{ B_{s}^{(k)}; \ga^{(ba)} \big\}  \Big)
 \pl{p=1}{k} \mc{F}^{(p)}\Big( \overleftarrow{ \bs{\ga}^{(pp-1)} } \cup \cdots \cup \overleftarrow{ \bs{\ga}^{(p1)} }  + \i \pi \ov{\bs{e}}_{\veps_{p}} ,
\bs{B}_p^{(k)}  \cup \bs{\ga}^{(kp)}  \cup \cdots \cup \bs{\ga}^{(p+1p)}   \big) \,.
\enq

In order to conclude, one still has to focus on the rewriting of the momentum  $\mc{P} \big( \{ A^{(s)}\}^{k}_{0} ; \{\bs{x}_a \}_{1}^{k} \big) $
as introduced in \eqref{defintion impulsion globale pour produit elements matrice}, this taken the partitioning
\eqref{ecriture sous partitions des Ap et Ak pour calcul multipoints} of the $A^{(p)}$s and the constraints
\label{ecriture contraintes Delta sur entrees As autres que k-1}-\label{ecriture contraintes Delta sur entrees As de type k-1}
which eventually impose that $A_s^{(p)}$s are given by \eqref{ecriture vecteurs Aps et Dp+11 apres resolution contrainte}.
Taken that $\ov{\bs{p}}\big( \bs{A}^{(s)} \big)$ is symmetric in respect to any permutation of the coordinates of $ \bs{A}^{(s)} $, it holds
\bem
 \mc{P} \big( \{ A^{(s)}\}^{k}_{0} ; \{\bs{x}_a \}_{1}^{k} \big) \; = \; \sul{p=1}{k-1} \sul{s=1}{p} \Big\{ \ov{\bs{p}} \, \big( \bs{A}^{(p)}_s  \big) \, + \,
 \ov{\bs{p}}\, \big(   \bs{\ga}^{(p+1s)}  \big)  \Big\} \cdot \bs{x}_{p+1 p}
\; - \; \sul{s=1}{k} \ov{\bs{p}}\, \big(  \bs{B}^{(k)}_s  \big)  \cdot \bs{x}_{k} \\
\;= \; \sul{p=1}{k-1} \sul{s=1}{p} \Big\{ \ov{\bs{p}} \big(  \bs{B}^{(k)}_s   \big) \, + \,
 \sul{u=p+1}{k}\ov{\bs{p}}\, \big( \bs{\ga}^{(us)}  \big)  \Big\} \cdot \bs{x}_{p+1 p}
\; - \; \sul{s=1}{k} \ov{\bs{p}}\, \big( \bs{B}^{(k)}_s \big)  \cdot \bs{x}_{k}  \\
\; = \; \sul{s=1}{k-1}\ov{\bs{p}}\, \big( \bs{B}^{(k)}_s  \big)  \cdot  \sul{p=s}{k-1} \bs{x}_{p+1 p}\; - \; \sul{s=1}{k} \ov{\bs{p}}\, \big( \bs{B}^{(k)}_s  \big)  \cdot \bs{x}_{k}
\, + \, \sul{u=2}{k} \sul{p=1}{u-1} \sul{s=1}{p} \ov{\bs{p}}\, \big(  \bs{\ga}^{(us)}   \big)   \cdot \bs{x}_{p+1 p}  \\
\; = \;  - \; \sul{s=1}{k} \ov{\bs{p}}\, \big( \bs{B}^{(k)}_s  \big)  \cdot \bs{x}_{s}
\, + \, \sul{u=2}{k} \sul{s=1}{u-1} \ov{\bs{p}}\, \big( \bs{\ga}^{(us)}  \big)   \cdot \sul{p=s}{u-1}   \bs{x}_{p+1 p}
\; = \; \msc{P}\big(  \{ B_s^{(k)} \}; \{\ga^{(ba)} \}; \{\bs{x}_s\} \big)  \;.
\end{multline}
as defined through \eqref{definition curly P}. This entails the claim. \qed

\subsection{The smeared integral representation}
\label{Subsection Smeared integral representation}

In order to state the next result, we need to introduce convenient multidimensional notations. Given $\bs{m} \in \mathbb{N}^{p}$ for some $p$
the multifactorial and the length of $\bs{m}$ are, respectively, defined as
\beq
\bs{m}! \; = \; \pl{s=1}{p} m_s!  \qquad \e{and} \qquad |\bs{m}|=\sul{a=1}{p}m_a\;.
\enq

\begin{prop}
\label{ecriture rep serie int mult tronquee pour fct multipts tronquee}
Let $g_1,\dots,g_k\in \mc{S}(\R^{1,1})$, let $G_k$ be as introduced in \eqref{definition fonction Gk}
and let $\bs{r}\, =\, \big( r_1,\dots, r_{k-1} \big) \in \mathbb{N}^{k-1}$.

Given $\bs{f}_{\e{vac}}$ as introduced in \eqref{definition vecteur f vacuum} and provided that each summand $\mc{I}_{ \bs{n} }\big[ G_k\big]$ is well-defined, it holds
\beq
\Big(\bs{f}_{\e{vac}},  \op{O}_1[g_1] \cdot  \op{O}^{(r_1)}_2[g_2] \cdots   \op{O}^{(r_{k-1})}_k[g_k]  \cdot \bs{f}_{\e{vac}} \Big)  \; = \;
\sul{ \bs{n} \in \mc{N}_{\bs{r}} }{}  \f{ \mc{I}_{ \bs{n} }\big[ G_k\big]   }{ \bs{n}! (2\pi)^{|\bs{n}|}}  \cdot \pl{b>a}{k} \ex{ - 2\i\pi \, n_{ba} \om_{ba}  }\;,
\enq
with
\beq
\om_{ba} \, = \, \sul{\ell=a+1}{b} \om_{\op{O}_{\ell}} \; .
\label{definition omega ba}
\enq
 Above one sums over integer valued vectors  $\bs{n}$  belonging  to $\mc{N}_{\bs{r}} \subset  \mathbb{N}^{ \f{k(k-1)}{2} } $:
\beq
\mc{N}_{\bs{r}}  \; = \; \bigg\{ \bs{n} \; = \, \big( n_{21}, n_{31}, n_{32}, n_{41}, \dots , n_{k k-1} \big) \; : \; \sul{u=p+1}{k} \sul{s=1}{p} n_{us} \, = \, r_p \quad p=1,\dots, k-1 \bigg\} \;.
\label{definition ensemble sommation restreint pour troncation fct a k pts}
\enq
Further, the summand takes the explicit form
\beq
\mc{I}_{ \bs{n} }\big[ G_k \big]  \; = \;
\lim_{\veps_1 \tend 0^+} \Int{  \R^{n_{21}}   }{} \dd^{ n_{21} } \ga^{(21)} \cdots
\lim_{\veps_{k-1} \tend 0^+} \hspace{-4mm} \Int{  \R^{n_{k1} + \dots + n_{kk-1} }  }{}  \hspace{-4mm} \dd^{ n_{k1} }  \ga^{(k1)} \cdots \dd^{ n_{kk-1} } \ga^{(kk-1)}
 \; \lim_{\veps_{k} \tend 0^+}   \Big( \mc{S}\cdot \mc{R}[ G_k ] \cdot  \mc{F}_{\e{tot};\bs{\veps}_{k} } \Big) \big(   \bs{\ga}     \big)    \; .
\label{definition sommant base dans serie int mult pour fct 2 pts}
\enq
The integrand contains three building blocks, each being a function of
\beq
\bs{\ga} \, = \, \big( \bs{\ga}^{(21)} ,\bs{\ga}^{(31)} ,\bs{\ga}^{(32)} , \dots , \bs{\ga}^{(kk-1)} \big) \in \Cx^{ n_{ \bs{\ga} } }
\qquad with \qquad  n_{ \bs{\ga} } \, = \,  \sul{b>a}{k} n_{ba}\;.
\label{definition vecteur gamma tot}
\enq
First of all, one has
\beq
\mc{S}\big(   \bs{\ga}     \big) \; = \;
  \pl{  \substack{ v >p \\ p \geq 3 } }{ k }  \pl{u>s}{p-1}
S\big(  \bs{\ga}^{(vu)}  \cup  \bs{\ga}^{(ps)}   \mid   \bs{\ga}^{(ps)}   \cup \bs{\ga}^{(vu)}   \big)  \;,
\label{definition facteur de diffusion complet k pts}
\enq
and
\beq
 \mc{R}[ G] \big(   \bs{\ga}    \big)   \; = \hspace{-2mm} \Int{ \big( \R^{1,1} \big)^{k} }{}  \hspace{-2mm} \pl{s=1}{k}\dd \bs{x}_s
\cdot  G\big(\bs{x}_1,\dots, \bs{x}_k \big) \cdot \pl{b>a}{k} \ex{ \i \ov{\bs{p}}(  \bs{\ga}^{(ba)} ) \cdot \bs{x}_{ba}  } \;.
\label{definition TF space-time de la fct test}
\enq
Finally, given $\bs{\veps}_{k}\, =\, \big(\veps_1,\dots,\veps_{k}\big)$,
\beq
 \mc{F}_{\e{tot};\bs{\veps}_{k} }\big(   \bs{\ga}     \big) \; = \;
 \pl{p=1}{k} \mc{F}^{(\op{O}_p)}\Big( \, \overleftarrow{ \bs{\ga}^{(pp-1)} } \cup \cdots \cup \overleftarrow{ \bs{\ga}^{(p1)} }  + \i \pi \ov{\bs{e}}_{\veps_{p}} ,
 \bs{\ga}^{(kp)}  \cup \cdots \cup \bs{\ga}^{(p+1p)}   \, \Big) \;,
\label{definition F tot eps}
\enq
in which $\ov{\bs{e}}_{\veps}$ is as defined in \eqref{definition ov bs eps avec indice regulateur}.

\end{prop}

\Proof

First of all, starting from \eqref{ecriture rep int fct k pts forme primordiale} specialised to $n_0=n_k=0$ and observing that, owing to the
symmetry properties \eqref{ecriture equation echange local pour M n m} of the individual building blocks constituting $\mc{G}_{\e{tot}}\Big( \big\{  \overrightarrow{A^{(s)}}   \big\}^{k}_{0} ; \bs{\veps}_k \Big)$,
\textit{c.f.} \eqref{definition Fourier direct de G a k veriable Minkowski}, \eqref{definition G produit noyaux integraux operateurs}
 and \eqref{definition G tot de eps}, $\mc{G}_{\e{tot}}$ is symmetric in each of the integration variables, what allows one to recast
\bem
\Big( \op{O}^{(0)}_1[g_1] \cdot  \op{O}^{(r_1)}_2[g_2] \cdots   \op{O}^{(r_{k-1})}_k[g_k]  \cdot \bs{f}_{\e{vac}} \Big)  \\
\; = \;
\lim_{\veps_1 \tend 0^+} \Int{  \R^{r_1}   }{} \f{ \dd^{ r_1 } \a^{(1)} }{ r_1! (2\pi)^{r_1} } \cdots \lim_{\veps_{k-1} \tend 0^+} \Int{  \R^{ r_{k-1} }   }{} \f{ \dd^{ r_{k-1} } \a^{(k-1)} }{ r_{k-1}!  (2\pi)^{r_{k-1}} }
\cdot \lim_{\veps_{k} \tend 0^+}
   \mc{G}\Big( \big\{  \bs{A}^{(s)}   \big\}^{k-1}_{0} ; \bs{\veps}_k \Big)  \cdot  \mc{R}[G_k]\Big(   \big\{  A^{(s)}     \big\}^{k-1}_{0} \Big)  \; ,
\end{multline}
with $A^{(0)}=\emptyset$. Then, by virtue of Proposition \ref{Proposition rep combinatoire pour G tot de epsk} and upon using the notations introduced there, provided that each summand is well defined, one gets
\bem
\Big( \op{O}^{(0)}_1[g_1] \cdot  \op{O}^{(r_1)}_2[g_2] \cdots   \op{O}^{(r_{k-1})}_k[g_k]  \cdot \bs{f}_{\e{vac}} \Big)
\;=\;    \pl{p=1}{k-1} \bigg\{   \sul{ \op{P}_p[  A^{(p)} ]  }{}
\pl{s=1}{p-1} \sul{ \sg_s^{(p-1)} \in \mf{S}_{  |  A^{(p-1)}_s |  } }{}   \bigg\} \cdot   \f{ \pl{ s = 2 }{ k } \ex{-2\i\pi \om_{\op{O}_s} r_{s-1} }  }
{  \pl{s=1}{k-1} \big\{ r_s! (2\pi)^{r_a} \big\}  } \\
\times \lim_{\veps_1 \tend 0^+} \Int{  \R^{r_1}   }{}  \dd  A_1^{(1)} \dd \ga^{(21)} \cdots \lim_{\veps_{k-1} \tend 0^+} \Int{  \R^{ r_{k-1} }   }{}
\dd  A_1^{(k-1)} \cdots A_{k-1}^{(k-1)} \dd \ga^{(k1)} \cdots  \dd \ga^{(kk-1)} \lim_{\veps_{k} \tend 0^+}
 \\
\times \pl{p=2}{k-1} \pl{s=1}{p-1} \De\Big( \bs{A}_s^{(p-1)}    \mid \overrightarrow{ \big( \bs{A}^{(p)}_s  \cup \bs{\ga}^{(p+1s)} \big) }^{ \sg_s^{(p-1)} } \Big)
\cdot \pl{s=1}{k-1} \De\big( \bs{A}_s^{(k-1)}  \mid \emptyset \Big)
\cdot \Big( \mc{S} \cdot \mc{R}[ G_k ]  \cdot \mc{F}_{\e{tot};\bs{\veps}_{k} } \Big)\big( \bs{\ga} \big) \;.
\label{ecriture dvpmt combinatoire fct multipoints sur base integrales en Asp et gammaba}
\end{multline}
We remind that the various collections of variables arising in the integration are subject to the constraints
\beq
| A_s^{(k-1)} | \, = \, 0 \, , \, s=1,\dots, k-1 \quad \e{and} \quad | A_s^{(p-1)} | \; = \; | A_s^{(p)} | \,  + \, | \ga^{(p+1s)} |\quad  \e{for} \quad
\left\{ \ba{ccc}    p & = & 2, \dots, k-1 \vspace{2mm} \\
                    s & = & 1,\dots, p-1
\ea \right. \;,
\enq
and that it also holds
\beq
\sul{s=1}{p} | A_s^{(p)} |  \, + \,  \sul{s=1}{p} | \ga^{(p+1s)} | \, = \, r_p \;.
\enq
At this stage, one may take successively the integrals over $A_1^{(1)}, A_s^{(2)}, \dots, A_s^{(k-2)}$. For convenience, we set
\beq
| \ga^{(ba)} | \, = \, n_{ba} \qquad  \e{and} \qquad  | A_s^{(p)} | = m_{ps} \;.
\enq
Once that the variables building the sets $A_s^{(p)}$ are all integrated, one may simplify the summation over the partitions and the permutations by simply evaluating their cardinality,
what yields
\bem
\Big( \op{O}^{(0)}_1[g_1] \cdot  \op{O}^{(r_1)}_2[g_2] \cdots   \op{O}^{(r_{k-1})}_k[g_k]  \cdot \bs{f}_{\e{vac}} \Big) \;=\;
 \sul{ \mc{C} }{}\pl{p=1}{k-1} \Bigg\{ \f{1}{r_p!} \cdot \f{ r_p!  }{ \pl{s=1}{p} ( n_{p+1s}! \cdot  m_{ps}! )    }  \cdot \pl{s=1}{p-1} m_{p-1s}!  \Bigg\} \\
 \times \pl{ s=2 }{ k } \ex{-2\i\pi \om_{\op{O}_s}r_{s-1} }  \cdot
\f{ \mc{I}_{ \bs{n} }\big[ G_k \big] }{ (2\pi)^{ |\bs{n}| } }  \;.
\end{multline}
There $\bs{n}\; = \;  \big( n_{21}, n_{31}, n_{32}, \dots , n_{k k-1} \big)$ and the summation over $m_{ba}, n_{ba}$ is subject to the constraints
\bem
\mc{C}\; = \; \Bigg\{ m_{ba}, n_{ba} \in \mathbb{N} \; : \;   \sul{s=1}{p} (n_{p+1s}+m_{ps}) \, =\, r_p \; , \; p=1,\dots, k-1 \;,   \\
m_{p-1s} \, = \, m_{ps} + n_{p+1s} \quad \e{for} \quad \left| \ba{c} p=2,\dots , k-1  \\  s=1,\dots, p-1 \ea \right.
\quad \e{and} \quad m_{k-1s} =0  \quad \e{for} \quad  s=1,\dots, p-1  \Bigg\} \;.
\end{multline}
It is important to note that, at this stage, the original object of interest already appears as a finite linear combination of
integrals $\mc{I}_{ \bs{n} }\big[ G_k \big]$ which are all well defined by assumption. Since passing from the integrals present in
\eqref{ecriture dvpmt combinatoire fct multipoints sur base integrales en Asp et gammaba} to a linear combination of  $\mc{I}_{ \bs{n} }\big[ G_k \big]$s
solely involves the action of Dirac masses, which is computed trivially, one gets that the individual terms appearing in the linear combination given in
\eqref{ecriture dvpmt combinatoire fct multipoints sur base integrales en Asp et gammaba} are also well defined. This thus justifies the finite sum-integral splitting
which was made in the first part of the proof.

Now, upon taking into account that $m_{k-1s} =0$, the combinatorial factor is readily reorganised into
\beq
\pl{p=1}{k-1} \Bigg\{ \f{1}{r_p!} \cdot \f{ r_p!  }{ \pl{s=1}{p} ( n_{p+1s}! \cdot  m_{ps}! )    }  \cdot \pl{s=1}{p-1} m_{p-1s}!  \Bigg\} \; = \; \pl{ b > a }{ k } \f{ 1 }{ n_{ba}! } \;.
\enq
The remaining constraints $m_{p-1s} \, = \, m_{ps} + n_{p+1s}$ may be solved, under the boundary condition $m_{k-1s} =0$ as
\beq
m_{ts} \; = \; \sul{p=t+2}{k}n_{ps} \qquad \e{for} \qquad s \in \intn{1}{t} \quad \e{and} \quad t \in \intn{1}{k-2} \;.
\enq
One thus gets that $\sul{s=1}{p}m_{ps} \, = \,  \sul{u=p+2}{k} \sul{s=1}{p} n_{us}$ what thus leads to the constraint
\beq
r_p \, = \,  \sul{u=p+1}{k} \sul{s=1}{p} n_{us}
\label{ecriture des entiers rp via les nab}
\enq
on the remaining summation integers $n_{ba}$. Finally, one observes that
\beq
\sul{\ell=1}{k-1} \om_{ \op{O}_{\ell+1} } r_{\ell} \; = \; \sul{b>a}{k} n_{ba} \sul{\ell=a+1}{b} \om_{ \op{O}_{\ell} } \;,
\enq
what entails the claim. \qed

\section{The \textit{per se} correlation function}
\label{Section Per se Correlation functions}

\subsection{Various auxiliary bounds}
\label{SubSection Auxiliary Bounds}

\begin{defin}

 Given a set of variables $\bs{z} \in \Cx^n$ and $\bs{\nu}=(\nu_1,\dots, \nu_n)\in \mathbb{N}^{n}$ we denote
\beq
\Dp{\bs{z}}^{\bs{\nu}} f (\bs{z}) \; = \; \pl{a=1}{n} \Dp{z_a}^{\nu_a} f(\bs{z}) \;.
\enq
Further,  given $\eta>0$, the open strip of width $\eta$ around the real axis is denoted as
\beq
\mc{S}_{\eta} \, = \, \big\{z \in \Cx \, : \, |\Im z | < \eta \big\} \;.
\enq
The ring of holomorphic functions on $U\subset \Cx^n$ open is denoted by $\mc{O}(U)$.

\end{defin}

\begin{lemme}
\label{Lemme decomposition FF en partie sing et reguliere}

Let $\bs{\ga}^{(ba)} \in \Cx^{n_{ba}}$ for any $k \geq b>a \geq 1$ and let
$\bs{\ga} \in \Cx^{ n_{\bs{\ga}} }$, $n_{\bs{\ga}}$ be as given through \eqref{definition vecteur gamma tot}.

Further, let
\beq
\bs{A}^{(p)} \; = \; \overleftarrow{ \bs{\ga}^{(p p-1)} } \cup \cdots \cup \overleftarrow{ \bs{\ga}^{(p  1)} } \qquad
\quad for \quad p=2,\dots, k
\label{definition vecteur Ap}
\enq
and let
\beq
\bs{B}^{(p)}  \; = \; \bs{\ga}^{( k p)}  \cup \cdots \cup \bs{\ga}^{(p + 1 p)}
\quad for \quad p=1,\dots, k-1 \;.
\label{definition vecteur Bp}
\enq
There exist $\eta, \eta^{\prime}>0$ and maps $\bs{\ga} \mapsto h_1(\bs{\ga})$,  $(\bs{\ga}, \veps) \mapsto h_p(\bs{\ga}, \veps)$
with  $p=2,\dots, k $, satisfying
\beq
\mc{F}^{(\op{O}_1)}\big(  \bs{B}^{(1)}   \big) \; = \; h_1\big(    \bs{\ga}  \big) \quad, \quad
\mc{F}^{(\op{O}_k)}\big( \bs{A}^{(k)} +\i\pi \ov{\bs{e}}_{\veps_{k}}  \big) \; = \; h_k\big(   \bs{\ga} , \veps_{k} \big)
\label{ecriture FF 1 et FF k en B1 et Ak}
\enq
and, for $p=2,\dots, k-1$, using the notations of \eqref{definition vecteur out plus contraction deux vars},
\beq
\mc{F}^{(\op{O}_p)}\big(  \bs{A}^{(p)}  \, + \, \i\pi \ov{\bs{e}}_{ \veps_{p} } ,  \bs{B}^{(p)}  \big) \; = \;
\f{  \pl{r>\ell}{ |A^{(p)} | } A_{r\ell}^{(p)} \cdot   \pl{r<\ell}{ |B^{(p)} | } B_{r\ell}^{(p)}   }
{ \pl{ r=1 }{ |A^{(p)} |  } \pl{ \ell=1 }{ |B^{(p)} |  }  \big( A_{r}^{(p)}  - B_{\ell}^{(p)}  - \i \veps_{p} \big)   }
\cdot h_p\big(  \bs{\ga}, \veps_{p}\big) \;.
\label{ecriture FF p dpt de Bp et Ap}
\enq
These maps are such that
\begin{itemize}

 \item $h_1 \in \mc{O}\Big( \mc{S}_{\eta}^{ n_{\bs{\ga}} } \Big)$;

 \item  pointwise in $\veps$ such that $|\veps| \,  < \, \eta^{\prime} $
$\bs{\ga} \mapsto h_p(\bs{\ga},\veps) \in \mc{O}\Big( \mc{S}_{\eta}^{ n_{\bs{\ga}} } \Big)$

\item for fixed $\bs{\ga} \in \mc{O}\Big( \mc{S}_{\eta}^{ n_{\bs{\ga}} } \Big)$,
$\veps \mapsto  h_p(\bs{\ga},\veps)$ is smooth for $|\veps|<\eta^{\prime}$.

\end{itemize}
Finally, for any $\bs{m} \in \mathbb{N}^{ n_{\bs{\ga}} }$ there exists $C>0$ such that uniformly in $  \bs{\ga}  \in \mc{S}_{ \tilde{\eta}}^{  n_{\bs{\ga}}  }$
with $0 < \tilde{\eta} < \eta $,
\beq
\bigg| \Dp{ \bs{\ga}  }^{\bs{m}  } \pl{a=1}{k} h_a\big(   \bs{\ga} , \veps_{a} \big)  \bigg| \; \leq \;
C \pl{ b > a }{ k } \pl{ j = 1 }{ n_{ba} } \Big| \cosh \Re\big[  \ga_{j}^{(ba)} \big] \Big|^{ n_{\bs{\ga}} +1 + \mf{w} } \qquad with \qquad
\mf{w}\; = \; \sul{a=1}{k}\mf{w}_{ \op{O}_a }\;.
\label{ecriture borne sur derivee produits des hp}
\enq
Above, it is to be understood that $ h_1\big(   \bs{\ga} , \veps \big) \equiv h_1\big(   \bs{\ga}  \big)$ and $\mf{w}_{ \op{O}_a }$ refers to the growth index of the operator $\op{O}_a$,
c.f. {\bf Bootstrap Axioms I-IV}.

\end{lemme}

\Proof

We first focus on establishing appropriate bounds for the form factor $\mc{F}^{(\op{O})}\Big(  \bs{\a}_p + \i\pi \ov{\bs{e}}_{\veps}, \bs{\vth}_q \Big)$
with $ \bs{\a}_p \in \Cx^{p}$ and $ \bs{\vth}_q \in \Cx^{q}$.
Recalling the $\mc{K}$ transform \eqref{definition K transformee fct p} representation of a form factor \eqref{solution eqns bootstrap via K transformee}
and using that for
\beq
\bs{\be}_n \, = \, \Big( \,  \bs{\a}_p + \i\pi \ov{\bs{e}}_{\veps},\bs{\vth}_q \Big)
\label{ecriture choix specifique betan coupe entre alphas et thetas}
\enq
one has the decomposition
\beq
\pl{k<s}{n} \sinh\big( \be_{ks} \big) \; = \; (-1)^{pq} \pl{k<s}{p} \sinh\big( \a_{ks} \big) \cdot \pl{k<s}{q} \sinh\big( \vth_{ks} \big)
\cdot \pl{k=1}{p} \pl{s=1}{q} \sinh \big( \a_k - \vth_s - \i\veps \big) \; ,
\enq
one is lead to the contour integral representation for the $\mc{K}$-transform \eqref{definition K transformee fct p}
\beq
  \mc{K}_{n}\big[ p_n^{(\op{O})}  \big]\big(   \bs{\a}_p + \i\pi \ov{\bs{e}}_{\veps}, \bs{\vth}_q \big) \; = \;
  \f{   \pl{k=1}{p} \big( \cosh \a_k \big)^{\mf{w}_{\op{O}}  +q+1 }  \pl{k=1}{q} \big( \cosh \vth_k \big)^{\mf{w}_{\op{O}}  +p+1 }   }{ \pl{k=1}{p} \pl{s=1}{q} \sinh \big( \a_k - \vth_s - \i\veps \big)  }
\cdot   \mc{U}_{n} \big(  \bs{\a}_p + \i\pi \ov{\bs{e}}_{\veps}, \bs{\vth}_q \big)
\label{ecriture rep int pour K transformee solution Pn}
\enq
valid provided that $\bs{\a}_p \in \mc{S}_{\eta}^p$, $\bs{\vth}_q \in \mc{S}_{\eta}^q$ for some $\eta>0$ and small enough.
Above, given sequences
\beq
\eta< \eta_1<\dots < \eta_p  \qquad  \e{and} \qquad  \eta< \eta_1^{\prime}<\dots < \eta_q^{\prime}
\enq
with $\eta_p, \eta_q^{\prime}>0$ and small enough, we agree upon
\beq
\mc{U}_{n} \big(   \bs{\a}_p + \i\pi \ov{\bs{e}}_{\veps}, \bs{\vth}_q \big) \; = \; \pl{a=1}{p} \Oint{  \Dp{}\mc{S}_{\eta_a} }{} \f{ \dd x_a   }{ 2\i\pi  }
\cdot  \pl{a=1}{q} \Oint{  \Dp{}\mc{S}_{\eta_a^{\prime} } }{} \f{ \dd y_a   }{ 2\i\pi  } u_n\big(  \bs{\a}_p ;  \bs{x}_p \mid  \bs{\vth}_q ;  \bs{y}_q \big) \;.
\label{ecriture rep int pour Un}
\enq
The integrand appearing above takes the explicit form
\bem
 u_n\big(  \bs{\a}_p ;  \bs{x}_p \mid \bs{\vth}_q ;  \bs{y}_q \big) \, = \,
\pl{a=1}{p} \Bigg\{ \f{ 1 }{    \sinh(x_a - \a_a) \, \big( \cosh x_a \big)^{\mf{w}_{\op{O}}  +q+1 }     } \Bigg\}  \\
\times \pl{a=1}{q} \Bigg\{ \f{ 1 }{   \sinh(y_a - \vth_a)  \, \big( \cosh y_a \big)^{\mf{w}_{\op{O}}  +p+1 }   }  \Bigg\}
\sul{  \bs{\ell}_n \in \{0,1\}^n}{} (-1)^{ \ov{\bs{\ell}}_n } \pl{k<s}{p} \bigg\{ 1 \, - \, \i \f{ \ell_{ks} \cdot \sin[2\pi \mf{b} ] }{ \sinh(x_{ks})  }  \bigg\}  \cdot
 \pl{k<s}{q} \bigg\{ 1 \, - \, \i \f{ \ell_{ks} \cdot \sin[2\pi \mf{b} ] }{ \sinh(y_{ks})  }  \bigg\}  \\
\times \pl{k=1}{p} \pl{s=1}{q} \Big\{ \sinh \big( x_k - y_s - \i\veps \big)  + \i \ell_{k (s+p)} \sin[2\pi \mf{b} ]  \Big\}
\cdot p_n^{(\op{O})}\big(   \bs{x}_p + \i\pi \ov{\bs{e}}_{\veps}, \bs{y}_q \mid \bs{\ell}_n \big) \;.
\end{multline}
 First of all, it follows from the bounds on $p_n^{(\op{O})}$ at $\infty$, \textit{c.f.} Proposition \ref{Proposition representation FF general en terme K transformee},
 that the contour integral defining $\mc{U}_{n}$ is well defined since
\beq
\Big| u_n\big(  \bs{\a}_p ;  \bs{x}_p \mid  \bs{\vth}_q ;  \bs{y}_q \big) \Big| \, \leq \, C
\pl{a=1}{p}   \bigg|  \f{ 1 }{     \cosh x_a      }  \bigg| \cdot \pl{a=1}{q}  \bigg|  \f{ 1 }{     \cosh y_a      }  \bigg| \;
\enq
for some $C>0$  and uniformly in $\big( \, \bs{\a}_p , \bs{\vth}_q \big) \in \mc{S}_{\eta}^p\times \mc{S}_{\eta}^q$ and $x_a \in  \mc{S}_{\eta_a}$, $y_a \in  \mc{S}_{\eta_a^{\prime}}$.
To check that \eqref{ecriture rep int pour K transformee solution Pn} does indeed hold, one takes the contour integral
definition $\mc{U}_n$ by the residues located inside each of the integration contours for $x_a$ and $y_a$ taken singly.
One should note that owing to $p_n^{(\op{O})}$ satisfying axiom $a)-d)$ stated in Proposition \ref{Proposition representation FF general en terme K transformee}
$u_n$ does not have poles at $x_a = x_b$, resp. $y_a=y_b$ so that its only poles inside the domain of integration for each variable
are at $x_a=\a_a$, $a=1,\dots, p$ and  $y_a=\vth_a$, $a=1,\dots, q$, what immediately leads to the claim.

It follows from the integral representation \eqref{ecriture rep int pour Un} that
$\big( \, \bs{\a}_p , \bs{\vth}_q \big) \mapsto \mc{U}_{n} \big(   \bs{\a}_p + \i\pi \ov{\bs{e}}_{\veps}, \bs{\vth}_q \big)$
is holomorphic on  $\mc{S}_{\eta}^p\times \mc{S}_{\eta}^q$. Moreover, derivations under the integral and straightforward bounds ensure that
for any $\bs{m}_p \in \mathbb{N}^p$ and $\bs{s}_q \in \mathbb{N}^q$, there exists $C>0$ such that
\beq
\Big| \Dp{ \bs{\a}_p }^{ \bs{m}_p } \cdot \Dp{ \bs{\vth}_q }^{ \bs{s}_q } \cdot \mc{U}_{n} \big(   \bs{\a}_p + \i\pi \ov{\bs{e}}_{\veps}, \bs{\vth}_q \big)  \Big| \; \leq \;
C
\enq
uniformly in $\veps$ small enough and throughout $\mc{S}_{\eta}^p\times \mc{S}_{\eta}^q$.

We now focus on estimating the growth of the remaining factor containing $\op{F}$. Recalling the Barnes function representation for $\op{F}$
given in \eqref{expression twpo body scattering via Barnes}, one may represent
\beq
\op{F}(\be)  \; = \; - \f{ \sin \big[ \tf{\i \be}{2} \big] }{  \pi }  \cdot \varpi_{\mf{b}}\Big( \f{\i \be }{2 \pi } \Big) \cdot \varpi_{ \hat{\mf{b}} }\Big( \f{\i \be }{2 \pi } \Big)
\label{ecriture representation explicite pour F}
\enq
with
\beq
\varpi_{\mf{b}}\big( \mf{z}  \big)  \; = \;  G\left( \ba{cccc} 1-\mf{b} - \mf{z} \, ,  &   1 + \mf{z} \,,  &   2 - \mf{b} + \mf{z}  \, ,  &   1 - \mf{z}    \\
  1 - \mf{z}      \, ,  &   1+  \mf{b} + \mf{z}   \, ,  &   1 + \mf{z}  \, ,  &  \mf{b} - \mf{z}   \ea \right) \;.
\enq

A direct calculation  building on \eqref{ecriture DA de rapport de G shifte sur G} yields that for $\be = \ga + \i \ups$ with $-\eta \leq \ups \leq \eta$ and
$\ga \tend +\infty$, one has
\beq
\varpi_{\mf{b}}\big( \mf{z}  \big)  \; = \;  \exp\bigg\{  \hat{\mf{b}} \, \big[ \i \pi \e{sgn}(\ga) + 2 \ln 2\pi + 2\i\pi \e{sgn}(\ga) \cdot \mf{z}  \big] \; + \; \e{O}\big( \mf{z}^{-1}  \big)  \bigg\}
\quad \e{where} \quad
\mf{z} \, = \, \f{\i \be }{2 \pi }
\enq
and with a remainder that is uniform in $\ups$ and differentiable to all orders.
From there one infers that
\beq
\op{F}(\ga \, + \, \i \ups ) \, = \, 1  \; + \; \e{O}\big( \mf{z}^{-1}  \big) \quad \e{as} \quad \ga \tend \pm \infty \;.
\label{ecriture DA pour facteur F}
\enq
Now, for the choice \eqref{ecriture choix specifique betan coupe entre alphas et thetas} of $\bs{\be}_n$, one has the decomposition
\beq
\pl{a<b}{n} \op{F}\big( \be_{ab} \big) \; = \; \pl{a<b}{p}  \a_{ab}  \cdot   \pl{a<b}{q} \vth_{ab} \cdot
\pl{a<b}{p} \f{ \op{F}\big( \a_{ab} \big) }{  \a_{ab}  } \cdot  \pl{a<b}{q} \f{ \op{F}\big( \vth_{ab} \big)  }{ \vth_{ab}  }
\cdot  \pl{a=1}{p} \pl{b=1}{q} \op{F}\big( \a_a -\vth_b +\i (\pi - \veps)  \big) \;.
\enq
Now, it follows from \eqref{ecriture representation explicite pour F} that $\be \tend \tf{ \op{F}(\be) }{ \be }$ is analytic on $\op{D}_{0,\tau}$ for some
$\tau>0$. Hence, uniformly in
\beq
|\be|=|\ga + \i\ups| \leq \tf{\tau}{2} \, \qquad \e{it} \; \e{holds}  \qquad \Big| \Dp{\ga}^m \cdot \f{ \op{F}(\be) }{ \be } \Big| \; \leq \; C \;,
\enq
for some $m$-dependent $C>0$. Further, given $|\ups| \leq \eta$ with $\eta$ small enough, and $|\be|=|\ga + \i\ups| \geq  \tf{\tau}{2}$ the denominator term $\tf{1}{\be}$ is non-zero and has bounded derivatives
as much as the numerator $\op{F}(\be)$ owing to the differentiability and uniformness of its asymptotic expansion \eqref{ecriture DA pour facteur F} as well as the fact that
$\op{F}(\be)$ is analytic in a strip of fixed with around $\R$. Hence, this reasoning ensures that
\beq
\max_{s \leq r } \sup_{ \ups \leq 2 \eta  } \sup_{ \ga \in \R} \Big| \Dp{\ga}^{s} \cdot \f{  \op{F}( \ga + \i\ups ) }{  \ga + \i\ups } \Big| \; \leq \; C \;.
\enq
Likewise,  the differentiability and uniformness of the asymptotic expansion \eqref{ecriture DA pour facteur F} yield
\beq
\max_{s, s^{\prime} \leq r } \sup_{ \ups \leq 2 \eta  } \sup_{ \ga, \ga^{\prime} \in \R}
\Big| \Dp{\ga}^{s} \cdot \Dp{\ga^{\prime}}^{s^{\prime}} \cdot  \op{F}\big( \ga-\ga^{\prime} + \i(\ups+\pi - \veps) \big)   \Big| \; \leq \; C \;.
\enq
From there, it follows that for any $\bs{m}_p \in \mathbb{N}^p$ and $\bs{s}_q \in \mathbb{N}^q$, there exists $C>0$ such that
\beq
\Bigg| \Dp{ \bs{\a}_p }^{ \bs{m}_p } \cdot \Dp{ \bs{\vth}_q }^{ \bs{s}_q } \cdot \bigg\{  \pl{a<b}{p} \f{ \op{F}\big( \a_{ab} \big) }{  \a_{ab}  } \cdot  \pl{a<b}{q} \f{ \op{F}\big( \vth_{ab} \big)  }{ \vth_{ab}  }
\cdot  \pl{a=1}{p} \pl{b=1}{q} \op{F}\big( \a_a -\vth_b +\i (\pi - \veps)  \big)   \bigg\} \Bigg| \; \leq \; C \;,
\enq
this uniformly in $\a_a \, =\, \ga_a + \i \ups_a$,  $\vth_a \, = \, \ga_a^{\prime} + \i \ups_a^{\prime}$ with $\ga_a, \ga^{\prime}_a \in \R$ and $|\ups_a|, |\ups^{\prime}_a|\,  \leq \, \eta$.

\noindent The above discussion thus provides one with the representation
\beq
\mc{F}^{(\op{O})}\big(  \bs{\a}_p + \i\pi \ov{\bs{e}}_{\veps}, \bs{\vth}_q \big) \; = \;
\f{  \pl{a<b}{p}  \a_{ab}  \cdot   \pl{a<b}{q} \vth_{ab} }{ \pl{k=1}{p} \pl{s=1}{q}  \Big\{ \a_k - \vth_s - \i\veps \Big\}  }
\cdot \mc{H}^{(\op{O})}\big(  \bs{\a}_p + \i\pi \ov{\bs{e}}_{\veps}, \bs{\vth}_q \big)
\;,
\enq
where
\bem
\mc{H}^{(\op{O})}\big(  \bs{\a}_p + \i\pi \ov{\bs{e}}_{\veps}, \bs{\vth}_q \big) \; = \;
\pl{k=1}{p} \big( \cosh \a_k \big)^{\mf{w}_{\op{O}}  +q+1 }  \pl{k=1}{q} \big( \cosh \vth_k \big)^{\mf{w}_{\op{O}}  +p+1 }
\pl{a<b}{p} \f{ \op{F}\big( \a_{ab} \big) }{  \a_{ab}  } \cdot  \pl{a<b}{q} \f{ \op{F}\big( \vth_{ab} \big)  }{ \vth_{ab}  } \\
\times \pl{k=1}{p} \pl{s=1}{q} \bigg\{ \f{ \a_k - \vth_s - \i\veps }{\sinh \big( \a_k - \vth_s - \i\veps \big) } \bigg\} \cdot  \pl{a=1}{p} \pl{b=1}{q} \op{F}\big( \a_a -\vth_b +\i (\pi - \veps)  \big)   \cdot
 \mc{U}_{n} \big(   \bs{\a}_p + \i\pi \ov{\bs{e}}_{\veps}, \bs{\vth}_q \big) \;.
\end{multline}
Since $x \mapsto \tf{ x }{ \sinh(x) }$ is bounded and analytic in a strip around the real axis and decays exponentially fast at $\Re(x) \tend \pm \infty$,
the previous bounds yield that for any $\bs{m}_p \in \mathbb{N}^p$ and $\bs{s}_q \in \mathbb{N}^q$, there exists $C>0$ such that
\beq
\Bigg| \Dp{ \bs{\a}_p }^{ \bs{m}_p } \cdot \Dp{ \bs{\vth}_q }^{ \bs{s}_q } \cdot \mc{H}^{(\op{O})}\big(  \bs{\a}_p + \i\pi \ov{\bs{e}}_{\veps}, \bs{\vth}_q \big) \Bigg|
\; \leq \; C  \pl{k=1}{p} \big| \cosh \Re[\a_k] \big|^{\mf{w}_{\op{O}} +q+1 } \cdot  \pl{k=1}{q} \big| \cosh \Re[\vth_k] \big|^{\mf{w}_{\op{O}} +p+1 } \;,
\label{ecriture bornes complete sur derivee de HO}
\enq
this uniformly in $\big( \bs{\a}_p  ,\bs{\vth}_q \big) \in  \mc{S}_{\eta}^{p}\times \mc{S}_{\eta}^{q}$.

Recalling the  vectors $\bs{A}^{(p)}$ and $\bs{B}^{(p)}$ introduced in \eqref{definition vecteur Ap}-\eqref{definition vecteur Bp},
one immediately infers \eqref{ecriture FF 1 et FF k en B1 et Ak}-\eqref{ecriture FF p dpt de Bp et Ap} with
\beq
h_1\big( \bs{\ga} \big) \; = \; \mc{H}^{(\op{O}_1)}\big(  \bs{B}^{(1)} \big)\; ,  \qquad
h_k\big( \bs{\ga}, \veps_{k} \big) \; = \; \mc{H}^{(\op{O}_k)}\big(  \bs{A}^{(k)}  + \i\pi \ov{\bs{e}}_{\veps_{k} } \big)
\enq
and, for $p=2,\dots, k-1$,
\beq
h_p\big(  \bs{\ga}, \veps_{p}  \big) \; = \; \mc{H}^{(\op{O}_p)}\big(   \bs{A}^{(p)}  + \i\pi \ov{\bs{e}}_{\veps_{p} }  \,  , \, \bs{B}^{(p)} \big)\; .
\enq
The bound \eqref{ecriture borne sur derivee produits des hp} then appears as a direct consequence of \eqref{ecriture bornes complete sur derivee de HO}.

\begin{lemme}
\label{Lemme borne sur derivees du facteur S total}

 Let $\bs{\ga}  \in \Cx^{n_{\bs{\ga}}}$, $n_{ \bs{\ga} }$ be as introduced in \eqref{definition vecteur gamma tot}, and $\mc{S}$
 as defined in \eqref{definition facteur de diffusion complet k pts}. Then, there exists $\eta >0$
 such that for any $\bs{m} \in \mathbb{N}^{  n_{ \bs{\ga} }  }$ there exists $C>0$ such that uniformly in $  \bs{\ga}  \in \mc{S}_{\eta}^{  n_{ \bs{\ga} }  }$,
\beq
\bigg| \Dp{ \bs{\ga}  }^{\bs{m}  } \mc{S}\big(   \bs{\ga}    \big) \bigg| \; \leq \;
C \;.
\label{ecriture borne sur derivee facteur S total pour k multipts}
\enq

\end{lemme}

\Proof

One observes that one has the product decomposition
\beq
 \mc{S}\big(   \bs{\ga}    \big) \; = \;
  \pl{  \substack{ v >p \\ p \geq 3 } }{ k }  \pl{u>s}{p-1} \pl{ j=1 }{ n_{vu} } \pl{ \ell=1 }{ n_{ps} }
S\big(  \ga^{(vu)}_j -  \ga^{(ps)}_{\ell}  \big) \;.
\enq
Further, one has the explicit expression
\beq
S\big( \ga  + \i\ups \big) \; = \; \f{  \sinh\big[ \tfrac{\ga}{2} + \i( \tfrac{\ups}{2}-\pi \mf{b} ) \big] \cosh\big[ \tfrac{\ga}{2} + \i( \tfrac{\ups}{2}+\pi \mf{b} ) \big]  }
{ \cosh\big[ \tfrac{\ga}{2} + \i( \tfrac{\ups}{2}-\pi \mf{b} ) \big] \sinh\big[ \tfrac{\ga}{2} + \i( \tfrac{\ups}{2}+\pi \mf{b} ) \big]   } \;.
\enq
The only singularities of the expression are simple poles, in the case of generic $\mf{b}$, which are located at
\beq
\ga\, = \, 0 \quad \e{and}\; \e{either} \quad \ups \,  = \,  -2\pi \mf{b} + 2\pi n \quad \e{or} \quad \ups \, = \,  2\pi \mf{b} + (2n+1)\pi  \quad \e{with} \quad n\in \mathbb{Z} \, .
\enq
Thus, provided that $|\ups|< \eta$ with $\eta$ small enough,
one has that for any $k\in \mathbb{N}$ there exists $C>0$ such that
\beq
\sup_{\ga \in \R} \Big| \Dp{\ga}^{k} S\big( \ga  + \i\ups \big) \Big| \; \leq \; C \;.
\enq
Thus by the multi-product Leibniz formula, \eqref{ecriture borne sur derivee facteur S total pour k multipts} follows. \qed

\begin{lemme}
\label{Lemme bornes sur derivees facteur TF de la smearing fct}

Let $G \in \mc{S}\big( (\R^{1,1})^k \big)$,  $n_{ \bs{\ga} } \in \mathbb{N}$ and
$\bs{\ga}  \in \Cx^{ n_{ \bs{\ga} } }$  be as introduced in \eqref{definition vecteur gamma tot}.
Finally, let $\mc{R}[G]$ be  as defined through \eqref{definition TF space-time de la fct test}. Then,
for any $\bs{m} \in \mathbb{N}^{ n_{ \bs{\ga} } }$ and $\bs{r} \in \mathbb{N}^{  n_{ \bs{\ga} } }$ there exists $C>0$ such that uniformly in $  \bs{\ga}  \in \R^{  n_{ \bs{\ga} } }$
\beq
\bigg| \Dp{ \bs{\ga} }^{\bs{m} }\mc{R}[ G]\big(   \bs{\ga}     \big) \bigg| \; \leq \;
C \, \norm{ G }_{ \bs{I}; \bs{m} } \,  \pl{b>a}{ k } \pl{s=1}{n_{ba} }  \bigg| \f{ 1 }{  \cosh\Re\big[ \ga_s^{(ba)} \big] }   \bigg|^{ r_s^{(ba)} } \;.
\label{ecriture borne sur derivee facteur R de G total pour k multipts}
\enq
Above, the $\bs{I}=\big( I_1,\dots, I_{k-1} \big)$ norm is defined in terms of the Minkowski coordinates of vectors $\bs{x}_k=\big( x_{k ; 0}, x_{k ; 1 } \big)$:
\beq
\norm{ G }_{ \bs{I} ;\bs{m} }  \; = \; \e{max}\bigg\{  \, \Big| \Big|  \big( 1+\sul{a=1}{k} ||\bs{x}_a|| \big)^{|\bs{m}|} \cdot  \prod_{\ell=1}^{k-1} \Dp{ x_{\ell ; 0} }^{ r_{\ell} }  G   \Big| \Big|_{ L^1\big( (\R^{1,1})^k \big)  }
                                                  \; : \; 0 \leq r_{\ell} \leq \sul{u=\ell}{k-1}I_{u}  \quad and \quad 0 \, \leq \, \sul{u=1}{k-1} r_u \, \leq \, \sul{u=1}{k-1}I_{u} \bigg\} \;.
\label{definition norme I de G}
\enq
Furthermore, the components of $\bs{I}$ appearing in \eqref{ecriture borne sur derivee facteur R de G total pour k multipts} take the form
\beq
I_{\ell} \; = \; \sul{ b = \ell + 1 }{ k } \sul{ j=1 }{ n_{ba} } \big( r_j^{(ba)} + |\bs{m}|  \big) \; .
\label{definition IEll}
\enq

Moreover, assume that $G \in \mc{C}^{\infty}_{\e{c}}\big( (\R^{1,1})^k \big)$. Then,
 there exists $\eta >0$ and small enough
 such that for any $\bs{m} \in \mathbb{N}^{ n_{ \bs{\ga} } }$ and $\bs{r} \in \mathbb{N}^{  n_{ \bs{\ga} } }$ there exists $C>0$ such that uniformly in $  \bs{\ga}  \in \mc{S}_{\eta}^{  n_{ \bs{\ga} } }$
 satisfying the constraint
\beq
\Im\Big[ \ov{\bs{p}}(  \bs{\ga}^{(ba)} ) \cdot \bs{x}_{ba}   \Big] \geq 0 \quad for \; any \quad \big( \bs{x}_1, \dots, \bs{x}_{k} \big) \in \e{supp}[G] \;,
\label{contrainte positivite PS Impulsion et difference position}
\enq
the bound \eqref{ecriture borne sur derivee facteur R de G total pour k multipts}
 holds as well.

\end{lemme}

\Proof

We first implement a change of coordinates $\big( \bs{x}_1,\dots, \bs{x}_k \big) \tend \big( \bs{y}_1,\dots, \bs{y}_k \big)$
in the integral representation \eqref{definition TF space-time de la fct test} for  $\mc{R}[G]$ defined as :
\beq
\bs{x}_s  \, = \, \sul{p=s}{k} \bs{y}_p \quad \e{for} \quad s=1,\dots, k \;,
\enq
so that $\bs{y}_k \, = \, \bs{x}_k$ and $\bs{y}_s \, = \, \bs{x}_s \, -\,  \bs{x}_{s+1}$ for $s=1,\dots, k-1$. This ensures that the change of coordinate map is a smooth diffeomorphism
from $ \big( \R^{1,1} \big)^{k} $ onto itself. Moreover, it holds
\beq
\det\Big[ D_{\bs{x}} \bs{y} \Big] \; = \; \det \left( \ba{cccccc}  D_{\bs{x}_1} \bs{y}_1 & 0 & \cdots   \\
                                                        D_{\bs{x}_2} \bs{y}_1 &  D_{\bs{x}_2} \bs{y}_2  &  0 & \cdots  \\
                                                              0 & \ddots &\ddots & 0   \\
                                                          \cdots  &  0 &    D_{\bs{x}_k} \bs{y}_{k-1} &  D_{\bs{x}_k} \bs{y}_k     \ea   \right) \; = \; 1 \;.
\enq
Next, for $b>a$ one has $\bs{x}_{ba} \; = \; - {\displaystyle \sul{p=a}{b-1}} \bs{y}_p$, so that
\beq
\sul{b>a}{k} \ov{\bs{p}}(  \bs{\ga}^{(ba)} ) \cdot \bs{x}_{ba} \; = \; - \sul{b=2}{k} \sul{a=1}{b-1}\sul{\ell = a}{ b-1 } \ov{\bs{p}}(  \bs{\ga}^{(ba)} ) \cdot \bs{y}_{\ell}
\; = \; - \sul{b=2}{k}\sul{\ell = 1}{ b-1 }  \bs{y}_{\ell}  \cdot  \sul{a=1}{\ell} \ov{\bs{p}}(  \bs{\ga}^{(ba)} )
\; = \; - \sul{\ell = 1}{ k-1 }  \bs{y}_{\ell}  \cdot \op{P}_{\ell}\big( \bs{\ga} \big) \;,
\enq
where
\beq
 \op{P}_{\ell}\big( \bs{\ga}  \big) \; = \; \sul{ b = \ell + 1 }{k}  \sul{ a = 1 }{ \ell } \ov{\bs{p}}(  \bs{\ga}^{(ba)} ) \; = \;
 \Big(  \op{P}_{\ell}^{(0)}\big( \bs{\ga}  \big) ,  \op{P}_{\ell}^{(1)}\big( \bs{\ga}  \big)  \Big) \;.
\enq
Thus, upon setting
\beq
f\big( \bs{y}_1,\dots, \bs{y}_k \big) \; = \; G\bigg( \sul{s=1}{k} \bs{y}_s, \sul{s=2}{k} \bs{y}_s , \dots, \bs{y}_k \bigg) \;,
\label{definition f via G dans estimee R de G}
\enq
one gets
\beq
\mc{R}[ G] \big(   \bs{\ga}    \big)   \; = \hspace{-2mm} \Int{ \big( \R^{1,1} \big)^{k} }{}  \hspace{-2mm} \pl{s=1}{k}\dd \bs{y}_s
\; f\big( \bs{y}_1,\dots, \bs{y}_k \big)  \pl{ \ell = 1 }{ k - 1 } \Big\{ \ex{   -\i  \bs{y}_{\ell}  \cdot \op{P}_{\ell}( \bs{\ga}  ) } \Big\} \;.
\enq
This representation is the starting point for establishing the desired bounds on the partial $\ga_{j}^{(ba)}$ derivatives of $\mc{R}[G]$.
This will be done by first evaluating the derivatives through the multi-dimensional Faa-di-Bruno formula.
For that purpose,  we introduce
\beq
  \mf{p}\big( \bs{\ga} \big) \; = \; -\sul{\ell=1}{k-1} \bs{y}_{\ell} \cdot  \op{P}_{\ell}\big( \bs{\ga} \big) \;.
\enq
Further,  given $\bs{m}\, = \, \big( \bs{m}^{(21)}, \bs{m}^{(31)}, \bs{m}^{(32)}, \dots, \bs{m}^{(kk-1)} \Big) \in \mathbb{N}^{ n_{\bs{\ga}} }$ with
$ \bs{m}^{(ba)} \, = \,  \big( m^{(ba)}_1,\dots, m^{(ba)}_{n_{ba}}  \big) \in \mathbb{N}^{n_{ba}}$, we set
\beq
 | \bs{m}   | \, = \, \sul{ b>a }{ k } \sul{ j=1 }{ n_{ba} } m_{j}^{(ba)} \;.
\enq
Next, we introduce the set
\bem
 \mc{C}_s( \bs{m}  , t) \; = \;  \Big\{ \big( k_1, \dots, k_s\big) \in \mathbb{N}^s \;, \; \big( \bs{\ell}_1,\dots, \bs{\ell}_s \big) \in   \big(  \mathbb{N}^{ |\bs{m} | } \big)^{s}  \; : \;
k_a >0 \; , \; \bs{0} \prec  \bs{\ell}_1 \prec \cdots  \prec  \bs{\ell}_s   \\
\quad \e{with} \quad \sul{a=1}{s} k_a \, = \, t \quad \e{and} \quad \sul{a=1}{s} k_a  \bs{\ell}_a \, = \, \bs{m}   \Big\}
\end{multline}
subordinate to the choice of $\bs{m} \in \mathbb{N}^{ n_{\bs{\ga}} }$ and $t \in \intn{ 1 }{ | \bs{m}  | }$. The
definition of the set makes use of the below order on $\mathbb{N}^{ | \bs{m}   | }$: given $\bs{\nu}, \bs{\mu}  \in \mathbb{N}^{ | \bs{m}   | } $ one has
\beq
\bs{\nu} \, \prec  \, \bs{\mu}  \quad \e{if} \; \e{either} \quad  |\bs{\nu}| \, <  \, |\bs{\mu}|
\enq
or, for some  $k \in \intn{ 1 }{  | \bs{m}   | - 1}$,
\beq
  |\bs{\nu}| \, =  \, |\bs{\mu}|  \quad \e{and} \quad
\nu_a \, = \, \mu_a \quad \e{for} \quad a=1,\dots,k \quad \e{while} \quad   \nu_{k+1} \, < \, \mu_{k+1} \;.
\enq
The multi-variable Faa-di-Bruno formula \cite{ConstantineSavitsFaaDiBrunoManyVarsVersion} leads to the explicit expression
\beq
 \f{1}{  \bs{m}  ! }\Dp{ \bs{\ga}  }^{\bs{m} } \cdot  \pl{ \ell = 1 }{ k - 1 } \Big\{  \ex{  -\i  \bs{y}_{\ell}  \cdot \op{P}_{\ell}( \bs{\ga} ) } \Big\}  \; = \;
\sul{ t =1  }{  | \bs{m}   | } (-\i)^{t} \sul{ s =1  }{  | \bs{m}   |  }  \sul{  \mc{C}_s( \bs{m}  , t) }{}
   \pl{j=1}{s}  \Bigg\{ \f{ \big[ \Dp{ \bs{\ga}  }^{ \bs{\ell}_j } \cdot \mf{p}\big( \bs{\ga} \big) \big]^{k_j} }{ k_j! \, ( \bs{\ell}_j !)^{k_j} }  \Bigg\}
\cdot  \pl{ \ell = 1 }{ k - 1 }  \Big\{ \ex{   -\i  \bs{y}_{\ell}  \cdot \op{P}_{\ell}( \bs{\ga} ) } \Big\} \;.
\label{ecriture Faa di Bruno multidimensionnel pour derivee facteur Fourier}
\enq
Therefore, setting
\beq
 f\big( \bs{y}_1,\dots, \bs{y}_k  \mid  \bs{m}  \big)  \; = \;  \bs{m}  !  \,  f\big( \bs{y}_1,\dots, \bs{y}_k   \big)
\sul{ t =1  }{  | \bs{m}   | } (-\i)^{t} \sul{ s =1  }{  | \bs{m}   |  }  \sul{  \mc{C}_s( \bs{m}  , t) }{}
   \pl{j=1}{s}  \Bigg\{ \f{ \big[ \Dp{ \bs{\ga}  }^{ \bs{\ell_j} } \cdot \mf{p}\big( \bs{\ga}  \big) \big]^{k_j} }{ k_j! ( \bs{\ell_j} !)^{k_j} }  \Bigg\}
\enq
one gets that
\beq
\Dp{ \bs{\ga}  }^{\bs{m}  } \cdot  \mc{R}[ G] \big(   \bs{\ga}     \big)   \; = \hspace{-2mm} \Int{ \big( \R^{1,1} \big)^{k} }{}  \hspace{-2mm} \pl{s=1}{k}\dd \bs{y}_s
\;  f\big( \bs{y}_1,\dots, \bs{y}_k  \mid  \bs{m}  \big)  \pl{ \ell = 1 }{ k - 1 } \Big\{ \ex{   -\i  \bs{y}_{\ell}  \cdot \op{P}_{\ell}( \bs{\ga}  ) } \Big\} \;.
\enq
Thus, integrating by parts $I_{\ell}$ times in respect to $y_{\ell;0}$ with $\ell=1,\dots, k-1$ yields that
\beq
\Dp{ \bs{\ga}  }^{\bs{m}  } \cdot  \mc{R}[ G] \big(   \bs{\ga}     \big)   \; = \; \pl{ \ell = 1 }{ k - 1 } \bigg\{  \f{  1 }{ \i \op{P}_{\ell}^{(0)}( \bs{\ga}  ) }  \bigg\}^{ I_{\ell} }
\hspace{-2mm} \Int{ \big( \R^{1,1} \big)^{k} }{}  \hspace{-2mm} \pl{s=1}{k}\dd \bs{y}_s
\; \pl{ \ell = 1 }{ k - 1 } \Big\{ \ex{  - \i  \bs{y}_{\ell}  \cdot \op{P}_{\ell}( \bs{\ga} ) } \Big\}
            \pl{ \ell = 1 }{ k - 1 }  \Dp{y_{\ell;0} }^{I_{\ell} } \cdot f\big( \bs{y}_1,\dots, \bs{y}_k  \mid  \bs{m}  \big) \;.
\enq
Here, we remind that $\bs{y}_{\ell} \, = \, \big(  y_{\ell;0}, y_{\ell;1} \big)$. Since $\mf{p}$ is given by a sum of functions of only one variable, the vectors $\bs{\ell}_j$ over which one sums in
\eqref{ecriture Faa di Bruno multidimensionnel pour derivee facteur Fourier} have necessarily only one non-zero component.

Now, decomposing into real and imaginary parts $\ga_{j}^{(ba)} \, = \, \chi_j^{(ba)} + \i\eta_j^{(ba)}$, for any $ \bs{s}_{k-1} \in \mathbb{N}^{k-1}$, one readily infers the upper bound
\bem
\Big| \Dp{ \bs{y}_{ k-1 ; 0} }^{ \bs{ s }_{k-1} } \Dp{ \ga_j^{(ba)} }^{\ell} \mf{p}\big( \bs{\ga}  \big) \Big| \; \leq \;
 \Big( 1+\max_{a \in \intn{1}{k-1}}\!\{\, ||\bs{y}_a||\}   \Big)  \cdot C \cdot \Big\{ \big| \cosh\big[ \ga_j^{(ba)} \big] \big| \vee   \big| \sinh\big[ \ga_j^{(ba)} \big] \big| \Big\} \\
\; \leq \;
\wt{C} \, \cdot \,  \Big( 1+ \max_{a \in \intn{1}{k-1}}\! \{ \, ||\bs{y}_a||\}  \Big)
\, \cdot  \, \Big\{  \cosh^2\big[ \chi_j^{(ba)} \big]    \cos^2\big[ \eta_j^{(ba)} \big] \, +  \,  \sinh^2\big[ \chi_j^{(ba)} \big]    \sin^2\big[ \eta_j^{(ba)} \big]  \Big\}^{\f{1}{2}} \\
\hspace{5mm}  \vee  \Big\{  \sinh^2\big[ \chi_j^{(ba)} \big]    \cos^2\big[ \eta_j^{(ba)} \big]  \, + \,   \cosh^2\big[ \chi_j^{(ba)} \big]    \sin^2\big[ \eta_j^{(ba)} \big]  \Big\}^{\f{1}{2}}
\; \leq \; \wt{C}^{\prime} \, \cdot  \, \cosh \Re\big[ \ga_j^{(ba)} \big] \;.
\end{multline}
Thus, by applying in each variable the higher order Leibnitz formula, one gets that for some $C>0$
\bem
 \Big| \pl{ \ell = 1 }{ k - 1 }  \Dp{y_{\ell;0} }^{I_{\ell} } \cdot f\big( \bs{y}_1,\dots, \bs{y}_k  \mid  \bs{m}  \big)   \Big| \; \leq \;
  \Big( \sul{a =1}{k-1} ||\bs{y}_a|| + 1 \Big)^{  | \bs{m}   |  }  \cdot \max_{ \substack{ b>a \\ j \in \intn{ 1 }{ n_{ba} } }  } \hspace{-2mm} \Big\{ \cosh \Re\big[ \ga_j^{(ba)} \big]    \Big\}^{  | \bs{m}   |  } \\
\times \sul{  s_1 =1 }{ I_{1} }  \cdots  \sul{  s_{k-1} =1 }{ I_{k-1} }  \,  \Big|  \prod_{\ell=1}^{k-1} \Dp{ y_{\ell ; 0} }^{ s_{\ell} }  f \big( \bs{y}_1,\dots, \bs{y}_k  \big)  \Big|  \; : \; 0 \leq s_{\ell} \leq I_{\ell}  \bigg\}
 \;.
\end{multline}
It follows directly from \eqref{definition f via G dans estimee R de G} that
\beq
 \Big| \Big|  \Big( \sul{a =1}{k-1} ||\bs{y}_a|| + 1 \Big)^{  | \bs{m}   |  } \prod_{\ell=1}^{k-1} \Dp{ y_{\ell ; 0} }^{ s_{\ell} }  f \big( \bs{y}_1,\dots, \bs{y}_k  \big)  \Big| \Big|_{ L^1\big( (\R^{1,1})^k \big)  }
\, \leq \, C \cdot
\norm{G}_{ \bs{I} ; \bs{m} }
\enq
for some $C>0$ and with $\norm{G}_{ \bs{I} ; \bs{m} }$ as defined in \eqref{definition norme I de G}.
Hence, all-in-all, when $G \in \mc{S}\big( (\R^{1,1})^k \big)$ and $\ga_{j}^{(ba)} \in \R$
or when  $G \in \mc{C}^{\infty}_{\e{c}}\big( (\R^{1,1})^k \big)$ and condition
 \eqref{contrainte positivite PS Impulsion et difference position} is fullfilled, the above  handlings yields, for some constant $C>0$
\beq
\bigg| \Dp{ \bs{\ga}  }^{\bs{m}  } \cdot  \mc{R}[ G] \big(   \bs{\ga}    \big)   \bigg| \; \leq  \; C \cdot \norm{G}_{\bs{I}}
\cdot \pl{ \ell = 1 }{ k - 1 } \bigg|  \f{ 1 }{ \op{P}_{\ell}^{(0)}( \bs{\ga} ) }  \bigg|^{ I_{\ell} } \cdot
\pl{  b>a }{ k}  \pl{j =1 }{ n_{ba} }    \Big( \cosh \Re\big[ \ga_j^{(ba)} \big]    \Big)^{  | \bs{m}   |  } \;.
\enq
It remains to lower bound the product of momentum related terms. One has that
\bem
\big| \op{P}_{\ell}^{(0)}( \bs{\ga}  ) \big| \, \geq \,  \big| \Re \op{P}_{\ell}( \bs{\ga}  ) \big|
\, \geq \, \sul{ b=\ell + 1 }{ k } \sul{ a=1 }{ \ell }  \sul{ j=1 }{ n_{ba} } m \cosh\big[ \chi_j^{(ba)} \big]    \cdot \big | \cos \big[ \eta_j^{(ba)} \big] \big| \\
\, \geq \, m \hspace{-2mm} \min_{  \substack{ b> a \\ j \in \intn{1}{  | \bs{m}   |  } } } \hspace{-2mm}  \big| \cos \big[ \eta_j^{(ba)} \big] \big| \cdot
 \cosh\big[ \chi_j^{(ba)} \big]  \, \geq \, c \cosh\Re\big[ \ga_j^{(ba)} \big]   \;,
\end{multline}
this provided that $|\Im\big[ \ga_j^{(ba)} \big]|$ is not too large. Thus, upon taking $I_{\ell}$ as given in \eqref{definition IEll}, one infers \eqref{ecriture borne sur derivee facteur S total pour k multipts}. \qed

\begin{lemme}
\label{Lemme decomposition produit SRF comme somme combinatoire avec poles explicites}

Let $n_{\bs{\ga}}$ and $\bs{\ga} \in \Cx^{ n_{\bs{\ga}} }$ be as defined in \eqref{definition vecteur gamma tot}
while $G_k, \mc{S}$, $\mc{R}[ G ]$  and $\mc{F}_{\e{tot};\bs{\veps}_{k} }$ be as given respectively in \eqref{definition fonction Gk},
\eqref{definition facteur de diffusion complet k pts}, \eqref{definition TF space-time de la fct test} and \eqref{definition F tot eps}.
Finally, let $\bs{A}^{(p)}$ and $\bs{B}^{(p)}$ be the vectors introduced in \eqref{definition vecteur Ap} and \eqref{definition vecteur Bp}
and $A^{(p)}$, $B^{(p)}$ the sets built out of their coordinates.
The following decomposition holds
\bem
\Big( \mc{S}\cdot \mc{R}[ G_k ] \cdot  \mc{F}_{\e{tot};\bs{\veps}_{k} } \Big) \big(   \bs{\ga}     \big) \; = \;
\pl{p=2}{k-1} \Bigg\{  \sul{ A^{(p)} \, = \, A^{(p)}_1 \underset{1}{\cup} A^{(p)}_2 }{}    \sul{ B^{(p)} \, = \, B^{(p)}_1 \underset{1}{\cup} B^{(p)}_2 }{}  \Bigg\}
\; \mc{H}_{\e{tot}}\Big( \bs{\ga}  \mid \big\{ \bs{A}_s^{(p)} , \bs{B}_s^{(p)} \big\}_{s=1, p=2}^{2\; \;\; , \;   k-1}  \, ; \, \bs{\veps}_{k} \Big) \\
\times \pl{p=2}{k-1} \pl{r=1}{ |A_1^{(p)}|  } \Bigg\{     \f{  1  }{   \big( \bs{A}_1^{(p)}\big)_{r} \, - \,  \big( \bs{B}_1^{(p)}\big)_{r}  - \i \veps_{p}    } \Bigg\} \;.
\label{decomposition produit SRF comme somme combinatoire avec poles explicites}
\end{multline}
Above, one sums over ordered in the first component partitions of the sets $A^{(p)}$ and $B^{(p)}$ with $p=2,\dots, k-1$ under the constraint
\beq
|A_1^{(p)}| \, = \, |B_1^{(p)}| \, = \, |A^{(p)}| \wedge |B^{(p)}|  \; \qquad for \qquad p=2,\dots, k-1 \;.
\enq
Below, $\eps\big( \, \bs{C}  \mid  \bs{C}_1 \cup  \bs{C}_2 \, \big)$ is a sign factor introduced in Definition \ref{definition singature vectorielle}
while
\bem
\mc{H}_{\e{tot}}\Big( \bs{\ga}  \mid \big\{ \bs{A}_s^{(p)} , \bs{B}_s^{(p)} \big\}_{s=1, p=2}^{2\; \;\; , \;   k-1}  \, ; \, \bs{\veps}_{k}  \Big) \; = \;
 \Big( \mc{S} \cdot \mc{R}[ G_k ] \Big)\big(  \bs{\ga}  \big) \cdot  \pl{p=1}{k} \Big\{ h_p\big( \bs{\ga} , \veps_{p} \big) \Big\}  \\
\times \pl{p=2}{k-1} \bigg\{  \eps\big( \, \bs{A}  \mid  \bs{A}_1 \cup  \bs{A}_2 \, \big) \cdot    \eps\big( \, \bs{B}  \mid  \bs{B}_1 \cup  \bs{B}_2 \, \big)
  \cdot \pl{r>\ell}{ |A_2^{(p)}| } \Big(  \big( \bs{A}_2^{(p)}\big)_{r} \, - \,  \big( \bs{A}_2^{(p)}\big)_{\ell}   \Big)
  \cdot \pl{r<\ell}{ |B_2^{(p)}| } \Big(  \big( \bs{B}_2^{(p)}\big)_{r} \, - \,  \big( \bs{B}_2^{(p)}\big)_{\ell}   \Big)   \bigg\} \;.
\label{ecriture explicite pour Htot}
\end{multline}
Moreover,   there exists $\eta >0$ and small enough
such that for any $\bs{m} \in \mathbb{N}^{  n_{\bs{\ga}} }$ there exists $C>0$, depending on $G$ such that uniformly in $  \bs{\ga} \in \mc{S}_{\eta}^{  n_{\bs{\ga}} }$
 satisfying \eqref{contrainte positivite PS Impulsion et difference position} and in $\bs{\veps}_{k}$ small enough
\beq
\bigg| \Dp{ \bs{\ga}  }^{\bs{m}  }\mc{H}_{\e{tot}}\Big( \bs{\ga}  \mid \big\{ \bs{A}_s^{(p)} , \bs{B}_s^{(p)} \big\}_{s=1, p=2}^{2\; \;\; , \;   k-1} \,; \, \bs{\veps}_{k} \Big)\bigg| \; \leq \;
C \,    \pl{b>a}{ k } \pl{s=1}{n_{ba} }  \bigg| \f{ 1 }{  \cosh\Re\big[ \ga_s^{(ba)} \big] }   \bigg| \;.
\label{ecriture borne sur facteur Htot}
\enq
Finally, $\mc{H}_{\e{tot}}$ is smooth pointwise in $\bs{\veps}_{k}$ small enough and pointwise in $\bs{\ga} \in \R^{ n_{\bs{\ga}} }$, $\mc{H}_{\e{tot}}$ is smooth in $\bs{\veps}_{k}$.
Moreover, in case $G$ has compact support, pointwise in $\bs{\veps}_{k}$ small enough, $\mc{H}_{\e{tot}}$ is holomorphic on $\mc{S}_{\eta}^{ n_{\bs{\ga}} }$
and pointwise in $\bs{\ga} \in \mc{S}_{\eta}^{ n_{\bs{\ga}} }$, $\mc{H}_{\e{tot}}$ is smooth in $\bs{\veps}_{k}$.

\end{lemme}

\Proof

Starting from the representation for individual form factors obtained in Lemma \eqref{Lemme decomposition FF en partie sing et reguliere}
and then implementing the expansion provided by Lemma \ref{Lemme decomposition sur pole simples Cauchy generalise}
relatively to each Cauchy determinant-like factor leads directly to \eqref{decomposition produit SRF comme somme combinatoire avec poles explicites}-\eqref{ecriture explicite pour Htot}.

\noindent The bounds on the multi-dimensional derivatives of $\mc{H}_{\e{tot}}$ are a direct consequence of
\begin{itemize}
\item the multi-dimensional Leibniz formula,
\item the fact that $ \eps\big( \, \bs{C}  \mid  \bs{C}_1 \cup  \bs{C}_2 \, \big)$ is a sign factor while the Vandermonde like products
are algebraic in the $\ga^{(ba)}_{j}$,
\item equation \eqref{ecriture borne sur derivee produits des hp} of Lemma \ref{Lemme decomposition FF en partie sing et reguliere},
\item equation \eqref{ecriture borne sur derivee facteur S total pour k multipts} of Lemma \ref{Lemme borne sur derivees du facteur S total}
\item and equation \eqref{ecriture borne sur derivee facteur R de G total pour k multipts} of Lemma \ref{Lemme bornes sur derivees facteur TF de la smearing fct}
where one should take
\beq
r_s^{(ba)} \, = \, n_{\bs{\ga}} + 2 + \mf{w}
\enq
with $\mf{w}$ as defined in \eqref{ecriture borne sur derivee produits des hp}.
\end{itemize}

This entails the claim. \qed

\begin{lemme}
\label{Lemme decomposition produit poles explicites en sous chaines}

 Let  $\bs{A}^{(p)}$ and $\bs{B}^{(p)}$ be the vectors introduced in \eqref{definition vecteur Ap} and \eqref{definition vecteur Bp}
and $A^{(p)}$, $B^{(p)}$ the sets built out of their coordinates. Assume one is given partitions
\beq
  A^{(p)} \, = \, A^{(p)}_1 \underset{1}{\cup} A^{(p)}_2  \qquad and \qquad  B^{(p)} \, = \, B^{(p)}_1 \underset{1}{\cup} B^{(p)}_2 \qquad for \qquad
p=2,\dots, k-1
\label{ecritrue partition ensembles Ap et Bp en deux ss ens ordonnes}
\enq
which satisfy the constraints $|A^{(p)}_1| \, = \, | B^{(p)}_1|\, = \, |A^{(p)}| \wedge |B^{(p)}|$ for $p=2,\dots, k-1$.
Then, there exist $n\in \mathbb{N}$ and sequences $\{\ell_u\}_{u=1}^{n}$, $\ell_u \in \intn{2}{k}$,
\beq
1 \,\leq \, a_0^{(u)}  \, < \,   a_1^{(u)}  \, < \,  \cdots  \, < \,   a_{\ell_u}^{(u)} \leq k
\qquad  and  \qquad j_s^{\, (u)} \in \intn{ 1 }{ n_{ a_s^{(u)} a_{s-1}^{(u)} } }
\label{ecriture proprietes des suites definissant les chaines de variables}
\enq
with $s=1,\dots, \ell_u$ and $u=1,\dots,n$  such that
\beq
 \pl{p=2}{k-1} \pl{r=1}{ |A_1^{(p)}|  } \Bigg\{     \f{  1  }{   \big( \bs{A}_1^{(p)}\big)_{r} \, - \,  \big( \bs{B}_1^{(p)}\big)_{\ell}  - \i \veps_{p}    } \Bigg\}
\; = \;  \pl{u=1}{n} \pl{r=1}{ \ell_u -1 } \Bigg\{  \f{ 1 }{   \ga_{j_r^{(u)}}^{ (a_r^{(u)} a_{r-1}^{(u)}) }  \, - \, \ga_{j_{r+1}^{(u)}}^{ (a_{r+1}^{(u)} a_{r}^{(u)}) } \, - \, \i \veps_{ a_r^{(u)} }   }  \Bigg\} \;.
\label{ecriture representation canonique produit facteurs singuliers}
\enq
Moreover, it holds that
\beq
\Big\{  \ga_{j_r^{\, (u)}}^{ (a_r^{(u)} a_{r-1}^{(u)}) }   \Big\}_{r=1}^{\ell_u} \, \cap \, \Big\{  \ga_{j_r^{\, (v)}}^{ (a_r^{(v)} a_{r-1}^{(v)}) }   \Big\}_{r=1}^{\ell_v} \; = \; \emptyset
\label{propriete de non intersection des variables de chaines singulieres}
\enq
as soon as $u\not= v$.

\end{lemme}

\Proof

First of all we observe,  \textit{c.f.} \eqref{definition vecteur Ap} and \eqref{definition vecteur Bp}, that a given variable
$\ga^{(ba)}_j$ appears exactly twice as a coordinate of the vectors $\bs{A}^{(p)}$, $\bs{B}^{(p)}$, namely
once as a coordinate of $\bs{A}^{(b)}$ and once as a coordinate of $\bs{B}^{(a)}$. This means that any
variable $\ga^{(ba)}_j$ may either appear once, twice or simply never in the \textit{lhs} product in \eqref{ecriture representation canonique produit facteurs singuliers}.

If all sets $A^{(p)}_1, B^{(p)}_1$, $p\in \intn{2}{k-1}$, are empty, then there is simply nothing to prove.
Else, we start by taking $p\in \intn{2}{k-1}$ minimal such that $|A^{(p)}_1|>0$ what thus implies that
\beq
|A^{(s)}_1| \, = \, |B^{(s)}_1| \, = \, 0 \qquad \e{for} \qquad s=2,\dots p-1 \;,
\enq
the above conditions being obviously empty if $p=2$.
Since $|A^{(p)}_1|>0$, there exists $s<p$ and $j\in \intn{1}{ n_{ps} }$ such that $\ga_{j}^{(ps)} \in A^{(p)}_1$, \textit{viz}.
there exists $r \in \intn{ 1 }{ | A^{(p)}_1| }$ such that
\beq
\ga_{j}^{(ps)} \, = \, \Big(  \bs{A}^{(p)}_1  \Big)_{r} \;.
\enq
By construction, \textit{c.f.} \eqref{definition vecteur Bp}, $\ga_{j}^{(ps)}$ appears as one of the coordinates
building up the vector $\bs{B}^{(s)}$. However, our choice of $p$ ensures that $|B^{(s)}_1| \, = \, 0$
since $s<p$. Therefore, the variable $\ga_{j}^{(ps)}$ cannot appear at any other place in the \textit{lhs}
product in \eqref{ecriture representation canonique produit facteurs singuliers}.
Now, since $|B_1^{(p)}|>0$, there exists $s^{\prime} > p $ and $j^{\prime}$ such that
\beq
\Big(  \bs{B}^{(p)}_1  \Big)_{r} \;= \; \ga_{j^{\prime}}^{(s^{\prime}p)}   \;.
\enq
We set
\beq
 a_{0}^{(1)} \, = \,  s \;, \quad  a_1^{(1)}  = p \; , \quad j^{\, (1)}_1 \, = \, j  \qquad \e{and}  \qquad
  a_2^{(1)}  \, = \,  s^{\prime} \;,  \quad j^{\, (1)}_2 \, = \, j^{\prime}    \;.
\enq
There are two options at this stage. Either, $\ga_{j^{\prime}}^{(s^{\prime}p)} \not\in  A^{(s^{\prime})}_1$, in which case
$\ga_{j^{\prime}}^{(s^{\prime}p)}$ does not appear anymore in the \textit{lhs} product in \eqref{ecriture representation canonique produit facteurs singuliers}.
Thus, the chain terminates $\ell_1=2$ and one repeats the reasoning relative to a reduced product involving the sets
\beq
\wt{A}^{\, (t)}_1 \; = \; A^{(t)}_1 \setminus \Big\{ \ga_{j_r^{\, (1)}}^{ (a_r^{(1)} a_{r-1}^{(1)}) } \Big\}_{r=1}^{ \ell_1 } \quad
\e{and} \qquad
\wt{B}^{(t)}_1 \; = \; B^{(t)}_1 \setminus \Big\{ \ga_{j_r^{\, (1)}}^{ (a_r^{(1)} a_{r-1}^{(1)}) } \Big\}_{r=1}^{ \ell_1 }
\enq
for $t=2,\dots,k-1$ and the associated vectors obtained by removing the coordinates $\ga_{j_r^{\, (u)}}^{ (a_r^{(u)} a_{r-1}^{(u)}) }$ -if these are present-
from the vectors $\bs{A}^{(t)}_1$ and $\bs{B}^{(t)}_1$.

Otherwise, $\ga_{j^{\prime}}^{(s^{\prime}p)} \in  A^{(s^{\prime})}_1$, which means that there exists $r^{\prime} \in \intn{1}{ |A^{(s^{\prime})}_1| }$
such that
\beq
\ga_{j}^{(s^{\prime}p)} \, = \, \Big(  \bs{A}^{(s^{\prime})}_1  \Big)_{ r^{\prime} } \;.
\enq
Then,  there exists $s^{\prime\prime} > s^{\prime} $ and $j^{\prime\prime}$ such that
\beq
\Big(  \bs{B}^{(s^{\prime})}_1  \Big)_{r} \;= \; \ga_{j^{\prime\prime}}^{(s^{\prime\prime}s^{\prime})}   \;.
\enq
One sets
\beq
  a_3^{(1)}  \, = \,  s^{\prime\prime} \;, \quad j^{\, (1)}_3 \, = \, j^{\prime \prime}    \;.
\enq
There are two options at this stage. Either, $\ga_{j^{\prime \prime}}^{(s^{\prime\prime} s^{\prime})} \not\in  A^{(s^{\prime\prime})}_1$, in which case
$\ga_{j^{\prime \prime}}^{(s^{\prime\prime} s^{\prime})} $ does not appear anymore in the \textit{lhs} product in \eqref{ecriture representation canonique produit facteurs singuliers}
and the chain terminates so that $\ell_1=3$. Otherwise, $\ga_{j^{\prime \prime}}^{(s^{\prime\prime} s^{\prime})}  \in  A^{(s^{\prime\prime})}_1$
and one continues the construction.

Eventually, one builds a sequence
\beq
 \ga_{j_r^{\, (1)}}^{ (a_r^{(1)} a_{r-1}^{(1)}) } \qquad \e{with} \qquad r=0,\dots, \ell_1 \;, \qquad  a_0^{(1)}<\cdots < a_{\ell_1}^{(1)} \qquad \e{and} \qquad
 j_r^{\, (1)} \in \intn{ 1 }{ n_{ a_r^{(1)} a_{r-1}^{(1)} } }
\enq
and such that
\beq
 \ga_{j_r^{\, (1)}}^{ (a_r^{(1)} a_{r-1}^{(1)}) }  \in A^{ (a_r^{(1)}) }_1\quad \e{for} \quad r=1,\dots, \ell_1-1
\qquad \e{and} \qquad
 \ga_{j_r^{\, (1)}}^{ (a_r^{(1)} a_{r-1}^{(1)}) }  \in B^{ (a_{r-1}^{(1)}) }_1\quad \e{for} \quad r=2,\dots, \ell_1
\enq
but
\beq
 \ga_{j_r^{\, (1)}}^{ (a_1^{(1)} a_{0}^{(1)}) }  \not\in B^{ (a_0^{(1)}) }_1 \qquad \e{and} \qquad
 \ga_{j_{\ell_1}^{\, (1)}}^{ (a_{\ell_1}^{(1)} a_{\ell_1-1}^{(1)}) }  \not\in A^{ (a_{\ell_u}^{(1)}) }_1  \;.
\enq
The sequence has to terminate as  $a_r^{(1)}$ is strictly increasing and belongs to $\intn{1}{k}$.
This being settled, one repeats the reasoning relative to a reduced product involving the sets
\beq
\wt{A}^{\, (t)}_1 \; = \; A^{(t)}_1 \setminus \Big\{ \ga_{j_r^{\, (1)}}^{ (a_r^{(1)} a_{r-1}^{(1)}) } \Big\}_{r=1}^{ \ell_1 } \quad
\e{and} \qquad
\wt{B}^{\, (t)}_1 \; = \; B^{(t)}_1 \setminus \Big\{ \ga_{j_r^{\, (1)}}^{ (a_r^{(1)} a_{r-1}^{(1)}) } \Big\}_{r=1}^{ \ell_1 }
\enq
for $t=2,\dots,k-1$ and the associated vectors obtained by removing the coordinates $\ga_{j_r^{\, (u)}}^{ (a_r^{(u)} a_{r-1}^{(u)}) }$ -if these are present-
from the vectors $\bs{A}^{ (t)}_1$ and $\bs{B}^{(t)}_1$. One repeats the process until all of the involved sets become empty, what must always happen  due to the finiteness of the
sets $|A^{(p)}|$ and $|B^{(p)}|$.

Finally, \eqref{propriete de non intersection des variables de chaines singulieres} follows from the very procedure which constructs the sequences
$j_r^{\, (u)}$ and $a_r^{(u)}$. \qed

\vspace{3mm}

\subsection{Closed representation for the $\bs{r}$-truncated multipoint functions}
\label{Subsection Closed reps for r truncated pt fcts}

We are now in position to prove the main technical result of the paper, which is the well definiteness of the multiple integral summands
$\mc{I}_{ \bs{n} }\big[ G_k \big]$ introduced in \eqref{definition sommant base dans serie int mult pour fct 2 pts}. The result
will be stated in two forms, depending on the support of the test functions $g_1,\dots, g_k$. In the case of general Schwartz functions,
we simply establish the well-definiteness in the sense of $-$ boundary valued multi-dimensional distribution
and is thus  less explicit, while for pair-wise mutually space-like separated supports, the result is explicit.

\begin{prop}
\label{Proposition Corr fct well defined through veps to zero limit}

Let $g_1,\dots, g_k$ be Schwartz functions on $\R^{1,1}$. Then, $\mc{I}_{ \bs{n} }\big[ G_k \big] $ given in \eqref{definition sommant base dans serie int mult pour fct 2 pts} is well defined,
viz. the $\bs{\veps}_k\tend \bs{0}^+$ limit exists, and the  conclusions of Proposition \ref{ecriture rep serie int mult tronquee pour fct multipts tronquee} are valid.
In particular, the truncated smeared correlation function
\beq
\Big( \bs{f}_{\e{vac}}, \op{O}_1[g_1] \cdot  \op{O}^{(r_1)}_2[g_2] \cdots   \op{O}^{(r_{k-1})}_k[g_k]  \cdot \bs{f}_{\e{vac}} \Big)
\enq
is a well-defined distribution on $\mc{S}\big( (\R^{1,1})^k \big)$.

\end{prop}

\Proof

Starting from the definition \eqref{definition sommant base dans serie int mult pour fct 2 pts}, one first decomposes the
product $\mc{S}\cdot \mc{R}[ G_k ] \cdot  \mc{F}_{\e{tot};\bs{\veps}_{k}}$ with the help of equation \eqref{decomposition produit SRF comme somme combinatoire avec poles explicites}
of Lemma \ref{Lemme decomposition produit SRF comme somme combinatoire avec poles explicites}. Then, for each given partition $A^{(p)}_1 \underset{1}{\cup} A^{(p)}_2$,
resp.  $ B^{(p)}_1 \underset{1}{\cup} B^{(p)}_2$,  of the sets $A^{(p)}$, resp. $B^{(p)}$,  $p=2,\dots, k-1$, as given in \eqref{ecritrue partition ensembles Ap et Bp en deux ss ens ordonnes}, one constructs
the independent chain factorisation as in \eqref{ecriture representation canonique produit facteurs singuliers} of Lemma \ref{Lemme decomposition produit poles explicites en sous chaines}.
Provided that each of the integrals is well-defined, a fact that we will establish below,
all of this allows one to recast $\mc{I}_{ \bs{n} }\big[ G_k \big]$ in the form
\bem
\mc{I}_{ \bs{n} }\big[ G_k \big]  \; = \; \pl{p=2}{k-1} \Bigg\{  \sul{ A^{(p)} \, = \, A^{(p)}_1 \underset{1}{\cup} A^{(p)}_2 }{}    \sul{ B^{(p)} \, = \, B^{(p)}_1 \underset{1}{\cup} B^{(p)}_2 }{}  \Bigg\}
\lim_{\veps_1 \tend 0^+} \Int{  \R^{n_{21}}   }{} \dd^{ n_{21} } \ga^{(21)} \cdots
\lim_{\veps_{k-1} \tend 0^+} \hspace{-4mm} \Int{  \R^{n_{k1} + \dots + n_{kk-1} }  }{}  \hspace{-4mm} \dd^{ n_{k1} }  \ga^{(k1)} \cdots \dd^{ n_{kk-1} } \ga^{(kk-1)} \\
\times  \lim_{\veps_{k } \tend 0^+}  \Bigg[
\mc{H}_{\e{tot}}\Big( \bs{\ga}  \mid \big\{ \bs{A}_s^{(p)} , \bs{B}_s^{(p)} \big\}_{s=1, p=2}^{2\; \;\; , \;   k-1}  \, ; \, \bs{\veps}_{k} \Big)
\pl{u=1}{n} \pl{r=1}{ \ell_u -1 } \Bigg\{  \f{ 1 }{   \ga_{j_r^{\,(u)}}^{ (a_r^{(u)} a_{r-1}^{(u)}) }  \, - \, \ga_{j_{r+1}^{\,(u)}}^{ (a_{r+1}^{(u)} a_{r}^{(u)}) } \, - \, \i \veps_{ a_r^{(u)} }   }  \Bigg\}  \Bigg]\;.
\label{reecriture In des Gk pour well definiteness Hilbert transforms}
\end{multline}
Here, we stress that the sequences $j_r^{\, (u)},a_r^{(u)} $ are as given in \eqref{ecriture proprietes des suites definissant les chaines de variables}
and \textit{do} depend on the partitions of $A^{(p)}, B^{(p)}$ considered.
Moreover, taken the non-intersection property \eqref{propriete de non intersection des variables de chaines singulieres}, each integration variable
$\ga_j^{(ba)}$ appears at most for \textit{one} value of $u$ in the outer product.

We now implement a permutation in the integration variables $ \bs{\ga} \, = \, \big( \bs{\ga}^{(21)},\dots, \bs{\ga}^{(kk-1)} \big) \, \hookrightarrow \,  \bs{\op{v}}\cup \bs{\gammaup}_2$,
with
\beq
\bs{\op{v}}  \, = \, \bs{\op{v}}^{(1)}\cup \cdots \cup \bs{\op{v}}^{(n)} \qquad \e{where}  \qquad
\bs{\op{v}}^{(u)} \, = \, \Big( \ga_{j_1^{\,(u)}}^{ (a_1^{(u)} a_{0}^{(u)}) } , \dots, \ga_{j_{\ell_u}^{\,(u)}}^{ (a_{\ell_u}^{(u)} a_{\ell_u-1}^{(u)}) } \Big) \in \mathbb{R}^{\ell_u}
\enq
and where $\bs{\gammaup}_2$ is the vector build from the remaining $\ga$-coordinates, taken in any order. For convenience, we set
\beq
\bs{\tau}^{(u)} \, = \, \Big( \veps_{a_{0}^{(u)}-1}, \dots, \veps_{a_{\ell_u-1}^{(u)}-1}\Big) \;.
\enq
This leads to
\bem
\mc{I}_{ \bs{n} }\big[ G_k \big]  \; = \; \pl{p=2}{k-1} \Bigg\{  \sul{ A^{(p)} \, = \, A^{(p)}_1 \underset{1}{\cup} A^{(p)}_2 }{}    \sul{ B^{(p)} \, = \, B^{(p)}_1 \underset{1}{\cup} B^{(p)}_2 }{}  \Bigg\}
  \Int{  \R^{ |\gammaup_2 | }   }{} \dd^{ |\gammaup_2 |  } \gammaup_2    \\
\times   \lim_{\bs{\veps}_{k} \tend \bs{0}^+} \pl{u=1}{n} \Bigg\{ \Int{ \R^{\ell_u} }{} \dd^{\ell_u} \op{v}^{(u)}
   \f{ \mc{H}_{\e{tot}}\Big( \bs{\ga}\big( \bs{\op{v}},\bs{\gammaup}_2\big)  \mid \big\{ \bs{A}_s^{(p)} , \bs{B}_s^{(p)} \big\}_{s=1, p=2}^{2\; \;\; , \;   k-1}  \, ; \, \bs{\veps}_{k} \Big) }
                        {  \pl{r=1}{ \ell_u -1 } \big[ \op{v}^{(u)}_{r} \, - \,\op{v}^{(u)}_{r+1} \, -\,  \i \tau^{(u)}_r  \big]  }  \Bigg\}   \;.
\end{multline}
Here we have moved $\lim_{\bs{\veps}_k\tend 0^+}$ jointly in front of the $\bs{\op{v}}$ integrals and through the $\gammaup_{2}$ integrals, what is licit if the convergence of the resulting integrals
is sufficiently uniform, a property we shall establish below.

Let $\msc{L}_{q}(x)$ be a $q^{\e{th}}$ antiderivative of $\tf{1}{x}$, \textit{viz}. $\msc{L}_{q}^{(q)}(x)=\tf{1}{x}$, \textit{e.g.} $\msc{L}_1(x)=\ln x$, $\msc{L}_2(x) = x \ln x - x $, \textit{etc}.
Then, for any $f\in \mc{S}(\R^{\ell})$, successive integrations by parts lead to
\bem
\Int{\R^{\ell}}{} \dd^{\ell} v f(\bs{v}) \pl{r=1}{\ell-1} \bigg\{  \f{ 1 }{v_r-v_{r+1}-\i \tau_r  } \bigg\}
\, =\, (-1)^{k_1} \Int{\R^{\ell}}{} \dd^{\ell} v   \pl{r=2}{\ell-1} \bigg\{  \f{ 1 }{v_r-v_{r+1}-\i \tau_r  } \bigg\} \cdot \msc{L}_{k_1}(v_1-v_2-\i \tau_1)  \Dp{v_1}^{k_1} f(\bs{v}) \\
\, = \, \pl{r=1}{s-1}\Big\{ (-1)^{k_r} \Big\} \Int{\R^{\ell}}{} \dd^{\ell} v   \pl{r=s}{\ell-1} \bigg\{  \f{ 1 }{v_r-v_{r+1}-\i \tau_r  } \bigg\} \cdot
\msc{L}_{ k_{s-1} }( v_{s-1}-v_{s}-\i \tau_{s-1} )  \Dp{v_{s-1}}^{k_{s-1}}  \cdots \msc{L}_{k_1}(v_1-v_2-\i \tau_1)  \Dp{v_1}^{k_1} f(\bs{v}) \\
\, = \,   \Int{\R^{\ell}}{} \dd^{\ell} v
\pl{r=1}{ \substack{ \curvearrowleft \\  \ell-1} } \bigg\{  (-1)^{k_r} \msc{L}_{ k_{r} }( v_{r}-v_{r+1}-\i \tau_{r} )  \Dp{v_{r}}^{k_{r}}   \bigg\}
\cdot f(\bs{v}) \;,
\end{multline}
where $\pl{r=1}{ \substack{ \curvearrowleft \\  u} } A_r \, = \, A_{u} \cdots A_1$. Upon taking $k_{r} = \ell-r $, \textit{i.e.} $k_r=k_{r+1}+1$, one
ensures that the joint action of the derivatives still leads to an at least piece-wise continuous integrand which has, at most, a polynomial growth in each variable
$v_a$ in what concerns the contributions of the antiderivatives $\msc{L}_q$.

This leads to the representation
\beq
\mc{I}_{ \bs{n} }\big[ G_k \big]  \; = \; \pl{p=2}{k-1} \Bigg\{  \sul{ A^{(p)} \, = \, A^{(p)}_1 \underset{1}{\cup} A^{(p)}_2 }{}    \sul{ B^{(p)} \, = \, B^{(p)}_1 \underset{1}{\cup} B^{(p)}_2 }{}  \Bigg\}
\Int{  \R^{ |\gammaup_2 | }   }{} \dd^{ |\gammaup_2 |  } \gammaup_2
  \lim_{\bs{\veps}_{k} \tend \bs{0}^+} \pl{u=1}{n} \Bigg\{ \Int{ \R^{\ell_u} }{} \dd^{\ell_u} \op{v}^{(u)} \Bigg\}  \mc{J}_{ \bs{n} }\Big[ G_k \mid \Big\{ \bs{A}^{(p)}_a, \bs{B}_a^{(p)} \Big\}\, ; \, \bs{\veps}_{k} \Big] \;,
\enq
where
\bem
\mc{J}_{ \bs{n} }\Big[ G_k \mid \Big\{ \bs{A}^{(p)}_a, \bs{B}_a^{(p)} \Big\} \, ; \, \bs{\veps}_{k} \Big]  \, = \,
\pl{u=1}{n}\pl{r=1}{ \substack{ \curvearrowleft \\  \ell_u-1} } \bigg\{  (-1)^{\ell_{u}-r} \msc{L}_{ \ell_{u}-r }( v_{r}^{(u)}-v_{r+1}^{(u)}-\i \tau_{r}^{(u)} )  \Dp{ v_{r}^{(u)} }^{ \ell_{u}-r }   \bigg\} \\
\times \mc{H}_{\e{tot}}\Big( \bs{\ga}\big( \bs{\op{v}},\bs{\gammaup}_2\big)  \mid \big\{ \bs{A}_s^{(p)} , \bs{B}_s^{(p)} \big\}_{s=1, p=2}^{2\; \;\; , \;   k-1}  \, ; \, \bs{\veps}_{k} \Big) \;.
\end{multline}
Invoking \eqref{ecriture borne sur facteur Htot} given in Lemma \eqref{Lemme decomposition produit SRF comme somme combinatoire avec poles explicites}, one
readily arrives to the bounds
\beq
\Big| \mc{J}_{ \bs{n} }\Big[ G_k \mid \Big\{ \bs{A}^{(p)}_a, \bs{B}_a^{(p)} \Big\} \,; \,  \bs{\veps}_{k} \Big]  \Big| \, \leq \,
\f{ \pl{u=1}{n}\pl{r=1}{  \ell_u-1 } \Big\{  \, \big( |v_r^{(u)}| + |v_{r+1}^{(u)}| \, \big)^{\ell_u} \cdot \ln \big| v_r^{(u)} - v_{r+1}^{(u)} \big| \Big\} }
{ \pl{a =1 }{ | \bs{\gammaup}_2 | } \big|   \cosh\big(   (\bs{\gammaup}_2)_a  \big)   \big| \cdot
\pl{a =1 }{ | \bs{\op{v}} | } \big|   \cosh\big(   (\bs{\op{v}})_a  \big) \big| } \;,
\enq
uniform in $||\bs{\veps}_k||$ small enough. The point-wise convergence
\bem
\mc{J}_{ \bs{n} }\Big[ G_k \mid \Big\{ \bs{A}^{(p)}_a, \bs{B}_a^{(p)} \Big\} \,; \,  \bs{\veps}_{k} \Big]  \, \limit{ \bs{\veps}_k }{ \bs{0}^+ }\, \\
\pl{u=1}{n}\pl{r=1}{ \substack{ \curvearrowleft \\  \ell_u-1} } \bigg\{  (-1)^{\ell_{u}-r} \msc{L}_{ \ell_{u}-r }( v_{r}^{(u)}-v_{r+1}^{(u)}-\i 0^+ )  \Dp{ v_{r}^{(u)} }^{ \ell_{u}-r }   \bigg\}
 \cdot \mc{H}_{\e{tot}}\Big( \bs{\ga}\big( \bs{\op{v}},\bs{\gammaup}_2\big)  \mid \big\{ \bs{A}_s^{(p)} , \bs{B}_s^{(p)} \big\}_{s=1, p=2}^{2\; \;\; , \;   k-1}  \, ; \, \bs{0} \Big)
\end{multline}
then allows one to conclude by invoking dominated convergence. \qed

\begin{prop}

Let $g_1,\dots, g_k$ be smooth, compactly supported on $\R^{1,1}$ and such that, for $a \not= b$,
\beq
\bs{x}_{ab}^2 \, <\, 0 \qquad for\; any \quad \bs{x}_a \in \e{supp}[g_a] \;\; with\; \; a\in \intn{1}{k}
\label{ecriture propriete pts ds support des gk}
\enq
and $\bs{x}_{a;1}>\bs{x}_{b;1}$ if $b>a$, i.e. $\bs{x}_{ab}$ is space-like with a strictly positive spatial coordinate.

Then, there exists $\eta>0$ and small enough such that for any sequence
\beq
\eta \,> \, \eta^{(k k-1)} \, > \, \cdots \, > \, \eta^{(k 1)} \,> \, \eta^{(k-1k-2)} \, > \, \cdots \, > \, \eta^{(2 1)} \, > \, 0 \; ,
\enq
it holds
\beq
\mc{I}_{ \bs{n} }\big[ G_k \big]  \; = \;  \pl{ b>a }{ k }
\Int{  \big\{ \R + \i \eta^{(ba)} \big\}^{n_{ba}}   }{} \hspace{-6mm} \dd^{ n_{ba} } \ga^{(ba)}
 \;  \Big( \mc{S}\cdot \mc{R}[ G_k ] \cdot  \mc{F}_{\e{tot}  } \Big) \big(   \bs{\ga}     \big)    \; ,
\label{ecriture compacte In de Gk avec contour integration dans plan complexe}
\enq
where
\beq
 \mc{F}_{\e{tot} }\big(   \bs{\ga}     \big) \; = \;
 \pl{p=1}{k} \mc{F}^{(\op{O}_p)}\Big( \, \overleftarrow{ \bs{\ga}^{(pp-1)} } \cup \cdots \cup \overleftarrow{ \bs{\ga}^{(p1)} }  + \i \pi \ov{\bs{e}} ,
 \bs{\ga}^{(kp)}  \cup \cdots \cup \bs{\ga}^{(p+1p)}   \, \Big) \;.
\label{definition F tot}
\enq
In particular, $\mc{I}_{ \bs{n} }\big[ G_k \big] $ is well-defined and the  conclusions of Proposition \ref{ecriture rep serie int mult tronquee pour fct multipts tronquee} are valid.

\end{prop}

\Proof

One starts from the representation \eqref{reecriture In des Gk pour well definiteness Hilbert transforms}.

For fixed $\veps_a>0$, $a=1,\dots, k$, owing to the property \eqref{ecriture propriete pts ds support des gk} and
the estimates \eqref{ecriture borne sur facteur Htot} provided by Lemma \ref{Lemme decomposition produit SRF comme somme combinatoire avec poles explicites}
one has the uniform in
\beq
\eta \, > \,  \Im\big[ \ga_j^{(cb)} \big] \, \geq \, \Im\big[ \ga_k^{(ba)} \big]  \, \geq \, 0 \qquad \e{for}\; \e{any} \quad  c \, > \, b \, > \, a \quad \e{and} \quad
j\in \intn{ 1 }{ n_{cb} } \quad \e{and} \quad k\in \intn{ 1 }{ n_{ba} }
\enq
with $\eta>0$ and small enough estimate
\bem
\Bigg|  \mc{H}_{\e{tot}}\Big( \bs{\ga}  \mid \big\{ \bs{A}_s^{(p)} , \bs{B}_s^{(p)} \big\}_{s=1, p=2}^{2\; \;\; , \;   k-1} \, ; \, \bs{\veps}_{k} \Big)
\cdot\pl{u=1}{n} \pl{r=1}{ \ell_u -1 } \Bigg\{  \f{ 1 }{   \ga_{j_r^{(u)}}^{ (a_r^{(u)} a_{r-1}^{(u)}) }  \, - \, \ga_{j_{r+1}^{(u)}}^{ (a_{r+1}^{(u)} a_{r}^{(u)}) } \, - \, \i \veps_{ a_r^{(u)} }   }  \Bigg\} \Bigg| \\
\; \leq \;    C    \pl{b>a}{ k } \pl{s=1}{n_{ba} }  \Bigg| \f{ 1 }{  \cosh\Re\big[ \ga_s^{(ba)} \big] }   \Bigg|^{ 2 }  \cdot
\pl{a<b<c}{ k } \pl{ k= 1 }{ n_{cb} } \pl{j =1 }{ n_{ba} }  \Bigg|  \f{  1  }{  \ga_{k}^{(cb)}   \, - \,  \ga_{j}^{(ba)}  + \i \veps_{b} }  \Bigg| \\
\; \leq \;    C    \pl{b>a}{ k } \pl{s=1}{n_{ba} }  \Bigg| \f{ 1 }{  \cosh\Re\big[ \ga_s^{(ba)} \big] }   \Bigg|^{ 2 } \cdot
\pl{a<b<c}{ k } \pl{ k= 1 }{ n_{cb} } \pl{j =1 }{ n_{ba} }  \Bigg|  \f{  1  }{  \Im\big[ \ga_{k}^{(cb)} \big]   \, - \,  \Im\big[ \ga_{j}^{(ba)} \big]  \, + \,  \veps_{b} }  \Bigg| \;.
\label{ecriture borne sur H tot plus poles a eps fini}
\end{multline}
This bound allows one to apply the unbounded contour variant of Morera's theorem
by first integrating over the contour
\beq
\intoo{-\infty}{+\infty} \,  \cup \,  \Big\{  \intoo{+\infty}{-\infty}  + \i \eta^{(k k-1)} \Big\} \; ,
\enq
successively for the variables $\ga_{n_{k k-1}}^{(k k-1)}, \dots , \ga_{n_{k 1}}^{(k k-1)}$ what allows one, each time, to trade the integration
over $\R$ into one over $\R + \i \eta^{(k k-1)}$. Then, one successively applies the same contour deformation for the
variables
\beq
\bs{\ga}^{(k k-2)}, \dots, \bs{\ga}^{(k 1)},  \bs{\ga}^{(k-1k-2)}, \dots ,  \bs{\ga}^{(2 1)}
\enq
by deforming the contour for $\bs{\ga}^{(ba)}$ from
\beq
\R^{n_{ba}} \qquad \e{on}\; \e{to} \qquad  \Big\{ \R+ \i \eta^{(ba)} \Big\}^{ n_{ba} } \;.
\enq
This yields
\bem
\mc{I}_{ \bs{n} }\big[ G_k \big]  \; = \;  \pl{p=2}{k-1} \Bigg\{  \sul{ A^{(p)} \, = \, A^{(p)}_1 \underset{1}{\cup} A^{(p)}_2 }{}    \sul{ B^{(p)} \, = \, B^{(p)}_1 \underset{1}{\cup} B^{(p)}_2 }{}  \Bigg\}
 \; \;  \lim_{\veps_1 \tend 0^+} \hspace{-4mm}  \Int{ \big\{ \R+ \i \eta^{(21)} \big\}^{ n_{21} }   }{} \hspace{-4mm}  \dd^{ n_{21} } \ga^{(21)} \cdots
\lim_{\veps_{k-1} \tend 0^+}    \pl{a=1}{k-1}\Int{ \big\{ \R+ \i \eta^{(ka)} \big\}^{ n_{ka} }   }{}  \hspace{-4mm} \dd^{ n_{ka} }  \ga^{(ka)}  \\
 \times \lim_{\veps_{k} \tend 0^+}  \mc{H}_{\e{tot}}\Big( \bs{\ga} \mid \big\{ \bs{A}_s^{(p)} , \bs{B}_s^{(p)} \big\}_{s=1, p=2}^{2\; \;\; , \;   k-1} \, ; \, \bs{\veps}_{k} \Big)
\cdot\pl{u=1}{n} \pl{r=1}{ \ell_u -1 } \Bigg\{  \f{ 1 }{   \ga_{j_r^{\, (u)}}^{ (a_r^{(u)} a_{r-1}^{(u)}) }  \, - \, \ga_{j_{r+1}^{\, (u)}}^{ (a_{r+1}^{(u)} a_{r}^{(u)}) } \, - \, \i \veps_{ a_r^{(u)} }   }  \Bigg\} \; .
\label{ecriture rep int pour In Gk avec contour deforme}
\end{multline}
At this stage, one observes that pointwise on the integration contour
\bem
 \mc{H}_{\e{tot}}\Big( \bs{\ga} \mid \big\{ \bs{A}_s^{(p)} , \bs{B}_s^{(p)} \big\}_{s=1, p=2}^{2\; \;\; , \;   k-1} \, ; \, \bs{\veps}_{k} \Big)
\cdot\pl{u=1}{n} \pl{r=1}{ \ell_u -1 } \Bigg\{  \f{ 1 }{   \ga_{j_r^{\, (u)}}^{ (a_r^{(u)} a_{r-1}^{(u)}) }  \, - \, \ga_{j_{r+1}^{\, (u)}}^{ (a_{r+1}^{(u)} a_{r}^{(u)}) } \, - \, \i \veps_{ a_r^{(u)} }   }  \Bigg\} \\
\limit{ \bs{\veps}_{k} }{ \bs{0} }
 \mc{H}_{\e{tot}}\Big( \bs{\ga}  \mid \big\{ \bs{A}_s^{(p)} , \bs{B}_s^{(p)} \big\}_{s=1, p=2}^{2\; \;\; , \;   k-1}  \, ; \, \bs{0}  \Big)
\cdot\pl{u=1}{n} \pl{r=1}{ \ell_u -1 } \Bigg\{  \f{ 1 }{   \ga_{j_r^{\, (u)}}^{ (a_r^{(u)} a_{r-1}^{(u)}) }  \, - \, \ga_{j_{r+1}^{\, (u)}}^{ (a_{r+1}^{(u)} a_{r}^{(u)}) }   }  \Bigg\}
\end{multline}
so that owing to the upper bound on the integration contour in \eqref{ecriture rep int pour In Gk avec contour deforme} valid uniformly in
\beq
|\veps_{a} |  ,  \qquad a=1,\dots, k \, , \; \e{small}\; \e{enough}
\enq
and which takes the explicit form
\bem
\Bigg|  \mc{H}_{\e{tot}}\Big( \bs{\ga}  \mid \big\{ \bs{A}_s^{(p)} , \bs{B}_s^{(p)} \big\}_{s=1, p=2}^{2\; \;\; , \;   k-1} \, ; \, \bs{\veps}_{k} \Big)
\cdot\pl{u=1}{n} \pl{r=1}{ \ell_u -1 }
\Bigg\{  \f{ 1 }{   \ga_{j_r^{\, (u)}}^{ (a_r^{(u)} a_{r-1}^{(u)}) }  \, - \, \ga_{j_{r+1}^{\, (u)}}^{ (a_{r+1}^{(u)} a_{r}^{(u)}) } \, - \, \i \veps_{ a_r^{(u)} }   }  \Bigg\} \Bigg| \\
\; \leq \;    C    \pl{b>a}{ k } \pl{s=1}{n_{ba} }  \Bigg| \f{ 1 }{  \cosh \Re\big[ \ga_s^{(ba)} \big] }   \Bigg|^{ 2 } \cdot
\pl{a<b<c}{ k }   \bigg(  \f{  1  }{  \eta^{(cb)} \, -  \, \eta^{(ba)}   }  \bigg)^{n_{cb} n_{ba}  } \;,
\end{multline}
one is in position to apply the dominated convergence theorem, what yields
\bem
\mc{I}_{ \bs{n} }\big[ G_k \big]  \; = \;  \pl{p=2}{k-1} \Bigg\{  \sul{ A^{(p)} \, = \, A^{(p)}_1 \underset{1}{\cup} A^{(p)}_2 }{}    \sul{ B^{(p)} \, = \, B^{(p)}_1 \underset{1}{\cup} B^{(p)}_2 }{}  \Bigg\}
 \; \;    \Int{ \big\{ \R+ \i \eta^{(21)} \big\}^{ n_{21} }   }{} \hspace{-4mm}  \dd^{ n_{21} } \ga^{(21)} \cdots
   \pl{a=1}{k-1}\Int{ \big\{ \R+ \i \eta^{(ka)} \big\}^{ n_{ka} }   }{}  \hspace{-4mm} \dd^{ n_{ka} }  \ga^{(ka)}  \\
 \times \mc{H}_{\e{tot}}\Big( \bs{\ga}  \mid \big\{ \bs{A}_s^{(p)} , \bs{B}_s^{(p)} \big\}_{s=1, p=2}^{2\; \;\; , \;   k-1} \, ; \, \bs{0} \Big)
\cdot\pl{u=1}{n} \pl{r=1}{ \ell_u -1 } \Bigg\{  \f{ 1 }{   \ga_{j_r^{\, (u)}}^{ (a_r^{(u)} a_{r-1}^{(u)}) }  \, - \, \ga_{j_{r+1}^{\, (u)}}^{ (a_{r+1}^{(u)} a_{r}^{(u)}) }    }  \Bigg\} \; .
\end{multline}
This justifies the fact that $\mc{I}_{ \bs{n} }\big[ G_k \big]$ is well defined, and thus validates the conclusions of Proposition \ref{ecriture rep serie int mult tronquee pour fct multipts tronquee}.
One now applies Lemma \ref{Lemme decomposition produit poles explicites en sous chaines} and then Lemma \ref{Lemme decomposition produit SRF comme somme combinatoire avec poles explicites}
backwards so as to re-sum, under the integral sign, the summations over the partitions of $A^{(p)}$ and $B^{(p)}$, hence leading directly to \eqref{ecriture compacte In de Gk avec contour integration dans plan complexe}.
\qed

We are now in position to state the main result of this work.
We first introduce the totally positively space-like subset of $(\R^{1,1})^k$
\beq
\mc{D}_{\e{space};+} \; = \; \Big\{  \big(\bs{x}_1,\dots, \bs{x}_k \big) \in  (\R^{1,1})^k \; : \;
\bs{x}_{ab}^2 \, <\, 0  \;\; \e{and} \;\;  \bs{x}_{a;1}>\bs{x}_{b;1}   \quad \e{for}\; \e{any} \quad 1\leq a < b \leq k \Big\}
\enq
\begin{theorem}
\label{Theoreme representation k pt fct totally spacelike}
The distribution on $\mc{C}^{\infty}_{\e{c}}\big(  \mc{D}_{\e{space};+} \big)$ induced by the truncated
$k$-point function
\beq
\Big( \bs{f}_{\e{vac}}, \op{O}_1[g_1] \cdot  \op{O}^{(r_1)}_2[g_2] \cdots   \op{O}^{(r_{k-1})}_k[g_k]  \cdot  \bs{f}_{\e{vac}}  \Big)
\enq
is given by a smooth function on $\mc{D}_{\e{space};+}$. Namely,  for any $g_1,\dots, g_k$ be smooth, compactly supported on $\R^{1,1}$ and such that, for $a \not= b$,
\beq
\bs{x}_{ab}^2 \, <\, 0 \qquad for \; any \quad \bs{x}_a \in \e{supp}[g_a] \;, \qquad a \in \intn{1}{k}\;,
\enq
and $\bs{x}_{a;1}>\bs{x}_{b;1}$ for $b>a$ it holds
\beq
\Big( \bs{f}_{\e{vac}} , \op{O}_1[g_1] \cdot  \op{O}^{(r_1)}_2[g_2] \cdots   \op{O}^{(r_{k-1})}_k[g_k]  \cdot \bs{f}_{\e{vac}} \Big) \;  = \;
\Int{ (\R^{1,1})^k }{} \pl{a=1}{k} \dd \bs{x}_a \cdot \pl{a=1}{k} g_a\big( \bs{x}_a \big)
\cdot \mc{W}_{ \bs{r}  }\big( \bs{x}_1, \dots, \bs{x}_k \big)
\enq
where
\bem
\mc{W}_{ \bs{r}  }\big( \bs{x}_1, \dots, \bs{x}_k \big) \; = \;
\sul{ \bs{n} \in \mc{N}_{\bs{r}} }{}  \f{1}{ \bs{n}! (2\pi)^{|\bs{n}|}  } \cdot \pl{b>a}{k} \bigg\{ \ex{ - 2\i\pi \, n_{ba} \om_{ba} }  \bigg\}
\pl{ b>a }{ k }
\Int{  \big\{ \R + \i \eta^{(ba)} \big\}^{n_{ba}}   }{} \hspace{-6mm} \dd^{ n_{ba} } \ga^{(ba)} \cdot  \pl{ b>a }{ k } \ex{ \i \ov{\bs{p}}(\bs{\ga}^{(ba)}) \cdot \bs{x}_{ba}}  \\
 \times \mc{S}\big(   \bs{\ga}     \big)  \cdot \pl{p=1}{k} \mc{F}^{(\op{O}_p)}\Big( \, \overleftarrow{ \bs{\ga}^{(pp-1)} } \cup \cdots \cup \overleftarrow{ \bs{\ga}^{(p1)} }  + \i \pi \ov{\bs{e}} ,
\bs{\ga}^{(kp)}  \cup \cdots \cup \bs{\ga}^{(p+1p)}   \, \Big)
 \; .
\label{ecriture expression explicite Wr D space de type +}
\end{multline}
Above, $\mc{N}_{\bs{r}}$ is as introduced in \eqref{definition ensemble sommation restreint pour troncation fct a k pts} and $\om_{ba}$
has been defined in \eqref{definition omega ba}.

\end{theorem}

Note that other representations leading to various other closed representations for the distributions restricted to other Weyl chambers of $(\R^{1,1})^k$
in terms of smooth functions may be obtained with the help of the master-representation provided in Proposition \ref{Proposition rep combinatoire pour G tot avec regularisation double + et -}.
However, the latter does not seem to lead to a set of closed representations, in terms of smooth functions, valid in patches that would cover $(\R^{1,1})^k$
with the exception of measure zero sets (such as for instance the null-cones $(\bs{x}_a-\bs{x}_{a+1})^2=0$)  where one expects singularities to arise.
Hence, we do not list all of these here as they can be readily deduced from  Proposition \ref{Proposition rep combinatoire pour G tot avec regularisation double + et -}
on a case-by-case study. In Sub-section \ref{SubSection Mixed rep} to come, we shall present one more representation
following from Proposition \ref{Proposition rep combinatoire pour G tot avec regularisation double + et -} that will be useful in establishing
the local commutativity property of the Wightman axioms in a publication to come.

Theorem \ref{Theoreme representation k pt fct totally spacelike} suggests a closed formula for the full, \textit{i.e.} non-truncated,
$k$-point correlation functions associated with mutually space-like separated points forming $\mc{D}_{\e{space};+}$. The latter would be obtained
by summing up \eqref{ecriture expression explicite Wr D space de type +} over $\bs{r}\in \mathbb{N}^{k-1}$. However, at this stage, such an expression
would only be formal in that one would have still to establish that the resulting series of multiple integrals is absolutely convergent. Such a result was obtained
for space-like separated two-point functions in \cite{KozConvergenceFFSeriesSinhGordon2ptFcts}. However the generalisation of that method
to the much more complex setting of multiple integrals defining summands for the $k$-point functions goes beyond the scope of the present work.
Here, we shall only state the result as a conjecture.

\begin{conj}

For any $g_1,\dots, g_k$ be smooth, compactly supported on $\R^{1,1}$ and such that, for $a \not= b$,
\beq
\bs{x}_{ab}^2 \, <\, 0 \qquad for \; any \quad \bs{x}_a \in \e{supp}[g_a] \;, \qquad a \in \intn{1}{k}\;,
\enq
and $\bs{x}_{a;1}>\bs{x}_{b;1}$ for $b>a$, the full $k$-point function
\beq
\Big(  \bs{f}_{\e{vac}} , \op{O}_1[g_1] \cdot  \op{O}_2[g_2] \cdots   \op{O}_k[g_k]  \cdot  \bs{f}_{\e{vac}}  \Big)
\enq
is represented by a smooth function
\beq
\Big( \bs{f}_{\e{vac}}, \op{O}_1[g_1] \cdot  \op{O}_2[g_2] \cdots   \op{O}_k[g_k]  \cdot \bs{f}_{\e{vac}} \Big) \;  =
\Int{ (\R^{1,1})^k }{} \pl{a=1}{k} \dd \bs{x}_a \cdot \pl{a=1}{k} g_a\big( \bs{x}_a \big)
\cdot \mc{W}\big( \bs{x}_1, \dots, \bs{x}_k \big)
\enq
where
\bem
\mc{W}\big( \bs{x}_1, \dots, \bs{x}_k \big) \; = \;
\sul{ \bs{n} \in \mathbb{N}^{\f{k(k-1)}{2}} }{}  \f{1}{ \bs{n}! (2\pi)^{|\bs{n}|} } \cdot \pl{b>a}{k} \exp\bigg\{ - 2\i\pi \, n_{ba} \om_{ba} \bigg\}
\pl{ b>a }{ k }
\Int{  \big\{ \R + \i \eta^{(ba)} \big\}^{n_{ba}}   }{} \hspace{-6mm} \dd^{ n_{ba} } \ga^{(ba)} \cdot  \pl{ b>a }{ k } \ex{ \i \ov{\bs{p}}(\bs{\ga}^{(ba)}) \cdot \bs{x}_{ba}}  \\
 \times \mc{S}\big(   \bs{\ga}     \big)  \cdot \pl{p=1}{k} \mc{F}^{(\op{O}_p)}\Big( \, \overleftarrow{ \bs{\ga}^{(pp-1)} } \cup \cdots \cup \overleftarrow{ \bs{\ga}^{(p1)} }  + \i \pi \ov{\bs{e}} ,
\bs{\ga}^{(kp)}  \cup \cdots \cup \bs{\ga}^{(p+1p)}   \, \Big)
 \; .
\end{multline}
The constants $\om_{ba}$ have been introduced in \eqref{definition omega ba}.

\end{conj}

\subsection{A mixed representation}
\label{SubSection Mixed rep}

We end this section by establishing an alternative representation for the truncated $k$-point function which allows for a different kind of contour deformations than in the one obtained
in Proposition \ref{Proposition Corr fct well defined through veps to zero limit}. This representation for the $\bs{r}$-truncated $k$ point function is well-tailored for proving
the local commutativity property of the \textit{per se} $k$-point correlation function, under the additional hypothesis of the convergence of the series over $\bs{r}\in \mathbb{N}^{k-1}$
of $\bs{r}$-truncated $k$ point functions. We shall address the verification of the Wightman axioms, starting from the expressions obtained in this work, in a separate publication.

\begin{prop}
	\label{Proposition alternate representation for the truncated}
Let $g_1,\dots, g_k \in \mc{C}_{\e{c}}^{\infty}(\R^{1,1})$ and pick $t \in \intn{1}{k}$ and $\bs{r}=(r_1,\dots, r_{k-1})\in \mathbb{N}^{k-1}$. Then, one has the below, well-defined, representation for the
$\bs{r}$-truncated $k$-point function which coincides with the one given in Proposition \ref{Proposition rep combinatoire pour G tot de epsk}.
\beq
\Big( \bs{f}_{\e{vac}}, \op{O}_1[g_1] \cdot  \op{O}^{(r_1)}_2[g_2] \cdots   \op{O}^{(r_{k-1})}_k[g_k]  \cdot \bs{f}_{\e{vac}} \Big)  \; = \;
\sul{ \bs{n} \in \mc{N}_{\bs{r}} }{}  \f{1}{ \bs{n}! (2\pi)^{|\bs{n}|} } \pl{a<b}{k} \Big\{ \ex{-2\i\pi n_{ba}\om_{ba}^{(t)}}  \Big\} \cdot \mc{I}_{\bs{n}}^{(t)}[G_k]
\enq
There, we agree upon $\om_{ba}^{(t)} \, = \, \sul{ \substack{\ell=a+1 \\ \not= t} }{b} \om_{\op{O}_{\ell}}$, and have set
\beq
 \mc{I}_{\bs{n}}^{(t)}[G_k]  =
\lim_{\veps_1 \tend 0^+} \hspace{-1mm} \Int{  \R^{n_{21}}   }{} \hspace{-1mm} \dd^{ n_{21} } \ga^{(21)} \cdots
\lim_{\veps_{k-1} \tend 0^+} \hspace{-5mm} \Int{  \R^{n_{k1} + \dots + n_{kk-1} }  }{}  \hspace{-6mm} \dd^{ n_{k1} }  \ga^{(k1)} \cdots \dd^{ n_{kk-1} } \ga^{(kk-1)}
\lim_{\veps_{k} \tend 0^+}   \Big( \mc{S}^{(t)}\cdot \mc{R}[ G_k ] \cdot  \mc{F}_{\e{tot};\bs{\veps}_{k} }^{(t)} \Big) \big(   \bs{\ga}     \big) \;,
\enq
with
\beq
\label{S product representation for alternating signs}
\mc{S}^{(t)}(\bs{\ga}) \, = \,
\mc{S}(\bs{\ga}) \cdot \pl{v=t+1}{k} \pl{u=1}{t-1} \left( \pl{s=1}{t-1} S\big(\bs{\ga}^{(ts)} \cup \bs{\ga}^{(vu)}   \mid \bs{\ga}^{(vu)} \cup \bs{\ga}^{(ts)} \big)
\cdot \pl{s=t+1}{k} S\big(\bs{\ga}^{(vu)} \cup \bs{\ga}^{(st)} \mid \bs{\ga}^{(st)} \cup \bs{\ga}^{(vu)} \big)\right)
\enq
$\mc{S}$  as defined in \eqref{definition facteur de diffusion complet k pts} and
\bem
\mc{F}_{\e{tot};\bs{\veps}_{k} }^{(t)} \big(   \bs{\ga}     \big) \; = \;
\pl{\substack{p=1 \\ p \neq t }}{k} \mc{F}^{(\op{O}_p)}\Big( \, \overleftarrow{ \bs{\ga}^{(pp-1)} } \cup \cdots \cup \overleftarrow{ \bs{\ga}^{(p1)} }  + \i \pi \ov{\bs{e}}_{\veps_{p}} ,
\bs{\ga}^{(kp)}  \cup \cdots \cup \bs{\ga}^{(p+1p)} \, \Big) \\
\times \mc{F}^{(\op{O}_t)}\Big( \, \bs{\ga}^{(kt)}  \cup \cdots \cup \bs{\ga}^{(t+1t)} ,
\overleftarrow{ \bs{\ga}^{(tt-1)} } \cup \cdots \cup \overleftarrow{ \bs{\ga}^{(t1)} }  - \i \pi \ov{\bs{e}}_{\veps_{t}}  \, \Big) \; .
\label{alternate form factor product}
\end{multline}
Finally,  $\mc{R}[G]$ is as defined through \eqref{definition TF space-time de la fct test} and $G_k$ as in \eqref{definition fonction Gk}.

\end{prop}

\Proof

Starting from the representation \eqref{ecriture rep int fct k pts forme la plus generale} and recasting the integrand by means of Proposition
\ref{Proposition rep combinatoire pour G tot cas avec une insertion + et que des -}, one may take explicitly the limits
$\veps_1^{\prime},\dots, \veps_{t-1}^{\prime}, \veps_t, \veps_{t+1}^{\prime},\dots, \veps_{k}^{\prime} \tend 0^+$
since nothing depends on these regulators. Then renaming $\veps_{t}^{\prime} \hookrightarrow \veps_t$
and following the very same steps and notations as in the proof of Proposition one infers the form of the multiple integral representation  upon observing that
\beq
\pl{ \substack{ s=1 \\ \not=t} }{k} \ex{ -2\i\pi \om_{\op{O}_s} |A^{(s-1)}|} \,= \,
\ex{2\i\pi \om_{\op{O}_t} r_{t-1} } \cdot \pl{s=1}{k} \ex{-2\i\pi \om_{\op{O}_s} r_{s-1} }  \, = \,
\pl{b>a}{k} \Big\{ \ex{-2\i\pi n_{ba} \sul{\ell=a+1}{k} \om_{\op{O}_{\ell}}  } \Big\}
\cdot \pl{u=t}{k} \pl{s=1}{t-1} \ex{2\i\pi \om_{\op{O}_t} n_{us} } \, = \, \pl{b>a}{k} \Big\{ \ex{-2\i\pi n_{ba}   \om_{ba}^{(t)}  } \Big\} \;.
\enq
There, we have used the expression \eqref{ecriture des entiers rp via les nab}.

Finally, the well definiteness of the $\bs{\veps}_k\tend \bs{0}^+$ limit is achieved exactly as in the proof of Proposition \ref{Proposition Corr fct well defined through veps to zero limit},
and we leave the details to the reader. This concludes the proof. \qed

\vspace{2mm}

\section{Conclusion}

This work provided various closed, rigorous, representation for the $\bs{r}$-truncated $k$ point functions in the quantum Sinh-Gordon field theory.
These lead to explicit representations for the \textit{per se} correlation functions of the theory upon summing them up over $\bs{r} \in \mathbb{N}^{k-1}$.
However, the question of convergence of such series is quite hard, and we plan to address it in a separate work. Nonetheless, such summations can already
be taken as quite serious conjectures for the closed expressions for the $k$-point functions in this integrable quantum field theory.
No closed result for any $k$-point function with $k \geq 3$ was ever obtained in the literature, even when disregarding the convergence issues.
We plan to study more precisely the expressions we obtain for the $k$ point functions in a separate publication where we will show that, if convergence is assumed,
the resulting expressions satisfy all of the Wightman axioms.

\section*{Acknowledgment}

The work of KKK and AS is supported by the ERC Project LDRAM : ERC-2019-ADG Project 884584. KKK acknowledges support from CNRS.
We thank F. Smirnov, Y. Potaux for stimulating discussions.

\appendix

\section{Auxiliary results}
\label{Appendix Section Auxiliary Results}
\subsection{Auxiliary identities}
\label{Appendix Subsection auxiliary identities}

\begin{defin}
\label{definition singature vectorielle}

Given a set $B=\{b_1,\dots, b_N\}$ and an ordered partition $B=B_1 \underset{12}{\cup} B_2$, let $\eps\Big( \, \bs{B}  \mid  \bs{B}_1 \cup  \bs{B}_2 \, \Big)$
denote the signature of the permutation which permutes the coordinates of the vector $\bs{B}=\big( b_1,\dots, b_N \big)$
into those obtained by the concatenation $\bs{B}_1 \cup  \bs{B}_2$.

\end{defin}

\begin{lemme}
\label{Lemme decomposition sur pole simples Cauchy generalise}

 Let $a_1,\dots, a_M$ and $b_1,\dots, b_N$ be two collections of mutually pairwise distinct complex numbers. Then, given $A=\{a_s\}_1^M$ and $B=\{b_{\ell}\}_{1}^{N}$ one has the combinatorial representation
\bem
\f{ \pl{r> \ell }{ M } a_{r\ell} \pl{r < \ell }{ M } b_{r\ell}    }{  \pl{r=1}{M} \pl{\ell}{N} (a_r   -   b_{\ell} ) } \; = \;
\sul{ B = B_1 \underset{1}{\cup} B_2   }{}\sul{ A = A_1 \underset{1}{\cup} A_2   }{}
\f{ \eps\,\big( \, \bs{A}  \mid  \bs{A}_1 \cup  \bs{A}_2 \, \big) \,  \eps\,\big( \, \bs{B}  \mid  \bs{B}_1 \cup  \bs{B}_2 \, \big)  }
{ ( |A| \wedge |B| )! } \\
\times \f{   \pl{ r<\ell }{ |B_2|}  \Big\{ \big( \bs{B}_2\big)_r \, - \,  \big( \bs{B}_2\big)_{\ell} \Big\} \cdot
                \pl{ r > \ell }{ |A_2|}  \Big\{ \big( \bs{A}_2\big)_r \, - \,  \big( \bs{A}_2\big)_{\ell} \Big\}  }{  \pl{r=1}{|A_1|} \Big\{ \big( \bs{A}_1\big)_r \, - \,  \big( \bs{B}_1\big)_r \Big\} } \;,
\label{ecriture decomposition Cauchy generalise sur somme permutation}
\end{multline}
where $x_{ab}=x_a-x_b$ and $\eps$ is as given in Definition \ref{definition singature vectorielle}. Finally, the summation over partitions is made under the constraint
\beq
|A_1| \, = \, |B_1| \, = \, |A| \wedge |B| \;,
\enq
which implies that one of the two partitions trivially reduce to a sum over the permutation group.

\end{lemme}

\Proof

Assume first that $M \geq N$ and introduce the integral
\beq
\mc{I}_M \; = \; \Int{  \Ga\big( \{a_r\} \big) }{} \f{ \dd^M z }{ (2\i\pi)^M } \cdot
\f{ \pl{r> \ell }{ M } z_{r\ell} \pl{r < \ell }{ M } b_{r\ell}    }{  \pl{r=1}{M} \Big\{ (z_r- a_r)  \pl{\ell=1}{N} (z_r   -   b_{\ell} ) \Big\} } \;,
\enq
where $ \Ga\big( \{a_r\} \big)$ is a collection of small index one loops around $a_1,\dots, a_{M}$ such that $b_1, \dots, b_N$ are located in their exterior.

It is direct to check by taking the resides at $z_r=a_r$ that
\beq
\mc{I}_M \; = \; \f{ \pl{r> \ell }{ M } a_{r\ell} \pl{r < \ell }{ M } b_{r\ell}    }{  \pl{r=1}{M} \pl{\ell=1}{N} (a_r   -   b_{\ell} ) } \; .
\enq
However, seen as a function of a single variable $z_r$, the integrand decays as $z_r^{M-N-2}$ at infinity, meaning that there is no residue at $\infty$
so that
\beq
\mc{I}_M \; = \; (-1)^M  \hspace{-3mm} \Int{  \Ga\big( \{b_r\} \big) }{} \hspace{-3mm} \f{ \dd^M z }{ (2\i\pi)^M } \cdot
\f{ \pl{r> \ell }{ M } z_{r\ell} \pl{r < \ell }{ M } b_{r\ell}    }{  \pl{r=1}{M} \Big\{ (z_r- a_r)  \pl{\ell=1}{N} (z_r   -   b_{\ell} ) \Big\} } \;,
\enq
where $ \Ga\big( \{b_r\} \big)$ is a collection of small, index one, loops around $b_1,\dots, b_N$ such that $a_1, \dots, a_M$ are located in the exterior.
Due to the presence of the Vandermonde determinant, the residue corresponding to two $z$-variables  evaluated at the same point $b_s$
vanishes. Thus, taking the integral by means of the residue at the poles $z_r   =   b_{\ell}$, $\ell=1,\dots, N$, amounts to
to picking an ordered partition $B = B_1 \underset{1}{\cup} B_2 $ with $|B_1|=M$ in which, elements of $B_1$ may be permuted in any order.
Then, one evaluates the residues at $z_r= (\bs{B}_1)_r, r=1,\dots, M$. All calculations done, this yields
\beq
\mc{I}_M \; = \; (-1)^M \sul{ B = B_1 \underset{1}{\cup} B_2   }{}
  \eps\,\big(  \bs{B}  \mid  \bs{B}_1 \cup  \bs{B}_2   \big)
\cdot  \f{   \pl{ r<\ell }{ |B_2|}  \Big\{ \big( \bs{B}_2\big)_r \, - \,  \big( \bs{B}_2\big)_{\ell} \Big\}   }
{  \pl{r=1}{|B_1|} \Big\{ \big( \bs{B}_1\big)_r \, - \,  a_r \Big\} } \;.
\enq
Then, performing the change of permutation $\sg \hookrightarrow \sg \circ \pi^{-1}$ and given
\beq
\bs{B}_1 \, = \,  \big( B_{\a_1}, \dots, B_{\a_M}  \big) \quad \e{with} \quad
1\leq \a_1 < \dots < \a_M \leq N\; ,  \quad \e{by} \, \e{setting} \quad \bs{B}_1^{\sg}  \, = \,  \big( B_{\a_{\sg(1)}}, \dots, B_{\a_{\sg(M)} }  \big) \;,
\enq
one has for every $\pi \in \mf{S}_{ |A_1| }$
\bem
\mc{I}_M \; = \; (-1)^M \hspace{-2mm} \sul{ B = B_1  \cup B_2   }{} \sul{ \sg \in \mf{S}_{ |B_1| } }{}
  \hspace{-2mm} \eps\,\big(   \bs{B}  \mid  \bs{B}_1^{\sg \circ \pi^{-1}} \! \! \cup  \bs{B}_2   \big)
\cdot  \f{   \pl{ r<\ell }{ |B_2|}  \Big\{ \big( \bs{B}_2\big)_r \, - \,  \big( \bs{B}_2\big)_{\ell} \Big\}   }
{  \pl{r=1}{|B_1|} \Big\{ \big( \bs{B}_1^{\sg}\big)_{\pi^{-1}(r)} \, - \,  a_r \Big\} }  \\
\; = \; (-1)^M \hspace{-2mm} \sul{ B = B_1  \underset{1}{\cup}  B_2   }{} \hspace{-2mm} (-1)^{\pi} \,
  \eps\,\big(  \bs{B}  \mid  \bs{B}_1  \cup  \bs{B}_2 \big)
\cdot  \f{   \pl{ r<\ell }{ |B_2|}  \Big\{ \big( \bs{B}_2\big)_r \, - \,  \big( \bs{B}_2\big)_{\ell} \Big\}   }
{  \pl{r=1}{|B_1|} \Big\{ \big( \bs{B}_1\big)_{r} \, - \,  a_{\pi(r)} \Big\} } \;.
\end{multline}
Thus summing the above over $\pi \in \mf{S}_{ |A_1| }$, dividing by $|A_1|! \, = \, ( |A| \wedge |B| )!$,
and observing that
\beq
\eps\,\big(  \bs{A}  \mid  \bs{A}_1^{\pi} \cup  \bs{A}_2 \big)= (-1)^{\pi} \quad \e{for} \quad  A_2=\emptyset \; ,
\enq
yields
\beq
\mc{I}_M \; =     \hspace{-2mm} \sul{ B = B_1  \underset{1}{\cup}  B_2   }{}  \sul{ A = A_1 \underset{1}{\cup} A_2   }{}
\f{ \eps\,\big(  \bs{A}  \mid  \bs{A}_1 \cup  \bs{A}_2  \big) \,  \eps\,\big(  \bs{B}  \mid  \bs{B}_1 \cup  \bs{B}_2  \big)  }
{ ( |A| \wedge |B| )! }
\cdot  \f{   \pl{ r<\ell }{ |B_2|}  \Big\{ \big( \bs{B}_2\big)_r \, - \,  \big( \bs{B}_2\big)_{\ell} \Big\}   }
{  \pl{r=1}{|B_1|} \Big\{ \big( \bs{A}_1\big)_{r} \, - \,  \big( \bs{B}_1\big)_{r} \Big\} } \;.
\enq
This exactly reproduces the formula \eqref{ecriture decomposition Cauchy generalise sur somme permutation} since one is in the setting where $|B| \geq |A|$
so that $A_2 = \emptyset$ meaning that the ordered product over $A_2$ elements reduces to $1$.

The reasonings in the case $M \geq N $ are similar and start from the integral identity
\beq
\mc{I}_M \; = \; \f{ \pl{r> \ell }{ M } a_{r\ell} \pl{r < \ell }{ M } b_{r\ell}    }{  \pl{r=1}{M} \pl{\ell=1}{N} (a_r   -   b_{\ell} ) }
\; =  \Int{  \Ga\big( \{b_r\} \big) }{} \hspace{-3mm} \f{ \dd^N z }{ (2\i\pi)^N } \cdot
\f{ \pl{r> \ell }{ M } a_{r\ell} \pl{r < \ell }{ N } z_{r\ell}    }{  \pl{r=1}{N} \Big\{ (z_r- b_r)  \pl{\ell=1}{M} (a_{\ell}   -  z_{r} ) \Big\} } \;,
\enq
Taking the integral by means of the residues located outside of the integration contour, one gets
\beq
\mc{I}_M \; = \;     \sul{ A = A_1 \underset{1}{\cup} A_2   }{}
 \eps\,\big( \bs{A}  \mid  \bs{A}_1 \cup  \bs{A}_2 \, \Big)
\cdot  \f{   \pl{ r>\ell }{ |A_2|}  \Big\{ \big( \bs{A}_2\big)_r \, - \,  \big( \bs{A}_2\big)_{\ell} \Big\}   }
{  \pl{\ell=1}{|A_1|} \Big\{ \big( \bs{A}_1\big)_{\ell} \, - \,  b_{\ell} \Big\} } \;.
\enq
Then, upon symmetrising as before, one gets
\beq
\mc{I}_M \; =     \hspace{-2mm} \sul{ B = B_1  \underset{1}{\cup}  B_2   }{}  \sul{ A = A_1 \underset{1}{\cup} A_2   }{}
\f{ \eps\,\big(  \bs{A}  \mid  \bs{A}_1 \cup  \bs{A}_2  \big) \,  \eps\,\big(  \bs{B}  \mid  \bs{B}_1 \cup  \bs{B}_2  \big)  }
{ ( |A| \wedge |B| )! }
\cdot  \f{   \pl{ r>\ell }{ |A_2|}  \Big\{ \big( \bs{A}_2\big)_r \, - \,  \big( \bs{A}_2\big)_{\ell} \Big\}   }
{  \pl{r=1}{|B_1|} \Big\{ \big( \bs{A}_1\big)_{r} \, - \,  \big( \bs{B}_1\big)_{r} \Big\} } \;.
\enq
This again reproduces  \eqref{ecriture decomposition Cauchy generalise sur somme permutation} taken that $|A| \geq |B|$
so that $B_2 = \emptyset$ meaning that the ordered product over $B_2$ elements reduces to $1$. \qed

\subsection{Special functions}
\label{Appendix Fonctions Speciales}

The Barnes $G$-function admits an integral representation involving the $\psi$-function, $\psi(z)=\ln^{\prime}\Ga(z)$,
\beq
G(z+1) \, = \, (2\pi)^{\f{z}{2}} \cdot \exp\bigg\{ - \f{z(z-1)}{2} \, + \, \Int{0}{z} t \psi(t) \dd t \bigg\} \; , \; \Re(z) >-1 \;. 
\enq
It is continued to the whole complex plane by means of the functional equation 
\beq
G(z+1)=\Ga(z) G(z) \;. 
\enq
The Barnes function is entire. Its zeroes are located at $-\mathbb{N}$ and $-n$ is a zero of $G$ of degree $n$. The Barnes function admits \cite{FerreiraLopezAEforBarnesFct} the large-$z$
asymptotic expansion which is valid uniformly on $|\e{arg}(z)|\leq \pi-\eps$ for any $\eps>0$ and fixed. Moreover, the remainder in this expansion
is infinitely differentiable, \textit{viz}. the control also holds for $G^{(k)}$ with any $k$ fixed, provided one differentiates the remainder an appropriate number of times.
The expansion takes the form
\beq
G(1+z) \; = \; \exp\bigg\{  z^2 \, \Big( \f{\ln z}{2} \, - \, \f{3}{4} \Big) \, +\, z \, \ln \sqrt{2\pi} \, - \, \f{\ln z}{12} + \zeta^{\prime}(-1) \, + \, \e{O}\Big( \tfrac{1}{z} \Big) \bigg\} \;.
\enq
It allows one to infer that for any $a\in \Cx$ fixed and $z\tend \infty$ with $|\e{arg}(z)|\leq \pi-\eps$, one has that
\beq
\f{ G(1+z+a) }{ G(1+z) } \; = \; \exp\bigg\{ a z \ln z \, - \,   a z  \, + \, \f{a^2}{2} \ln z  \, + \, a \ln \sqrt{2\pi} \, + \, \e{O}\Big( \tfrac{1}{z} \Big) \bigg\} \;.
\label{ecriture DA de rapport de G shifte sur G}
\enq

\section{Master representation for multi-point densities and applications}
\label{Appendix Section Master representation multi pt fcts}

\subsection{The Master representation}
\label{Appendix SubSection Master representation}

In Proposition \ref{Proposition rep combinatoire pour G tot de epsk},
we have establish one kind of representation, based on Lemma  \ref{Lemma rep direct pour noyau integral generalise secteur n m particules},
for the regularised multi-point generalised density $\mc{G}_{\e{tot}}[G] \Big( \big\{  \bs{A}^{(s)}   \big\}^{k}_{0} ; \bs{\veps}_k \Big)$
defined in \eqref{definition G tot de eps}. However, one may obtain more general representations for the multi-point generalised density by building on the representation of the
individual integral kernels $\mc{M}^{(\op{O})}_{n;m}\big(  \bs{\alpha}_n ; \bs{\be}_{m} \big)$ provided by \eqref{ecriture representation combinatoire pour MO via reduction mixte entre 1 et 2}
of Lemma \ref{Lemme ecriture generale densite FF}.  The latter provides an additional regularisation and allows one to obtain the below from of a smeared field action.
Given $\bs{\alpha}_n \in \R^{n}$ and $\bs{\be}_{m} \in \R^m$, we denote $B=\{\be_a\}_{1}^{m}$,  $A=\{\a_{a}\}_1^{n}$ and consider the partition $A=A_1\cup A_2$
Then, one has
\beq
\op{M}_{\op{O}}^{(m)} \big[g , f^{(m)}\big]  \big(  \bs{\a}_n \big)   \; =    \lim_{ \veps, \veps^{\prime} \tend 0^+} \hspace{-1mm}
  \Int{ \R_{>}^m }{}    \f{ \dd^m \be }{ (2\pi)^m }
\mc{M}^{(\op{O})}_{ |A| ; |B| }\big(  \bs{A}     ; \bs{B}  \big)_{\veps, \veps^{\prime} }  \cdot \mc{R}[g]\big(    \bs{A} ,  \bs{B} \big)  \cdot  f^{(m)}\big(   \bs{B}   \big) \;.
\enq
The momentum regulator $\mc{R}[g]\big(    \bs{A} ,  \bs{B} \big)$ is as introduced in \eqref{definition facteur regularisant espace reel}
while  the regularised kernel is expressed as
\bem
\mc{M}^{(\op{O})}_{ |A| ; |B| }\big(  \bs{A}     ; \bs{B}  \big)_{\veps, \veps^{\prime} }   \; =
 \sul{   A_1 = C_1 \cup C_2  }{}
\sul{   A_2 = D_1 \cup D_2  }{} \hspace{-3mm} \ex{-2  \i\pi |A_1 | \om_{\op{O}}    }
  \sul{ B =  \underset{13}{\cup}\, _{a=1}^{3} B_a  }{}   \hspace{-2mm}  \De\big( \bs{C}_1\cup \bs{D}_1 \mid \bs{B}_1 \cup \bs{B}_3 \big)
  \;  \\
 \times\op{S}\Big( \, \overleftarrow{\bs{A}}  \mid \overleftarrow{ \bs{D}_1 }\cup  \overleftarrow{ \bs{D}_2 }  \cup   \overleftarrow{ \bs{C}_2}\cup  \overleftarrow{ \bs{C}_1}  \Big)
\cdot \op{S}\big(   \bs{B} \mid \bs{B}_1 \cup \bs{B}_2\cup  \bs{B}_3 \big)
\cdot \mc{F}^{(\op{O})} \Big( \,  \overleftarrow{ \bs{C}_2} + \i \pi \ov{\bs{e}}_{\veps} ,  \bs{B}_2, \overleftarrow{ \bs{D}_2} - \i \pi \ov{\bs{e}}_{\veps^{\prime}}  \Big) \;.
%
%
%
%
%
%
%
%
%
%
%
%
%
\end{multline}
We remind that the overall shift regulator $\ov{\bs{e}}_{\veps}$ has been introduced in \eqref{definition ov bs eps avec indice regulateur}.
Further, the partitions involved in the expression for the regularised kernel have their cardinalities constrained as
$ |\bs{C}_1| \,= \, |\bs{B}_1| $ and $ |\bs{D}_1| \,= \,  |\bs{B}_3|$.

\noindent Given $r_0, \dots, r_{k} \in \mathbb{N}$, the above yields
\bem
\Big( \op{O}^{(r_0)}_1[g_1] \cdot  \op{O}^{(r_1)}_2[g_2] \cdots   \op{O}^{(r_{k-1})}_k[g_k]  \cdot \bs{f}_{r_k} \Big) \big( \bs{\a}^{(0)}_{r_0} \big) \; = \;
\lim_{\veps_1,\veps_1^{\prime}  \tend 0^+}  \Int{  \R_{>}^{r_1}   }{} \f{ \dd^{ r_1 } \a^{(1)} }{ (2\pi)^{r_1} }\cdots
\lim_{\veps_{k-1}^{\prime}, \veps_{k-1} \tend 0^+} \Int{  \R_{ > }^{ r_{k-1} }   }{ } \hspace{-2mm} \f{ \dd^{ r_{k-1} } \a^{ (k-1) } }{ (2\pi)^{r_{k-1} }}  \\
 \times \lim_{\veps_{k}^{\prime}, \veps_{k} \tend 0^+} \Int{  \R_{>}^{r_k}   }{} \hspace{-1mm} \f{ \dd^{ r_k } \a^{(k)} }{ (2\pi)^{r_k} }   f^{(r_k)} \big( \bs{A}^{(k)} \big) \cdot
  \mc{G}\big( \big\{  \bs{A}^{(s)}   \big\}^{k}_{0} ; \bs{\veps}_k, \bs{\veps}^{\prime}_k \big)  \cdot  \mc{R}[ G_k]\big(   \big\{  A^{(s)}     \big\}^{k}_{0} \big)  \; .
\label{ecriture rep int fct k pts forme la plus generale}
\end{multline}
Above, $\bs{f}_{r}$ has been defined in \eqref{definition vecteur fr dans secteur r particule pure} while $G_k$ and $\mc{R}[G]$ are as given in
\eqref{definition fonction Gk} and \eqref{definition Fourier direct de G a k veriable Minkowski}. The remaining building block of the integrand takes the form
\beq
 \mc{G}\big( \big\{  \bs{A}^{(s)}   \big\}^{k}_{0} ; \bs{\veps}_k, \bs{\veps}^{\prime}_k \big)  \; = \;
 \mc{M}^{(\op{O}_1)}_{r_0;r_1}\big(  \bs{\alpha}_{r_0}^{(0)} ; \bs{\alpha}_{r_1}^{(1)} \big)_{\veps_1, \veps_{1}^{\prime}} \cdot
 \mc{M}^{(\op{O}_2)}_{r_1;r_2}\big(  \bs{\alpha}_{r_1}^{(1)} ; \bs{\alpha}_{r_2}^{(2)} \big)_{\veps_2, \veps_{2}^{\prime}}  \cdots
\mc{M}^{(\op{O}_k)}_{r_{k-1};r_k}\big(  \bs{\alpha}_{r_{k-1}}^{(k-1)} ; \bs{\alpha}_{r_k}^{(k)} \big)_{\veps_k, \veps_{k}^{\prime}} \;.
\label{definition G regularise plus et moins}
\enq
There, the regularised kernels are subordinate to the fixed partitions $\bs{A}^{(\ell-1)} \, =\,  \bs{A}^{(\ell-1)}_1  \cup \bs{A}^{(\ell-1)}_2$.

\begin{prop}
  \label{Proposition rep combinatoire pour G tot avec regularisation double + et -}

  Let $A^{(0)}=\emptyset$ and $A^{(s)}=\{ \a_a^{(s)} \}_{1}^{r_s}$ for $s=1,\dots, k$.
\beq
\mc{G}_{\e{tot}}[G] \Big( \big\{  \bs{A}^{(s)}   \big\}^{k}_{0} ; \bs{\veps}_k , \bs{\veps}_k^{\prime} \Big)  \; = \;  \mc{G}\Big( \big\{  \bs{A}^{(s)}   \big\}^{k}_{0} ; \bs{\veps}_k, \bs{\veps}^{\prime}_k \Big)
\cdot  \mc{R}[ G_k]\big(   \big\{  A^{(s)}     \big\}^{k}_{0} \big)
\label{ecriture G tot de eps et eps prime k}
\enq
  with $\mc{G}$ as in \eqref{definition G regularise plus et moins} and $\mc{R}[G]$ as in  \eqref{definition facteur regularisant espace reel}.
Then, given a partitioning $A^{(s)} = A_1^{(s)}\cup A_2^{(s)}$ for $s=1,\dots, p-1$, one has the expansion
\bem
\mc{G}_{\e{tot}}[G] \Big( \big\{  \bs{A}^{(s)}   \big\}^{k}_{0} ; \bs{\veps}_k , \bs{\veps}_k^{\prime} \Big) \; = \;
\pl{s=1}{k-1} \ex{-2\i\pi \om_{\op{O}_{s+1}}|A_1^{(s)}| } \cdot   \pl{p=1}{k-1} \bigg\{   \sul{ \op{P}_p[  A^{(p)}_1, A^{(p)}_2 ]  }{}
\pl{a=1}{2^{p-1}-1} \sul{ \sg_{a}^{(p-1)} \in \mf{S}_{  |  X_{a,+}^{(p-1)} |  } }{} \sul{ \sg_{a+2^{p-1}}^{(p-1)} \in \mf{S}_{  |  X_{a,-}^{(p-1)} |  } }{}   \bigg\} \sul{ \op{P}_k[  A^{(k)} ]  }{}
 \\
\times \pl{a=1}{2^{k-1}-1} \bigg\{ \De\big( \bs{X}_{a,+}^{(k-1)}  \mid \bs{E}_{a,+}^{(k)}  \big) \cdot \De\big( \bs{X}_{a,-}^{(k-1)}  \mid \bs{E}_{a,-}^{(k)}  \big)  \bigg\}
\cdot \pl{p=2}{k-1} \pl{a=1}{2^{p-1}-1} \bigg\{
\De\Big( \bs{X}_{a,+}^{(p-1)}  \mid \overrightarrow{ \big( \bs{X}_{a,+}^{(p)} \cup \bs{X}_{a,-}^{(p)} \cup \bs{W}_{a,+}^{(p)} \cup \bs{W}_{a,-}^{(p)}  \big)}^{ \sg_{a}^{(p-1)} }  \Big)  \\
\times \De\Big( \bs{X}_{a,-}^{(p-1)}  \mid
                        \overrightarrow{ \big( \bs{X}_{a+2^{p-1},+}^{(p)} \cup \bs{X}_{a+2^{p-1},-}^{(p)} \cup \bs{W}_{a+2^{p-1},+}^{(p)} \cup \bs{W}_{a+2^{p-1},-}^{(p)}  \big)}^{ \sg_{a+2^{p-1} }^{(p-1)} }  \Big) \bigg\}
\\
\times \msc{S}_{\e{tot}}\Big( \big\{ \Xi_{\eps}^{(p)}; W_{a,\eps}^{(p)} \big\}  \Big)  \cdot
\pl{p=1}{k} \bigg\{  \mc{F}^{(\op{O}_p)} \Big( \overleftarrow{ \bs{\ga}_+^{(pp-1)} } \cup \dots \cup  \overleftarrow{ \bs{\ga}_+^{(p1)} }+ \i \pi \ov{\bs{e}}_{\veps_p}, \\
\bs{\Xi}_{+}^{(p)} \cup \bs{\Xi}_{-}^{(p)} \cup \bs{\ga}_{+}^{(kp)} \cup \bs{\ga}_{-}^{(kp)} \cup \dots \cup \bs{\ga}_{+}^{(p+1p)} \cup \bs{\ga}_{-}^{(p+1p)},
\overleftarrow{ \bs{\ga}_-^{(pp-1)} }  \cup \dots \cup   \overleftarrow{ \bs{\ga}_-^{(p1)} }   - \i \pi \ov{\bs{e}}_{\veps_p^{\prime}} \Big)  \bigg\}
\cdot  \msc{R}_{\e{glob}}[ G]\Big(   \{ \Xi_{\eps}^{(a)} \}; \{\ga_{\eps}^{(ba)} \}  \Big)   \;.
\label{formule pour reecriture action mixte + et -}
\end{multline}
The sums run through partitions  $\op{P}_p[  A^{(p)}_1, A^{(p)}_2 ]$ of $A^{(p)}_1$ and $A^{(p)}_2$ which, for  $p \in \intn{1}{ k-1 }$ take the form
\beq
A^{(p)}_1 \,  = \,  \bigcup\limits_{a=1}^{2^p-1}\Big\{ W_{a,+}^{(p)} \cup  X_{a,+}^{(p)}   \Big\} \; , \qquad
A^{(p)}_2 \, = \,  \bigcup\limits_{a=1}^{2^p-1}\Big\{ W_{a,-}^{(p)} \cup  X_{a,-}^{(p)}   \Big\}
\label{ecriture sous partitions des Ap pour calcul multipoints}
\enq
and  index-ordered partitions $\op{P}_k[  A^{(k)} ]$ of $A^{(k)}$:
\beq
A^{(k)} \, = \, E_{0}^{(k)}  \bigcup\limits_{\eps = \pm} \bigcup\limits_{1, \cdots , 2^{k-1}-1 }^{ 2^{k-1}-1 } \hspace{-2mm} E_{s,\eps}^{(k)} \;.
\label{ecriture sous partitions des Ak pour calcul multipoints}
\enq
Further, one sums over permutations
\beq
\sg_{a}^{(p-1)}  \in \mf{S}_{  |  X_{a,+}^{(p-1)}  | } \; , \quad  \sg_{2^{p-1}+a}^{(p-1)} \in \mf{S}_{  |  X_{a,-}^{(p-1)}  | }
\label{definition permutations porusommations de type + et - jointe}
\enq
with $a \in \intn{1}{ 2^{p-1}-1}$ and $p \in \intn{2}{k-1}$.  The cardinalities of the sets building up the partitions of the $A^{(p)}_a$s are constrained as follows
\beqa
\big|  X_{a,+}^{(p-1)} \big| & = & \big| X_{a,+}^{(p)}  \cup X_{a,-}^{(p)}  \cup W_{a,+}^{(p)} \cup W_{a,-}^{(p)} \big|  \vspace{2mm}  \label{ecriture contrainte partitions Xa+} \vspace{4mm} \\
\big|  X_{a,-}^{(p-1)} \big| & = & \big| X_{a+2^{p-1},+}^{(p)}  \cup X_{a+2^{p-1},-}^{(p)}  \cup W_{a+2^{p-1},+}^{(p)} \cup W_{a+2^{p-1},-}^{(p)} \big|
\label{ecriture contrainte partitions Xa-}
\eeqa
this for $a \in \intn{1}{2^{p-1}-1}$ and $p \in \intn{1}{k-1}$. Finally, the last constraint reads
and
\beq
\big|  X_{a,\eps}^{(k-1)} \big| \; = \, \big| E_{a,\eps}^{(k)}  \big| \qquad  for  \qquad  a \in \intn{1}{2^{k-1}-1 } \quad and  \quad \eps = \pm \;.
\label{ecriture contrainte partitions Ea+ et Ea-}
\enq
Next,
\beq
\msc{R}_{\e{glob}}[ G]\Big(   \{ \Xi_{\eps}^{(a)} \}; \{\ga_{\eps}^{(ba)} \}  \Big)   \; = \hspace{-2mm} \Int{ \big( \R^{1,1} \big)^{k} }{}  \hspace{-2mm} \pl{s=1}{k}\dd \bs{x}_s
\cdot  G\big(\bs{x}_1,\dots, \bs{x}_k \big) \cdot
\ex{\i \msc{P}_{\e{glob}}\big(  \{ \Xi_{\eps}^{(a)} \}; \{\ga_{\eps}^{(ba)} \}; \{\bs{x}_s\} \big)  } \; ,
\enq
where we have used the shorthand notation
\beq
\msc{P}_{\e{glob}}\big(  \{ \Xi_{\eps}^{(a)} \}; \{\ga_{\eps}^{(ba)}\}; \{\bs{x}_s\} \big) \; = \;\sul{\eps = \pm}{} \bigg\{     \sul{b>a}{k}  \ov{\bs{p}} \, \big( \bs{\ga}_{\eps}^{(ba)}  \big)
\cdot \bs{x}_{ba}  \, - \, \sul{a=1}{k}   \ov{\bs{p}} \, \big( \bs{\Xi}^{(a)}_{\eps} \big)  \cdot \bs{x}_p \bigg\}
\label{definition curly P global}
\enq
and introduced the auxiliary sets
\beq
\ga_{\eps}^{(ba)} = \bigcup_{s=0}^{2^{b-a-1}-1} W_{2^{a-1} (2s+1) ,\eps}^{(b-1)} \qquad with \quad  1 < a < b \leq k  \;,
\label{definition des ensembles gamma ba eps}
\enq
and for $p=1,\dots, k-1$
\beq
\Xi^{(p)}_{\eps} \; =   \hspace{-2mm} \bigcup_{s=0}^{2^{k-p-1}-1} \hspace{-2mm} E_{2^{p-1} (2s+1) ,\eps}^{(k)} \quad while \quad \Xi^{(k)}_{+}   \; = \; E_{0}^{(k)}  \quad and  \quad \Xi_{-}^{(k)}=\emptyset \;.
\label{definition des ensembles Xi p eps}
\enq
Finally, one has the product representation
\bem
\msc{S}_{\e{tot}}\Big( \big\{ \Xi_{\eps}^{(p)}; W_{a,\eps}^{(p)} \big\}  \Big) \; = \; \op{S}\big(  \bs{A}^{(k)} \mid \bs{C}_1^{(k-1)} \cup  \bs{\Xi}_+^{(k)} \cup  \bs{D}_1^{(k-1)}\big)
 \pl{p=1}{k-1}  \Bigg\{ \op{S}\Big( \,  \bs{C}_1^{(p)} \cup  \bs{\ga}_+^{(p+11)} \cup \dots \cup \bs{\ga}_+^{(p+1p)} \cup \bs{\ga}_-^{(p+11)} \cup \dots \cup \bs{\ga}_-^{(p+1p)}    \\
\cup  \bs{D}_1^{(p)}
\mid \bs{C}_1^{(p-1)} \cup  \bs{\Xi}_{+}^{(p)} \cup \bs{\Xi}_{-}^{(p)} \cup \bs{\ga}_{+}^{(kp)} \cup \bs{\ga}_{-}^{(kp)} \cup \dots
 \cup \bs{\ga}_{+}^{(p+1p)} \cup \bs{\ga}_{-}^{(p+1p)} \cup \bs{D}_1^{(p-1)}  \Big) \Bigg\}\;.
\label{definition matrice Stot cas + et - mixte}
\end{multline}
It is understood that $\bs{C}_1^{(0)}, \bs{D}_1^{(0)}$ are absent from the concatenation while  $\bs{C}_1^{(p)}, \bs{D}_1^{(p)}$ correspond to vectors built from the below sets with any choice of
ordering for their coordinates:
\beqa
C_1^{(p)} & = &  \bigcup\limits_{\eps=\pm}^{}   \bigcup\limits_{v=p+1}^{k-1}  \bigcup\limits_{a=1}^{2^p-1} \bigcup\limits_{s=0}^{2^{v-p-1}-1} \hspace{-2mm} W_{a+s 2^{p+1}  ; \eps}^{(v)} \quad
\bigcup\limits_{\eps=\pm}^{}  \bigcup\limits_{a=1}^{2^p-1} \bigcup\limits_{s=0}^{2^{k-p-2}-1} E_{a+ s 2^{p+1}  ; \eps}^{(k)}   \;,  \label{ecriture C1p apres resolution recurrence} \\
D_1^{(p)} & = &  \bigcup\limits_{\eps=\pm}^{}   \bigcup\limits_{v=p+1}^{k-1}  \bigcup\limits_{a=1}^{2^p-1} \bigcup\limits_{s=0}^{2^{v-p-1}-1} \hspace{-2mm}  W_{a+(2s+1) 2^{p}  ; \eps}^{(v)}  \;
\bigcup\limits_{\eps=\pm}^{}  \bigcup\limits_{a=1}^{2^p-1} \bigcup\limits_{s=0}^{2^{k-p-2}-1} E_{a+(2s+1)2^{p}  ; \eps}^{(k)}  \;.
\label{ecriture D1p apres resolution recurrence}
\eeqa
if $p=1,\dots, k-2$, while
\beq
C_1^{(k-1)} \, = \,  \bigcup\limits_{a=1}^{2^{k-1}-1} E_{a ; +}^{(k)}  \quad and \quad
D_1^{(k-1)} \, = \,  \bigcup\limits_{a=1}^{2^{k-1}-1} E_{a ; -}^{(k)}   \;.
\label{ecriture C1k-1 et D1k-1 apres resolution recurrence}
\enq

\end{prop}

\Proof

We first focus on re-expressing, in a more convenient way,  the product of regularised kernels $ \mc{M}^{(\op{O}_{\ell})}_{r_{\ell-1};r_{\ell}}$
\beq
\wt{\mc{G}}\big( \big\{  \bs{A}^{(s)}   \big\}^{k}_{0} ; \bs{\veps}_k, \bs{\veps}^{\prime}_k \big) \; = \;
\pl{s=2}{k} \Big\{ \ex{2\i\pi \om_{\op{O}_s}|A^{(s-1)}| } \Big\} \cdot  \mc{G}\big( \big\{  \bs{A}^{(s)}   \big\}^{k}_{0} ; \bs{\veps}_k, \bs{\veps}^{\prime}_k \big) \, ,
\enq
with $\mc{G}$ as given through \eqref{definition G regularise plus et moins}, .

Each of the building blocks may be represented through  \eqref{ecriture representation combinatoire pour MO via reduction mixte entre 1 et 2} what yields to
\bem
\wt{\mc{G}}\big( \big\{  \bs{A}^{(s)}   \big\}^{k}_{0} ; \bs{\veps}_k, \bs{\veps}^{\prime}_k \big) \; = \;
\pl{p=1}{k-1} \Bigg\{   \sul{ A_1^{(p)}= C_1^{(p)} \cup C_2^{(p)}  }{}  \sul{ A_2^{(p)} = D_1^{(p)} \cup D_2^{(p)}  }{} \hspace{-3mm} \ex{ 2 \i\pi \om_{\op{O}_{p+1}} |A_2^{(p)}| }  \Bigg\}
\pl{p=1}{k} \Bigg\{ \sul{ A^{(p)}= \underset{13}{\cup}\, _{a=1}^{3} B_a^{(p)} }{}  \Bigg\} \\
\times \pl{p=1}{k} \De\big( \bs{C}_1^{(p-1)} \cup \bs{D}_1^{(p-1)}  \mid \bs{B}_1^{(p)} \cup \bs{B}_3^{(p)} \big)
\pl{p=1}{k-1}\op{S}\Big( \, \overleftarrow{ \bs{A}^{(p)} }  \mid \overleftarrow{ \bs{D}_1^{(p)} }\cup  \overleftarrow{ \bs{D}_2^{(p)} } \cup \overleftarrow{\bs{C}_2^{(p)}} \cup  \overleftarrow{ \bs{C}_1^{(p)} } \Big) \\
\times   \pl{p=1}{k}\op{S}\big(  \bs{A}^{(p)} \mid \bs{B}_1^{(p)} \cup  \bs{B}_2^{(p)} \cup  \bs{B}_3^{(p)}\big)  \cdot
\pl{p=1}{k}\mc{F}^{(\op{O}_{p})}\Big( \,  \overleftarrow{ \bs{C}_2^{(p-1)}} + \i \pi \ov{\bs{e}}_{\veps_{p}} ,  \bs{B}_2^{(p)}, \overleftarrow{ \bs{D}_2^{(p-1)}} - \i \pi \ov{\bs{e}}_{\veps_{p}^{\prime}}  \Big) \;.
\label{ecriture somme originelle sur partitions pour G a deux regulateur eps}
\end{multline}
Above, we agree that
\beq
C_{1}^{(0)} \; = \; C_{2}^{(0)} \; = \;D_{1}^{(0)} \; = \; D_{2}^{(0)} \; = \; \emptyset
\label{ecriture ensemble vides CaDa index 0}
\enq
The sub-partitions that are summed over are constrained as
\beq
 |C_1^{(p-1)}| \, = \, |B_1^{(p)}| \;, \quad |D_1^{(p-1)}| \, = \, |B_3^{(p)}| \qquad p=1,\dots, k \, .
\label{ecriture contrainte cardinatlite Ba en termes des D1 et C1}
\enq
At this stage, we implement of change of summation over the partitionings by introducing finer ones. Namely, for $p=1,\dots, k-1$,
we consider partitions of $A_1^{(p)}$ and $A_2^{(p)}$ into $2^p-1$ sets as described in \eqref{ecriture sous partitions des Ap pour calcul multipoints} and partitions of $A^{(k)}$ into $2^{k}-1$ sets as given in \eqref{ecriture sous partitions des Ak pour calcul multipoints}.
These finer partitions are constrained according to \eqref{ecriture contrainte partitions Xa+}, \eqref{ecriture contrainte partitions Xa-} and \eqref{ecriture contrainte partitions Ea+ et Ea-}.

Given these data, we define
\beq
C_1^{(p)} \; = \; \bigcup\limits_{a=1}^{2^p-1}X_{a,+}^{(p)} \; , \qquad  C_2^{(p)} \; = \; \bigcup\limits_{a=1}^{2^p-1} W_{a,+}^{(p)} \qquad \e{for} \qquad p=1 , \dots, k-1 \;,
\label{definition partitions Ca(p) pour cas + et -}
\enq
as well as
\beq
D_1^{(p)} \; = \; \bigcup\limits_{a=1}^{2^p-1}X_{a,-}^{(p)} \; , \qquad  D_2^{(p)} \; = \; \bigcup\limits_{a=1}^{2^p-1} W_{a,-}^{(p)} \qquad \e{for} \qquad p=1 , \dots, k-1 \;.
\label{definition partitions Da(p) pour cas + et -}
\enq
Further, introducing permutations $\sg_{a}^{(p-1)}  \in \mf{S}_{  | \cup_{\eps} \{  X_{a,\eps}^{(p)} \cup W_{a,\eps}^{(p)}\} | } $
for $a \in \intn{1}{2^p-1}\setminus \{2^{p-1}\}$, we set, for $p=1, \dots, k-1$
\beq
B_1^{(p)} \; = \; \bigcup\limits_{a=1}^{2^{p-1}-1}  \bigg\{ \bigcup\limits_{\eps=\pm} \Big\{ X_{a,\eps}^{(p)} \cup W_{a,\eps}^{(p)}   \Big\} \bigg\}^{ \sg_{a}^{(p-1)} },
\qquad  B_2^{(p)} \; = \;   \bigcup\limits_{\eps=\pm} \Big\{ X_{2^{p-1},\eps}^{(p)} \cup W_{2^{p-1},\eps}^{(p)} \Big\} \;,
\label{definition partitions B1(p) B2(p) pour cas + et -}
\enq
as well as
\beq
B_3^{(p)} \; = \; \bigcup\limits_{a=1}^{2^{p-1}-1} \bigg\{ \bigcup\limits_{\eps=\pm} \Big\{ X_{a+2^{p-1},\eps}^{(p)} \cup W_{a+2^{p-1},\eps}^{(p)}   \Big\}\bigg\}^{ \sg_{a + 2^{p-1} }^{(p-1)} }\; .
\label{definition partitions B3(p) pour cas + et -}
\enq
Finally, we set
\beq
B_1^{(k)} \; = \; \bigcup\limits_{ \substack{ a=1 \\ 1,\cdots, 2^{k-1}-1}  }^{ 2^{k-1}-1} E_{a,+}^{(k)},  \qquad B_2^{(k)} \; = \; E_{0}^{(k)},
\qquad \e{and}   \qquad B_3^{(k)} \; = \; \bigcup\limits_{ \substack{ a=1 \\ 1,\cdots, 2^{k-1}-1}  }^{2^{k-1}-1} E_{a,-}^{(k)}\;.
\label{definition partitions des Baks}
\enq

The original summation over partitions given in \eqref{ecriture somme originelle sur partitions pour G a deux regulateur eps} is equivalent to
summing up over partitions given in \eqref{ecriture sous partitions des Ap pour calcul multipoints}, \eqref{ecriture sous partitions des Ak pour calcul multipoints},
permutations \eqref{definition permutations porusommations de type + et - jointe} and cardinality constraints \eqref{ecriture contrainte partitions Xa-}-\eqref{ecriture contrainte partitions Ea+ et Ea-},
with the "original" partitions $C_a^{(p)}, D_a^{(p)}$ and $B^{(p)}_a$ reconstructed as \eqref{definition partitions Ca(p) pour cas + et -}, \eqref{definition partitions Da(p) pour cas + et -},
\eqref{definition partitions B1(p) B2(p) pour cas + et -} and \eqref{definition partitions B3(p) pour cas + et -}.
This equivalence is established quite analogously to the one given in the proof of Proposition \ref{Proposition rep combinatoire pour G tot de epsk}, so we do not reproduce
it here. Still, we point out that the correspondence is achieved in such a way that for $p = 2, \dots, k-1$ the Dirac masses factorise as
\beq
\De\big( \bs{C}_1^{(p-1)}  \mid \bs{B}_1^{(p)}  \big) \; = \;
\pl{a=1}{2^{p-1}-1}  \De\Big( \bs{X}_{a,+}^{(p-1)}  \mid \overrightarrow{ \big( \bs{W}_{a,+}^{(p)} \cup \bs{W}_{a,-}^{(p)} \cup  \bs{X}_{a,+}^{(p)} \cup \bs{X}_{a,-}^{(p)}   \big)}^{ \sg_a^{(p-1)} }  \Big)
\label{ecriture contraintes Delta de type +}
\enq
and
\beq
\De\big( \bs{D}_1^{(p-1)}  \mid \bs{B}_3^{(p)}  \big) \; = \;
\pl{a=1}{2^{p-1}-1}  \De\Big( \bs{X}_{a,-}^{(p-1)}  \mid
      \overrightarrow{ \big(  \bs{W}_{a+2^{p-1},+}^{(p)} \cup \bs{W}_{a+2^{p-1},-}^{(p)} \cup \bs{X}_{a+2^{p-1},+}^{(p)} \cup \bs{X}_{a+2^{p-1},-}^{(p)}    \big)}^{ \sg_{a+2^{p-1}}^{(p-1)} }  \Big) \;.
\label{ecriture contraintes Delta de type -}
\enq
Moreover, a similar factorisation holds for the Dirac masses involving the $k^{\e{th}}$ variables
\beq
\De\big( \bs{C}_1^{(k-1)}  \mid \bs{B}_1^{(k)}  \big) \; = \; \pl{a=1}{2^{k-1}-1}  \De\big(  \bs{X}_{a,+}^{(k-1)}  \mid \bs{E}_{a,+}^{(k)}   \big)
\qquad \e{and}   \qquad
\De\big( \bs{D}_1^{(k-1)}  \mid \bs{B}_3^{(k)}  \big) \; = \;  \pl{a=1}{2^{k-1}-1}  \De\big(  \bs{X}_{a,-}^{(k-1)}  \mid \bs{E}_{a,-}^{(k)}   \big) \;.
\label{ecriture contraintes Delta extérieures mixtes}
\enq
This change of summation variables leads to the expression
\bem
\wt{\mc{G}}\big( \big\{  \bs{A}^{(s)}   \big\}^{k}_{0} ; \bs{\veps}_k, \bs{\veps}^{\prime}_k \big) \; = \;
\pl{p=1}{k-1} \bigg\{ \ex{ 2 \i\pi \om_{\op{O}_{p+1}} |A_2^{(p)}| } \hspace{-2mm} \sul{ \op{P}_p[  A^{(p)} ]  }{}
\pl{a=1}{2^{p-1}-1} \sul{ \sg_{a}^{(p-1)} \in \mf{S}_{  |  X_{a,+}^{(p-1)} |  } }{} \sul{ \sg_{a+2^{p-1}}^{(p-1)} \in \mf{S}_{  |  X_{a,-}^{(p-1)} |  } }{}   \bigg\} \sul{ \op{P}_k[  A^{(k)} ]  }{}
 \\
\pl{\ups=\pm}{}  \Bigg\{  \pl{a=1}{2^{k-1}-1} \!\! \De\big(  \bs{X}_{a,\ups}^{(k-1)}  \mid \bs{E}_{a,\eps}^{(k)}   \big) \cdot
\pl{p=2}{k-1} \pl{a=1}{ 2^{p-1}-1 } \!\! \De\Big( \bs{X}_{a,\ups}^{(p-1)}  \mid
          \overrightarrow{ \big( \bs{W}_{a+t_{p,\ups},+}^{(p)} \cup \bs{W}_{a+t_{p,\ups},-}^{(p)} \cup \bs{X}_{a+t_{p,\ups},+}^{(p)}
                                                          \cup \bs{X}_{a+t_{p,\ups},-}^{(p)}    \big)}^{ \sg_{a+t_{p,\ups}}^{(p-1)} }  \Big) \Bigg\} \\
\times \mc{W}\bigg( \Big\{ \big\{ \bs{C}_a^{(p)}, \bs{D}_a^{(p)} \big\}_{a=1}^{2} \Big\}_{p=1}^{k-1}; \Big\{\big\{ \bs{B}_a^{(p)} \big\}_{a=1}^{3} \Big\}^{k}_{p=1} \bigg)   \;.
\label{ecriture représentation hybride pour wt G apres changement parametrisation partition}
\end{multline}
There, $t_{p,+}=0$ and $t_{p,-}=2^{p-1}$.
The sets $C_a^{(p)}$s, $D_a^{(p)}$s  and $B_a^{(p)}$s should be now understood as built as in \eqref{definition partitions Ca(p) pour cas + et -}, \eqref{definition partitions Da(p) pour cas + et -},
\eqref{definition partitions B1(p) B2(p) pour cas + et -} and \eqref{definition partitions B3(p) pour cas + et -}. Finally, one has
\bem
\mc{W}\Big( \big\{ C_a^{(p)}, D_a^{(p)}, B_a^{(p)} \big\} \Big) \, = \,
\pl{p=1}{k-1}\op{S}\Big( \, \overleftarrow{ \bs{A}^{(p)} }  \mid \overleftarrow{ \bs{D}_1^{(p)} }\cup  \overleftarrow{ \bs{D}_2^{(p)} } \cup \overleftarrow{\bs{C}_2^{(p)}} \cup  \overleftarrow{ \bs{C}_1^{(p)} } \Big)
\cdot  \pl{p=1}{k}\op{S}\big(  \bs{A}^{(p)} \mid \bs{B}_1^{(p)} \cup  \bs{B}_2^{(p)} \cup  \bs{B}_3^{(p)}\big) \\
\times  \pl{p=1}{k}\mc{F}^{(\op{O}_p)} \Big( \overleftarrow{ \bs{C}_2^{(p-1)} } + \i \pi \ov{\bs{e}}_{\veps_p},  \bs{B}_2^{(p)}, \overleftarrow{\bs{D}_2^{(p-1)} } - \i \pi \ov{\bs{e}}_{\veps_p^{\prime}} \Big) \;.
\label{definition sommant W hybride}
\end{multline}

The partitioning that we have just introduced allows one for a direct resolution of the constraints imposed by the Dirac masses. We will
establish the precise formulae by induction. The $\De$-enforced constraints at $p=k$ gives
\beq
\bs{X}_{a,\eps}^{(k-1)}  \, = \,  \bs{E}_{a,\eps}^{(k)}  \qquad \e{for} \qquad a=1,\dots 2^{k-1}-1 \quad \e{and}  \quad \eps = \pm \;.
\enq
Further, substituting $\bs{X}_{a,+}^{(k-1)} $ and $\bs{X}_{a,-}^{(k-1)}$ in the next constraint yields
\beq
 \bs{X}_{a,+}^{(k-2)} \; = \;  \overrightarrow{ \big( \bs{W}_{a,+}^{(k-1)} \cup \bs{W}_{a,-}^{(k-1)} \cup \bs{E}_{a,+}^{(k)} \cup \bs{E}_{a,-}^{(k)}   \big)}^{ \sg_{a}^{(k-2)} }   \qquad a=1, \dots, 2^{k-2}-1 \;.
\enq
Here we stress that one should understand the resulting vector as being obtained from a direct concatenation of the four vector entries
followed by a global permutation of the entries of the resulting vector. This produces the output vector.

One gets a similar result for the other kind of constraints:
\beq
\bs{X}_{a,-}^{(k-2)} \; = \;  \overrightarrow{ \big(  \bs{W}_{s+2^{k-2},+}^{(k-1)} \cup \bs{W}_{s+2^{k-2},-}^{(k-1)} \cup \bs{E}_{s+2^{k-2},+}^{(k-1)} \cup \bs{E}_{s+2^{k-2},-}^{(k-1)}  \big)}^{ \sg_{a+2^{p-1}}^{(k-2)} }
\qquad a=1, \dots, 2^{k-2}-1 \;.
\enq
Note that, formally speaking, one has $\bs{X}_{a,-}^{(k-2)}  = \bs{X}_{a+2^{k-2},+}^{(k-2)}$, \textit{i.e.} one obtains the expression for $\bs{X}_{a,-}^{(k-2)}$ by shifting $a$
in the expression for $ \bs{X}_{a,+}^{(k-2)}$ by $2^{p-1}$.

Now, we state the induction hypothesis of rank $r$. There exists a change of variables in respect to the summed permutations, such that for any $ p \in \intn{r}{k-2}$,
and modulo the Dirac mass reduction constraints, it holds
\beq
 \bs{X}_{a,+}^{(p)} \; = \; \e{Vect}\Bigg\{    \Bigg[  \bigcup\limits_{\eps=\pm} \bigcup\limits_{v=p+1}^{k-1} \bigcup\limits_{s=0}^{2^{v-p-1}-1}
                                 \bs{W}^{(v)}_{a+s2^{p+1},\eps}   \Bigg] \bigcup \Bigg[ \bigcup\limits_{\eps=\pm} \bigcup\limits_{s=0}^{2^{k-p-2}-1} \bs{E}_{a+s2^{p+1},\eps}^{(k)}  \Bigg] \Bigg\}^{\sg^{(p)}_a}
\label{ecriture expression reduite pour X a + p}
\enq
while $ \bs{X}_{a,-}^{(p)}$ is obtained by a shift of the above expression by $2^{p}$, \textit{viz}.
\beq
 \bs{X}_{a,-}^{(p)} \; = \;  \bs{X}_{a+2^{p},+}^{(p)} \; = \; \e{Vect}\Bigg\{  \Bigg[  \bigcup\limits_{\eps=\pm}\bigcup\limits_{v=p+1}^{k-1} \bigcup\limits_{s=0}^{2^{v-p-1}-1}
                                 \bs{W}^{(v)}_{a+(2s+1)2^{p},\eps} \Bigg] \bigcup
                                        \Bigg[ \bigcup\limits_{\eps=\pm} \bigcup\limits_{s=0}^{2^{k-p-2}-1} \bs{E}_{a+(2s+1)2^{p},\eps}^{(k)}  \Bigg] \Bigg\}^{\sg^{(p)}_{a+2^p}} \;.
\label{ecriture expression reduite pour X a - p}
\enq
The concatenation of vectors is done by using the increasing union on the vectors, \textit{viz}.
\beq
\bigcup\limits_{\eps=\pm}  \bigcup\limits_{v=p+1}^{k-1} \bigcup\limits_{s=0}^{2^{v-p-1}-1} \bs{W}^{(v)}_{a+s2^{p+1},\eps} \; = \;
 \bs{W}^{(p+1)}_{a,+} \cup \bs{W}^{(p+1)}_{a+2^{p+1},+} \cup \bs{W}^{(p+2)}_{a,+} \cup \cdots \cup \bs{W}^{(k-1)}_{a+2^{k-1}-2^{p+1},-} \;.
\enq

The induction hypothesis is true for $p=k-2$. Assuming that it is true for $r$, we observe that the $\De$-constraints of $+$ type enforce
\bem
\bs{X}_{a,+}^{(r-1)}  \, =\,   \e{Vect}\Bigg(  \bs{W}_{a,+}^{(r)} \bigcup \bs{W}_{a,-}^{(r)} \bigcup
\e{Vect}\Bigg\{    \Bigg[  \bigcup\limits_{\eps=\pm} \bigcup\limits_{v=r+1}^{k-1} \bigcup\limits_{s=0}^{2^{v-r-1}-1}
                                 \bs{W}^{(v)}_{a+s2^{r+1},\eps}   \Bigg] \bigcup \Bigg[ \bigcup\limits_{\eps=\pm} \bigcup\limits_{s=0}^{2^{k-r-2}-1} \bs{E}_{a+s2^{r+1},\eps}^{(k)}  \Bigg] \Bigg\}^{\sg^{(r)}_a} \\
\bigcup \e{Vect}\Bigg\{  \Bigg[  \bigcup\limits_{\eps=\pm}\bigcup\limits_{v=r+1}^{k-1} \bigcup\limits_{s=0}^{2^{v-r-1}-1}
                                 \bs{W}^{(v)}_{a+(2s+1)2^{r},\eps} \Bigg] \bigcup
                                        \Bigg[ \bigcup\limits_{\eps=\pm} \bigcup\limits_{s=0}^{2^{k-r-2}-1} \bs{E}_{a+(2s+1)2^{r},\eps}^{(k)}  \Bigg] \Bigg\}^{\sg^{(r)}_{a+2^r}}    \Bigg)^{ \sg_a^{(r-1)} } \;.
\label{ecriture grosse formule recursive pour X a + rang r-1}
\end{multline}
Observe that in terms of set - \textit{i.e.} upon forgetting the ordered nature of the concatenations imposed by the vectors- the above unions may be recast as
\beq
X_{a,+}^{(r-1)}  \, =\,  \bigcup\limits_{\eps=\pm} \Bigg[   W_{a,\eps}^{(r)}
  \bigcup\limits_{v=r+1}^{k-1} \bigcup\limits_{s=0}^{2^{v-r-1}-1}
                               \Big\{   W^{(v)}_{a+s2^{r+1},\eps}  \cup W^{(v)}_{a+(2s+1)2^{r},\eps} \Big\}
                                    \,   \bigcup  \,   \bigcup\limits_{s=0}^{2^{k-r-2}-1} \Big\{ E_{a+s2^{r+1},\eps}^{(k)} \cup E_{a+(2s+1)2^{r},\eps}^{(k)} \Big\}  \Bigg]
\enq
Since
\beq
\bigcup\limits_{s=0}^{2^{v-r-1}-1} \Big\{ \{ a+2 s2^{r} \} \cup \{ a+(2s+1)2^{r} \} \Big\} \;  = \; \bigcup\limits_{s=0}^{2^{v-r}-1} \{ a+ s 2^{r} \} \;,
\enq
this yields
\beq
X_{a,+}^{(r-1)}  \, =\,  \bigcup\limits_{\eps=\pm} \Bigg[
  \bigcup\limits_{v=r}^{k-1} \bigcup\limits_{s=0}^{2^{v-r}-1}
                                  W^{(v)}_{a+s2^{r},\eps}
                                     \,  \bigcup \,   \bigcup\limits_{s=0}^{2^{k-r-1}-1}  E_{a+s2^{r},\eps}^{(k)}   \Bigg] \;.
\label{ecriture formule recursive r moins 1 pour Xa +}
\enq
The very same handlings establish that $X_{a,-}^{(r-1)}=X_{a+2^{r-1},+}^{(r-1)}$, where the \textit{rhs} is to be understood as
given by the above expression. To prove the induction hypothesis at rank $r-1$, one needs to raise these to the level of vectors, it is just enough
to change the outer permutation $\sg_{a}^{(r-1)} \hookrightarrow \sg_{a}^{(r-1)}\circ \tau $ in \eqref{ecriture grosse formule recursive pour X a + rang r-1}
in such a way that $\tau$ reorganises the coordinates of the set-elements so that eventually these produce the concatenated chain of vectors
\beq
 \Bigg[  \bigcup\limits_{\eps=\pm} \bigcup\limits_{v=p+1}^{k-1} \bigcup\limits_{s=0}^{2^{v-p-1}-1}
                                 \bs{W}^{(v)}_{a+s2^{p+1},\eps}   \Bigg] \bigcup \Bigg[ \bigcup\limits_{\eps=\pm} \bigcup\limits_{s=0}^{2^{k-p-2}-1} \bs{E}_{a+s2^{p+1},\eps}^{(k)}  \Bigg] \;.
\enq
The reasoning is word-for-word analogous for $\bs{X}_{a,-}^{(r-1)}$ with the sole exception that one performs $\sg_{a+2^{r-1}}^{(r-1)} \hookrightarrow \sg_{a+2^{r-1}}^{(r-1)}\circ \tau $ in
the $\De$-constraint defining that vector.

Once the reduction is solved, it is a matter of direct calculations to obtain the expressions for the sets $C^{(p)}_a, D_a^{(p)}$ and $B^{(p)}_a$.
One readily establishes by induction on $k$, that given $k \in \intn{1}{p}$, it holds
\beq
\intn{ 1 }{ 2^p - 1 } \, = \, \bigcup\limits_{a=1}^{k}  \bigcup\limits_{s=0}^{2^{p-a}-1} \{ 2^{a-1}(2s+1) \}   \;  \cup   \; \bigcup\limits_{s=0}^{2^{p-k}-1} \{ 2^{k}s \} \;.
\enq
In particular, for $k=p$, one gets $\intn{ 1 }{ 2^p - 1 }  \, = \,  \bigcup\limits_{a=1}^{p} \bigcup\limits_{s=0}^{2^{p-a}-1} \{ 2^{a-1}(2s+1) \} $, what implies that
\beq
C_2^{(p)}\,= \,  \bigcup\limits_{a=1}^{p} \bigcup\limits_{s=0}^{2^{p-a}-1} W_{2^{a-1}(2s+1) ; +}^{(p)} \; = \;  \bigcup\limits_{a=1}^{p} \ga_{+}^{(p+1a)}
\qquad \e{and} \qquad
D_2^{(p)}\,= \,  \bigcup\limits_{a=1}^{p} \bigcup\limits_{s=0}^{2^{p-a}-1} W_{2^{a-1}(2s+1) ; -}^{(p)} \; = \;  \bigcup\limits_{a=1}^{p} \ga_{-}^{(p+1a)} \;.
\label{ecriture C2p et D2p via gammas}
\enq
Here, we remind that $\ga_{\eps}^{( b a)}$ have been introduced in \eqref{definition des ensembles gamma ba eps}. The expressions for $C_1^{(p)}$
and $D_{1}^{(p)}$ do not seem to reduce in terms of $\ga_{\eps}^{(ba)}$ only. These sets take the form given in \eqref{ecriture C1p apres resolution recurrence}
-\eqref{ecriture D1p apres resolution recurrence} for $p \in \intn{1}{k-2}$ and \eqref{ecriture C1k-1 et D1k-1 apres resolution recurrence} for $p=k-1$.

In what concerns the $B$-ensembles, we first start with the $B_{a}^{(k)}$s. By construction \eqref{definition partitions des Baks} and
the definition of $\Xi^{(k)}_{\eps}$ \eqref{definition des ensembles Xi p eps}, one has that $B_2^{(k)}=\Xi_{+}^{(k)}$, however,
$B_1^{(k)}$ and $B_3^{(k)}$ do not recast in terms of $\Xi_{\eps}^{(p)}$ solely. The same holds true for $B_1^{(p)}$ and $B_3^{(p)}$s.
However, starting from the definition \eqref{definition partitions B1(p) B2(p) pour cas + et -}, observing from
\eqref{definition des ensembles gamma ba eps} that $\ga_{\eps}^{(p+1p)}=W_{2^{p-1},\eps}^{(p)}$, and substituting the expressions for the $X^{(p)}_{2^{p-1},\eps}$
given in \eqref{ecriture expression reduite pour X a + p}-\eqref{ecriture expression reduite pour X a - p}, one gets
\bem
\bigcup\limits_{\eps=\pm}^{} X_{2^{p-1};\eps}^{(p)}  \; = \; \bigcup\limits_{\eps=\pm}^{}
\Bigg[   \bigcup\limits_{v=p+1}^{k-1} \bigcup\limits_{s=0}^{2^{v-p-1}-1} \hspace{-2mm} \Big\{  W_{2^{p-1}+2s2^{p}  ; \eps}^{(v)}   \cup W_{2^{p-1}+(2s+1)2^{p}  ; \eps}^{(v)} \Big\}
\bigcup\limits_{s=0}^{2^{k-p-2}-1}  \Big\{  E_{2^{p-1}+2s2^{p}  ; \eps}^{(k)}   \cup E_{2^{p-1}+(2s+1)2^{p}  ; \eps}^{(k)} \Big\}   \Bigg] \\
 \; = \; \bigcup\limits_{\eps=\pm}^{}
\Bigg[   \bigcup\limits_{v=p+1}^{k-1} \bigcup\limits_{s=0}^{2^{v-p}-1} \hspace{-2mm}   W_{2^{p-1}+s2^{p}  ; \eps}^{(v)}
\bigcup\limits_{s=0}^{2^{k-p-1}-1}     E_{2^{p-1}+s2^{p}  ; \eps}^{(k)}      \Bigg]
\; = \; \bigcup\limits_{\eps=\pm}^{} \bigg\{ \Xi_{\eps}^{(p)}  \bigcup\limits_{b=p+1}^{k}  \ga_{\eps}^{(b+1p)}   \bigg\} \;.
\end{multline}
Thus, putting these together yields
\beq
B^{(p)}_2 \; = \;\bigcup\limits_{\eps=\pm}^{} \bigg\{ \Xi_{\eps}^{(p)} \bigcup\limits_{b=p+1}^{k}  \ga_{\eps}^{(bp)} \bigg\} \;, \qquad p\in \intn{1}{k-1} \quad \e{and} \quad
B_2^{(k)}=\Xi_{+}^{(k)} \;.
\label{ecriture B2p via gammas et Xis}
\enq

We are now in position to finally produce the reduced forms for the momentum contributions $\mc{P} \big( \{ A^{(s)}\}^{k}_{0} ; \{\bs{x}_a \}_{1}^{k} \big) $ \eqref{defintion impulsion globale pour produit elements matrice}
arising in $\mc{R}[G]$ \eqref{definition Fourier direct de G a k veriable Minkowski}   as well as the
form factor and $S$-matrix contributions that build up the $\mc{W}$ factor defined in \eqref{definition sommant W hybride}.

We first focus on the expression for the momentum which, when $A^{(0)}=\emptyset$, can be recast as
\bem
\mc{P} \big( \{ A^{(s)}\}^{k}_{0} ; \{\bs{x}_a \}_{1}^{k} \big)_{\mid A^{(0)}=\emptyset} \; = \; \sul{p=2}{k} \Big\{ \ov{\bs{p}} \, \big( \bs{A}^{(p-1)}  \big) \, - \,
\ov{\bs{p}} \, \big( \bs{A}^{(p)}  \big)  \Big\} \cdot \bs{x}_{p}  \, - \, \ov{\bs{p}} \, \big( \bs{A}^{(1)}  \big) \cdot \bs{x}_1 \\
\;= \; \sul{p=2}{k} \Big\{ \ov{\bs{p}} \, \big( \bs{C}_1^{(p-1)}  \big) \, + \, \ov{\bs{p}} \, \big( \bs{C}_2^{(p-1)}  \big) \, + \, \ov{\bs{p}} \, \big( \bs{D}_1^{(p-1)}  \big)
\, + \, \ov{\bs{p}} \, \big( \bs{D}_2^{(p-1)}  \big)
\, - \, \ov{\bs{p}} \, \big( \bs{B}_1^{(p)}  \big) \, - \, \ov{\bs{p}} \, \big( \bs{B}_2^{(p)}  \big) \, - \, \ov{\bs{p}} \, \big( \bs{B}_3^{(p)}  \big)   \Big\} \cdot \bs{x}_{p}
\, - \, \ov{\bs{p}} \, \big( \bs{B}^{(1)}_2  \big) \cdot \bs{x}_1
\end{multline}
Above, we have replaced the vectors $\bs{A}^{(p-1)}$ by using the set decomposition
\beq
A^{(p-1)} \, = \, C_1^{(p-1)} \cup C_2^{(p-1)} \cup D_1^{(p-1)} \cup D_2^{(p-1)} \;,
\enq
for $p=2,\dots, k$ and the vectors $\bs{A}^{(p)}$  by using the set decomposition $A^{(p)} \, = \, B_1^{(p)} \cup B_2^{(p)} \cup B_3^{(p)}$ for $p=1,\dots, k$,
with the peculiarity that $B_1^{(1)}=B_{3}^{(1)}=\emptyset$, see \eqref{ecriture ensemble vides CaDa index 0}-\eqref{ecriture contrainte cardinatlite Ba en termes des D1 et C1}.
We remind that  $\ov{\bs{p}}\big( \bs{A} \big)$, \textit{c.f.} \eqref{definition de bar vect p},
is a set function and thus symmetric in respect to any permutation of the coordinates of $ \bs{A}$.  The orderings of the coordinates of the vectors appearing above are thus irrelevant.
One may now implement the reduction enforced by the $\De$-constraints in \eqref{ecriture somme originelle sur partitions pour G a deux regulateur eps}:
$\bs{C}_1^{(p-1)} \, = \,  \bs{B}_1^{(p)}$ and $ \bs{D}_1^{(p-1)} \, = \,  \bs{B}_3^{(p)}$. This leads to
\beq
\mc{P} \big( \{ A^{(s)}\}^{k}_{0} ; \{\bs{x}_a \}_{1}^{k} \big)_{\mid A^{(0)}=\emptyset} \; = \;
\sul{p=2}{k} \Big\{  \ov{\bs{p}} \, \big( \bs{C}_2^{(p-1)}  \big) \, + \, \ov{\bs{p}} \, \big( \bs{D}_2^{(p-1)}  \big)
\, - \, \ov{\bs{p}} \, \big( \bs{B}_2^{(p)}  \big) \Big\} \cdot \bs{x}_{p}
                                          \, - \, \ov{\bs{p}} \, \big( \bs{B}_2^{(1)}  \big)  \cdot \bs{x}_{1} \;.
\enq
One may insert the parameterisation of $B_2^{(p)}$ \eqref{ecriture B2p via gammas et Xis} and $C_2^{(p)}, D_2^{(p)}$ \eqref{ecriture C2p et D2p via gammas} and in terms of the "final" variables
$\ga^{(ba)}_{\eps}$ and $\Xi_{\eps}^{(p)}$, what yields
\bem
\mc{P} \big( \{ A^{(s)}\}^{k}_{0} ; \{\bs{x}_a \}_{1}^{k} \big)_{\mid A^{(0)}=\emptyset}
\, = \, \sul{\eps = \pm}{} \sul{p=2}{k}   \Biggr\{ \sul{a=1}{p-1} \ov{\bs{p}} \, \big( \bs{\ga}_{\eps}^{(pa)}  \big)  \, - \, \sul{b=p+1}{k} \ov{\bs{p}} \, \big( \bs{\ga}_{\eps}^{(bp)}  \big)
 \, - \, \ov{\bs{p}} \, \big( \bs{\Xi}_{\eps}^{(p)}  \big) \Biggr\} \cdot \bs{x}_{p}   \\
 \, - \,    \sul{\eps = \pm}{} \Bigg\{ \sul{b=2}{k} \ov{\bs{p}} \, \big( \bs{\ga}_{\eps}^{(b1)}  \big)  \, +\, \ov{\bs{p}} \, \big( \bs{\Xi}_{\eps}^{(1)}  \big)  \Bigg\} \cdot \bs{x}_{1}
\, .
\end{multline}
It is readily seen that the above expression reduces to $\msc{P}_{\e{glob}}\big(  \{ \Xi_{\eps}^{(p)} \}; \{\ga^{(ba)}_{\eps} \}; \{\bs{x}_s\} \big) $ as given in  \eqref{definition curly P global}.

We now turn on to rewriting $\mc{W}$ introduced in \eqref{definition sommant W hybride}. From its very structure, it is clear that this is a symmetric function
of the coordinates of the vectors $\bs{C}_{2}^{(p)}$, $\bs{D}_{2}^{(p)}$, $p=1,\dots, k-1$, and $\bs{B}_{2}^{(p)}$, $p=1,\dots,k$. One may thus build these vectors
from their associated sets any way one likes, provided that the ordering is inserted consistently in each of its appearances in the formula.
We thus choose
\beqa
\bs{C}_2^{(p-1)} & \hookrightarrow & \bs{\ga}_+^{(p1)} \cup \dots \cup \bs{\ga}_+^{(pp-1)} \; ,   \\
\bs{D}_2^{(p-1)} &  \hookrightarrow & \bs{\ga}_-^{(p1)} \cup \dots \cup \bs{\ga}_-^{(pp-1)} \; ,   \\
\bs{B}_2^{(p)} \hspace{2mm} & \hookrightarrow & \bs{\Xi}_{+}^{(p)} \cup \bs{\Xi}_{-}^{(p)} \cup \bs{\ga}_{+}^{(kp)} \cup \bs{\ga}_{-}^{(kp)} \cup \dots \cup \bs{\ga}_{+}^{(p+1p)} \cup \bs{\ga}_{-}^{(p+1p)} \; .
\eeqa
This leads to the substitutions
\beq
\mc{F}^{(\op{O}_1)} \Big(    \bs{B}_2^{(1)}  \Big)  \; \hookrightarrow  \;
\mc{F}^{(\op{O}_1)} \Big(   \bs{\Xi}_{+}^{(1)} \cup \bs{\Xi}_{-}^{(1)} \cup \bs{\ga}_{+}^{(k1)} \cup \bs{\ga}_{-}^{(k1)} \cup \dots \cup \bs{\ga}_{+}^{(21)} \cup \bs{\ga}_{-}^{(21)}   \Big)
\enq
and
\bem
\mc{F}^{(\op{O}_p)} \Big( \overleftarrow{ \bs{C}_2^{(p-1)} } + \i \pi \ov{\bs{e}}_{\veps_p},  \bs{B}_2^{(p)}, \overleftarrow{\bs{D}_2^{(p-1)} } - \i \pi \ov{\bs{e}}_{\veps_p^{\prime}} \Big)
 \; \hookrightarrow  \;
\mc{F}^{(\op{O}_p)} \Big( \overleftarrow{ \bs{\ga}_+^{(pp-1)} } \cup \dots \cup  \overleftarrow{ \bs{\ga}_+^{(p1)} }+ \i \pi \ov{\bs{e}}_{\veps_p},\\
\bs{\Xi}_{+}^{(p)} \cup \bs{\Xi}_{-}^{(p)} \cup \bs{\ga}_{+}^{(kp)} \cup \bs{\ga}_{-}^{(kp)} \cup \dots \cup \bs{\ga}_{+}^{(p+1p)} \cup \bs{\ga}_{-}^{(p+1p)},
\overleftarrow{ \bs{\ga}_-^{(pp-1)} }  \cup \dots \cup   \overleftarrow{ \bs{\ga}_-^{(p1)} }   - \i \pi \ov{\bs{e}}_{\veps_p^{\prime}} \Big) \;.
\end{multline}
Taken altogether, this reproduces the product over form factors in \eqref{formule pour reecriture action mixte + et -}.

Finally, it remains to focus on the product of $\op{S}$-matrices arising in \eqref{definition sommant W hybride}. By using the
$\De$-constraints \eqref{ecriture somme originelle sur partitions pour G a deux regulateur eps} and symmetry properties of the $\op{S}$-matrices, one has

\bem
\pl{p=1}{k-1}\op{S}\Big( \, \overleftarrow{ \bs{A}^{(p)} }  \mid \overleftarrow{ \bs{D}_1^{(p)} }\cup  \overleftarrow{ \bs{D}_2^{(p)} } \cup \overleftarrow{\bs{C}_2^{(p)}} \cup  \overleftarrow{ \bs{C}_1^{(p)} } \Big)
\cdot  \pl{p=1}{k}\op{S}\big(  \bs{A}^{(p)} \mid \bs{B}_1^{(p)} \cup  \bs{B}_2^{(p)} \cup  \bs{B}_3^{(p)}\big)  \\
\; = \;\op{S}\big(  \bs{A}^{(k)} \mid \bs{C}_1^{(k-1)} \cup  \bs{B}_2^{(k)} \cup  \bs{D}_1^{(k-1)}\big) \cdot
\pl{p=2}{k-1}\op{S}\big(  \bs{C}_1^{(p)} \cup  \bs{C}_2^{(p)}  \cup    \bs{D}_2^{(p)}  \cup \bs{D}_1^{(p)}    \mid \bs{C}_1^{(p-1)}  \cup  \bs{B}_2^{(p)} \cup  \bs{D}_1^{(p-1)}   \big)  \\
\times \op{S}\big(  \bs{C}_1^{(1)} \cup  \bs{C}_2^{(1)}  \cup    \bs{D}_2^{(1)}  \cup \bs{D}_1^{(1)}    \mid \bs{B}_2^{(1)}    \big)  \;.
\end{multline}
It is direct to check that the above product is invariant under any permutations of the coordinates of the vectors $\bs{C}_1^{(p)}, \bs{D}_1^{(p)}$ with $p=1,\dots, k-1$.

This entails the claim. \qed

\subsection{Proposition \ref{Proposition rep combinatoire pour G tot de epsk} revisited}
\label{SousSection Preuve alternative representation totalement -}

We provide a short proof of Proposition \ref{Proposition rep combinatoire pour G tot de epsk} by building on the results of
Proposition \ref{Proposition rep combinatoire pour G tot avec regularisation double + et -}.

\Proof
Starting from the representation given in Proposition \ref{Proposition rep combinatoire pour G tot avec regularisation double + et -}, one sets $A_{2}^{(p)}=\emptyset$
for $p=1,\dots, k-1$. Owing to the decompositions \eqref{ecriture sous partitions des Ap pour calcul multipoints}, this choice implies that
\beq
W_{a,-}^{(p)}\,= \, X_{a,-}^{(p)}\, = \, \emptyset \quad \e{for} \quad a=1,\dots, 2^p-1\quad \e{and} \quad p=1,\dots, k-1 \;,
\enq
Further, it holds that
\beq
W_{a,+}^{(p)}\,= \, X_{a,+}^{(p)}\, = \, \emptyset \quad \e{unless} \quad a=2^{s} \quad \e{for} \quad s=0,\dots, p-1 \; \e{and} \quad p=1,\dots, k-1 \;.
\enq
Indeed, from the very construction it holds that $A_1^{(1)}=W_{1,+}^{(1)}\cup X_{1,+}^{(1)}$, so the claim holds for $p=1$. Assume that it holds for some $p-1$.
Then, owing to the constraint \eqref{ecriture contrainte partitions Xa-} and the fact that $X_{a,-}^{(p-1)}=\emptyset$, one infers that
\beq
W_{a,+}^{(p)}\,= \, X_{a,+}^{(p)}\, = \, \emptyset \quad \e{for} \quad a=1+2^{p-1},\dots, 2^p-1  \;.
\enq
Further, the constraints \eqref{ecriture contrainte partitions Xa+} and the fact that $X_{a,+}^{(p-1)}=\emptyset$ for $a \not=2^{s} \quad \e{with} \quad s=0,\dots, p-2$
prove the induction hypothesis for $p$. Finally, using the constraint \eqref{ecriture contrainte partitions Ea+ et Ea-}, one infers that
\beq
E_{a,+}^{(k)}\, = \, \emptyset \quad \e{unless} \quad a=2^{s} \quad \e{with} \quad s=0,\dots, k-2 \quad \e{and} \quad
E_{a,-}^{(k)} \, = \, \emptyset \quad \e{for} \, \e{any} \, a\;.
\enq
This leads to
\beq
\ga^{(ba)}_- \, = \, \emptyset \;, \quad \ga^{(ba)}_+ \, =\, W^{(b-1)}_{2^{a-1},+} \;, \quad \Xi_{-}^{(p)}=\emptyset \;, \quad
\Xi_{+}^{(s)}\, =\, E^{(k)}_{2^{s-1},+}  \;,
\enq
and we remind that $\Xi_{+}^{(k)}\, = \, E_0^{(k)}$. Moreover, due to cardinality constrains, all permutations but $\sg_{2^{s}}^{(p-1)}$, $p \in \intn{1}{k-1}$
and $s\in \intn{0}{p-2}$ trivialise.

In order to recover the sets that were summed over in \eqref{ecriture formule combinatoire pour G tot totalement moins}, we proceed to the relabelling:
\beq
\ga^{(ba)}_+ \, \hookrightarrow \, \ga^{(ba)} \;, \quad A_{s}^{(p)}  \, \hookrightarrow \, X^{(p)}_{2^{s-1},+}
       \;, \quad \Xi_{+}^{(s)}  \, \hookrightarrow \, B^{(k)}_{s} \;.
\enq
Upon these relabelling, the summations over partitions in \eqref{formule pour reecriture action mixte + et -} directly reduce to those in  \eqref{ecriture formule combinatoire pour G tot totalement moins}
of Proposition \ref{Proposition rep combinatoire pour G tot de epsk}.

Further, the above relabelling immediately leads to the reduction
\beq
\msc{P}_{\e{glob}}\big(  \{ \Xi_{\eps}^{(a)} \}; \{\ga_{\eps}^{(ba)}\}; \{\bs{x}_s\} \big) \, \hookrightarrow \;
\msc{P}\big(  \{ B_s^{(k)} \}; \{\ga^{(ba)} \}; \{\bs{x}_s\} \big)
\enq
where the two quantities are defined respectively in \eqref{definition curly P global}, \eqref{definition curly P}. Likewise, one gets
\bem
\pl{p=1}{k} \bigg\{  \mc{F}^{(\op{O}_p)} \Big( \overleftarrow{ \bs{\ga}_+^{(pp-1)} } \cup \dots \cup  \overleftarrow{ \bs{\ga}_+^{(p1)} }+ \i \pi \ov{\bs{e}}_{\veps_p},
\bs{\Xi}_{+}^{(p)} \cup \bs{\Xi}_{-}^{(p)} \cup \bs{\ga}_{+}^{(kp)} \cup \bs{\ga}_{-}^{(kp)} \cup \dots \cup \bs{\ga}_{+}^{(p+1p)} \cup \bs{\ga}_{-}^{(p+1p)},\\
\overleftarrow{ \bs{\ga}_-^{(pp-1)} }  \cup \dots \cup   \overleftarrow{ \bs{\ga}_-^{(p1)} }   - \i \pi \ov{\bs{e}}_{\veps_p^{\prime}} \Big)  \bigg\}  \\
 \hookrightarrow \; \pl{p=1}{k} \mc{F}^{(\op{O}_p)}\Big( \overleftarrow{ \bs{\ga}^{(pp-1)} } \cup \cdots \cup \overleftarrow{ \bs{\ga}^{(p1)} }  + \i \pi \ov{\bs{e}}_{\veps_{p}} ,
\bs{B}_p^{(k)}  \cup \bs{\ga}^{(kp)}  \cup \cdots \cup \bs{\ga}^{(p+1p)}   \Big)
\end{multline}
Finally, to relate $\msc{S}_{\e{tot}}\Big( \big\{ \Xi_{\eps}^{(p)}; W_{a,\eps}^{(p)} \big\}  \Big) $ defined in \eqref{definition matrice Stot cas + et - mixte}
to $\msc{S}\Big( \big\{ B_{s}^{(k)}; \ga^{(ba)} \big\}  \Big) $ defined in \eqref{definition matrice S cas totalement -},
one observes that the $D_a^{(p)}$ partitions arising in \eqref{ecriture tilge G via somme sur les C et D partitions}
and parameterised through \eqref{definition partitions Da(p) pour cas + et -} are all empty by construction. This then leads to
\beq
C_1^{(p)}\,= \, C_1^{(p)}\cup D_1^{(p)} \, = \, \bigcup\limits_{\eps = \pm} \bigg\{ \bigcup\limits_{b=p+2}^{k} \bigcup\limits_{a=1}^{p} \ga_{\eps}^{(ba)}   \bigcup\limits_{a=1}^{p} \Xi_{\eps}^{(a)} \bigg\}
\, \hookrightarrow \, \bigcup\limits_{b=p+2}^{k} \bigcup\limits_{a=1}^{p} \ga^{(ba)}   \bigcup\limits_{a=1}^{p} B_a^{(k)}  \;.
\label{ecriture valeur des ensembles C1p union D1p}
\enq
It remains to choose the associated vectors as
\beq
\bs{C}_1^{(p)} \, = \,  \bs{B}_1^{(k)} \cup  \bs{\ga}^{(k 1)}  \cup  \cdots \cup \bs{\ga}^{( p+2 1)}
\cup  \bs{B}_2^{(k)} \cup  \bs{\ga}^{(k 2)}  \cup  \cdots \cup \bs{\ga}^{( p+2 2)}   \cup \cdots \cup
\bs{B}_{p-1}^{(k)} \cup  \bs{\ga}^{(k p)}  \cup  \cdots \cup \bs{\ga}^{( p+2  p)}
\enq
where it is understood that the $\bs{\ga}^{(ba)}$ insertions are absent when $p=k-1$. Then one observes that with such an identification,
 the factors building up $\msc{S}_{\e{tot}}\Big( \big\{ \Xi_{\eps}^{(p)}; W_{a,\eps}^{(p)} \big\}  \Big) $
in \eqref{definition matrice Stot cas + et - mixte} reduces to \eqref{ecriture substitution pour S Ak slash D1k D2k}-\eqref{ecriture substitution pour S D1p+1 C2p slash D1p D2p}.
Since the rest of the reductions follows word for word the handlings outlined in the proof of Proposition  \ref{Proposition rep combinatoire pour G tot de epsk},
this entails the claim. \qed

\subsection{An intermediate decomposition for the multi-point function}
\label{Appendix SubSection Intermediate Decomposition Multi pt fcts}

\begin{prop}
\label{Proposition rep combinatoire pour G tot cas avec une insertion + et que des -}

Let $t\in \intn{1}{k}$ and $A^{(0)}=A^{(k)}=\emptyset$ then  $\mc{G}_{\e{tot}}[ G] \big( \big\{  \bs{A}^{(s)}   \big\}^{k}_{0} ; \bs{\veps}_k, \bs{\veps}_k^{\prime} \big) $
introduced in \eqref{ecriture G tot de eps et eps prime k} admits the representation
\bem
 \mc{G}_{\e{tot}}[ G] \Big( \big\{  \bs{A}^{(s)}   \big\}^{k}_{0} ; \bs{\veps}_k, \bs{\veps}_k^{\prime} \Big) \; = \;
 \pl{ \substack{ s=1 \\  \not= t} }{k} \ex{-2\i\pi \om_{\op{O}_s}|A^{(s-1)}| } \cdot   \pl{p=1}{k-1} \bigg\{   \sul{ \op{P}_p[  A^{(p)} ]  }{}
\, \pl{s=1}{p-1} \sul{ \sg_s^{(p-1)} \in \mf{S}_{  |  A^{(p-1)}_s |  } }{}   \bigg\}
\Big( \mc{S}^{(t)} \, \mc{F}^{(t)}_{\e{tot};{\bs{\veps}_k}^{(t)} } \, \mc{R}[G]\Big) \big( \bs{\ga}   \big) \\
\times \pl{p=2}{k-1} \pl{s=1}{p-1} \De\Big( \bs{A}_s^{(p-1)}  \mid \overrightarrow{ \big( \bs{A}^{(p)}_s  \cup \bs{\ga}^{(p+1s)} \big) }^{ \sg_s^{(p-1)} } \Big)
\cdot \pl{s=1}{k-1} \De\big( \bs{ A}_s^{(k-1)}  \mid \emptyset  \big)   \;.
\end{multline}
Above, $\mc{S}^{(t)}$, $\mc{F}^{(t)}_{\e{tot};{\bs{\veps}_k} } $ and $\mc{R}[G]$ are as defined through
\eqref{S product representation for alternating signs}, \eqref{alternate form factor product} and \eqref{definition TF space-time de la fct test}
and $\bs{\ga}$ has been introduced in \eqref{definition vecteur gamma tot}.
Next, we agree upon
\beq
\bs{\veps}^{(t)}_k= \big(\veps_1,\dots, \veps_{t-1}, \veps_t^{\prime}, \veps_{t+1}, \veps_{k} \big) \, .
\enq
Finally, the sums run through partitions  $\op{P}_p[  A^{(p)} ]$ of $A^{(p)}$ such that
\beq
 A^{(p)} \, = \, \bigcup\limits_{s=1}^{p}A_s^{(p)}  \, \bigcup\limits_{s=1}^{p}\ga^{(p+1s)} \quad p=1 , \dots, k-1  \;.
%
\enq
 The summations are constrained so that $\big|  A^{(k-1)}_s \big| \; = \,0 $ for $s=1,\dots, k-1$ and
\beq
 \big|  A^{(p-1)}_s \big| \; = \, \big| A^{(p)}_s  \cup \ga^{(p+1s)} \big| \qquad for \quad s=1,\dots, p-1\;,  \quad p=1,\dots, k-1 \;.
\enq
\end{prop}

\Proof
One starts with the decomposition provided by Proposition \ref{Proposition rep combinatoire pour G tot avec regularisation double + et -} and chooses the system of
partitions as
\beq
A_{2}^{(p)}=\emptyset \quad \e{for} \quad p \in \intn{1}{k-1}\setminus\{t-1\} \quad \e{and} \quad A_1^{(t-1)} \, = \, \emptyset \;.
\label{ecriture choix des ensembles partitiones}
\enq
Owing to the decompositions \eqref{ecriture sous partitions des Ap pour calcul multipoints}, this choice implies that
\beq
W_{a,-}^{(p)}\,= \, X_{a,-}^{(p)}\, = \, \emptyset \quad \e{for} \quad a=1,\dots, 2^p-1\quad \e{and} \quad p=1,\dots, k-1 \;, p \neq t-1 \;,
\label{ecriture vanishing des W et X - p}
\enq
and
\beq
W_{a,+}^{(t-1)}\,= \, X_{a,+}^{(t-1)}\, = \, \emptyset \quad \e{for} \quad a=1,\dots, 2^{t-1}-1 \;.
\label{ecriture vanishing des W et X + t}
\enq
By following the exact same reasoning as in the proof of Proposition \ref{Proposition rep combinatoire pour G tot de epsk}
given in Subsection \ref{SousSection Preuve alternative representation totalement -},
one shows that
\beq
W_{a,+}^{(p)}\,= \, X_{a,+}^{(p)}\, = \, \emptyset \quad \e{unless} \quad  a  =  2^{s} \quad \e{with} \quad
\left\{ \ba{ccc}   s & = & 0,\dots, p-1  \vspace{1mm} \\
                    p & = & 1,\dots, t-2  \ea \right.  \;.
\label{ecriture vanishing des W et X + p avant t}
\enq
The choice \eqref{ecriture choix des ensembles partitiones} along with \eqref{ecriture vanishing des W et X - p}-\eqref{ecriture vanishing des W et X + p avant t},
allows one to infer from the constraints \eqref{ecriture contrainte partitions Xa-} that
\beq
W_{a,-}^{(t-1)}\,= \, X_{a,-}^{(t-1)}\, = \, \emptyset \quad \e{unless} \quad a=2^{s} \quad \e{with} \quad s=0,\dots, t-2 \;.
\enq
We now establish by induction that, when $p = t,\dots,k-1$, one has
\beq
W_{a,+}^{(p)}\,= \, X_{a,+}^{(p)}\, = \, \emptyset \quad \e{unless} \quad a = 2^{s} + 2^{t-1}  \mathds{1}_{s \leq t-2}\quad \e{with} \quad
\left\{ \ba{ccc}   s& = & 0,\dots, p-1  \vspace{1mm} \\
       p & = & t,\dots, k-1 \ea \right. \;.
\enq
This is once again proven by recursion. For $p=t$, this just comes from the reading out of the empty sets from the constraints \eqref{ecriture contrainte partitions Xa+}-\eqref{ecriture contrainte partitions Xa-}
at $p=t$ and observing that  $W_{2^{t-1},+}^{(t)}\,= \, X_{ 2^{t-1} ,+}^{(t)}$ may also be non-empty.
Now assume that the claim holds up to some $p \geq t$. Then, the constraints \eqref{ecriture contrainte partitions Xa-}
and the fact that $X_{a,-}^{(p)}=\emptyset$ ensure that
\beq
W_{a,+}^{(p+1)}\,= \, X_{a,+}^{(p+1)}\, = \, \emptyset \quad \e{for} \quad a=1+2^{p},\dots, 2^{p+1}-1  \;.
\enq
Further, the constraints \eqref{ecriture contrainte partitions Xa+} adjoined to
$X_{a,+}^{(p)}=\emptyset$ for $a \not = 2^{s} + \mathds{1}_{s\leq t-2}2^{t-1}$ with $s=0,\dots, p-1$ and
the potential non-emptiness of  $W_{2^{p},+}^{(p+1)}\,= \, X_{ 2^{p} ,+}^{(p+1)}$ lead to the claim.

We now introduce the relabelling for the non-empty sets for $a<b$
\beq
\ga^{(ba)} \, = \,
\begin{cases}
W^{(b-1)}_{2^{a-1},+} \;, &    b < t \vspace{1mm}  \\
W^{(t-1)}_{2^{a-1},-} \;,  &    b = t  \vspace{1mm} \\
W^{(b-1)}_{ 2^{a-1} + 2^{t-1} \mathds{1}_{a\leq t-1 }  ,+} \;, &   b \geq  t+1
\end{cases}
\qquad \e{and} \qquad
A_{s}^{(p)} \, = \,
\begin{cases}
X^{(p)}_{2^{s-1},+} \;, &  p \leq t-2  \vspace{1mm} \\
X^{(t-1)}_{2^{s-1},-} \;,  &  p = t-1  \vspace{1mm} \\
X^{(p)}_{2^{s-1} + 2^{t-1} \mathds{1}_{s \leq t-1},+} \;, &   t \leq p \leq k-2
\end{cases}
\; ,
\enq
where $s\in \intn{1}{p}$ and since $A_{s}^{(k-1)}=\emptyset$ for any $s \in \intn{1}{k-1}$.
Note that this change of sets re-expresses the $\ga_{\eps}^{(ba)}$ as introduced through \eqref{definition des ensembles gamma ba eps}
in terms of the $\ga^{(ba)}$ as
\beq
\ga^{(ba)}_+  \, = \, \ga^{(ba)} \quad \e{for} \quad  a<b\, , \; \; b \in \intn{2}{k}\setminus\{t\}  \quad \e{and} \quad
\ga^{(ta)}_-  \, = \, \ga^{(ta)} \, , \; a \in \intn{1}{t-1} \;.
\enq
All other $\ga_{\eps}^{(ba)}$ being empty.

This identification immediately reduces the product over form factors in \eqref{formule pour reecriture action mixte + et -}
to $\mc{F}^{(t)}_{\e{tot};{\bs{\veps}_k}^{(t)} }$ \eqref{alternate form factor product}. Likewise, it is direct to check starting from \eqref{definition curly P global}
that
\beq
\msc{P}_{\e{glob}}\big(  \{ \Xi_{\eps}^{(a)} \}; \{\ga_{\eps}^{(ba)}\}; \{\bs{x}_s\} \big) \;\hookrightarrow \;   \sul{b>a}{k}  \ov{\bs{p}} \, \big( \bs{\ga}^{(ba)}  \big)
\cdot \bs{x}_{ba}   \;.
\enq
Reaching the closed form for $\mc{S}^{(t)}$ as given in \eqref{S product representation for alternating signs} demands more
work. Recall the $C_1^{(p)}$ and $D_{1}^{(p)}$ partitions introduced in
\eqref{definition partitions Ca(p) pour cas + et -} and \eqref{definition partitions Da(p) pour cas + et -}
and which appear in vector form as the building blocks of $\msc{S}_{\e{tot}}\Big( \big\{ \Xi_{\eps}^{(p)}; W_{a,\eps}^{(p)} \big\}  \Big)$ introduced in
\eqref{definition matrice Stot cas + et - mixte}. One starts from the observation given in \eqref{ecriture valeur des ensembles C1p union D1p} that, under the present
reduction of summation variables, one has for $p\in \intn{1}{k-2}$
\beq
 C_1^{(p)}\cup D_1^{(p)} \, = \,  \bigcup\limits_{b=p+2}^{k} \bigcup\limits_{a=1}^{p} \ga^{(ba)}   \;.
\enq
Furthermore, by construction $C_{1}^{(k-1)}=D_{1}^{(k-1)}=\emptyset$. Moreover, it holds that $C_{1}^{(t-1)}=\emptyset$ while $D_{1}^{(p)}=\emptyset$ for $p\not= t-1$. Recalling that $\msc{S}_{\e{tot}}$
is invariant under any permutation of the entries of the vectors $\bs{C}_1^{(p)}$ and $\bs{D}_1^{(p)}$, one may just as well set in the formula for $\msc{S}_{\e{tot}}$
\beq
\bs{C}_1^{(p)} \, = \, \bs{U}^{(p)} \quad  \e{for} \quad  p\in \intn{1}{k-2}\setminus \{t-1\} \quad \e{and} \quad \bs{D}_1^{(t-1)} \, = \, \bs{U}^{(t-1)}
\enq
where
\beq
\bs{U}^{(p)}  \,= \,  \bs{\ga}^{(k 1)}  \cup  \cdots \cup \bs{\ga}^{( p+2 1)}
\cup   \bs{\ga}^{(k 2)}  \cup  \cdots \cup \bs{\ga}^{( p+2 2)}   \cup \cdots \cup
\bs{\ga}^{(k p)}  \cup  \cdots \cup \bs{\ga}^{( p+2  p)}
\enq
for $p\in \intn{1}{k-2}$ and $\bs{U}^{(k-1)}  \,= \, \bs{\emptyset}$. This substitution recasts $\msc{S}_{\e{tot}}$ as
\bem
\msc{S}_{\e{tot}}\Big( \big\{ \Xi_{\eps}^{(p)}; W_{a,\eps}^{(p)} \big\}  \Big) \; = \;
 \pl{ \substack{ p=1  \\ \not= t, t-1} }{k-1}
 \Bigg\{ \op{S}\Big( \,  \bs{U}^{(p)} \cup  \bs{\ga}^{(p+11)} \cup \dots \cup \bs{\ga}^{(p+1p)}
\mid \bs{U}^{(p-1)} \cup   \bs{\ga}^{(kp)}  \cup \dots  \cup \bs{\ga}^{(p+1p)}   \Big) \Bigg\}  \\
\times \op{S}\Big( \,  \bs{U}^{(t)} \cup  \bs{\ga}^{(t+11)} \cup \dots \cup \bs{\ga}^{(t+1t)}
\mid  \bs{\ga}^{(kt)} \cup \dots  \cup \bs{\ga}^{(t+1t)} \cup \bs{U}^{(t-1)}  \Big)  \\
\times \op{S}\Big( \,  \bs{\ga}^{(t1)} \cup \dots \cup \bs{\ga}^{(t t-1)} \cup  \bs{U}^{(t-1)}
\mid \bs{U}^{(t-2)} \cup \bs{\ga}^{(kt-1)}  \cup \dots \cup \bs{\ga}^{(t t-1)}   \Big) \; = \;
\mc{S}_1\cdot \mc{S}_2 \cdot \mc{S}_3(\bs{\ga})\;.
\end{multline}
Above, we have introduced
\beq
\mc{S}_1 (\bs{\ga}) \; = \; \pl{   p=1   }{k-1}
 \Bigg\{ \op{S}\Big( \,  \bs{U}^{(p)} \cup  \bs{\ga}^{(p+11)} \cup \dots \cup \bs{\ga}^{(p+1p)}
\mid \bs{U}^{(p-1)} \cup   \bs{\ga}^{(kp)}  \cup \dots  \cup \bs{\ga}^{(p+1p)}   \Big) \Bigg\} \; = \; \mc{S}(\bs{\ga})
\enq
with $\mc{S}$ as defined in \eqref{definition facteur de diffusion complet k pts} and upon using the chain of reductions outlined in the proof of Proposition
\ref{Proposition rep combinatoire pour G tot de epsk}.
In its turn, ones has
\bem
 \mc{S}_2(\bs{\ga}) \, = \, \op{S}\Big( \,  \bs{\ga}^{(t1)} \cup \dots \cup \bs{\ga}^{(t t-1)} \cup  \bs{U}^{(t-1)} \mid
\bs{U}^{(t-1)}  \cup  \bs{\ga}^{(t1)} \cup \dots \cup \bs{\ga}^{(t t-1)}  \, \Big)  \\
\, = \,  \pl{a=1}{t-1} \pl{b=t+1}{k} \pl{c=1}{t-1}
 \op{S}\Big(\,  \bs{\ga}^{(ta)} \cup  \bs{\ga}^{(b c)}  \mid  \bs{\ga}^{(b c)} \cup  \bs{\ga}^{(ta)} \, \Big) \,.
\end{multline}
Finally, it holds
\bem
 \mc{S}_3(\bs{\ga}) \, = \, \op{S}\Big( \,  \bs{\ga}^{(t1)} \cup \dots \cup \bs{\ga}^{(t t-1)} \cup  \bs{U}^{(t-1)}
\mid   \bs{U}^{(t-1)} \cup \bs{\ga}^{(t1)} \cup \dots \cup \bs{\ga}^{(t t-1)}   \Big)  \\
\; = \;\pl{b=t+1}{k} \pl{c=1}{t-1}  \pl{a=t+1}{k}
 \op{S}\Big(\,  \bs{\ga}^{(bc)} \cup  \bs{\ga}^{(at)}  \mid  \bs{\ga}^{(at)} \cup  \bs{\ga}^{(bc)} \, \Big) \;.
\end{multline}
Hence, all-in-all, one recovers $\mc{S}^{(t)}$ introduced in
\eqref{S product representation for alternating signs}. This entails the claim. \qed

\end{document}